\documentclass{aa} 
\usepackage{graphicx}
\usepackage{txfonts}
\usepackage{longtable}
\usepackage{subfig}
\usepackage{float}
\usepackage{booktabs}
\usepackage{natbib}
\bibpunct{(}{)}{;}{a}{}{,} % to follow the A&A style

\begin{document}
   \title{X-ray spectral variability of LINERs selected from the Palomar sample}

   \subtitle{}

 \author{Hern\'{a}ndez-Garc\'{i}a, L.\inst{1}; Gonz\'{a}lez-Mart\'{i}n, O.\inst{2}$^,$\inst{3}; Masegosa, J.\inst{1}; M\'{a}rquez, I. \inst{1}
          }

   \institute{Instituto de Astrof\'{i}sica de Andaluc\'{i}a, CSIC, Glorieta de la Astronom\'{i}a, s/n, 18008 Granada, Spain\\
              \email{lorena@iaa.es}
         \and
             Instituto de Astrof\'{i}sica de Canarias (IAC), C/ V\'{i}a Lactea, s/n, 38205 La Laguna, Tenerife, Spain \\
        \vspace*{-0.35cm}
         \and
             Departamento de Astrof\'{i}sica, Universidad de La Laguna (ULL), 38205 La Laguna, Tenerife, Spain \\
             }

   \date{Received XXXX; accepted YYYY}

\authorrunning{Hern\'{a}ndez-Garc\'{i}a et al.}
\titlerunning{X-ray variability in LINERs}

% \abstract{}{}{}{}{} 
% 5 {} token are mandatory
 
  \abstract
   {Variability is a general property of active galactic nuclei (AGN). At X-rays, the way in which these changes occur is not yet clear. In the particular case of low ionisation nuclear emission line region (LINER) nuclei, variations on months/years timescales have been found for some objects, but the main driver of these changes is still an open question.}
   {The main purpose of this work is to investigate the X-ray variability in LINERs, including the main driver of such variations, and to search for eventual differences between type 1 and 2 objects. }
   {We use the 18 LINERs in the Palomar sample with data retrieved from \emph{Chandra} and/or \emph{XMM}--Newton archives corresponding to observations gathered at different epochs. All the spectra for the same object are simultaneously fitted in order to study long term variations. The nature of the variability patterns are studied allowing different parameters to vary during the spectral fit.
Whenever possible, short term variations from the analysis of the light curves and UV variability are studied. }
   {Short term variations are not reported in X-rays. Three LINERs are classified as non-AGN candidates in X-rays, all of them being \emph{Compton}-thick candidates; none of them show variations at these frequencies and two of them vary in the UV. Long term X-ray variations can be analysed in 12 out of 15 AGN candidates, about half of them showing variability (7 out of the 12). At UV frequencies, most of the AGN candidates with available data are variable (five out of six). Thus, 13 AGN candidates are analysed at UV and/or X-rays, ten of them variable at least in one energy band. None of the three objects that do not vary in X-rays have UV available data. Therefore, long timescale variability is very common in LINERs. The main driver of these X-ray variations is related to changes in the nuclear power, while changes in absorptions are found only for NGC\,1052. We do not find any difference between type 1 and 2 LINERs, nor in the number of variable cases (three out of five type 1s and four out of seven type 2s) nor in the nature of the variability pattern. We find indications of an anticorrelation between the slope of the power law, $\Gamma$, and the Eddington ratio.}


   \maketitle
%
%________________________________________________________________

\begin{table*}
\begin{center}
\caption{\label{properties} General properties of the sample galaxies.}
\begin{tabular}{lccccccccccc} \hline
\hline
Name    & RA & DEC &  Dist.$^1$  & N$_{Gal}$ & $m_B$ & Morph.  & Optical & X-ray & Jet & Ref.$^2$ \\
 & (J2000)  & (J2000) & (Mpc) & ($10^{20}$ cm$^{-2}$) & & Type & class. & class. & \\
(1) & (2) & (3) & (4) & (5) & (6) & (7) & (8) & (9) & (10) & (11)   \\  \hline
NGC~315   &   00 57 48.88 & +00 21 08.8  & 59.60 & 5.88 & 12.20 & E &		 L1.9 & AGN	  & Y & (1) \\
NGC~1052  &    02 41 04.80 & +08 15 20.8  & 19.48 & 3.07  & 11.44 & E & 	 L1.9 & AGN	  & Y & (2) \\
NGC~1961  &  05 42 04.6 & +69 22 42	&  56.20 & 8.28  &  11.01 &  SAB(rs)c &  L2 & AGN	  & Y & (3) \\
NGC~2681  &    08 53 32.73 & +51 18 49.3  &  15.25 & 2.45 & 11.15 & S0-a(s) &	 L1.9 & AGN*	  & N & (1) \\
NGC~2787  &   09 19 18.56 & +69 12 12.0  & 10.24 & 4.32 & 11.60 &  S0-a(sr) &	 L1.9 & AGN	  & N & (1) \\
NGC~2841  &    09 22 02.63 & +50 58 35.5  & 16.62 & 1.45 & 10.06  & Sb(r) &	 L2 & AGN	  & N & (1) \\
NGC~3226$^3$  &	 10 23 27.01 & +19 53 54.7  & 29.84 & 2.14  & 12.34 &  E &	 L1.9 & AGN	  & N & (1) \\
NGC~3608  &   11 16 58.96 & +18 08 54.9  & 24.27 & 1.49 & 11.57 &  E &  	 L2/S2: & Non-AGN* &  N & (1) \\
NGC~3718  &   11 32 34.8 & +53 04 05	 & 17.00 & 1.08 & 11.19      & SB(s)a &  L1.9 & AGN	  & Y & (1) \\
NGC~4261  &    12 19 23.22 & +05 49 30.8  & 31.32 & 1.55 & 11.35 &  E & 	 L2 & AGN	  & Y & (4) \\
NGC~4278  &   12 20 06.83 & +29 16 50.7  & 15.83 & 1.77 & 11.04  & E &  	 L1.9 & AGN	  & Y & (5) \\
NGC~4374  &   12 25 03.74 & +12 53 13.1  & 17.18 & 2.60 & 10.11 &  E &  	 L2 & AGN*	  & Y & (6) \\
NGC~4494  &   12 31 24.03 & +25 46 29.9  & 13.84 & 1.52 & 10.68 &  E &  	 L2:: & AGN	  & N & (1) \\
NGC~4636  &   12 42 49.87 & +02 41 16.0  & 16.24 & 1.81 & 10.43  & E &  	 L1.9 & Non-AGN*  &  Y & (7) \\
NGC~4736  &   12 50 53.06 & +41 07 13.6  & 5.02 & 1.44 & 8.71 &  Sab(r) &	 L2 & AGN	  & N & (1) \\
NGC~5195  &   13 29 59.6 & +47 15 58	 &  7.91 & 1.56 & 10.38 & IA &  	 L2: & AGN	  & N & (8) \\
NGC~5813  &   15 01 11.26 & +01 42 07.1  & 30.15 & 4.21 & 11.48 & E &		 L2: & Non-AGN*   &  Y & (9) \\ 
NGC~5982  &   15 38 39.8 & +59 21 21	 & 41.22 & 1.82 & 12.05 & E &		 L2:: & AGN	  & N & (10) \\ 
\hline
\end{tabular}
\caption*{  (Col. 1) Name, (Col. 2) right ascension, (Col. 3) declination, (Col. 4) distance, (Col. 5) galactic absorption, (Col. 6) aparent magnitude in the Johnson filter B from \cite{ho1997}, (Col. 7) galaxy morphological type from \cite{omaira2009b}, (Col. 8) optical classification from \cite{ho1997}, (Col. 9) X-ray classification from \cite{omaira2009a}, where the * represent \emph{Compton}--thick candidates from \cite{omaira2009b}, (Col. 10) evidence of radio jet, and (Col. 11) references for radio data. }
\end{center}
\vspace*{-0.5cm}
\footnote*{All distances are taken from NED and correspond to the average redshift-independent distance estimates.}
\footnote*{References: (1) \cite{nagar2005}; (2) \cite{vermeulen2003}; (3) \cite{krips2007}; (4) \cite{birkinshawdavies1985}; (5) \cite{giroletti2005}; (6) \cite{xu2000}; (7) \cite{giacintucci2011}; (8) \cite{houlvestad2001}; (9) \cite{randall2011}; (10) \cite{vrtilek2013}.
\footnote*{We rejected the long term variability analysis (i.e., comparison of spectra at different epochs) of NGC\,3226 because \emph{XMM}--Newton data may be contaminated by emission from NGC\,3227 (see HG13), whereas we have mantained the UV and short term analyses (i.e., light curves).} }
\end{table*}

\section{\label{intro}Introduction}

Active galactic nuclei (AGN) are divided into two classes depending on the width of optical Balmer permitted spectral lines, which can be broad (type 1) or narrow (type 2). From the viewpoint of the unified model (UM) of AGN \citep{antonucci1993,urrypadovani1995}, the difference between type 1 and 2 objects is due to orientation effects relative to the obscuring medium, where a direct view into the black hole (type 1) or a view through the absorbing material (type 2) give rise to a variety of subtypes between both classes.
In the case of low luminosity nuclear emission line regions (LINER), it is tempting to view them as a scaled-down version of Seyfert galaxies. However, different physical properties (e.g., black hole masses or luminosities) have been inferred \citep[][]{eracleous2010b, masegosa2011}, and the way to introduce them in the UM is still controversial \citep{ho2008}. \cite{ho1997} optically classified a variety of LINERs as 1.9 or 2.0 types, while pure type 1 like Seyfert 1--1.5 objects have not been found.

X-ray data offer the most reliable probe of the high-energy spectrum, providing many AGN signatures \citep{donofriomarzianisulentic2012}. 
AGN are detected as a point-like source at hard X-rays. This method was applied for LINERs in a number of publications
\citep[e.g.,][]{satyapal2004, satyapal2005, dudik2005, ho2008}. The most extensive work was carried out by \cite{omaira2009a}. They analyzed 82 LINERs with \emph{Chandra} and/or \emph{XMM}-Newton data and found that 60\% of the sample show a compact nuclear source in the 4.5-8 keV band; a multiwavelenght analysis resulted in about 80\% of the nuclei showing evidence of AGN-related properties. Moreover, their result is a low limit since \emph{Compton}-thick (CT) objects (i.e., $N_H > 1.5 \times 10^ {24} cm^ {-2}$) were not taken into account.

Variability is one of the main properties characterizing AGN \citep{bradley1997}. In the case of LINERs, the first clear evidence showing variability was the work by \cite{maoz2005} at UV frequencies. In X-rays, the study of variability can be accomplished by the comparison of spectra at different epochs, what can account for long term variations.
This was done for LINERs by different authors. \cite{pian2010} and \cite{younes2011} showed that long term variability is common in type 1 LINERs. \cite{omaira2011b} studied a type 2 LINER also showing long term variations. In a previous paper, we studied long term spectral variability in six type 1 and 2 LINERs, where spectral and flux variations were found at long timescales in four
objects \citep[][hereinafter HG13]{lore2013}. 
These spectral variations may be related to the soft excess, the absorber, and/or the nuclear power.

For one epoch observation, when high signal to noise data are available, short timescale variations can be investigated through the power spectral density (PSD) analysis of the light curve \citep{lawrence1987, omairavaughan2012}.
Using PSDs, \cite{omairavaughan2012} found that variable LINERs amount to 14\%, in comparison to the 79\% found for Seyfert galaxies.

On the other hand, the normalised excess variance, $\sigma_{NXS}^2$, is the most straightforward method to search for short term variations \citep{nandra1997, vaughan2003}.
This quantity can be understood as a proxy of the area below the PSD shape, and can be used to search for short term variations, with the advantage that high quality data are not required for its calculation. In HG13 we did not find short timescale variations in six LINERs. 

The aim of this paper is to study the main driver of the X-ray variability in LINERs. We analyzed the X-ray variability in the largest sample of LINERs studied for this purpose. 
This paper is organized as follows: the sample and the data are presented in Sect. \ref{sample}. The reduction of the data is explained in Sect. \ref{reduction}. A review of the methodolody is described in Sect. \ref{method}, where individual and simultaneous spectral fittings, comparisons of different appertures, flux variability at X-ray and UV frequencies, and short term variability are explained. The results from this analysis are given in Sect. \ref{results}, which are discussed in Sect. \ref{discusion}. Finally, our main results are summarized in Sect. \ref{conclusion}.

%__________________________________________________________________

\section{\label{sample}Sample and data}

We used the Palomar Sample \citep{ho1997}, which is the largest sample of nearby galaxies with optical spectra, containing HII nuclei, Seyferts, LINERs, and transition objects. It includes measurements of
the spectroscopic parameters for 418 emission-line nuclei. Since we are interested in LINERs, objects clasified as L1, L1:, L1::, L2, L2:, L2:: and L/S\footnote{Quality ratings as described by \cite{ho1997} are given by ``:" and ``::" for uncertain and highly uncertain classification, respectively.} were taken into account. This sample contains 89 LINERs, 22 type 1 and 67 type 2. Note that along this paper, we divide the objects into two groups, type 1 (1.9) and type 2 (2.0), in accordance with the classification by \cite{ho1997}.

We made use of all the publicly available \emph{XMM}--Newton and \emph{Chandra} data up to October 2013. Initially, 63 objects had either \emph{Chandra} or \emph{XMM}--Newton observations by the date of the sample selection. LINERs with only one available observation were rejected from the sample (28 objects). Objects affected by a pileup fraction of 10\% or more were excluded (four objects, and one observation of another object).
The pileup fractions were estimated using the simulation software {\sc pimms}\footnote{http://heasarc.gsfc.nasa.gov/docs/software/tools/pimms.html} version 4.6. We used the 0.5--2 keV and 2--10 keV fluxes, the best fit model, and the redshift to evaluate its importance. Only two objects in the final sample are affected by a pileup fraction of 6\% (obsID. 2079 of NGC\,4494, and obsID. 5908 of NGC\,4374). As shown later in the results (see Sect. \ref{ind}), this does not have consequences in the variability studies.
To guarantee a proper spectral fitting, observations with less than 400 number counts were also excluded (12 objects, and 18 observations). ObsID 011119010 and 13814 of NGC\,4636 and NGC\,5195, respectively, met these criteria but a visual inspection showed low number counts in the hard band and were rejected from the sample.
Finally, NGC\,4486 was rejected because it is well known that this source is dominated by the jet emission \citep{harris2003, harris2006, harris2009, harris2011}.

The final sample of LINERs contains 18 objects, eight type 1 and 10 type 2.   
Table \ref{properties}
shows the general properties of the galaxies. 
This sample covers the same range in total apparent blue magnitudes than all LINERs in \cite{ho1997} sample, with $B_T$ from 8.7 to 12.3, included in Col. 6. The X-ray classification from \cite{omaira2009a} divides the objects into AGN candidates (when a point-like source is detected in the 4.5--8.0 keV energy band) and non-AGN  candidates (otherwise).
Evidence of jet structure at radio frequencies is provided in Col. 10. 
Table \ref{obs} shows the log of the valid observations, where the observational identification (Col. 3), dates (Col. 4), extraction radius (Col. 5), and the net exposure time (Col. 6) are presented. Number of counts and
hardness ratios, defined as HR = (H-S)/(H+S)\footnote{H is the
number of counts in the hard (2--10 keV) band and S is the number of counts in the soft (0.5--2 keV) band} are also included in Cols. 7 and 8, respectively. Finally, UV luminosities from the optical monitor (OM) and its corresponding filter are given in Cols. 9 and 10.

\section{\label{reduction}Data reduction}

\subsection{Chandra data}

\emph{Chandra} observations were obtained with the ACIS instrument \citep{garmire2003}. The data reduction and analysis were carried out in a systematic, uniform way using CXC
\emph{Chandra} Interactive Analysis of Observations (CIAO\footnote{http://cxc.harvard.edu/ciao4.4/}), version 4.3. Level 2 event data were extracted with the 
{\sc acis-process-events} task. We first cleaned the data from background flares (i.e., periods of high background) using the 
{ \sc lc\_clean.sl}\footnote{http://cxc.harvard.edu/ciao/ahelp/lc\_clean. html} task, which removes periods of anomalously low (or
high) count rates from light curves from source-free background
regions of the CCD. This routine calculates a mean rate from
which it deduces a minimum and maximum valid count rate,
and creates a file with the periods that are considered by the
algorithm to be good.

Nuclear spectra were extracted from a circular region centered on the positions given by NED\footnote{http://ned.ipac.caltech.edu/}. These positions were visually inspected to make sure that the coordinates match the X-ray source position. We chose circular
radii, aiming to include all possible photons, while excluding other sources or background effects. The radii are in the
range between $r_{Chandra}$ = 1.5--5.0 $\arcsec$ (or 3--10 pixels, see Table \ref{obs}). The background selection was made taking circular regions between 5--10$\arcsec$ apertures free of sources in
the same chip as the target and close to the source to minimize effects related to the
spatial variations of the CCD response.
We used the {\sc dmextract} task to extract the spectra of
the source and the background regions. The response matrix
file (RMF) and ancillary reference file (ARF) were generated for
each source region using the {\sc mkacisrmf} and {\sc mkwarf} tasks,
respectively. The spectra were
binned to have a minimum of 20 counts per spectral bin, to be
able the use of the $\chi^2$-statistics, that was made with the {\sc grppha}
task included in {\sc ftools}.

\subsection{XMM--Newton data}

All \emph{XMM}--Newton observations were obtained from the EPIC pn
camera \citep{struder2001}. The data were reduced in a systematic, uniform way
using the Science Analysis Software (SAS\footnote{http://xmm.esa.int/sas/}), version 11.0.0.
Before extracting the spectra, good-time intervals were selected (i.e., flares were excluded). The method we used for this purpose maximizes the signal to noise ratio of the net source spectrum by applying a different constant count rate threshold on the
single-events, E $>$ 10 keV field-of-view background light curve.
The nuclear positions were taken from NED and visually inspected to check that they match the X-ray nuclear positions. As a sanity check, the {\sc eregionanalyse} task was used to compare wether our visual selection deviates from this selection. This task was applied to three objects with low number counts in the sample (namely NGC\,1961, NGC\,3608, and NGC\,5982) and relative diferences $< 1\%$ were obtained. The extraction region was determined through circles of $r_{XMM}$ = 15--35 $\arcsec$ (i.e., 300--700 px) radius and
the background has been determined with an algorithm that selects
the best circular region around the source, that is free of other
sources and as close as possible to the nucleus. This automatic
selection was checked manually to ensure the best selection for
the backgrounds.

We extracted the source and background regions with the
{\sc evselect} task. RMFs were generated using the {\sc rmfgen} task,
and the ARFs were generated using the {\sc arfgen} task. We then
grouped the spectra to obtain at least 20 counts per spectral bin
using the {\sc grppha} task, as is required to use the $\chi^2$-statistics.

\subsection{Light curves}

Light curves in the 0.5--10 keV, 0.5--2.0 keV and 2.0 --10.0 keV energy bands of the source and
background were extracted using the {\sc dmextract} and
{\sc evselect} tasks for \emph{XMM}--Newton and \emph{Chandra}, respectively, with a 1000s
bin. We studied only those light curves with exposure times longer than 30 ksec. For light curves with longer exposure times, we divided them in segments of 40 ksec. Thus, in some cases more than one segment are obtained from the same light curve. The light curve from the source was manually screened
for high background and flaring activity, i.e., when
the background light curve showed flare-like events and/or
prominent decreasing/increasing trends. After this process the total useful observation time is usually lower, thus only light curves with more than a total of 30 ksec are used for the analysis. The light curves are shown
in Appendix \ref{lightcurves}, where the solid line represents the mean value
of the count rate and the dashed lines represent 1$\sigma$ standard
deviation. Note that these values will not be used for the variability analysis, but for a visual inspection of the data.

\section{\label{method}Methodology}

The methodology is explained in HG13, whereas it differs in the treatment of the short-term variability (see Sect. \ref{short}). For clarity, we recall the procedure in the following. 

\subsection{Individual spectral analysis}

An individual spectral analysis allowed us to select the best fit model for each data set. We used XSPEC\footnote{http://heasarc.nasa.gov/xanadu/xspec/} version 12.7.0 to fit the data with five different models:

\begin{itemize}
\item[$\bullet$] ME : $e^{N_{Gal} \sigma (E)} \cdot e^{N_{H} \sigma (E(1+z))}[N_{H}] \cdot MEKAL[kT, Norm]$
\vspace*{0.2cm}
\item[$\bullet$]
PL : $e^{N_{Gal} \sigma (E)} \cdot e^{N_{H} \sigma (E(1+z))}[N_{H}] \cdot Norm e^{-\Gamma}[\Gamma, Norm]$
\vspace*{0.2cm}
\item[$\bullet$] 2PL : $e^{N_{Gal} \sigma (E)} \big( e^{N_{H1} \sigma (E(1+z))}[N_{H1}] \cdot Norm_1 e^{-\Gamma}[\Gamma, Norm_1] + e^{N_{H2} \sigma (E(1+z))}[N_{H2}] \cdot Norm_2 e^{-\Gamma}[\Gamma, Norm_2]\big)$
\vspace*{0.2cm}
\item[$\bullet$] MEPL : $e^{N_{Gal} \sigma (E)} \big(e^{N_{H1} \sigma (E(1+z))}[N_{H1}] \cdot MEKAL[kT, Norm_1] + e^{N_{H2} \sigma (E(1+z))}[N_{H2}] \cdot Norm_2 e^{-\Gamma}[\Gamma, Norm_2]\big)$
\vspace*{0.2cm}
\item[$\bullet$] ME2PL : $e^{N_{Gal} \sigma (E)} \big( e^{N_{H1} \sigma (E(1+z))}[N_{H1}] \cdot Norm_1 e^{-\Gamma}[\Gamma, Norm_1] + MEKAL[kT] + e^{N_{H2} \sigma (E(1+z))}[N_{H2}] \cdot Norm_2 e^{-\Gamma}[\Gamma, Norm_2]\big)$
\end{itemize}

\noindent where $\sigma (E)$ is the photo-electric cross-section, $z$ is the redshift, and $Norm_i$ are the normalizations of the power law or the thermal component (i.e., $Norm_1$ and $Norm_2$). For each model, the parameters varying are written in brackets.
The Galactic absoption, $N_{Gal}$, is included in each model and fixed to the predicted value (Col. 5 in Table \ref{properties}) using the {\sc nh} tool within {\sc ftools} \citep{dickeylockman1990, kalberla2005}.

The $\chi^2/d.o.f$ and F-test were used to select the simplest model which better represents the data.

\subsection{\label{simult} Simultaneous spectral analysis}

We simultaneously fitted the spectra for each object with the same model. The baseline model was obtained from the individual fittings. For each galaxy, the initial values for the parameters were set to those obtained for the spectrum with the largest number counts.

The simultaneous fit was done in three steps:

\begin{itemize}
\item[0.] SMF0 (Simultaneous fit 0) : The same model was used with all parameters linked to the same value to fit every spectra of the same object, i.e., the non-variable case.
\item[1.] SMF1 : Using SMF0 as the baseline for this step, we let the parameters $N_{H1}$, $N_{H2}$, $\Gamma$, $Norm_1$, $Norm_2$, and $kT$ vary individually. 
The best fit was selected for $\chi^2_r$ closest to unity that improved SMF0 (using the F-test).
\item[2.] SMF2 : Using SMF1 as the baseline for this step (in case SMF1 did not fit the data well), we let two parameters vary, the one that varied in SMF1 along 
with any of the other parameters of the fit. $\chi^2_r$ and F-test were again used to confirm an improvement of the fit.
\end{itemize}

Whenever possible, this method was applied to the data coming from the two instruments separately. When data from \emph{Chandra} and \emph{XMM}--Newton were used together, 
an additional analysis was performed to make sure the sources included in the larger aperture did not produce the observed variability.
A spectrum of an annular region was then extracted from \emph{Chandra} data, with $r_{ext}$ = $r_{XMM}$ and $r_{int}$ = $r_{Chandra}$. 
 We recall that the PSF of \emph{Chandra} is energy dependent and therefore the annular region might be affected by contamination from the source photons at high energies. We have estimate this contribution by simulating the PSF of the sources in our sample using ChaRT\footnote{http://cxc.harvard.edu/chart/} and MARX\footnote{http://space.mit.edu/ASC/MARX/}. A monochromatic energy of 8 keV was used and the ray density was obtained for each observation individually. We find that the highest contribution from the source photons at 8 keV is 7\%. Note that this contribution is at high energies (i.e., the contribution is lower at lower energies) and does not affect our results (see Sect. \ref{ind}).
The data used for comparisons are marked with $c$ in Table \ref{obs}.
When the contamination by the annular
region to the \emph{Chandra} data with the $r_{XMM}$ aperture emission was higher than 50\% in the 0.5-10.0 keV energy band, we did not consider the joint analysis since the accuracy
of the derived parameters could be seriously affected. For lower contamination levels, we considered that \emph{Chandra} data can be used to estimate the contribution
of the annular region to the \emph{XMM}--Newton spectrum. The ring from \emph{Chandra} data was fitted with the five models explained above. The resulting model was incorporated
(with its parameters frozen) in the fit of the \emph{XMM}--Newton nuclear spectrum, so we were able to extract the parameters of the nuclear emission. When multiple observations of the same object and instrument
were available, we compared the data with similar dates (see Table \ref{obs}).

\subsection{Flux variability}

X-ray luminosities for the individual and simultaneous fits were computed using XSPEC for the soft and hard bands. Distances were taken from NED and correspond to the average redshift-independent distance estimate for each object when available (or to the redshift estimated distance otherwise) and are listed in Table \ref{properties}.

UV luminosities were obtained (when available) from the optical monitor (OM) onboard \emph{XMM}--Newton simultaneously to X-ray data. Whenever possible, measurements from different filters 
were retrieved. We recall that UVW2 is centered at 1894$\AA$ (1805-2454) $\AA$, UVM2 at 2205$\AA$ (1970-2675) $\AA$, and UVW1 at 2675$\AA$ (2410-3565) $\AA$. 
In the case of NGC\,4736 we used data from the U filter (centered at 3275$\AA$ (3030-3890) $\AA$) because measures from other filters were not available.
We used the OM observation FITS source lists (OBSMLI)\footnote{ftp://xmm2.esac.esa.int/pub/odf/data/docs/XMM-SOC-GEN-ICD-0024.pdf} to obtain the photometry. When OM data were not available, we searched for UV information
in the literature. It is worth to notice that in this case the X-ray and UV data might not be simultaneous (see Appendix \ref{indivnotes}).

We assumed an object to be variable when:

\begin{equation}
 L_{max} - L_{min} > 3 \times  \sqrt{(errL_{max})^ 2+(errL_{min})^ 2} 
\end{equation}

\noindent  where $L_{max}$ and $L_{min}$ are the maximum and minimum luminosities of an object and $errL_{max}$ and $errL_{min}$ the errors of the measurements. We notice that this relation is used to determine if an object is variable, and not as an error estimation.

\begin{figure*}
\centering

\subfloat{\includegraphics[width=0.30\textwidth]{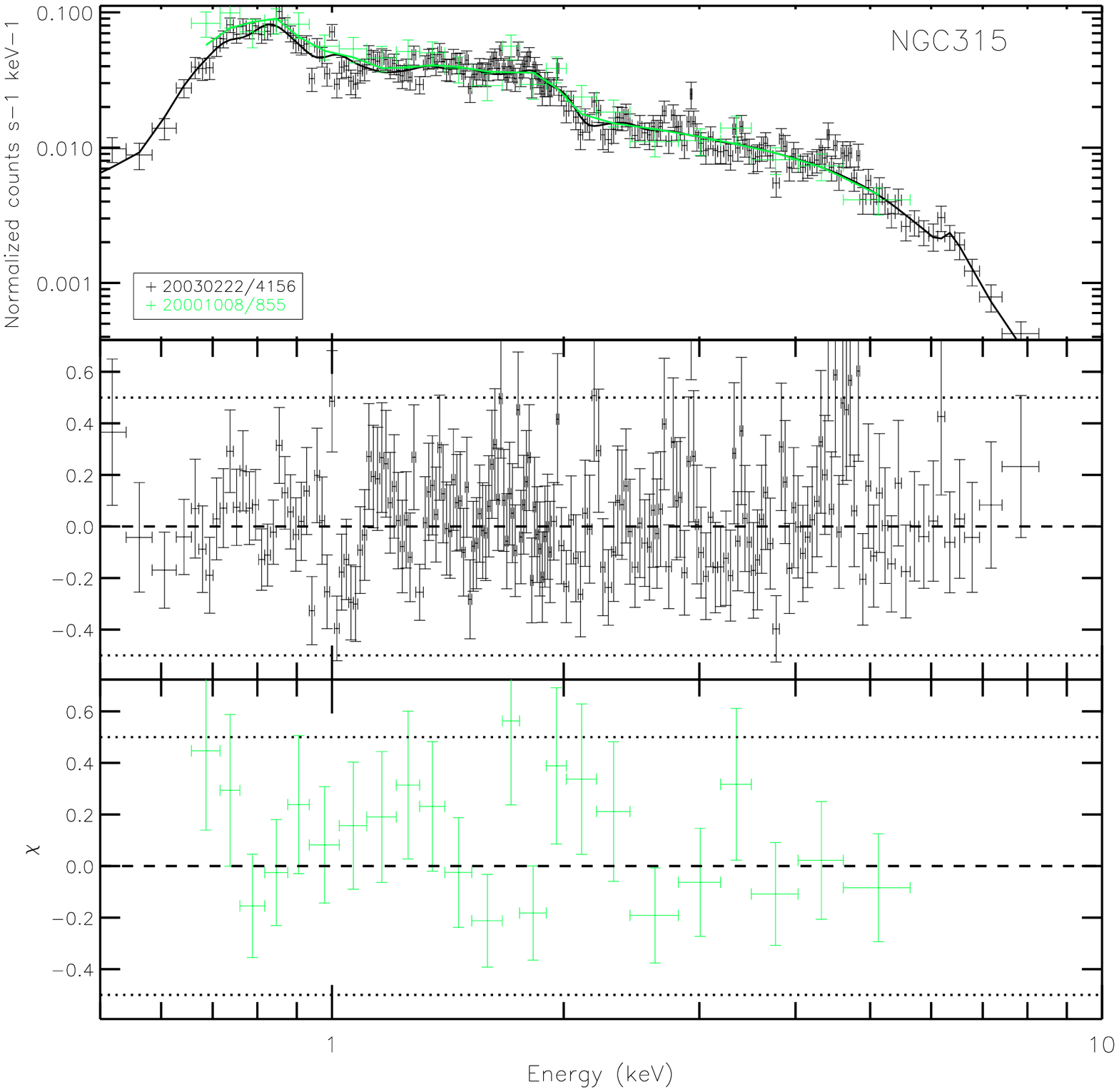}}
\subfloat{\includegraphics[width=0.30\textwidth]{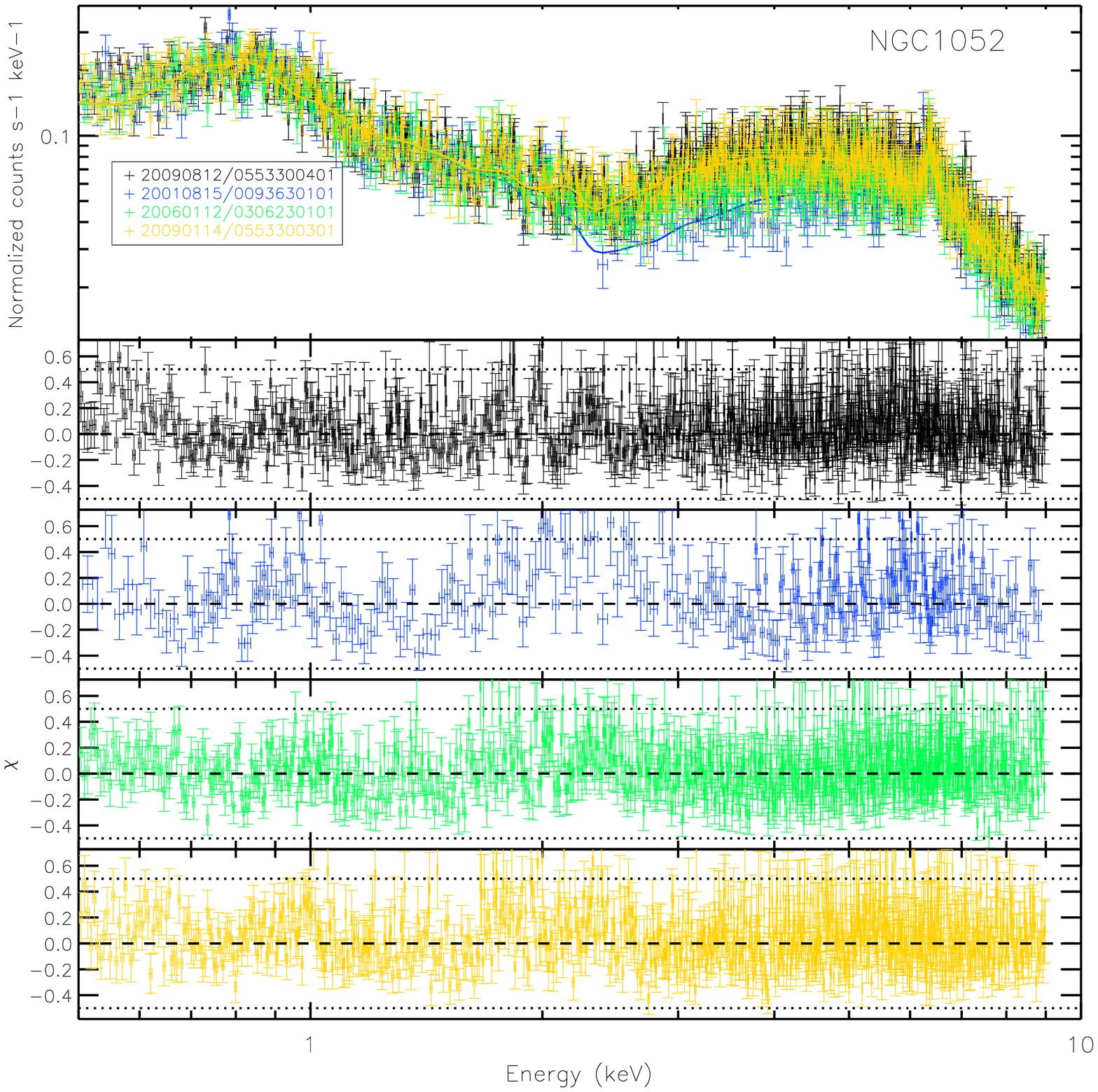}}
\subfloat{\includegraphics[width=0.30\textwidth]{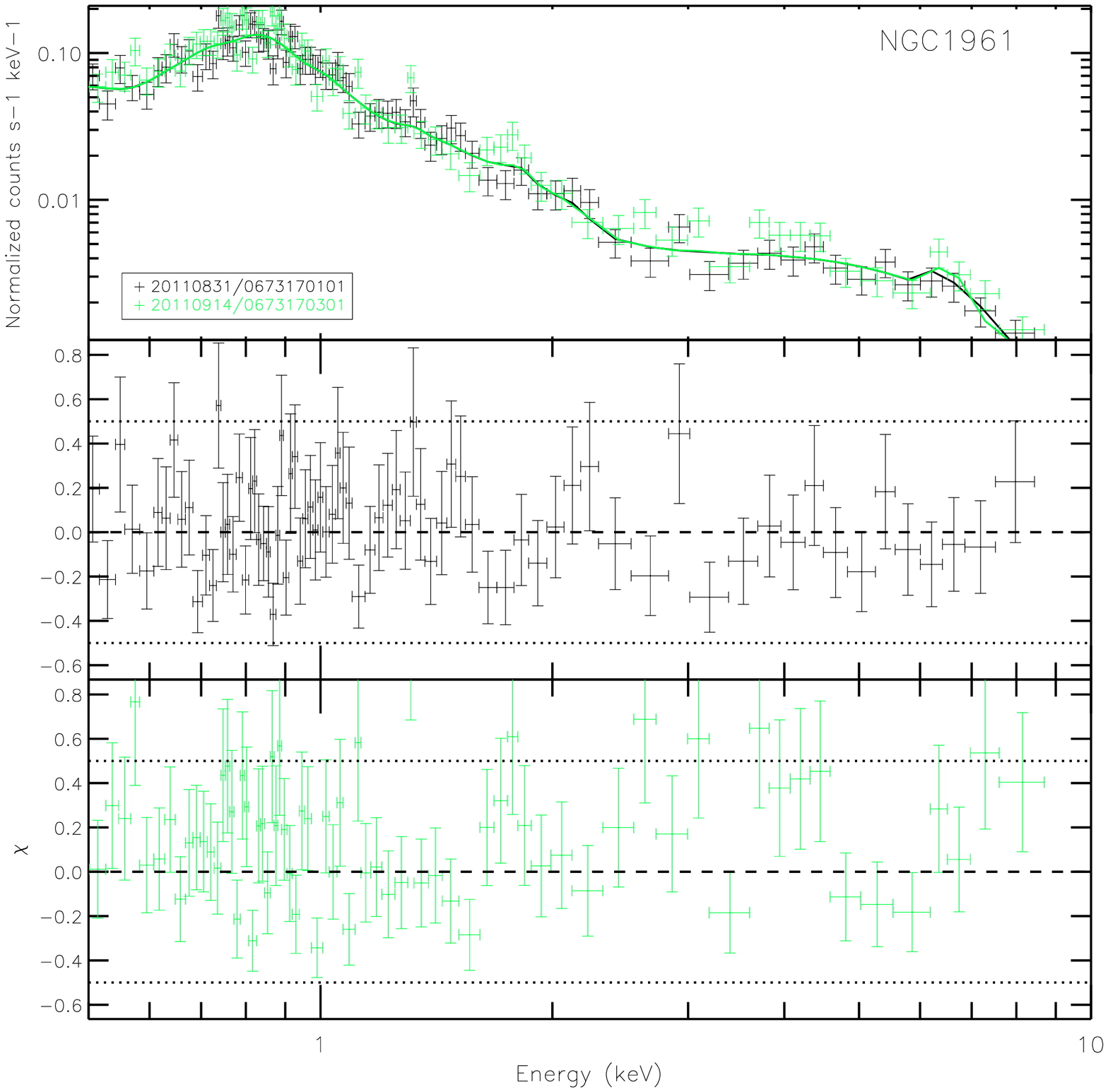}}

\subfloat{\includegraphics[width=0.30\textwidth]{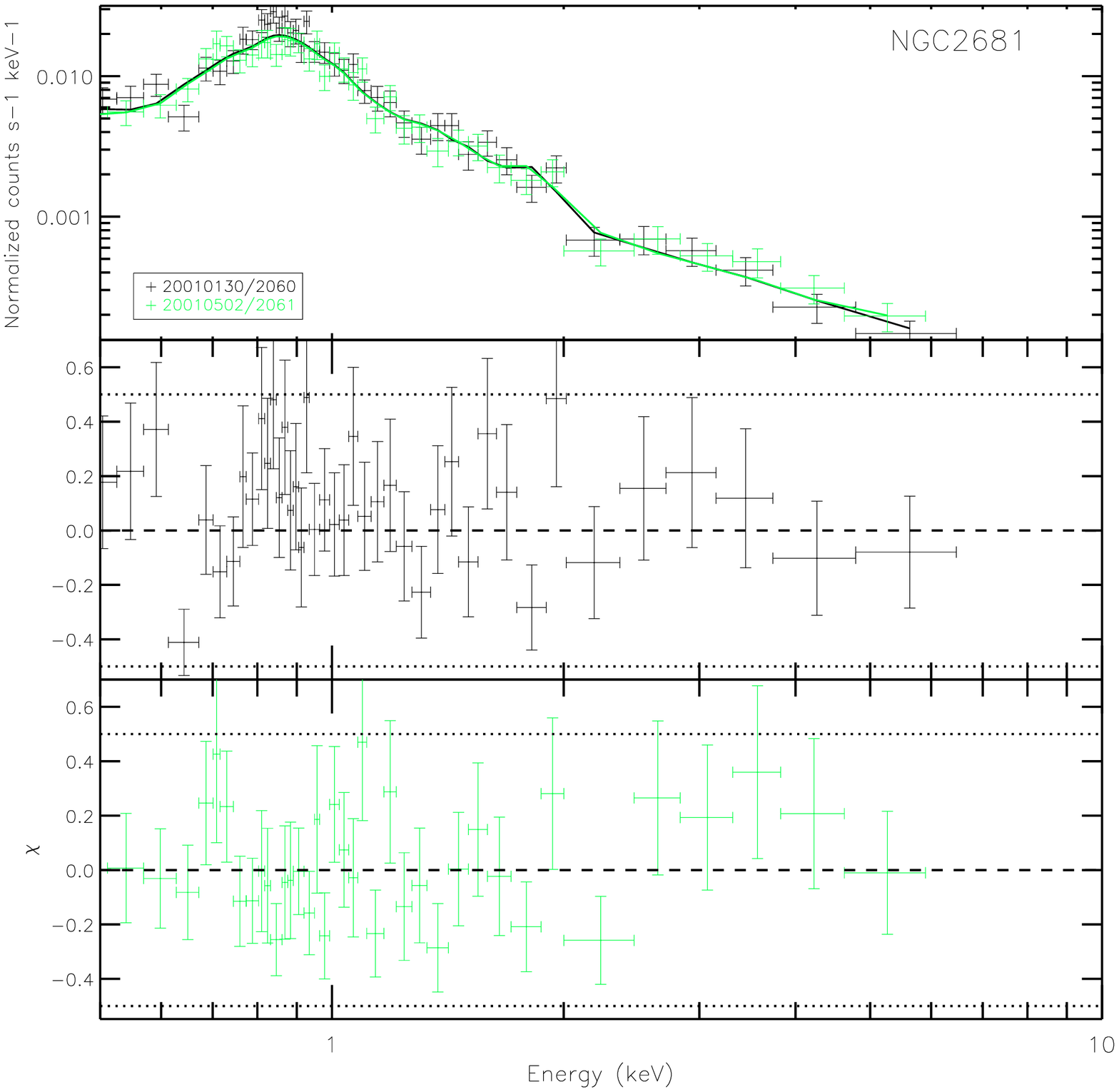}}
\subfloat{\includegraphics[width=0.30\textwidth]{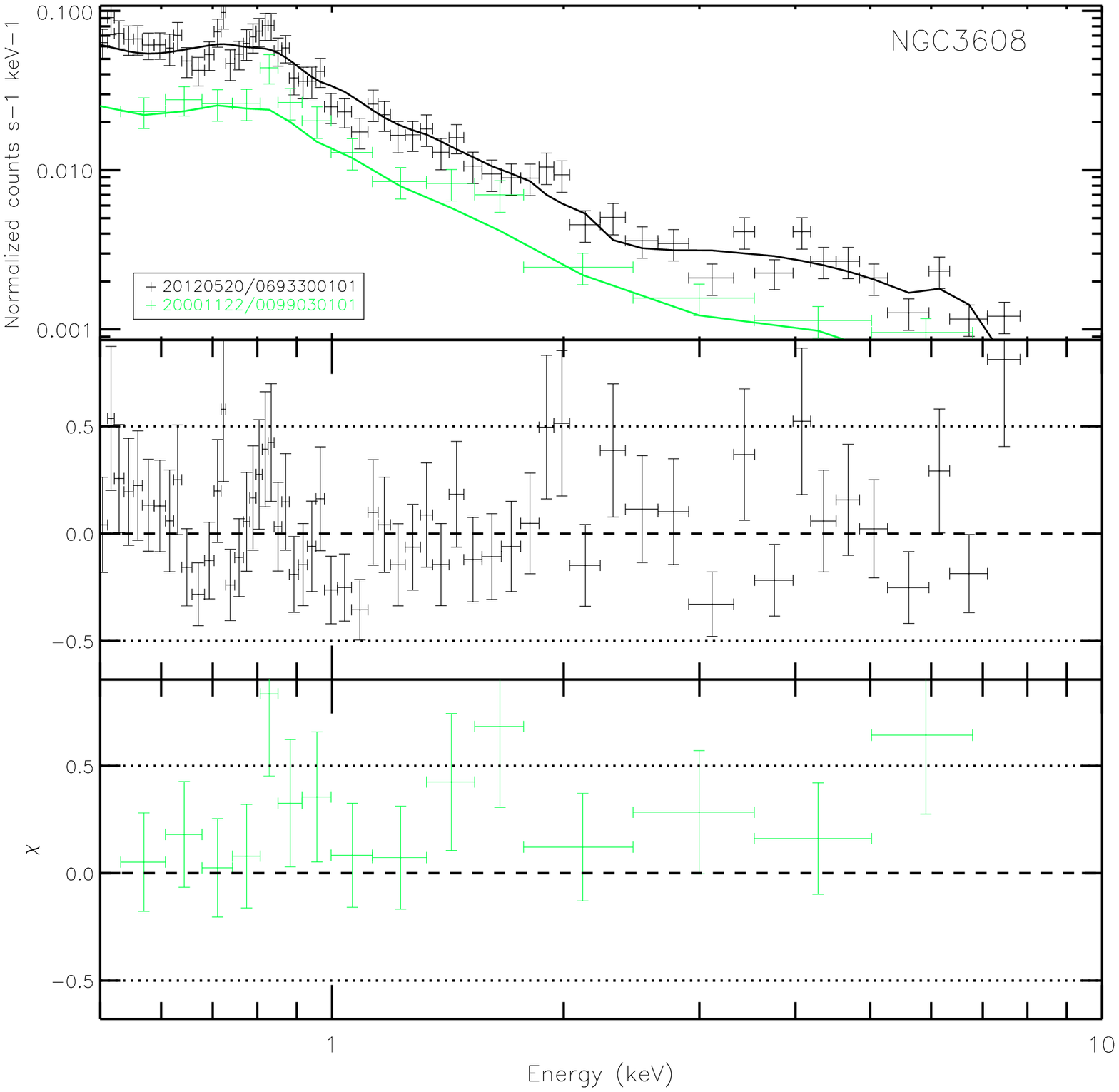}}
\subfloat{\includegraphics[width=0.30\textwidth]{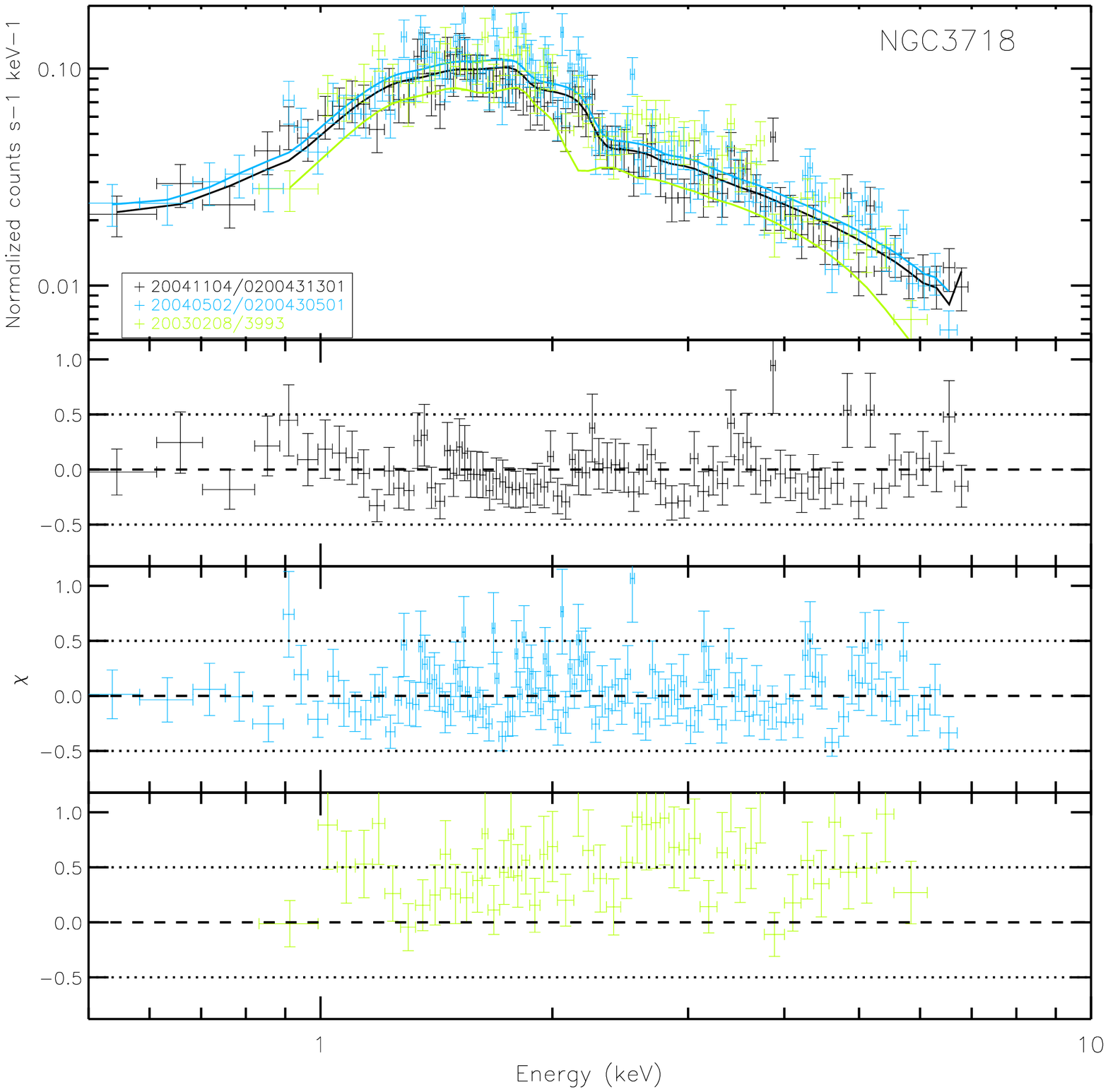}}

\subfloat{\includegraphics[width=0.30\textwidth]{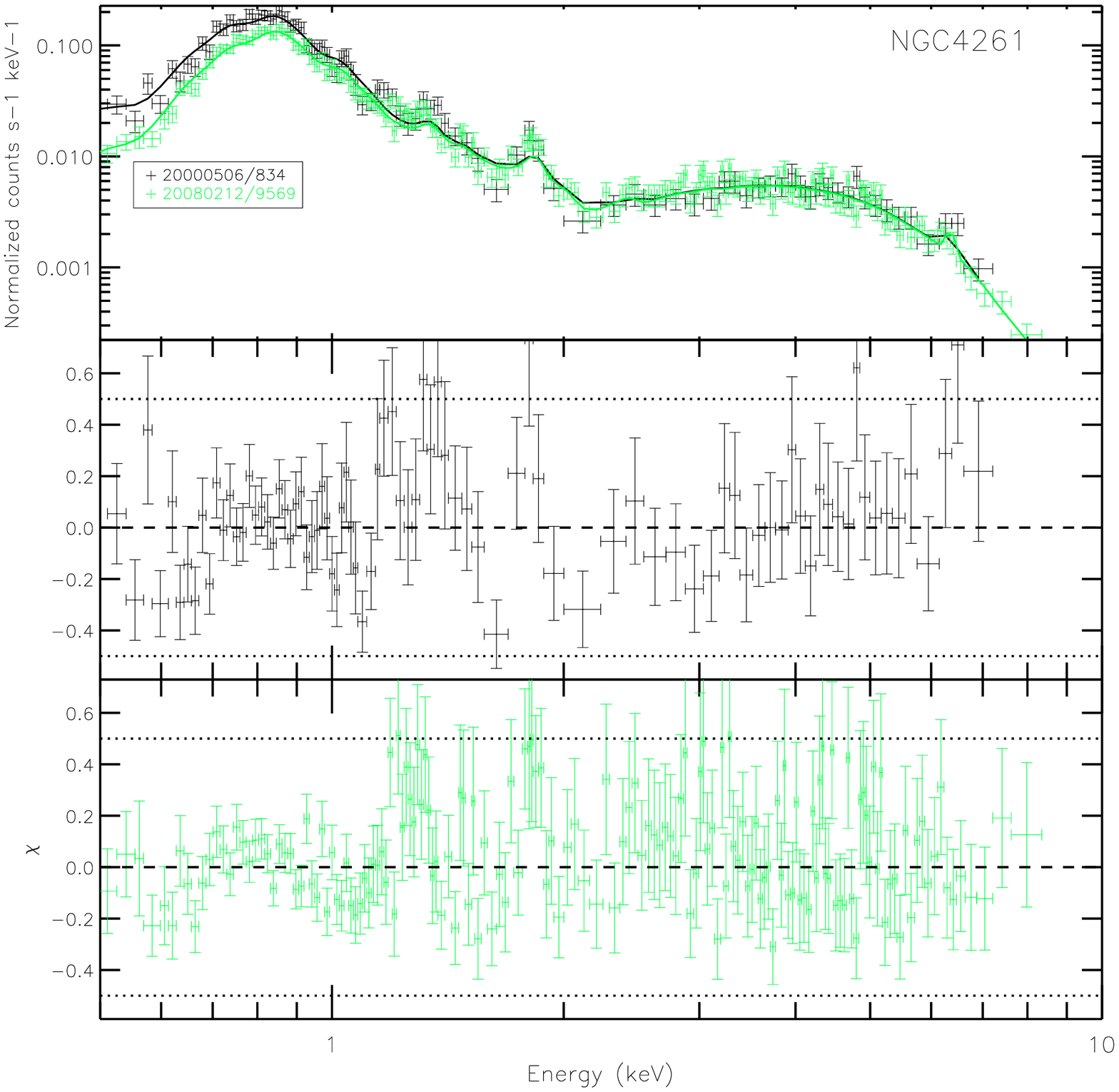}}
\subfloat{\includegraphics[width=0.30\textwidth]{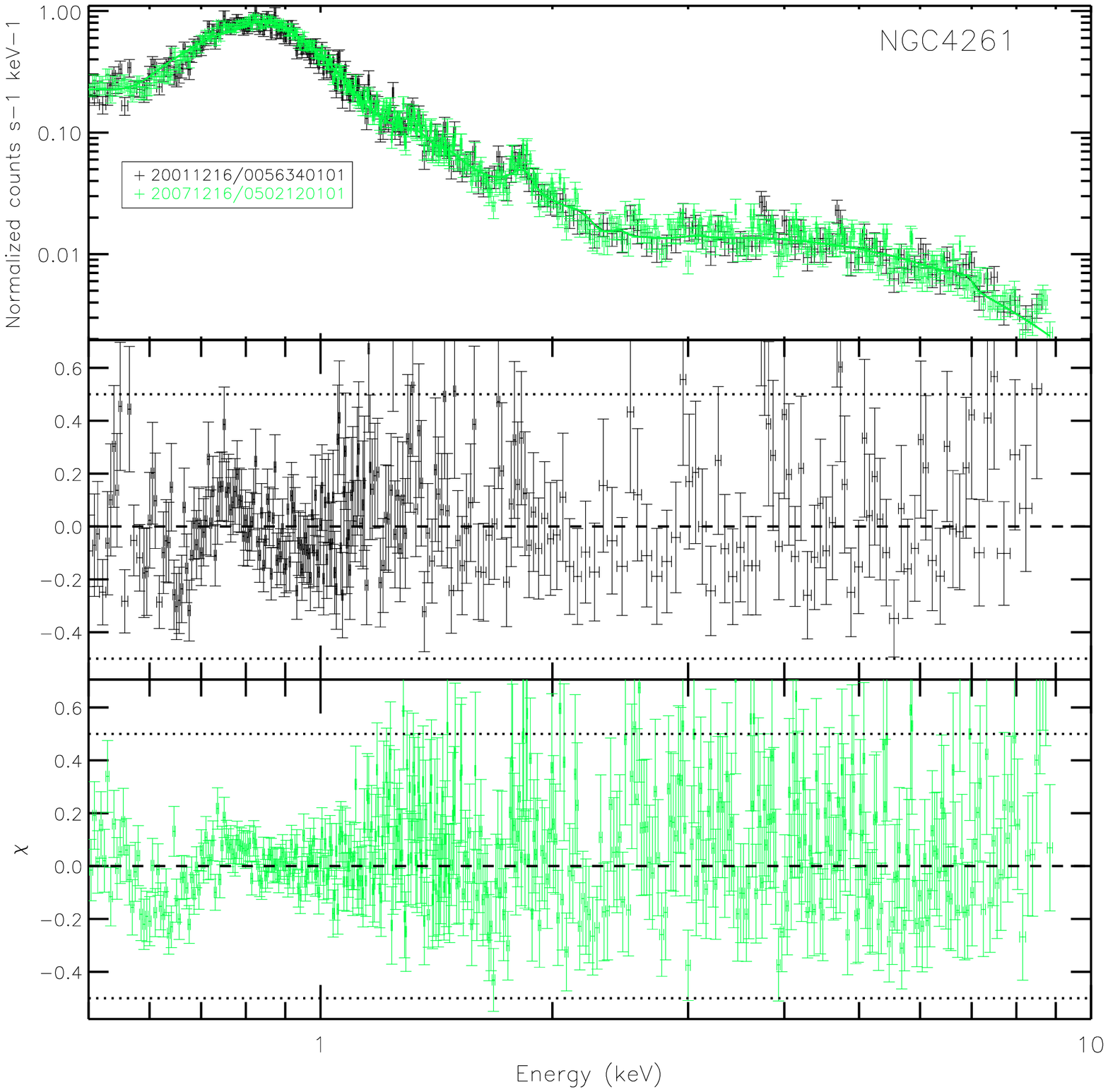}}
\subfloat{\includegraphics[width=0.30\textwidth]{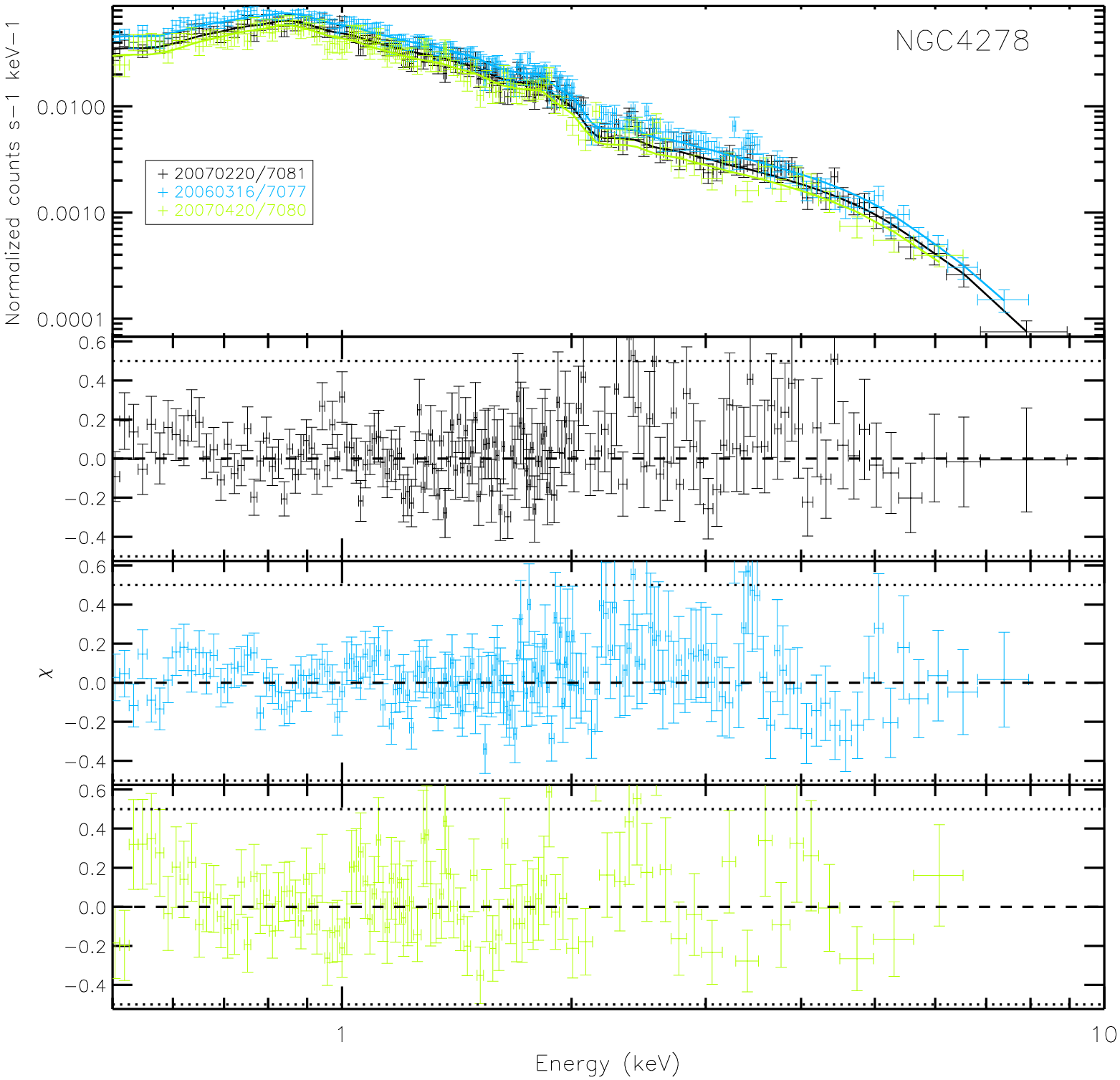}}

\subfloat{\includegraphics[width=0.30\textwidth]{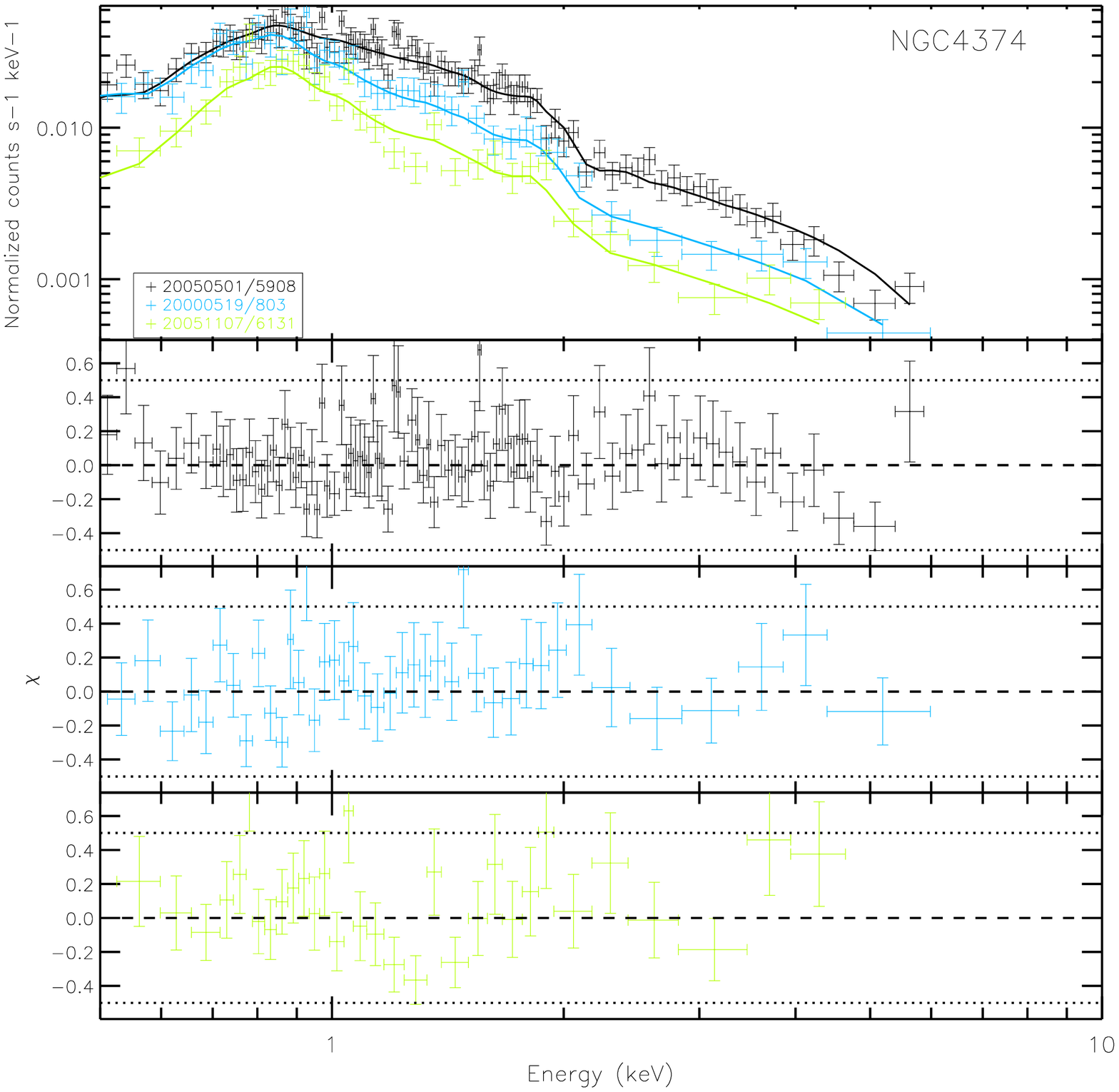}}
\subfloat{\includegraphics[width=0.30\textwidth]{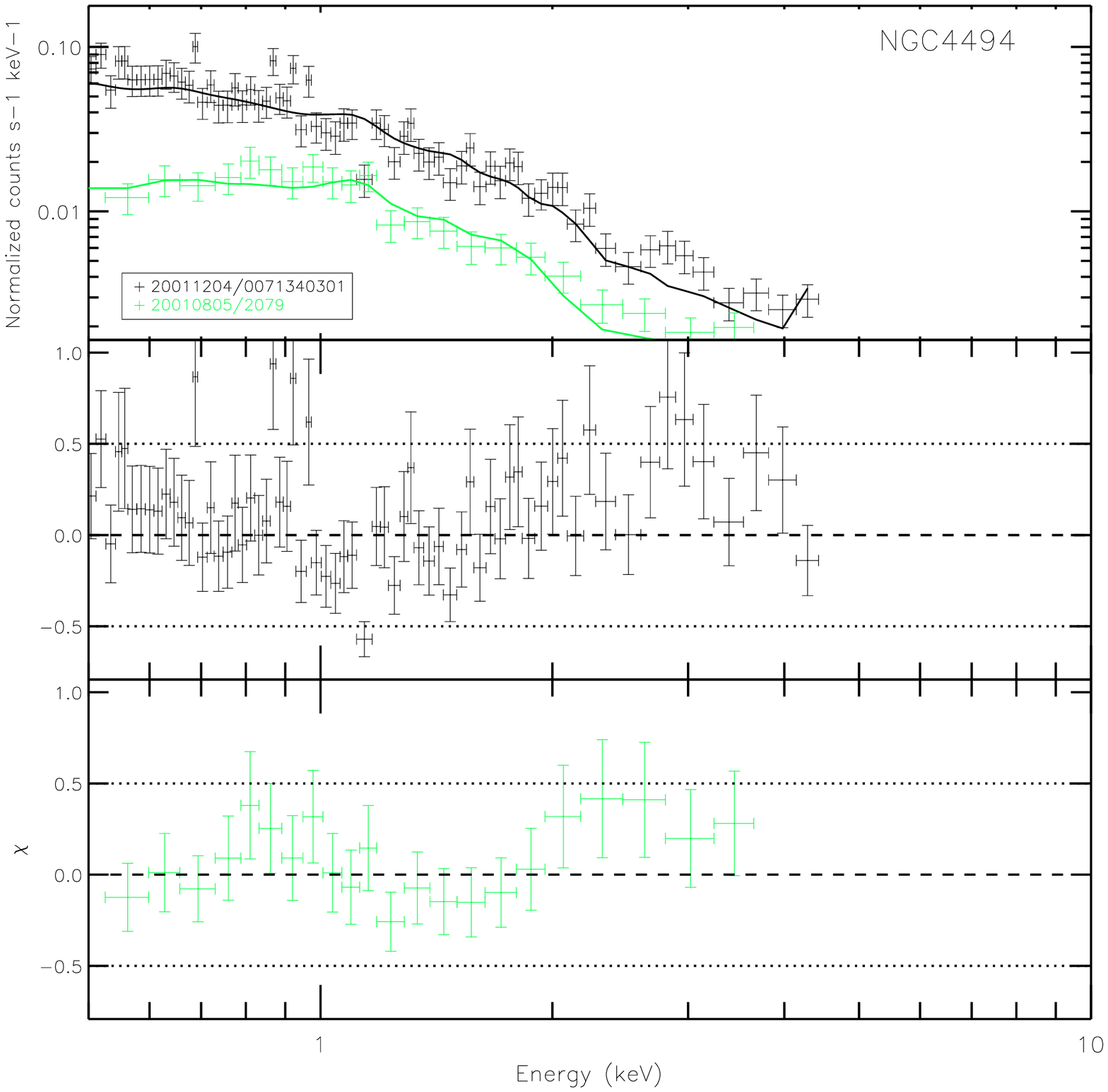}}
\subfloat{\includegraphics[width=0.30\textwidth]{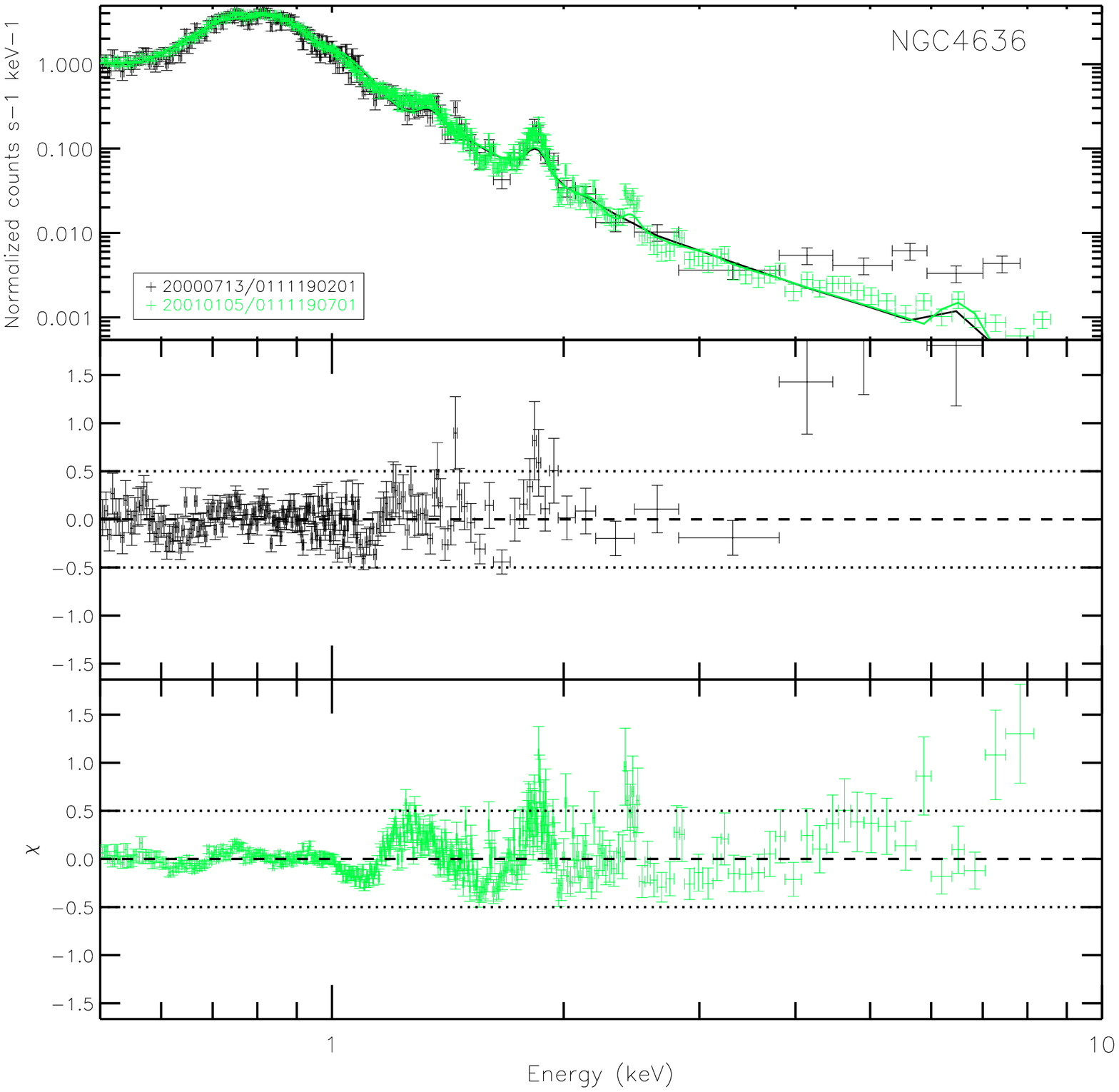}}

\caption{For each object, (top): simultaneous fit of X-ray spectra; (from second row on): residuals. The legends contain the date (in the format yyyymmdd) and the obsID. Details are given in Table \ref{obs}.}
\label{bestfig}
\end{figure*}

\begin{figure*}
\setcounter{figure}{0}
\centering
\subfloat{\includegraphics[width=0.30\textwidth]{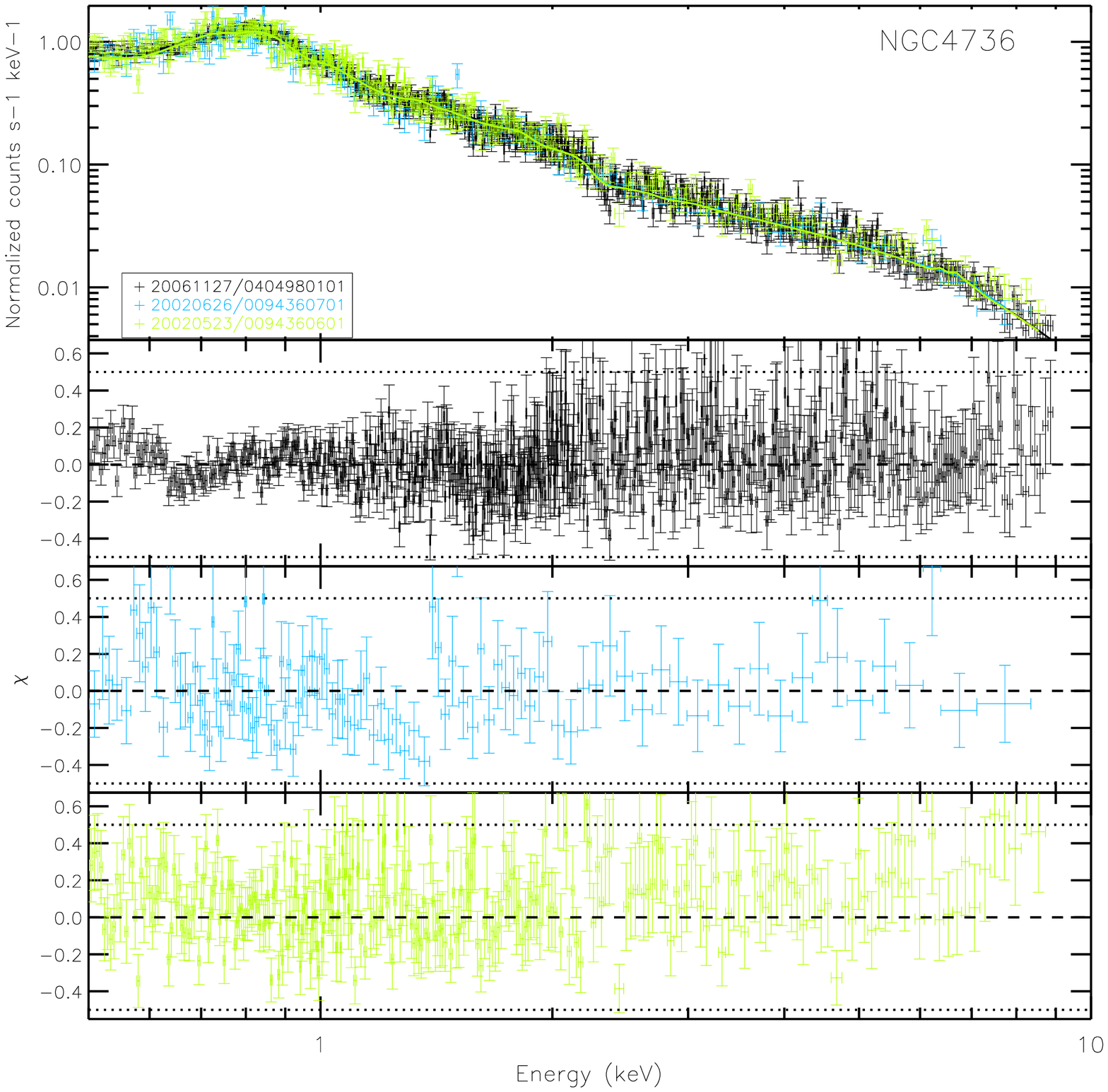}}
\subfloat{\includegraphics[width=0.30\textwidth]{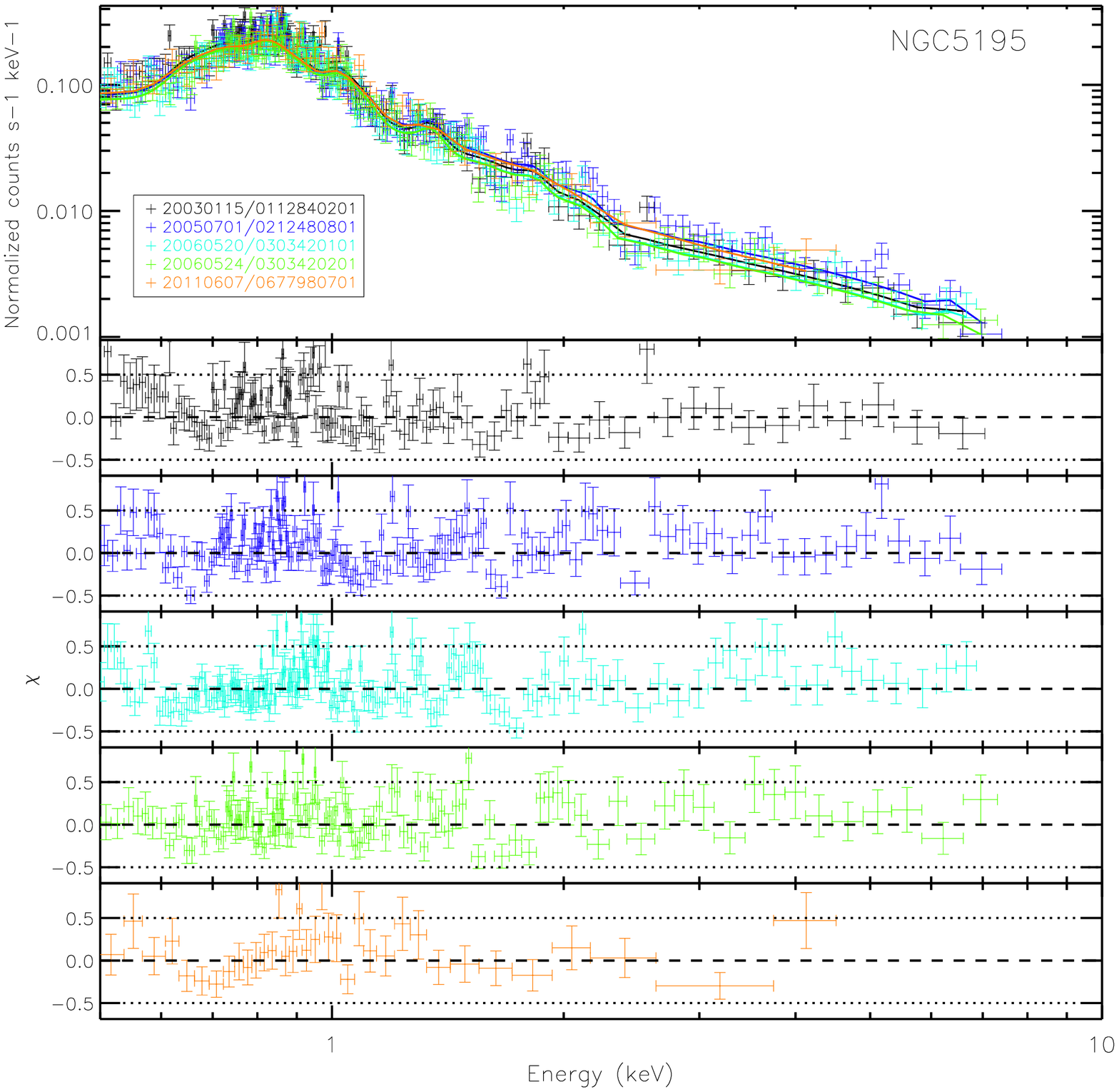}}
\subfloat{\includegraphics[width=0.30\textwidth]{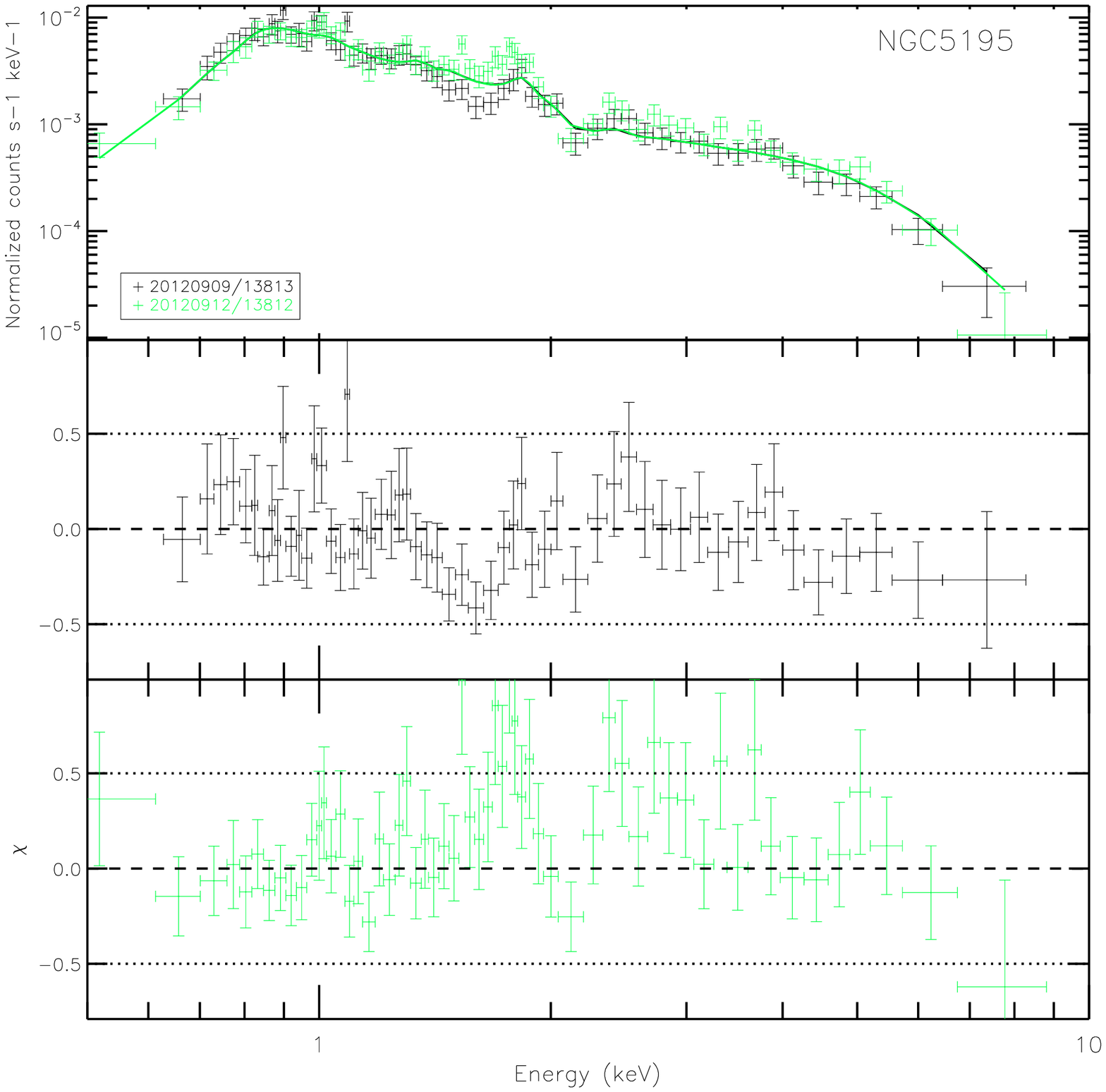}}

\subfloat{\includegraphics[width=0.30\textwidth]{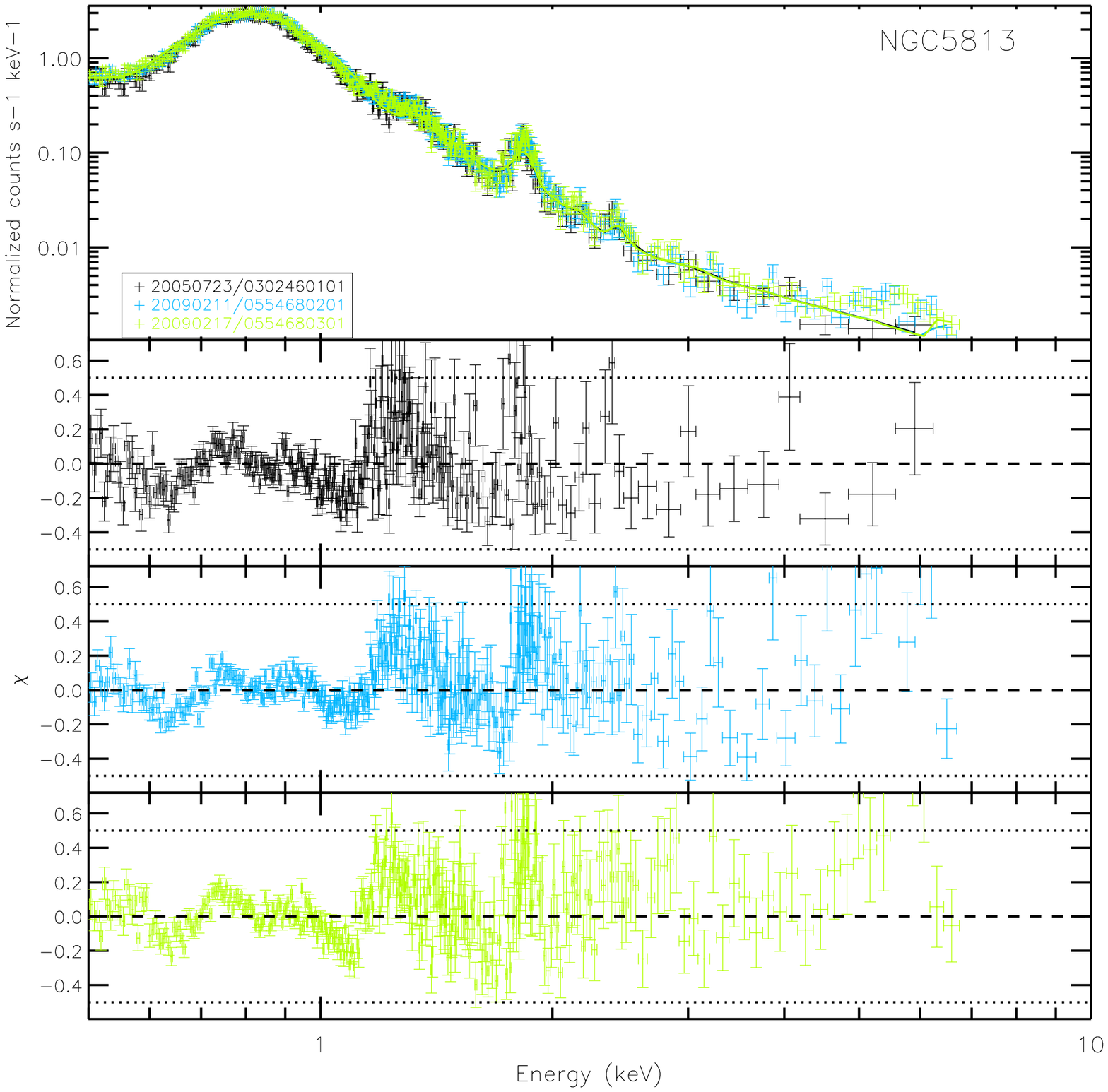}}
\subfloat{\includegraphics[width=0.30\textwidth]{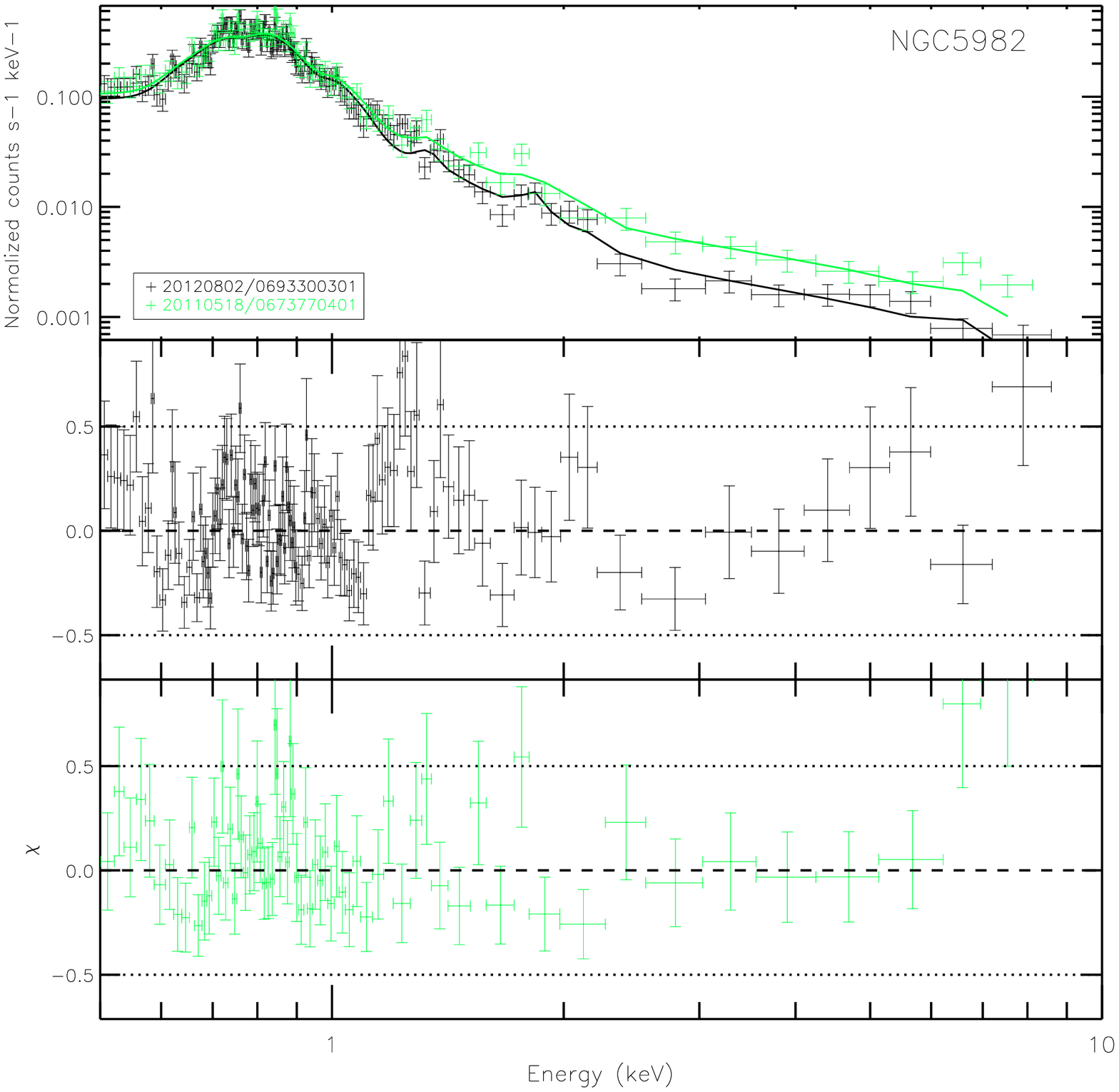}}

\caption{(Cont.)}
\end{figure*}

\subsection{\label{short}Short timescale variability}

Assuming a constant count rate for the whole observation in the 0.5--10 keV energy band, we calculated $\rm{\chi^2/d.o.f}$ as a proxy to the variations. We considered the source to be variable if the count rate differed from the average above 3$\rm{\sigma}$ (or 99.7\% probability). 

To compare the variability amplitude of the light curves between observations, we calculated the normalized excess variance, $\rm{\sigma_{NXS}^2}$, for each light curve segment with 30-40 ksec. This magnitude is related to the area below the power spectral density (PSD) shape.
We followed the prescriptions given by \cite{vaughan2003} to estimate $\rm{\sigma_{NXS}^2}$ and its error, $\rm{err(\sigma_{NXS}^2)}$ \citep[see also][]{omaira2011a}:

\begin{equation} \sigma_{NXS}^2 = \frac{S^2 - <\sigma_{err}^2>}{<x>^2} \end{equation}

\begin{equation} err(\sigma_{NXS}^2) = \sqrt{\frac{2}{N} \Big(\frac{<\sigma_{err}^2>}{<x>^2}\Big)^2 + \frac{<\sigma_{err}^2>}{N} \frac{4\sigma_{NXS}^2}{<x>^2} }, \end{equation}

\noindent where $x$, $\sigma_{err}^2$ and $N$ are the count rate, its error and the number of points in the light curve, respectively, and $S^2$ is the variance of the light curve:

\begin{equation} S^2 = \frac{1}{N-1} \sum_{i=1}^N (x_i - <x>)^2 . \end{equation}

When $\rm{\sigma_{NXS}^2}$ was negative or compatible with zero within the errors, we estimated the 90\% upper limits using Table 1 in \cite{vaughan2003}. We assumed a PSD slope of -1, the upper limit from \cite{vaughan2003} and we added the value of 1.282err($\rm{\sigma_{NXS}^2}$) (to take into account the uncertainity due to the experimental Poisson fluctuations) to the limit. For a number of segments, N, obtained from an individual light curve, an upper limit for the normalised excess variance was calculated as:

\begin{equation}
\sigma_{NXS}^2 = \frac{\sqrt{\sum_{i=1}^N \sigma_{NXS_i}^2}}{N}
\end{equation}

In case N segments were obtained for the same light curve and at least one was consistent with being variable, we calculated the normalised weighted mean
and its error as the weighted variance.

\section{\label{results}Results}

In this section we present the results on the variability in LINERs of all the sources
individually (Sect. \ref{ind}) and the general results (Sects. \ref{spectral}, \ref{flux}, and \ref{lightcurve}). This includes short and long term variations in X rays and long term variations at UV frequencies. 
The summary of the results obtained for the variability is given in Table \ref{variab}.
Notes and comparisons with previous works for individual
objects are included in Appendix \ref{indivnotes}.

\subsection{\label{ind}Individual objects}

\begin{figure*}
\centering
\subfloat{\includegraphics[width=0.30\textwidth]{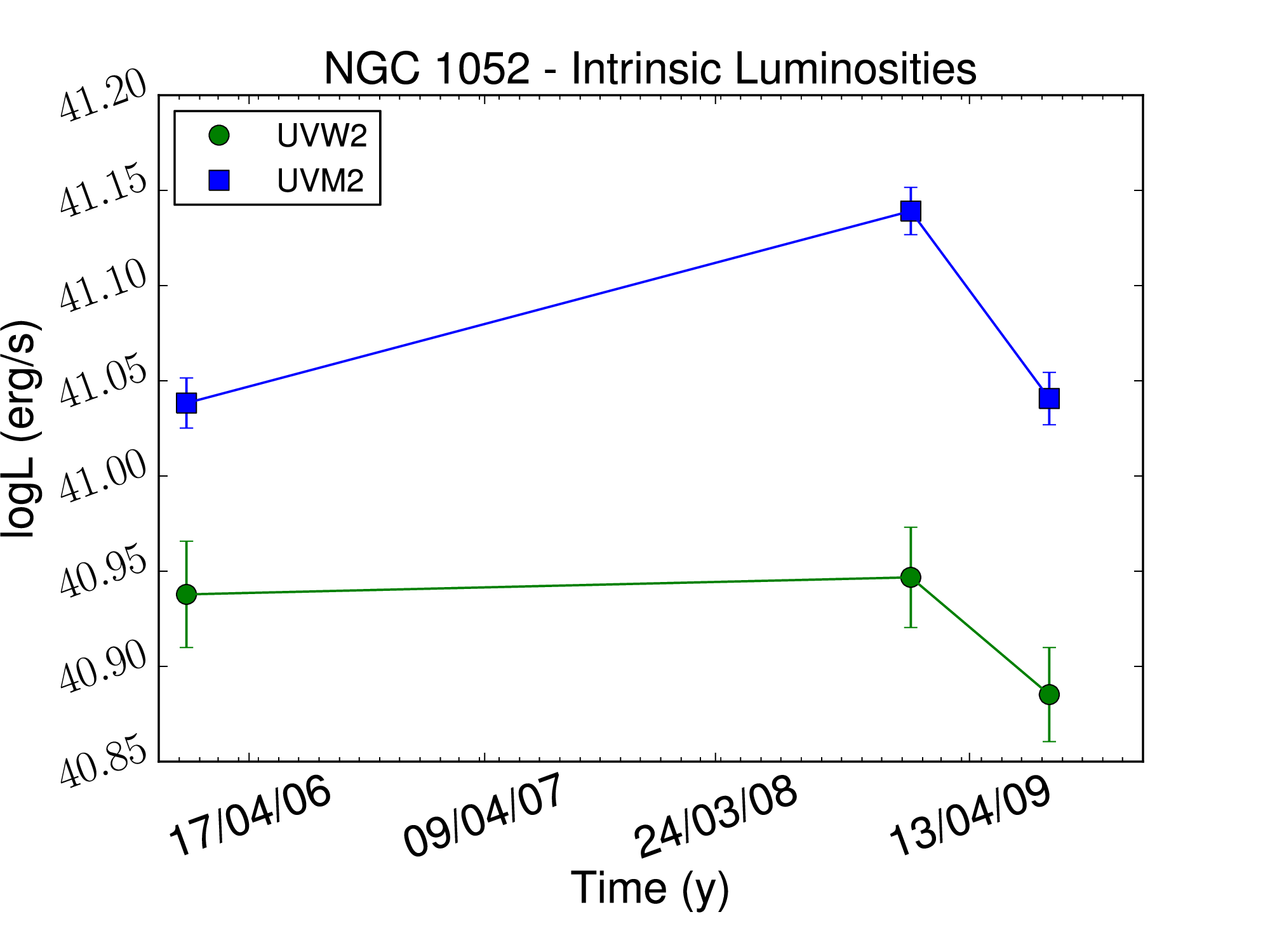}}
\subfloat{\includegraphics[width=0.30\textwidth]{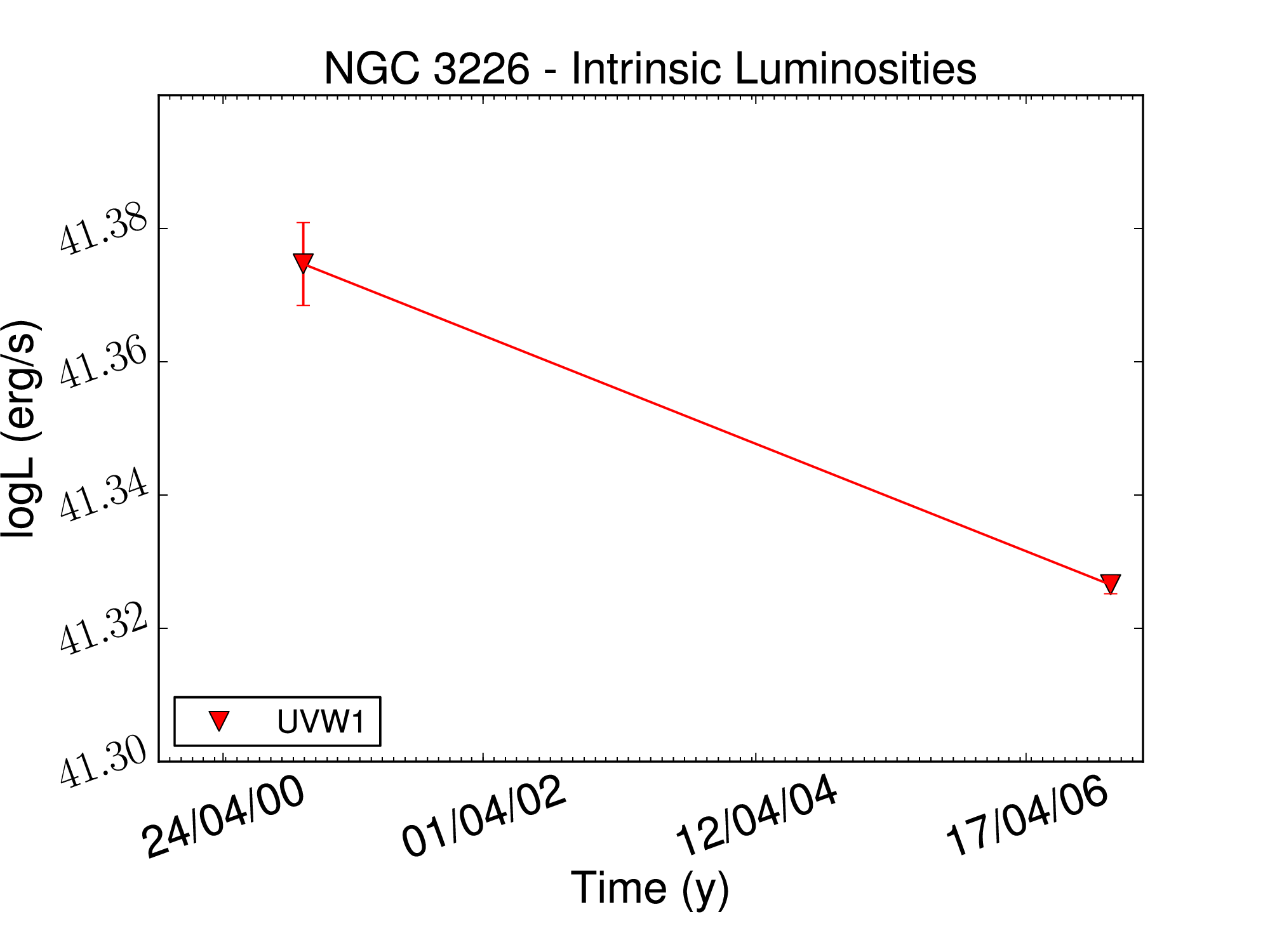}}
\subfloat{\includegraphics[width=0.30\textwidth]{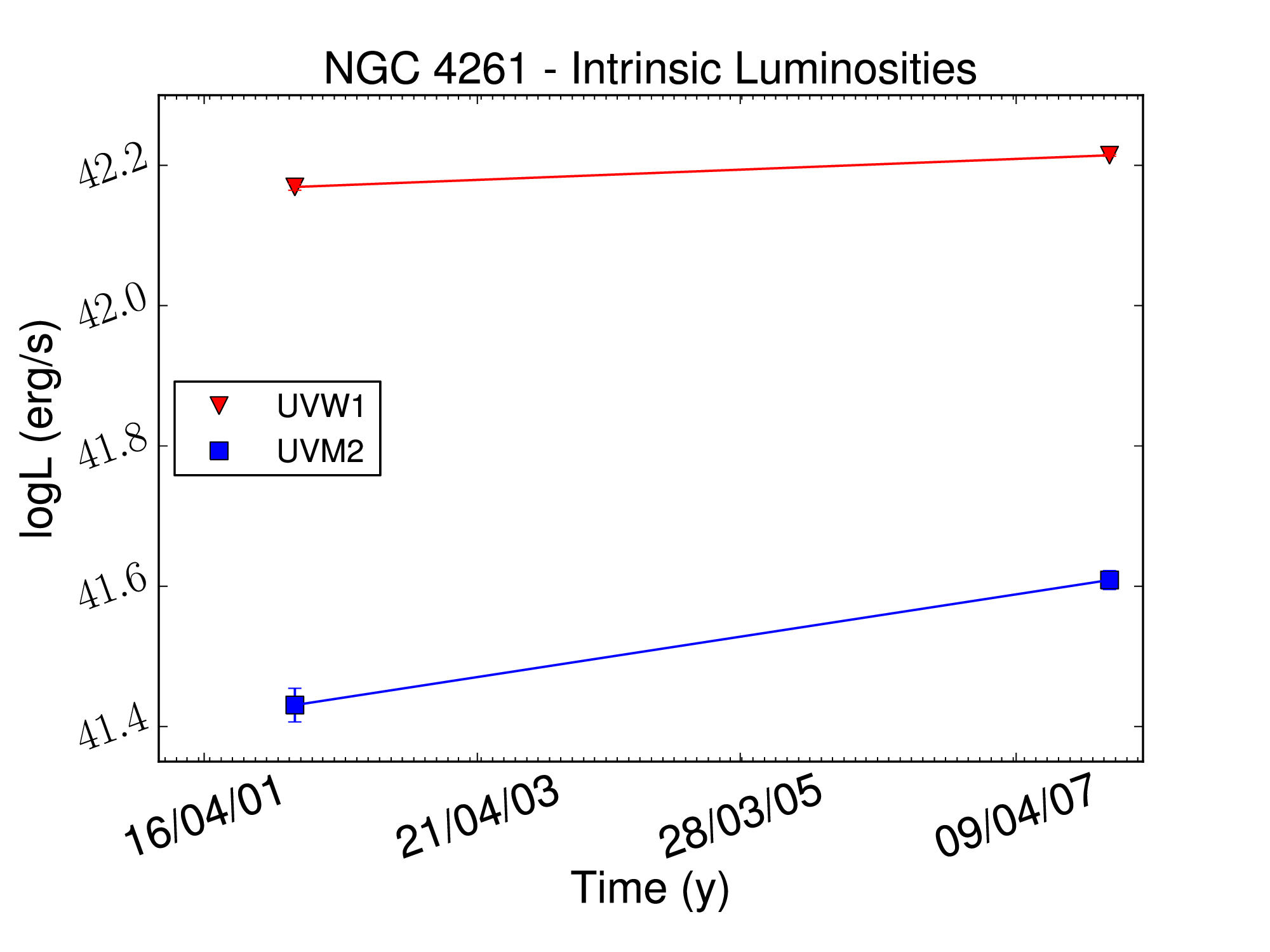}}

\subfloat{\includegraphics[width=0.30\textwidth]{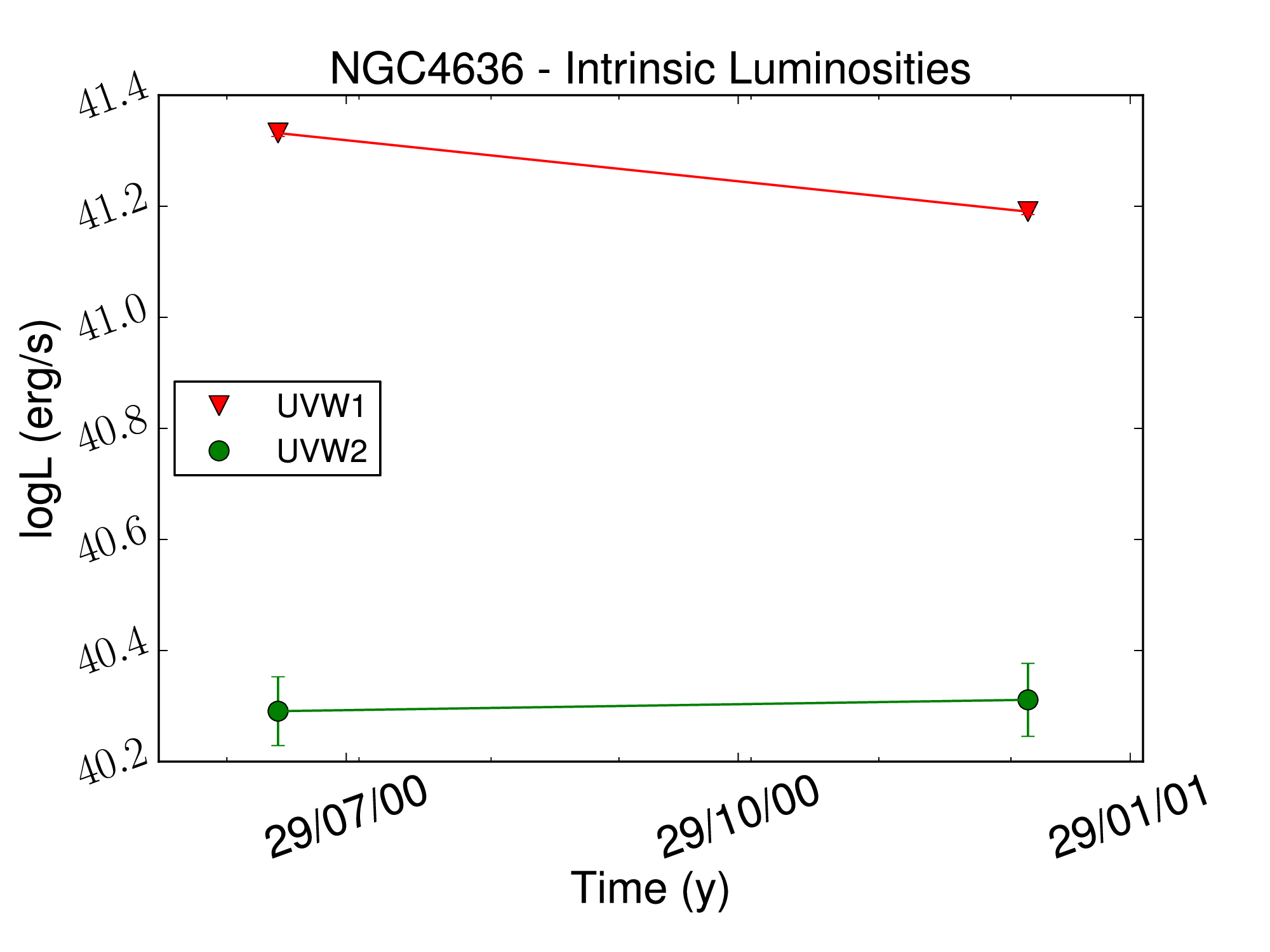}}
\subfloat{\includegraphics[width=0.30\textwidth]{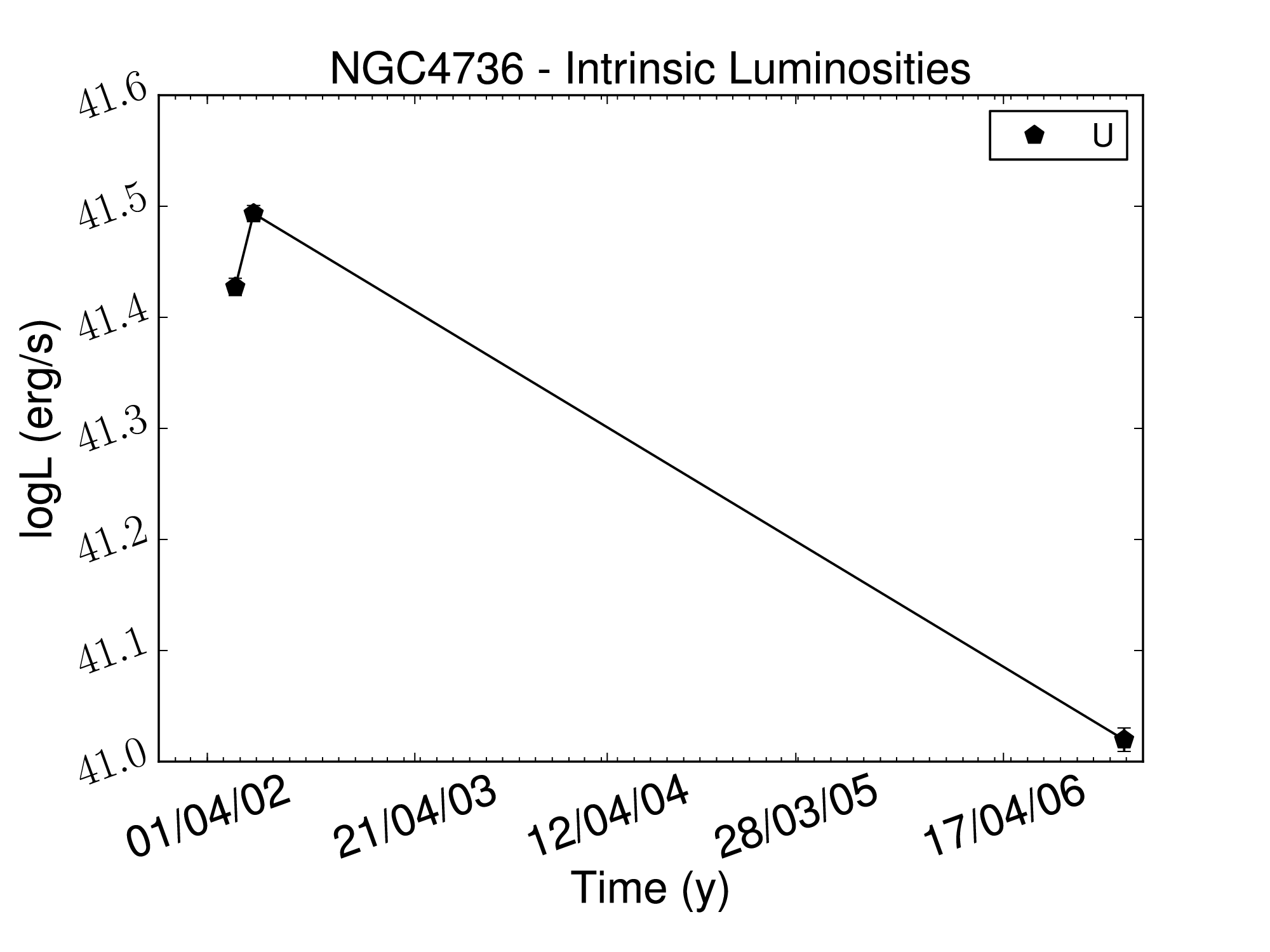}}
\subfloat{\includegraphics[width=0.30\textwidth]{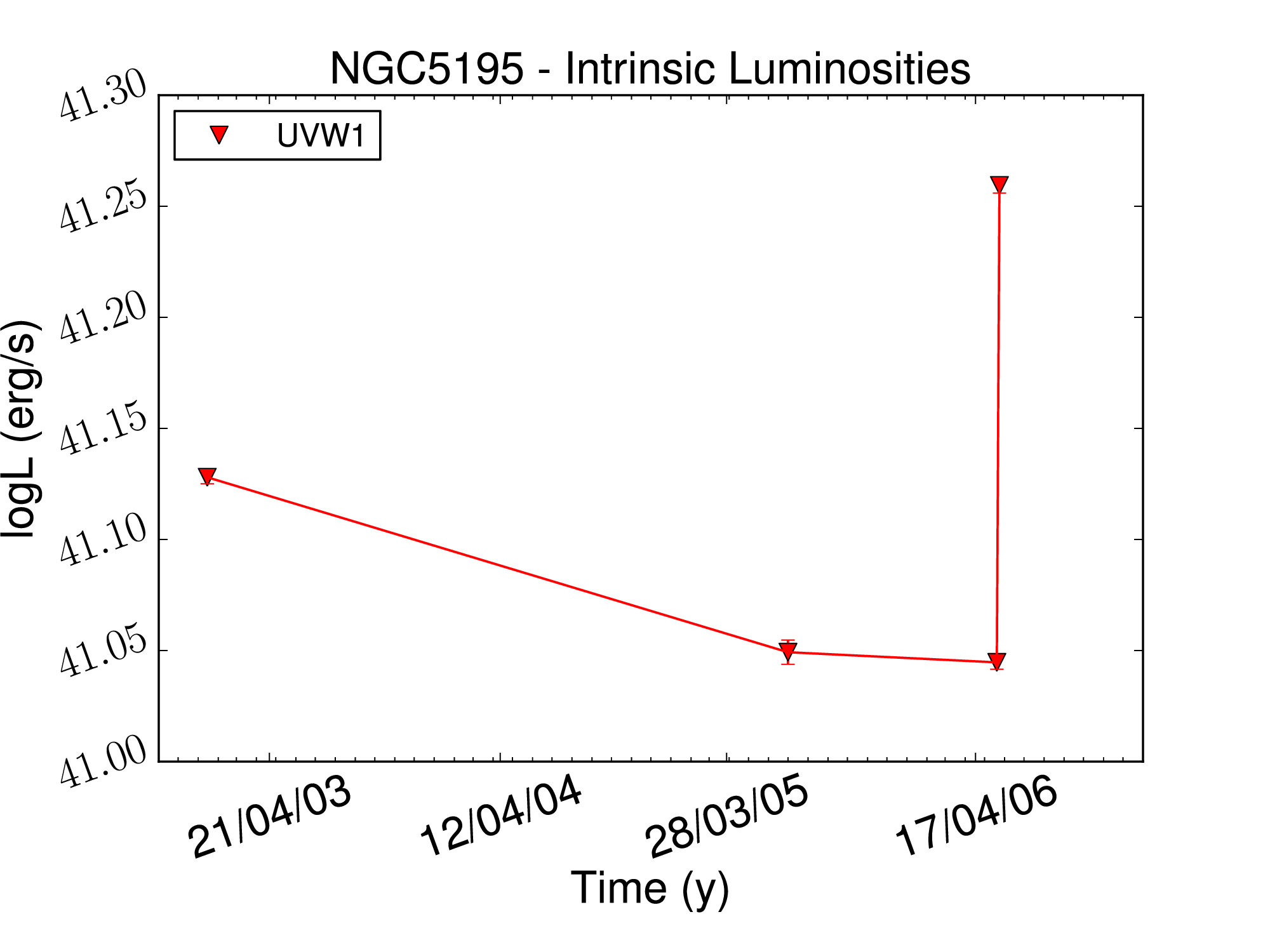}}

\subfloat{\includegraphics[width=0.30\textwidth]{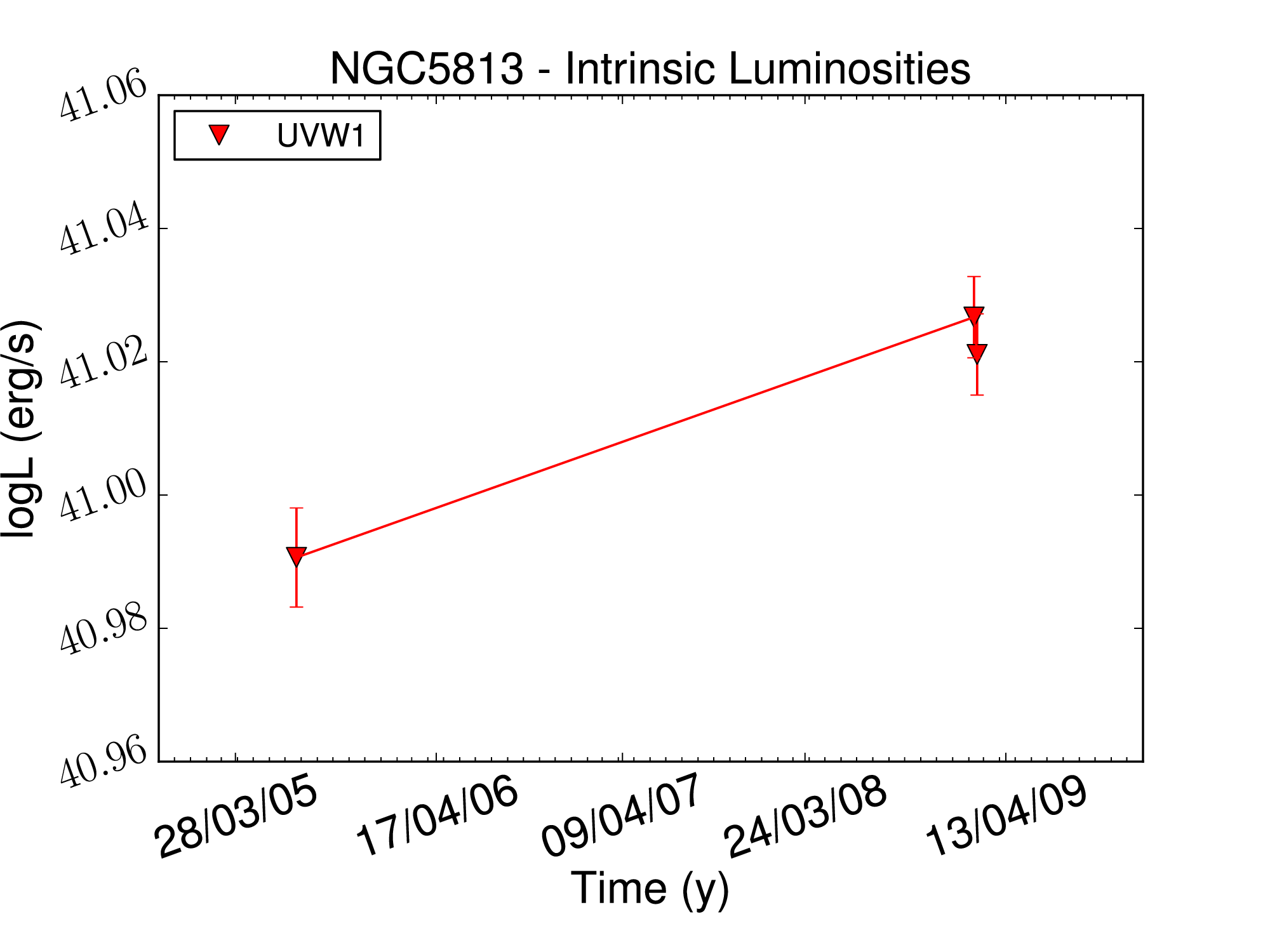}}
\subfloat{\includegraphics[width=0.30\textwidth]{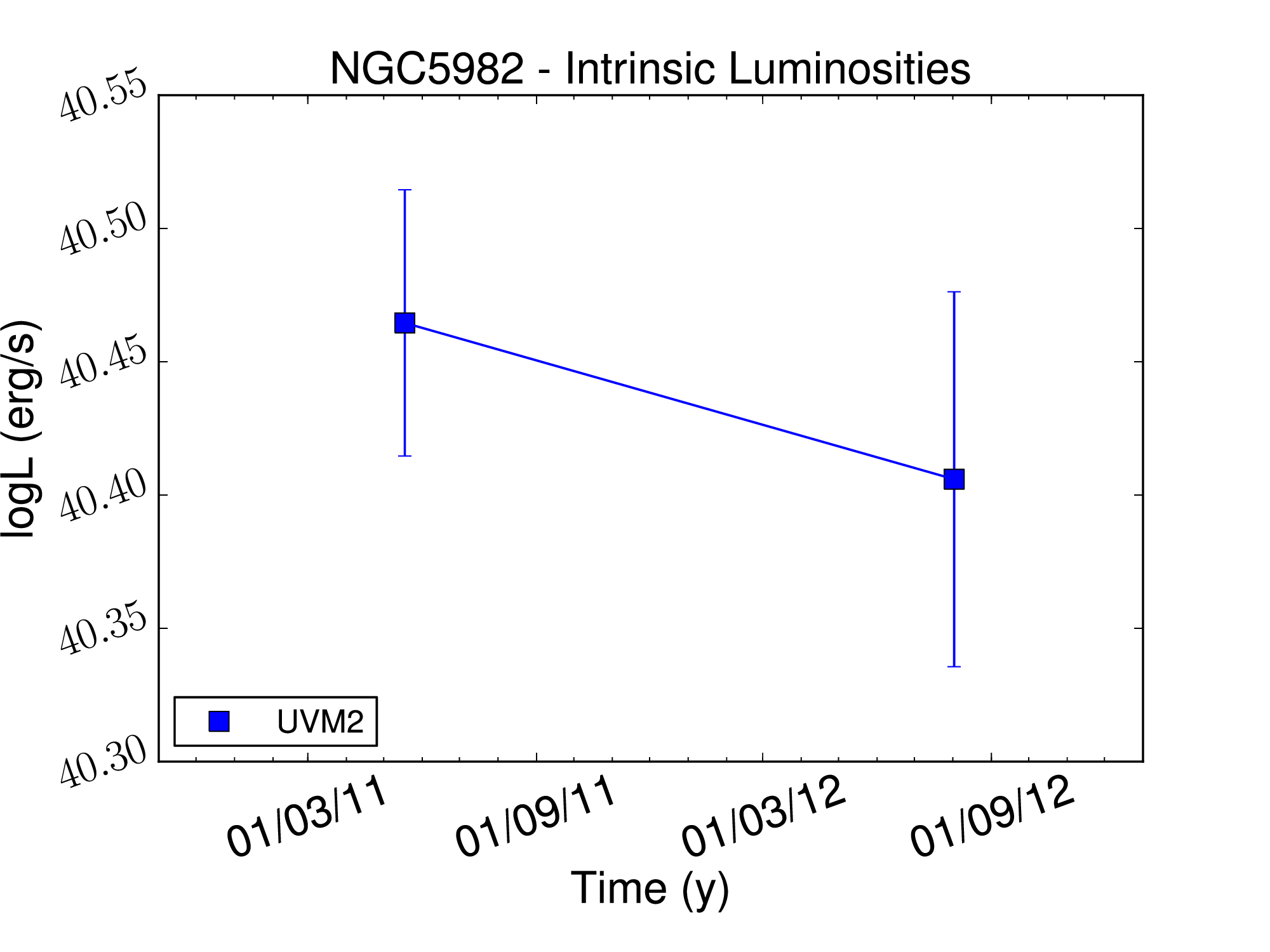}}
\caption{UV luminosities obtained from the data with the OM camera onboard \emph{XMM}--Newton, when available. Different filters have been used; UVW1 (red triangles), UVW2 (green circles), UVM2 (blue squares), U (black pentagons).}
\label{luminUVfig}
\end{figure*}

\begin{figure*}
\centering
\subfloat{\includegraphics[width=0.30\textwidth]{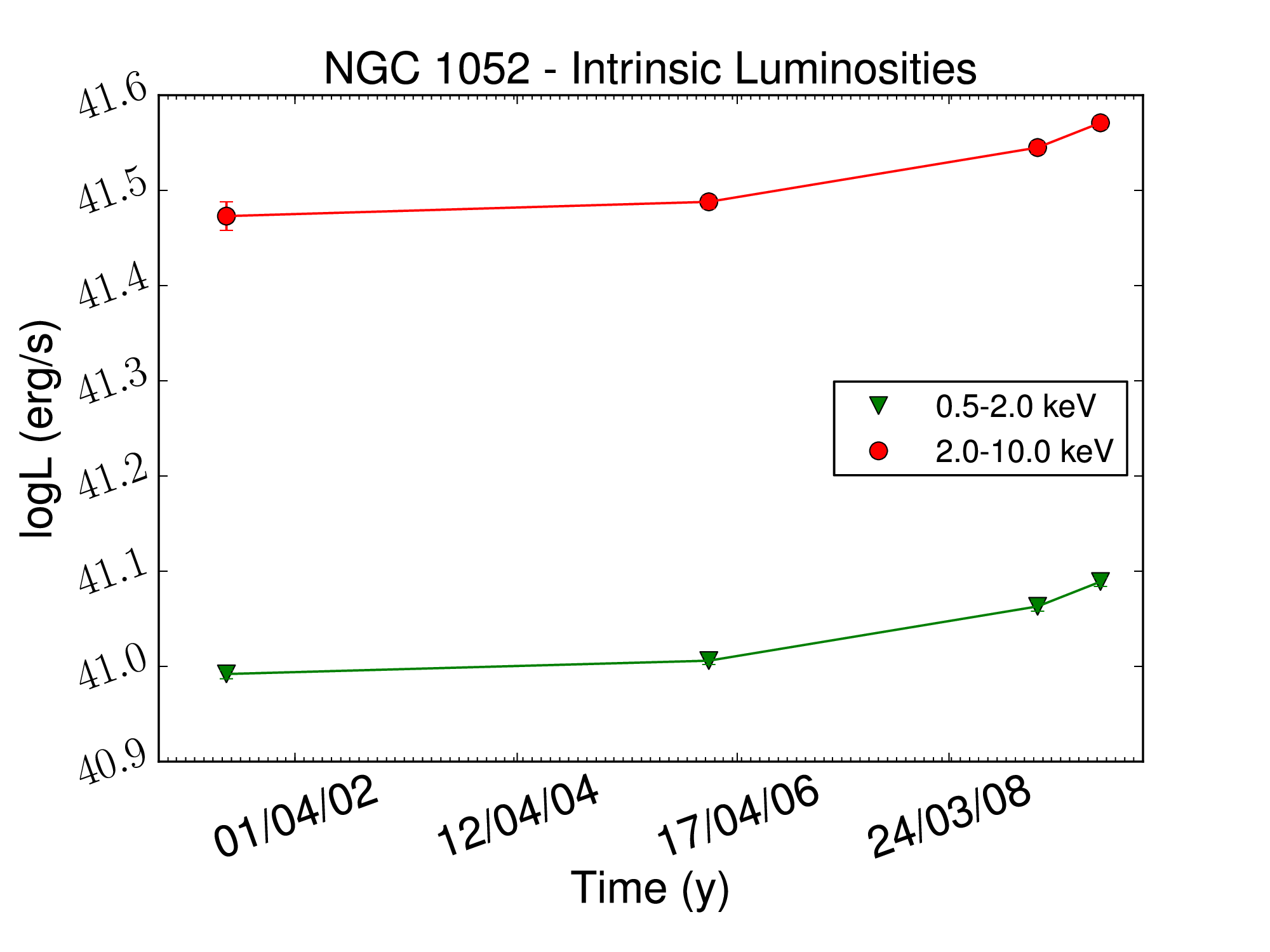}}
\subfloat{\includegraphics[width=0.30\textwidth]{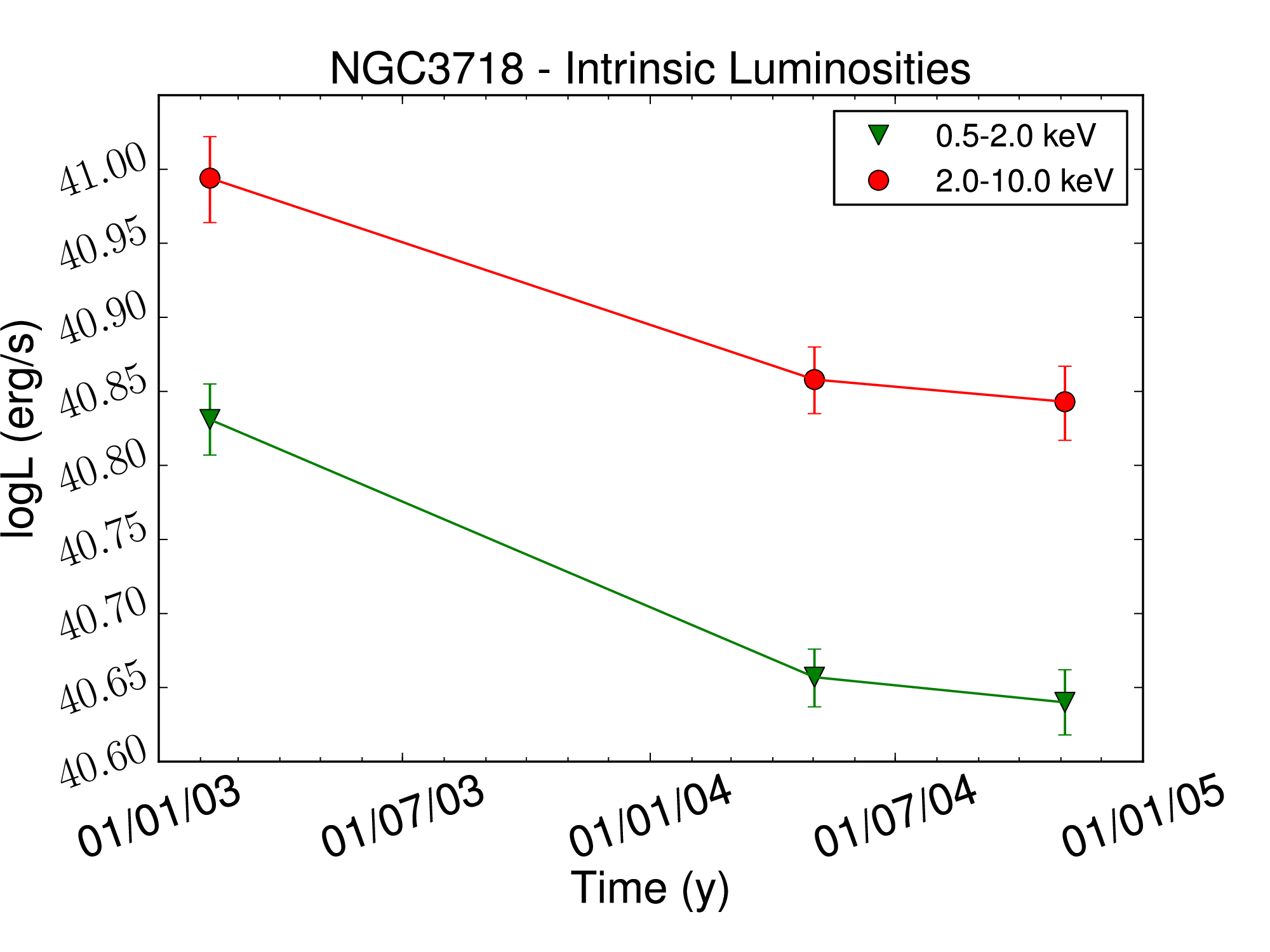}}
\subfloat{\includegraphics[width=0.30\textwidth]{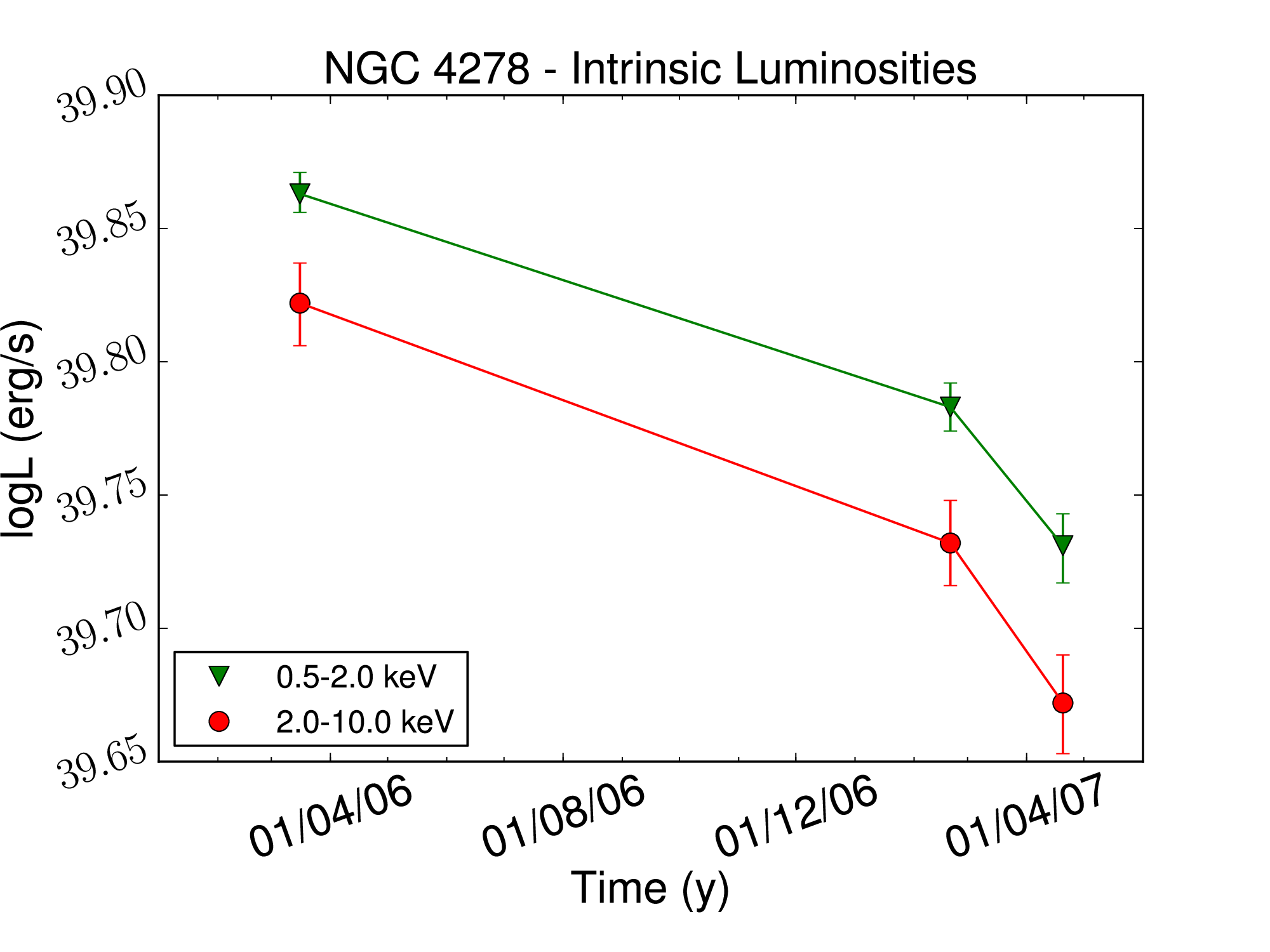}}

\subfloat{\includegraphics[width=0.30\textwidth]{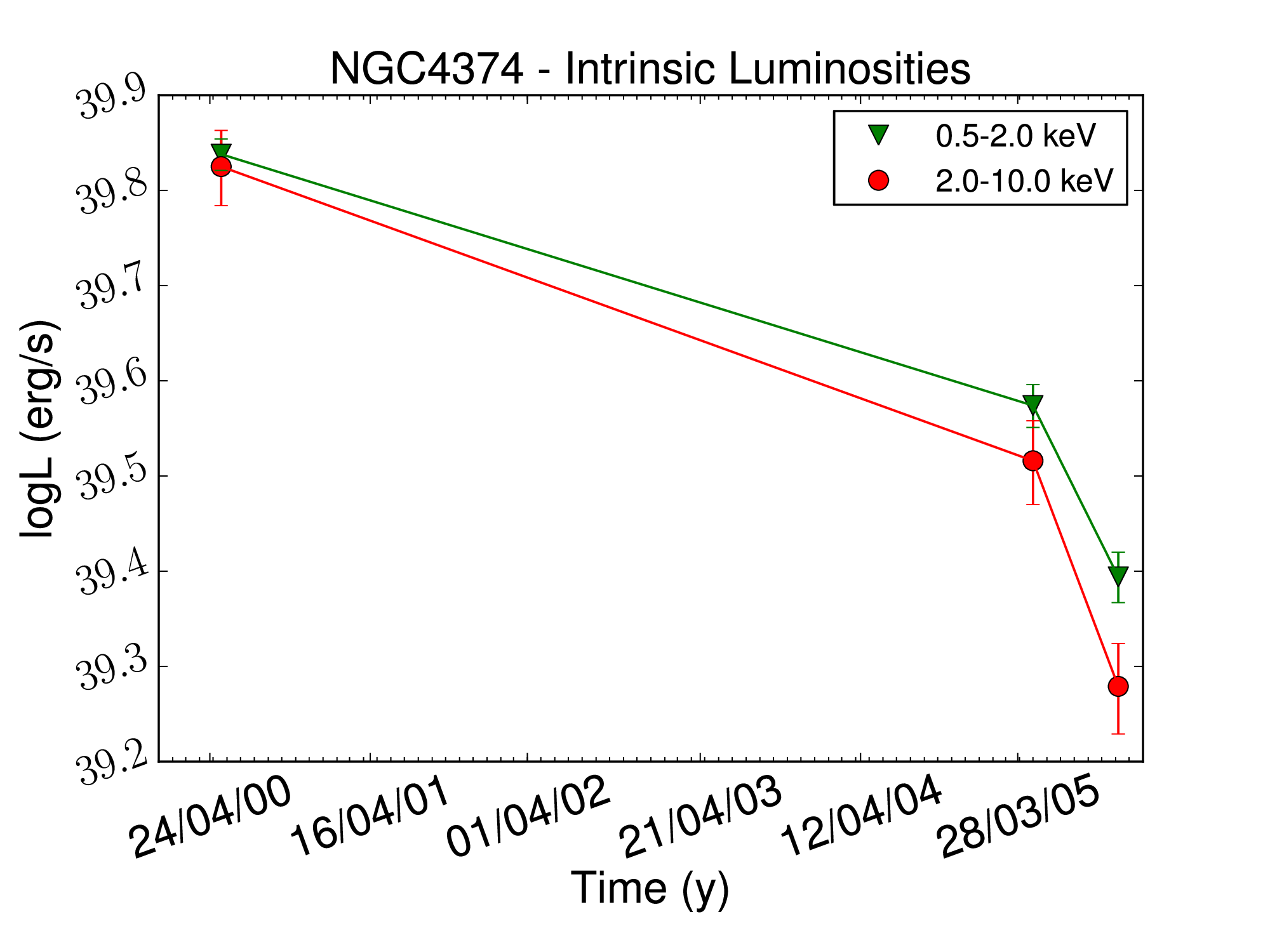}}
\subfloat{\includegraphics[width=0.30\textwidth]{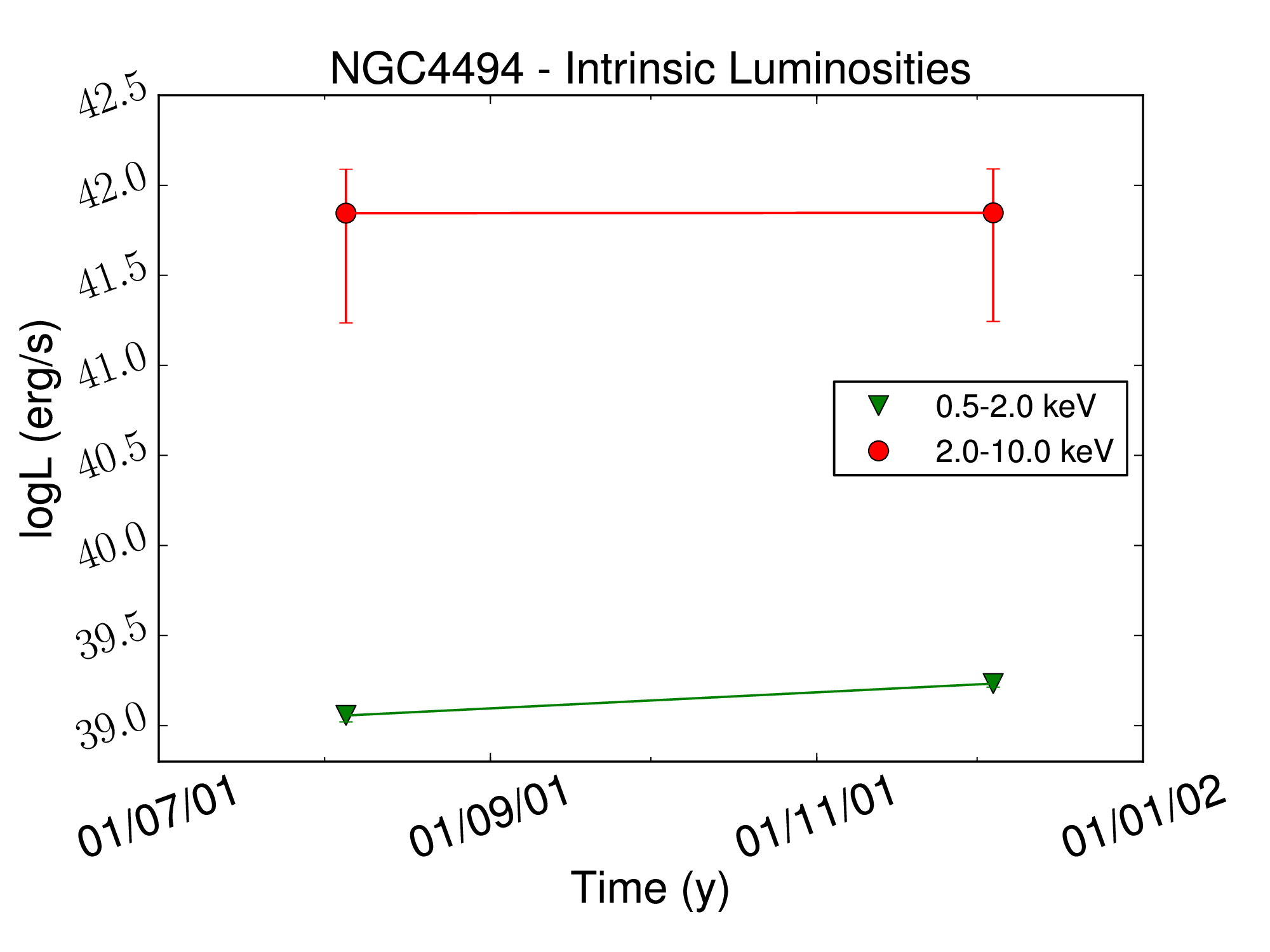}}
\subfloat{\includegraphics[width=0.30\textwidth]{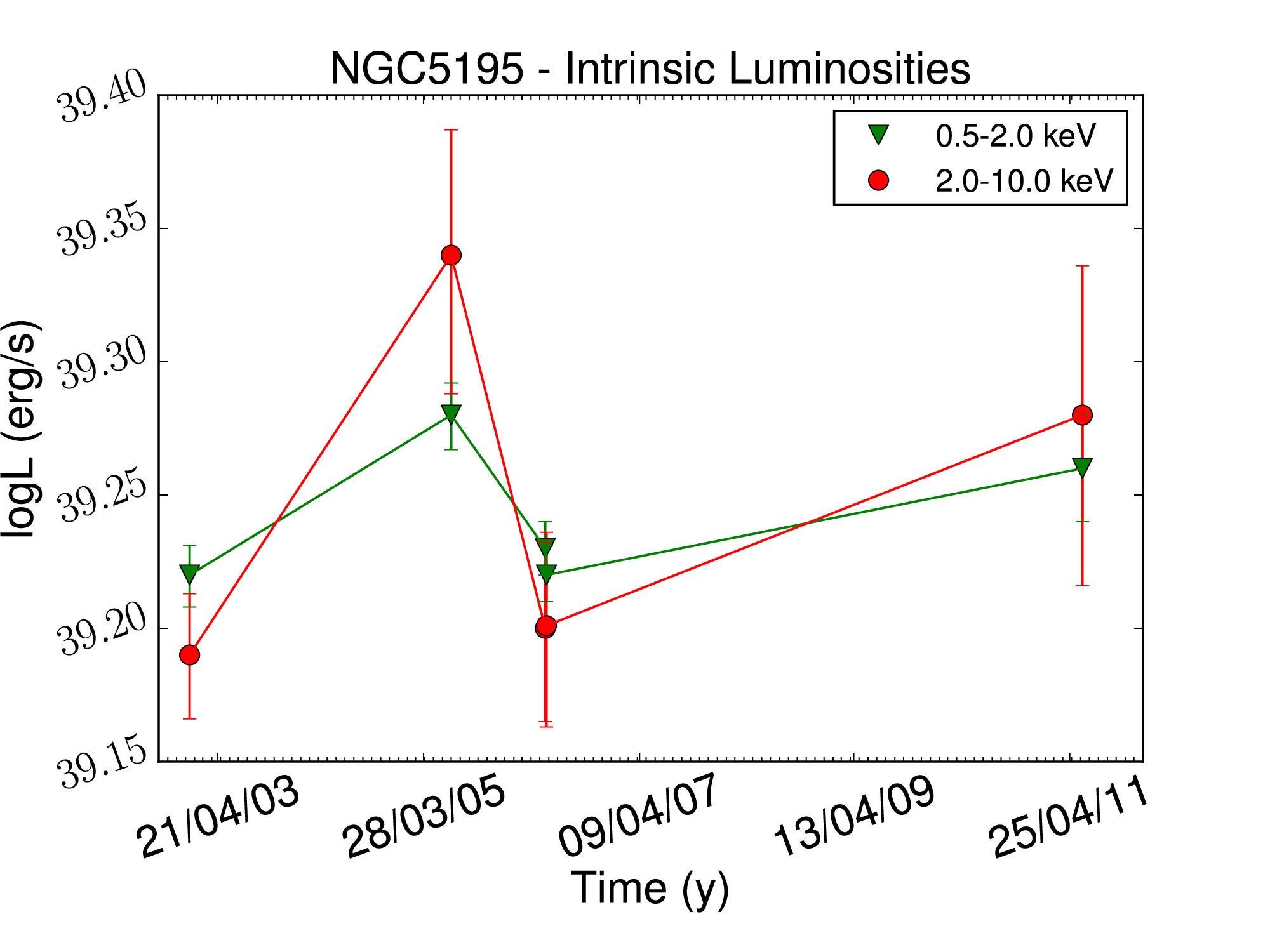}}

\subfloat{\includegraphics[width=0.30\textwidth]{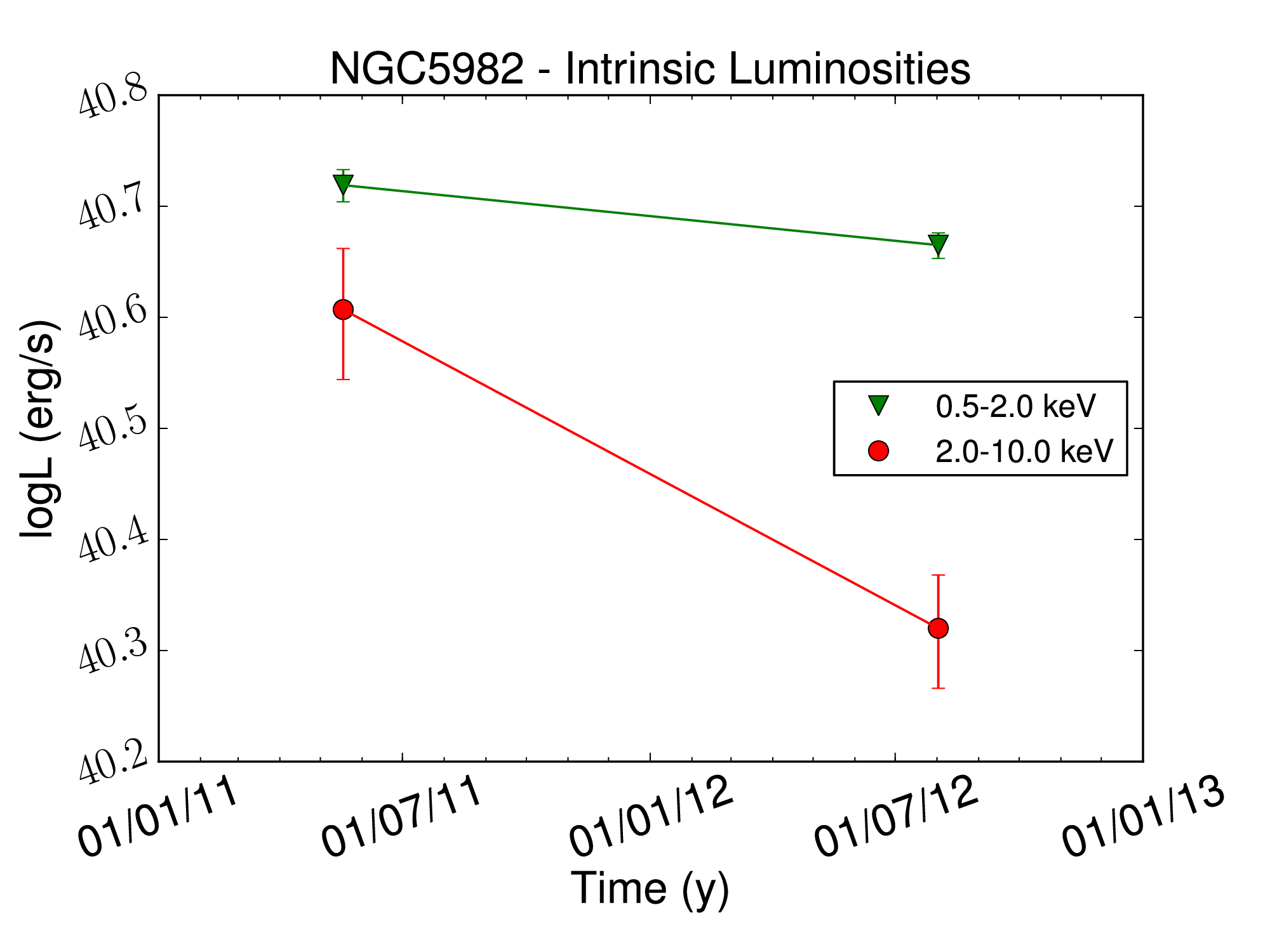}}
\caption{Intrinsic luminosities calculated for the soft (0.5-2.0 keV, green triangles) and hard (2.0-10.0 keV, red circles) energies in the simultaneous fitting, only for the variable objects.}
\label{luminXfig}
\end{figure*}

To be concise, we stand out the peculiarities of each source. 
For details on the data and results, we refer the reader to the following tables and figures: the observations used in the analysis (Table \ref{obs}), variations of the hardness ratio, HR, only compared for data from the same instrument 
(Col. 8 in Table \ref{obs}), UV luminosities when simultaneous
data from the OM monitor were available in more than one date (Col. 9 in Table \ref{obs} and Fig. \ref{luminUVfig}),
individual and simultaneous best fit and the parameters varying in the model (Table \ref{bestfit} and Fig. \ref{bestfig}), X-ray
flux variations (Table \ref{lumincorr} and Fig. \ref{luminXfig}), the analysis of the annular region when data from \emph{Chandra} and \emph{XMM}--Newton are
used together (Table \ref{annulus}), and the simultaneous fittings of these observations (Table \ref{simultanillo}), and short-term variability from the analysis of the light curves 
(Table \ref{estcurvas} and Appendix \ref{lightcurves}). When short term variations are not detected, upper limits of $\sigma_{NXS}^2$ are calculated.

\begin{itemize}
\item \underline{NGC\,315}: From the simultaneous analysis of \emph{Chandra} data, variations are not found (i.e., SMF0) in a three years period. The annular region contributes with 3\% in \emph{Chandra} data. When comparing with \emph{XMM}--Newton data, variations of the parameters do not improve the fit within the five years period. The analysis from one of the \emph{Chandra} light curves shows variations in the hard band at 1.6$\sigma$ confidence level.
\item \underline{NGC\,1052}: SMF2 is used to fit its \emph{XMM}--Newton data, with variations of $Norm_2$ (49\%) and $N_{H2}$ (31\%) over a period of eight years. Flux variations of 20\% are obtained for soft and hard energies in the same period. Since the annular region contributes with 10\% in \emph{Chandra} data, \emph{Chandra} and \emph{XMM}--Newton data are compared, without changes in one year period. Short term variations are not detected. UV variations from the UVW2 (13\%) and UVM2 (21\%) are found. 
\item \underline{NGC\,1961}: \emph{XMM}--Newton data do not show variations in one month period (i.e., SMF0). UV data are available, but the nucleus of the galaxy is not detected.
\item \underline{NGC\,2681}: The SMF0 results for \emph{Chandra} data with no improvement when vary the parameters. Thus, the object do not vary in a period of four months. Short term variations are not detected.
\item \underline{NGC\,2787}: One observation per instrument is available. When they are compared, the emission from the annular region contributes with 53\% in \emph{Chandra} data. Thus, we do not perform a simultaneous fit and we can not use this object to discuss long term variations. Short term variations are not detected.
\item \underline{NGC\,2841}: One observation per instrument is available. When they are compared, the emission from the annular region contributes with 60\% in \emph{Chandra} data. Thus, we do not perform a simultaneous fit and we can not use this object to discuss long term variations. In this case \emph{Chandra} image reveals at least three X-ray sources within the annular region (see Appendix \ref{Ximages}).
\item \underline{NGC\,3226}: Long-term X-ray variations from this source are not taken into account due to possible contamination from NGC\,3227. We refer the reader to HG13 for details. The analysis from the \emph{Chandra} light curve shows variations in the soft and total bands below 2$\sigma$ confidence level. UV variations amounts to 11\% in the UVW1 filter.
\item \underline{NGC\,3608}: SMF0 is used to fit the \emph{XMM}--Newton data, with no variations in a 12 years period.
\item \underline{NGC\,3718}: We jointly fit \emph{Chandra} and \emph{XMM}--Newton data since emission from the annular region is negligible. The best representation of the data needs $Norm_2$ vary (37\%), i.e., SMF1 is used. This implies a change in luminosity of 35\% (29\%) at soft (hard) energy in one years period. In the UV data, the nucleus of the galaxy is not detected.
\item \underline{NGC\,4261}: The simultaneous fit without allowing to vary any parameter (i.e., SMF0) results in a good fit both in \emph{Chandra} and \emph{XMM}--Newton data over a period of eight and six years, respectively.
Simultaneous fit of \emph{Chandra} and \emph{XMM}--Newton (the annular region contributes 37\% in \emph{Chandra} data) do not show changes.
Short term variations are not detected.
Considering the UV range, variations amount to 9\% in the UVW1 filter and 34\% in the UVM2 filter.
\item \underline{NGC\,4278}: The best fit for \emph{Chandra} data is SMF1 with $Norm_2$ varying (30\%) in a one year period. An X-ray intrinsic luminosity variation at soft (hard) energy of 26\% (29\%) is found. The contribution of the annular region in \emph{Chandra} data amounts to 38\%. When comparing \emph{XMM}--Newton and \emph{Chandra} data, a variation in the normalization of the PL (15\%) along two years is found. Short term variations are not detected.
\item \underline{NGC\,4374}: SMF1 is used for the simultaneous fit with \emph{Chandra} data, with variations of $Norm_2$ (73\%) in a period of five years. Flux variation of  64\% (71\%) in the soft (hard) band during the same period are found. Data from different instruments are not compared because the annular region contributes 84\% in \emph{Chandra} data.
Short term variations in the soft and total bands are found from one \emph{Chandra} observation below 2$\sigma$ confidence level.
\item \underline{NGC\,4494}: The simultaneous fit is jointly performed for \emph{Chandra} and \emph{XMM}--Newton data (the contribution of the annular region is 21\% in \emph{Chandra} data) up to 4.5 keV, because \emph{Chandra} data show a low count-rate at harder energies. We use SMF1, obtaining the best representation of the data set when $Norm$ varies (33\%). Flux variation of 31\% (35\%) is obtained for the soft (hard) energy in four months period.
\item \underline{NGC\,4636}: SMF0 is used to fit \emph{XMM}--Newton data, i.e., variations are not found over six months period. Note that $\chi^2_r \sim 2$. Unfortunately, none of the proposed models are good enough to improve the final fit. 
From the analysis of the light curves, short term variations in the hard band are obtained from one \emph{XMM}--Newton observation at 1.4$\sigma$ confidence level. In the UV, variations of 28\% are obtained from the UVW1 filter.
\item \underline{NGC\,4736}: Variations are not found from \emph{XMM}--Newton data, i.e., SMF0 is used in a period of four years. \emph{Chandra} and \emph{XMM}--Newton data are not compared since the emission from the annular region contributes with 84\% in \emph{Chandra} data.
From the analysis of the light curves, variations in the soft, hard, and total bands are obtained below 2$\sigma$ confidence level in all cases. At UV frequencies, a variation of 66\% is obtained from the U filter.
\item \underline{NGC\,5195}: SMF1 is used for \emph{XMM}--Newton data, where the best representation let $Norm_2$ vary (20\%). A flux variation of 9\% (19\%) in the soft (hard) energy band is found in a period of eight years. 
In the case of \emph{Chandra} data, no variations are found from the simultaneous fit (i.e., SMF0) in a three days period. The annular region contributes with 74\% in \emph{Chandra} data, so the data from the two instruments are not compared.
The analysis of one \emph{Chandra} light curve reveals short term variations in the soft and total bands below 2$\sigma$ confidence level. At UV frequencies, variation of 16\% are found with the UVW1 filter.
\item \underline{NGC\,5813}: For \emph{Chandra} data we make the simultaneous analysis up to 4 keV because of the low count rate at harder energies. For both \emph{Chandra} and \emph{XMM}--Newton data SMF0 is used, with no improvement of the fit when varying the parameters in a period of six and four years, respectively. Note that $\chi^2_r \sim 2$ in \emph{XMM}--Newton data. Unfortunately, none of the proposed models are good enough to improve the final fit. The data from the two instruments are not compared since the annular region contributes with 100\% in \emph{Chandra} data. Short term variations are not detected.
In the UV, OM observations with the UVW1 filter are used, which show variations of 8\% in a period of four years. 
\item \underline{NGC\,5982}: SMF1 is used to fit \emph{XMM}--Newton data, with the best representation being when varying $Norm_2$ (50\%). Flux variations of 11\% (49\%) in the soft (hard) band are obtained in a period of one year. UV variations are not found.
\end{itemize}

\onecolumn
\tiny
\begin{table*}
\begin{center}
\caption{\label{variab} Results of the variability analysis. } 
\begin{tabular}{lcccccccccc} \hline \hline
Name & Type & log ($L_{soft}$) & log ($L_{hard}$) & log ($M_{BH}$) & log ($R_{Edd}$) & & Variability & & T & HR  \\ \cline{7-9} 
 & & (0.5-2 keV) & (2-10 keV) & & & SMF0 &  SMF1 & SMF2 & (Years) & \\ 
(1) & (2) & (3)          & (4) & (5) & (6) & (7) & (8) & (9) & (10) & (11) \\ \hline
NGC\,315 (C) & AGN & 41.38 & 41.58 & 8.65 & -3.67 & MEPL & - & - & 3 & \\
			 & L1.9 & 0\%  & 0\%   &      &       &      &   &   &   & 8\% \\
NGC\,1052 (X) & AGN       & 41.04  & 41.52  & 8.07 &  -3.14   &  ME2PL & $Norm_2$  & $N_{H2}$  &   8 &   \\
 & L1.9 & 20\% & 20\% & & & & 49\% & 31\% & & 33\%  \\
NGC\,1961 (X) & AGN & 41.20 & 41.23 & 8.67 & -4.03 & ME2PL & - & - & 0.08 &  \\

			  & L2 & 0\%  &  0\%  &      &       &       &   &   &      & 0\% \\
NGC\,2681* (C) & AGN & 39.02 & 38.93 & 7.07 & -4.73 & MEPL & - & - & 0.4 & \\
		  & L1.9 & 0\%  & 0\%   &      &       &      &   &   &     & 4\% \\
NGC\,3608* (X) & Non-AGN & 40.32 & 40.24 & 8.06 & -4.41 & ME2PL & - & - & 12 & \\
			 & L2/S2: & 0\%    & 0\%   &      &        &      &   &   &    & 4\% \\
NGC\,3718 (C+X) & AGN & 40.76 & 40.99 & 7.85 & -3.60 & 2PL & $Norm_2$ & - & 1 & \\
				& L1.9 & 35\% & 29\% &       &       &     & 37\%     &   &   & 14\% \\
NGC\,4261 (X) & AGN       & 40.98  & 41.02  & 8.96 &  -4.54   &  ME2PL &  - &  - & 8 &   \\
 & L2 & 0\% & 0\% & & & & & & &  0\%  \\
\hspace*{1.0cm}  (C) & &  & &  &     &  ME2PL &  - &  - & 6 &  \\
 &  & 0\% & 0\% & & & & & & &  19\%  \\
NGC\,4278 (C) & AGN       & 39.80  & 39.75  & 8.46 &  -5.30  &  MEPL    &  $Norm_2$ & - & 1  &  \\
 & L1.9 & 26\% & 29\% & & & & 30\% & & & 4\%  \\
NGC\,4374* (C) & AGN & 39.64 & 39.59 & 8.74 & -5.79 & MEPL & $Norm_2$ & - & 5 & \\
              & L2 & 64\%   & 71\%  &      &        &      & 73\%    &   &   & 12\% \\
NGC\,4494 (X,C) & AGN & 39.13 & 39.37 & 7.64 & -4.84 & PL & $Norm$ & - & 0.3 & \\
                & L2:: & 31\% & 35\% &        &       &    & 33\% &     &    & \\
NGC\,4636* (X) & Non-AGN & 40.86 & 39.81 & 8.16 & -5.00 & MEPL & - & - & 0.5 & \\
              & L1.9 & 0\%       & 0\% &       &       &      &  &       &     & 14\%  \\
NGC\,4736 (X) & AGN & 39.61 & 39.73 & 6.98 & -3.84 & MEPL & - & - & 4 & \\
			  & L2 & 0\%    & 0\%   &      &       &      &   &   &   & 5\% \\
NGC\,5195 (X) & AGN & 39.24 & 39.25 & 7.59 & -4.86 & MEPL & $Norm_2$ & - & 8 & \\
			  & L2: &  9\%  & 19\%   &      &       &       & 20\%    &   &   & 3\% \\
\hspace*{1.0cm} (C) & & 38.56 & 38.61  & & & MEPL & - & - & 0 &  \\
              &     &    0\% & 0\% & & & & & & & 2\% \\
	  
NGC\,5813* (X) & Non-AGN & 41.33 & 40.30 & 8.42 & -4.72 & MEPL & - & - & 4 & \\ 
			  & L2:           & 0\% & 0\%  &      &       &      &   &   &   & 1\% \\
\hspace*{1.0cm} (C) &         & 39.68 & 39.07 & 8.42 & -5.95 & MEPL & - & - & 6 & \\
			  &            & 0\% & 0\%  &      &       &      &   &   &   & 5\% \\
NGC\,5982 (X) & AGN & 40.70 & 40.49 & 8.44 & -4.71 & MEPL & $Norm_2$ & - & 1 & \\
			  & L2:: & 11\% & 49\% &     &       &      & 50\%     &   &   & 10\%  \\ 

\hline
\end{tabular}
\caption*{{\bf Notes.} (Col. 1) Name (the asterisks represent \emph{Compton}--thick objects), and the instrument (C: \emph{Chandra} and/or X: \emph{XMM}--Newton) in parenthesis, (Col. 2) X-ray and optical types, (Col. 3 and 4) logarithm of the soft (0.5-2 keV) and hard (2-10 keV) X-ray luminosities, where the mean was calculated for objects showing variability, and percentages in flux variations, (Col. 5) black-hole mass in logarithmical scale, determined using the correlation between stellar velocity dispersion (from HyperLeda) and black hole mass \citep{tremaine2002}, (Col. 6) Eddington ratio, $L_{bol}/L_{Edd}$, calculated from \cite{eracleous2010b} using $L_{bol}=33L_{2-10 keV}$, (Col. 7) best fit for SMF0, (Col. 8) parameter varying in SMF1, with the percentage of variation, (Col. 9) parameter varying in SMF2, with the percentage of variation, (Col. 10) the sampling timescale, and (Col. 11) variations in the hardness ratios.}
\end{center}
\end{table*} 
\normalsize
\twocolumn

\subsection{\label{spectral}Long term X-ray spectral variability}

A first approximation to the spectral variations can be estimated from the hardness ratios (HR). 
Following the results in HG13, an object can be considered as variable when HR varies more than 20\%. 
One out of the 14 objects in our sample is variable according to this criterium using the HR measurements with the same instrument (namely NGC\,1052). 
Since we are mainly doubling the sample number, we conclude that the result obtained in HG13 is a consecuence of low number statistics. However, no clear relation can be invoked between variable objects and a minimum in HR variations (see Table \ref{variab}).

\emph{Chandra} and \emph{XMM}--Newton data are available together for the same object in 12 cases. We recall that comparison of the data from the two instruments is done only when the emission from the annular region with $r_{ext}$ = $r_{XMM}$ and $r_{int}$ = $r_{Chandra}$ contributes less than 50\% in \emph{Chandra} data with the $r_{XMM}$ aperture.
In the case of NGC\,3718 the extranuclear contamination is null, so the simultaneous analysis is performed without any prior analysis of the extended emission.
In six cases a simultaneous fit with data from the two instruments is not conducted. 
In five objects (NGC\,315, NGC\,1052, NGC\,4261, NGC\,4278, and NGC\,4494) the extranuclear contamination is taken into account for the simultaneous fit following the methodology described in Sect. \ref{simult}.

None of the three non-AGN candidates show variations
(one type 1 and two type 2).
Seven out of the 12 AGN candidates (three out of five type 1, and four out of seven type 2) show spectral variations. 
We find no variations in the spectral index, $\Gamma$, in any of the objects in the sample.
In all cases $Norm_2$ is responsible for these variations (between 20-73\%). In one case (NGC\,1052, type 1) variation in $N_{H2}$ (31\%) is required along with variations in $Norm_2$. 
These variations are found irrespectively of the LINER type (see Table \ref{variab}).

\subsection{\label{flux}UV and X-ray long term flux variability}

Since variations in $Norm_2$ naturally imply changes in the flux, at X-ray frequencies all the objects showing spectral variations also show flux variability (see Sect. \ref{spectral}). This means that none of the non-AGN candidates show flux 
variations, whereas seven out of the 12 AGN candidates do.  
Variations from 9 to 64\% (19 to 71\%) are obtained in the soft (hard) band (see Table \ref{variab}).
Soft and hard X-ray luminosities are listed in Table \ref{lumincorr}, and 
presented in Fig. \ref{luminXfig} for objects with flux variations.

In eight out of the 18 cases, data at UV frequencies are provided by the OM onboard \emph{XMM}--Newton at different epochs (simultaneously with X-ray data). Two of them are non-AGN candidates (the type 1 NGC\,4636 and the type 2 NGC\,5813). Both show variations in the UVW1 filter, while NGC\,4636 does not vary in the UVW2 filter.  
Five out of the six AGN candidates show UV variability in at least one filter (except NGC\,5982, type 2). 
Two variable objects are type 1 
(NGC\,1052 varies about 20\% in the UVM2 and UVW2 filters, and NGC\,3226\footnote{We recall that NGC\,3226 varies at UV frequencies, but long term variations in X-rays were rejected for the analysis.} shows 11\% variation in the UVW1 filter), and three are type 2
(NGC\,4261 shows 10\% (33\%) variations in the UVW1 (UVM2) filter, NGC\,4736 varies 66\% in the U filter, and NGC\,5195 varies 51\% in the UVW1
filter). In summary, three out of four type 1 and the two type 2 AGN candidates are variable objects at UV frequencies.
Their UV luminosities are presented in Table \ref{obs} and Fig. \ref{luminUVfig}.

Comparing X-ray and UV flux variations, two non-AGN candidates show UV variations but do not show X-ray variations (one type 1 and one type 2). From the AGN candidates, three show X-ray and UV flux variations (two type 1 and one type 2), and three type 2s show variations only in one of the frequencies (two in the UV, and one in X-rays).

Taking into account UV and/or X-ray variations, ten out of 13 AGN candidates are variable (four out of six type 1, and six out of seven type 2).
We note that the three objects showing no variations at X-ray frequencies (NGC\,315, NGC\,2681, and NGC\,1961) do not have UV data in more than one epoch.

\subsection{\label{lightcurve}Short term variability}

According to the values of $\sigma_{NXS}^2$, four objects show positive values within the errors in the soft and total bands, one type 1 (NGC\,3226), and three type 2 (NGC\,4374, NGC\,4736, NGC\,5195).
We obtain $\sigma_{NXS}^2$ values above zero for three objects in the hard band, two type 1 (NGC\,315, and NGC\,4636), and one type 2 (NGC\,4736). 
However, all the measurements are consistent with zero at 2$\sigma$ level. 
For the remaining light curves we estimate upper limits for the normalised excess variance.
Therefore, we can not confirm short term variability in our sample.
The light curves are in Appendix \ref{lightcurves}, and their statistics in Table \ref{estcurvas}.

%
%______________________________________________________________

\section{\label{discusion}Discussion}

\subsection{Long term variations}

We analysed three non-AGN candidates (one type 1 and two type 2), and none of them show X-ray spectral nor flux variations. An additional source, NGC\,5846 (type 2), was studied in HG13, that does not show variations either. All of them are classified as CT candidates by \cite{omaira2009b}. In these objects, the nuclear obscuration is such that X-ray emission can not be directly observed, i.e., the view of their nuclear emission is suppressed below $\sim$ 10 keV \citep{maiolino1998}. If this is the case, spectral variations might not be detected, in accordance with our results. Indications to classify objects as an AGN candidate can be found at other frequencies, for example with radio data. 
At these frequencies, a compact, flat spectrum nuclear source can be considered as AGN signature \citep{nagar2002,nagar2005}.
\cite{omaira2009a} collected multiwavelenght properties of 82 LINERs. In their sample, 18 objects are classified in X-rays as non-AGN candidates and have nuclear radio cores detected. From these, 14 are classified as CT candidates. Thus, it might be possible that the AGN in CT objects are not seen at X-ray frequencies (and therefore its X-ray classification is non-AGN), whereas at radio frequencies the AGN can be detected.
In the four non-AGN candidates studied in HG13 and this work, radio cores are detected in three objects and, in fact, evidence of jet structures are reported in the literature (NGC\,4636, \citealt{giacintucci2011}; NGC\,5813, \citealt{randall2011}; NGC\,5846, \citealt{filho2004}), suggesting their AGN nature.
Moreover, in this work UV variability is found for NGC\,4636 and NGC\,5813, whereas UV data is not available to study long term variations in the other two cases.
A nuclear counterpart was not detected for NGC\,3608 with \emph{VLA} by \cite{nagar2005}.
Therefore, the X-ray variable nature of CT objects can not be dismissed, but variability analyses at higher energies should be performed.

Among the 12 AGN candidates in our sample, two objects are proposed to be CT candidates \cite[NGC\,2681 and NGC\,4374,][]{omaira2009b}. 
In these objects a point-like source at hard energies is detected, what might indicate that part of the AGN continuum is still contributing below 10 keV. 
Hence, variations in the nuclear continuum may be observed, as is the case of NGC\,4374.
An example of a confirmed CT type 2 Seyfert that shows spectral variations with \emph{XMM}--Newton data is Mrk 3 \citep{guainazzi2012}.
%The classification of CT in \cite{omaira2009b} is based in three indicators. The first condition is $\Gamma < 1.2$, that their data agreed within uncertainities. However, we have found $\Gamma > 1.2$ for NGC\,4374 for all the observations (see Table \ref{bestfit}). The second criterium was $log(F_x(2--10 keV)/F([OIII])) < 0.5$, whereas NGC\,4374 showed a value of 0.44 and was classified as CT?. Recently, \cite{ford2013} found a small amount of star formation in this galaxy, so we have estimated the $L_{[OIII]}$ based on the star formation rate using the relation in \cite{moustakas2006}. We obtained a value two orders of magnitude lower than that estimated by \cite{ho1997} and, therefore, we cannot conclude that star formation is contributing to the $L_{[OIII]}$. The third indicator is the equivalent width of the iron line, that was compatible with being CT. We cannot obtain this measure since the line is not observed in the X-ray spectrum. Then, we cannot exclude NGC\,4374 from its CT classification, but we have found that at least one criterium contradicts it.

Only in one case (NGC\,1052) variations in $N_{H2}$ are needed along with those in $Norm_2$ (see below). 
 Variations in the column density have been extensively observed in type 1 Seyferts
(e.g., NGC\,1365, \citealt{risaliti2007}; NGC\,4151, \citealt{puccetti2007}; Mrk\,766, \citealt{risaliti2011}; Swift J2127.4+5654, \citealt{sanfrutos2013}). 
%However, an essential difference between Seyfert galaxies and NGC\,1052 is the presence of a jet in the latter \citep{kadler2004a}. 
 \cite{brenneman2009} studied a 101 ksec observation of NGC\,1052 from \emph{Suzaku} data and did not find short-term variations.
%Recently, \cite{connolly2014} studied the type 1.8 Seyfert NGC\,1365 with \emph{Swift} data and found an anticorrelation between the absorbing column and the normalisation of the power %law. They interpreted such behavior as an X-ray wind of absorbing material getting away from the accretion disc. We plotted the same relation for the parameters in the simultaneous fit of %NGC\,1052 in Fig. \ref{NGC1052normnh}, where this behavior can be appreciated. However, since only four observations have been studied we cannot confirm this result. 
%Note also that such a trend is not observed in any of the sources in our sample, even if we consider the individual observations. 
%This implies that NGC\,1052 probably behaves differently from the other objects in the sample. 
NGC\,1052 also shows variations at UV frequencies, as shown by \cite{maoz2005} and we confirm in this work. 
However, the LINER nature of this source has been discussed in the literature;
\cite{pogge2000} studied 14 LINERs with \emph{HST} data and only NGC\,1052 shows clear evidence for an ionization cone, analogous to those seen in Seyferts. 
From a study where artificial neural networks (ANN) were used to classify X-ray spectra, NGC\,1052 seems to be associated to type 1 Seyfert galaxies in X-rays \citep{omaira2014}.
The fact that the observed variations in NGC\,1052 are similar to those seen in type 1 Seyfert galaxies agrees well with this galaxy resembling Seyferts
at X-ray frequencies.

Spectral variations do not necessarily imply flux variations. For example, if variations in the column density, $N_H$, were found alone, flux variations would not be present. However, all the results reported in the literature for LINERs show spectral variations related to flux variability.
Variations in the normalisation of the power law, i.e. $Norm_2$, are found in all the variable sources in our sample.
Variations of other components, as the soft emission (NGC\,4102, \citealt{omaira2011b}; NGC\,4552, HG13) or the slope of the power law (NGC\,7213, \citealt{emmanoulopoulos2012}) are reported in the literature.
The variations in $\Gamma$ found by \cite{emmanoulopoulos2012} are small and were obtained on average every two days from 2006 to 2009. On the contrary, the observations we use in the present analysis were obtained with separations of months, and therefore it might be that if these variations occur in LINERs we are not able to detect them.
The most natural explanation for the variations in $Norm_2$ is that the AGN continuum is changing with time. 
The first conclusion derived from this result is that the X-ray emission in these variable LINERs is AGN-like.
Moreover, these kind of variations are common in other AGN \citep[e.g.,][]{turner1997}. Thus, even if the sample is not large enough to be conclusive, from the point of view of the X-ray variability, LINERs are similar to more powerful AGN. This is confirmed by the characteristic timescales derived from our analysis (see Sect. \ref{variabtimes}).

Our results show that UV and X-ray variations are not simultaneous (see Sect. \ref{flux}). This means that some X-ray variable sources are not UV variable, and vice versa. The most illustrative case is NGC\,5195, that changes 39\% at UV frequencies but do not vary in X-rays in the same period (see Tables \ref{obs} and \ref{lumincorr}, and Figures \ref{luminUVfig} and \ref{luminXfig}).
The most accepted scenario assumes that the X-ray emission is produced by a disc-corona system, where UV photons from the inner parts of the accretion disc are thermally Comptonized and scattered into the X-rays by a hot corona surrounding the accretion disc \citep{haardt1991}.  
In this case we expect that X-ray and UV emissions reach us at different times, due to the time that light takes to travel from one place to another.
These time lags will depend on the sizes of the BH, the disc and the corona, so that at larger sizes, the greater the time lags we expect.
For example, \cite{degenaar2014} conducted a multi-wavelenght study of the X-ray binary (XRB) Swift J1910.2-0546 and find time lags between X-ray and UV frequencies of $\sim$ 8 days. They argue that the changes may be related to the accretion morphology, perhaps due to a jet or a hot flow. 
LINERs have larger sizes than XRBs and, therefore, larger time lags are expected. Thus, the mismatch between UV and X-ray variabilities might be due to these time lags. 
Simultaneous X-ray and UV studies monitoring the sources would be useful to measure these time lags, and also to calculate the sizes of the variable regions.

\subsection{\label{variabtimes}Variability timescales}

\begin{figure}
\centering
\subfloat{\includegraphics[width=0.52\textwidth]{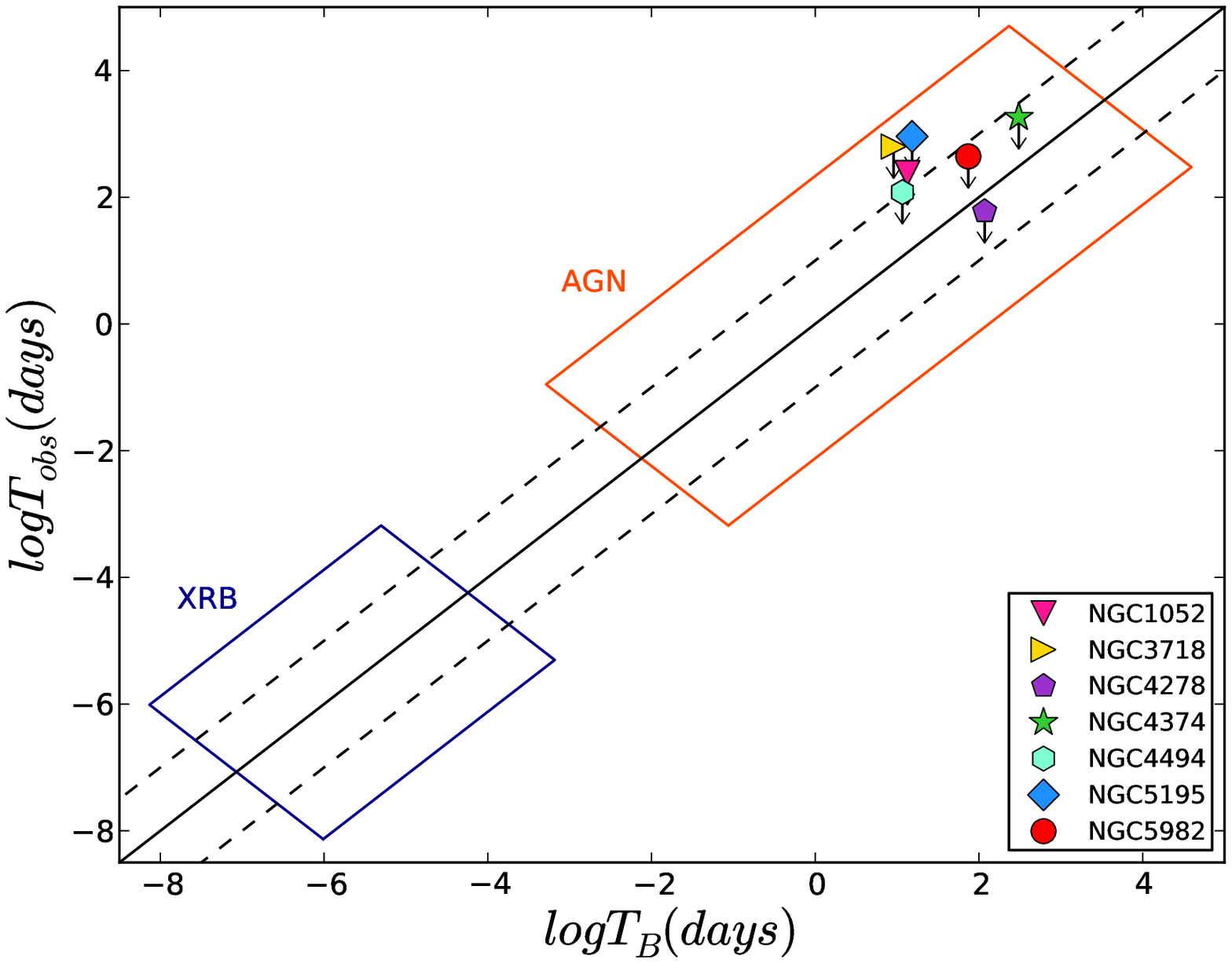}}
\caption{Observed variability timescale, $T_{obs}$, against the predicted value, $T_B$, from \cite{omairavaughan2012}. The solid line represents the 1:1 relationship, and the dashed lines the errors. Only variable objects are represented. The big orange rectangle represents the location of AGN, and the small blue rectangle the location of XRB as in \cite{mchardy2006}.}
\label{tiempos}
\end{figure}

Short variability timescales ($\sim$ few days) are found in the literature.
\cite{pian2010} and \cite{younes2011} from their analysis of type 1 LINERs find short term variations in two objects, i.e., four in total. Two objects are common with our sample (NGC\,3226 and NGC\,4278). However, in HG13 we did not find short term variations, nor in these two nor in the other objects in the sample.
\cite{omairavaughan2012} reported two out of 14 variable LINERs, one of them common with those reported by \cite{pian2010}.
In the present paper, we study short term variability from the analysis of the light curves for a total of 12 objects in three energy bands (soft, hard, total). Six objects show $\sigma_{NXS}^2 > 0$ in at least one of the three bands. However, we note that these variations are below 2$\sigma$ confidence level. None of the objects show a value consistent with being greater than zero above 3$\sigma$, and therefore we can not confirm short term variations. 

\cite{mchardy2006} reported a relation between the bend timescale for variations (i.e., the predicted timescale, $T_B$), black hole mass and bolometric luminosity.
This $T_B$ corresponds to a characteristic frequency, $\nu_B$, of the power spectral density (PSD), happening when the spectral index of the power law bends from $\sim 1$ to $ \sim 2$.
 \cite{omairavaughan2012} updated this relation as:

\begin{equation}
\label{equation}
log(T_B) = Alog(M_{BH}) + Blog(L_{bol}) + C
\end{equation}

\noindent where $A=1.34 ^+_- 0.36$, $B=-0.24 ^+_- 0.28$, $C=-1.88 ^+_-0.36$, and $T_B$, $M_{BH}$, and $L_{bol}$ are in units of $days$, $10^6 M_{\odot}$, and $10^{44} erg/s$, respectively.
Using Eq. \ref{equation} we plot the observed timescales of the variability, $T_{obs}$, against the predicted timescales, $T_B$, for the sources showing variations in our sample (Fig. \ref{tiempos}). The observed timescales are computed from the shortest periods in which variations are observed, and are represented as upper limits. 
It is important to note that the timescales between the observations probably differs from the predicted timescales. This is obvious since the observations were obtained randomly at different epochs. All the variable objects are compatible with the 1:1 relation (represented by a solid line, and dashed lines are the errors), although most of them have larger $T_{obs}$ than predicted. In Fig. \ref{tiempos}, we also plot the location of AGN (big orange rectangle) and X-ray binaries (XRB, small blue rectangle) as in \cite{mchardy2006}. It can be observed that LINERs are located in the upper part of the relation together with the most massive AGN, due to the strong dependence of $T_B$ with the $M_{BH}$. Note here that while $T_B$ represents the bending frequency of the PSD, $T_{obs}$ is a direct measure of changes in the spectral shape. This implies that the variability timescales are often shorter than the timescales between the observations (except for NGC\,4278) for our sample. All the variable objects are then consistent with the relation reported by \cite{omairavaughan2012}.

On the other hand, five objects in our sample do not show variations (and are not represented in Fig. \ref{tiempos}).
For these (NGC\,315, NGC\,1961, NGC\,2681, NGC\,4261, and NGC\,4494), we obtain $T_{B}$ ($T_{obs}$) $\sim$ 77 (873), 100 (14), 3 (92), 273 (2830), and 12 (120) days, respectively.
In the case of NGC\,1961, $T_B > T_{obs}$, so variations between the observations are not expected. Under Eq. \ref{equation}, we would expect variations from the other sources.
It could be possible that we do not detect variations because observations were taken at random. However, it could also be possible that these objects do not follow Eq. \ref{equation}.

Our results are consistent with the scaling relation found by \cite{mchardy2006} and \cite{omairavaughan2012} because, according to the BH mass and accretion rates of LINERs, variations of the intrinsic continuum are expected to be of large scales. Therefore, LINERs would follow the same relation than other AGN and XRBs (see Fig. \ref{tiempos}). We recall that $T_B$ has a strong dependence on $M_{BH}$, while the dependence with $L_{bol}$ (and with the accretion rate) is much lower \citep{mchardy2009}, and hence it prevents us to obtain useful information related to accretion physics.

Despite LINERs and more powerful AGN are located in the same plane, different authors have pointed out that the accretion mechanism in LINERs could be different from that happening in more powerful AGN \citep[e.g.,][]{gucao2009,younes2011}. When a source accretes at a very low Eddington rate ($R_{Edd} < 10^{-3}$), the accretion is dominated by radiatively inefficient accretion flows \citep[RIAF,][]{narayanyi1994,quataert2004}. 
Such flows are though to be present in XRB, which are closer accreting black holes that can be easily studied. It is well known that XRB show different X-ray emission states which are differenciated by their spectral properties  \citep[e.g.,][]{remillard2006}.
In comparison to XRB, LINERs should be in the ``low/hard" state or, if the Eddington ratio is too low, in the``quiescent" state, while more powerful AGN should be in the ``high/soft" state. 
%In XRBs the hard state is characterized by a hard PL component ($\Gamma \sim 1.7$), the accretion disc is faint and cool, and there is the presence of a quasi-steady radio jet \citep{remillard2006}. 

%These states can also be differenciated by their PSD. The ``high/soft" state is characterized by a broken power-law, where the break occur when the slope passes from $\alpha \sim$ 2 to 1. In the ``low/hard" state, the PSD is more complex and shows at least two breaks in the power-law. The first one occurs at high frequencies, passing from $\alpha \sim$ 2 to 1, and the additional break when passing from $\alpha \sim$ 1 to 0 at lower frequencies. The main problem to characterize AGN in the hard state is that the frequency coverage is not sufficient to distinguish it from the soft state.

%When \cite{mchardy2006} first proposed the relation presented in Eq. \ref{equation}, the timescale associated to the break from $\alpha \sim$ 2 to 1 was used as a characteristic timescale, which is included in both states. Therefore, it is valid for objects in the ``soft'' and ``hard'' states, avoiding us to obtain useful information related to accretion physics.

An anticorrelation between the slope of the power law, $\Gamma$, and the Eddington ratio, $R_{Edd}$, is expected from RIAF models. \cite{qiao2013a} theoretically investigated this correlation for XRB and found that advection dominated accretion flow (ADAF)\footnote{The RIAF model is an updated version of the ADAF model.} models can reproduce it well. In these models the X-ray emission is produced by Comptonization of the synchrotron and bremsstrahlung photons. Later, they studied low luminosity AGN (LLAGN) in the framework of a disc evaporation model (inner ADAF plus an outer truncated accretion disc) and found that it can also reproduce the anticorrelation \citep{qiao2013b}. 
%\cite{xu2011} argued in favour of RIAF models as the accretion mechanism in LLAGN, where the reason of the change between luminous and low luminous AGN may be due to a higher contribution of the outer thin disc plus hot corona from the ADAF in LLAGN.

\begin{figure}[h]
\centering
\subfloat{\includegraphics[width=0.5\textwidth]{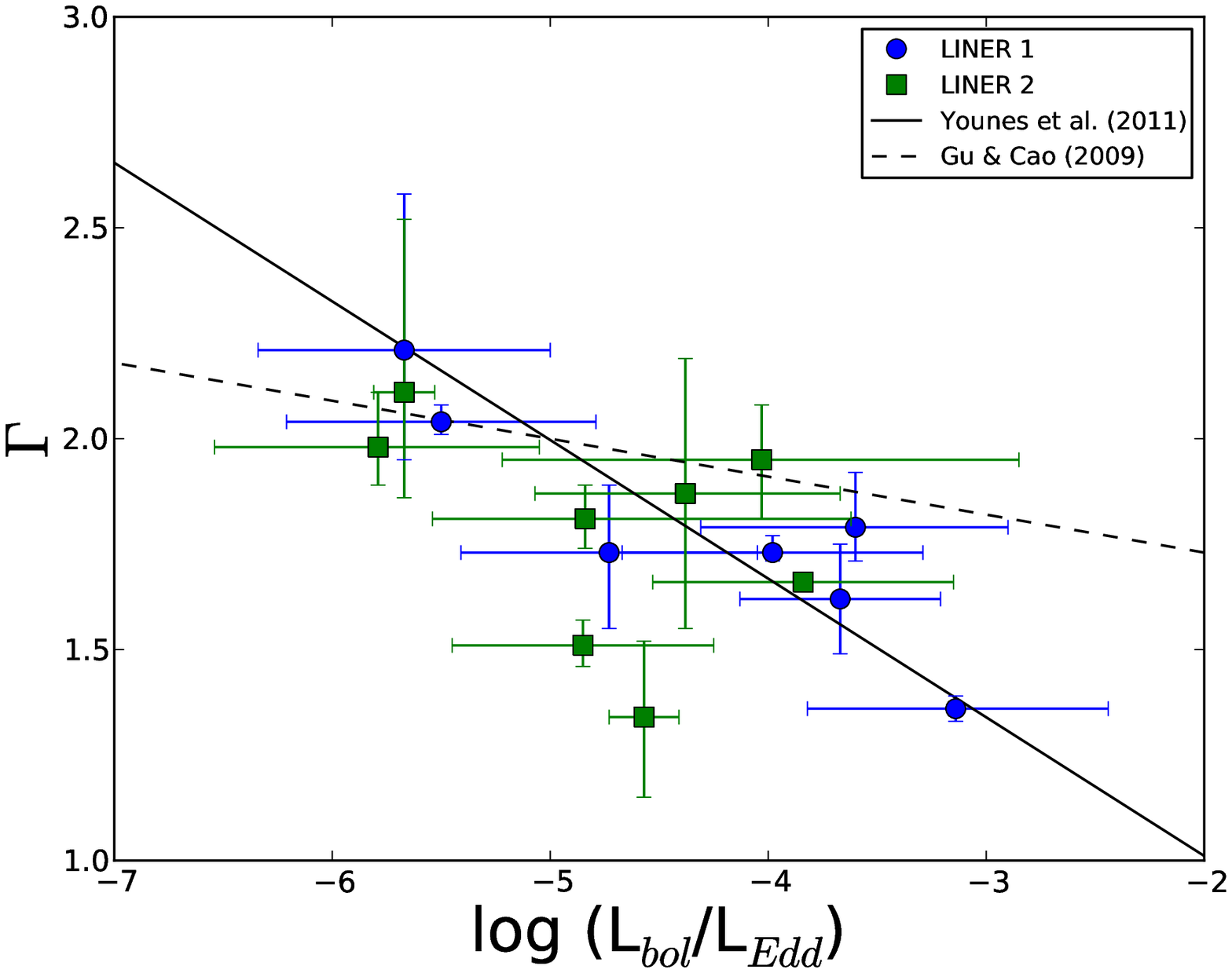}}
\caption{Spectral index, $\Gamma$, versus the Eddington ratio, $R_{Edd}=log(L_{bol}/L_{Edd})$. Type 1 (blue circles) and type 2 (green squares) LINERs are distinguished. The solid and dashed lines represent the relations given by \cite{younes2011} and \cite{gucao2009}, respectively, shifted to the same bolometric correction (see text).}
\label{gammaRedd}
\end{figure}

Some efforts have been done to observationally investigate that relationship for LLAGN. \cite{gucao2009} used a sample of 55 LLAGN (including 27 LINERs and 28 Seyferts) and found a regular anticorrelation between $\Gamma$ and $R_{Edd}$. However, when LINERs were considered alone, they did not find a strong correlation. Later, \cite{younes2011} studied a sample of type 1 LINERs and found a statistically significant anticorrelation. In HG13 we found that the seven LINERs studied in the sample fitted well in the relation given by \cite{younes2011}.
Here we plot the same relation in Fig. \ref{gammaRedd}, where the values of $\Gamma$ and $L_{2-10 keV}$ were obtained from the simultaneous fittings, since $\Gamma$ did not vary in any of the objects (see Table \ref{variab}). In the cases of NGC\,2787 and NGC\,2841 the individual fit from \emph{Chandra} data was used because the simultaneous fit can not be performed (see Sect. \ref{ind}). $R_{Edd}$ were calculated following the formulation given in \cite{eracleous2010}, assuming $L_{bol}=33L_{2-10 keV}$. The solid and dashed lines in Fig. \ref{gammaRedd} are the relations given by \cite{younes2011} and \cite{gucao2009}, respectively, corrected to our $L_{bol}$. 
The coefficient of the Pearson correlation is r=-0.66, with a coefficient of determination of p=0.008. This might suggest that RIAF models apply to LINERs, indicating an inefficient accretion disc, in contrast to the efficient accretion discs found for more powerful AGN.
A larger sample of LINERs would be useful to be conclusive.

\subsection{Type1/Type 2 variable AGN candidates}

Some studies at X-ray frequencies have shown the variable nature of LINERs. 
Type 1 LINERs were studied by \cite{pian2010}, \cite{younes2011}, \cite{emmanoulopoulos2012}, and HG13. \cite{pian2010} studied four sources with \emph{Swift} and find variations in two of them. \cite{younes2011} detected long term variations in seven out of the nine sources in their sample from \emph{XMM}--Newton and \emph{Chandra} data. \cite{emmanoulopoulos2012} studied one object with Rossi X-ray Timing Explorer (RXTE) and find a `harder when brighter' behaviour. In HG13 we analysed three AGN candidates using \emph{Chandra} and/or \emph{XMM}--Newton data and find variations in all the three.

In the present analysis our sample contains five type 1 AGN candidates, three of them consistent with being variable. From the sample by \cite{younes2011}, five objects are in common with our sample, four of them with similar results and only differing in the case of NGC\,315 (see Appendix \ref{indivnotes} for notes and comparisons).

Type 2 objects were studied by \cite{omaira2011b} and HG13. 
\cite{omaira2011b} used \emph{Suzaku}, \emph{Chandra} and \emph{Swift} data to study one object and find variations in the thermal component. In HG13 we reported long term variations in one of the two AGN candidates. 

In the present analysis our sample contains seven type 2 AGN candidates, four of them consistent with being variable.

Taking into account the present analysis and the studies listed above from the literature, 14 out of 22 LINERs are X-ray variable objects (eight out of 13 type 1, and six out of nine type 2). Therefore, there is no significant difference in the proportion of variable objects in X-rays in terms of the optical type1/type2 classification. Given that the observed variations are intrinsic to the sources, similar proportions were expected in view of the UM, in good agreement with our results.

Similar behaviour is found at UV frequencies. \cite{maoz2005} were the first authors to show that UV variability is common in LINERs, and to demostrate the presence of a nonstellar component at these frequencies. From their sample of 17 LINERs, only three do not show neither short-term ($<$1 yr) nor long-term ($>$1 yr) variability. From these, all the seven type 1s and seven out of 10 type 2 objects show variations.

When data from the OM onboard \emph{XMM}--Newton are available, we searched for UV variability. Taking into account only AGN candidates, five out of six objects show variations (two type 1 and three type 2). This gives strenght to the variable nature of LINERs.  
In common to our sample, \cite{maoz2005} already showed the variable nature of NGC\,1052 and NGC\,4736. 
Thus, taking into account the present analysis and the study by \cite{maoz2005}, 17 out of 21 LINERs are variable at UV frequencies (the eight type 1, and nine out of 13 type 2).
As previously noted by \cite{maoz2005}, the fact that type 2 LINERs show variations at UV frequencies suggests that the UM may not always apply to LINERs. 
It has been suggested that the broad line region (BLR) and the torus, responsible for obscuring the continuum that is visible in type 1 AGN, dissapear at low luminosities \citep{elitzur2006,elitzur2009}. The dissapearence of the torus could in principle explain why type 2 LINERs do vary in the UV, because the naked AGN is directly seen at these frequencies.

%\begin{equation} \label{elitzur}
%L < 5 \times 10^{39} \Big(\frac{M_{BH}}{M_{\odot}}\Big)^{2/3} erg \hspace*{0.1cm} s^{-1}
%\end{equation} 

We find that some objects show variations in X-rays but do not vary at UV frequencies, or vice versa. This means that the percentage of variable objects is higher if we take into account different frequencies. From all the 34 studied LINERs at UV and X-ray frequencies in the present and other works, 27 LINERs show variations in at least one energy band (13 out of 16 type 1, and 14 out of 18 type 2). Thus, the percentage of type 1 and 2 variable objects is similar.
Consequently, variability is very common in LINER nuclei, a property that share with other AGN.

\section{\label{conclusion}Conclusions}

Using \emph{Chandra} and \emph{XMM}--Newton public archives we performed a spectral and flux, short and long term variability analysis of 18 LINERs in the Palomar sample. The main results of this study can be summarized as follows:

\begin{enumerate}
\item Seven out of 12 AGN candidate LINERs show long term spectral variability, while the three non-AGN candidates do not. In two cases the simultaneous fit was not possible due to strong external contamination, and in one case the long term analysis was rejected due to possible contamination of a companion galaxy.
\item No significant difference in the proportion of X-ray variable nuclei (type 1 or 2) are found.
\item The main driver of the spectral variations is the change in the normalization of the power law, $Norm_2$; only for NGC\,1052 this is accompained by variations in the column density, $N_{H2}$.
\item UV variations are found in five out of six AGN candidates. The two non-AGN candidates also show variations.
\item Short term variations are not found.

\end{enumerate}

From X-ray and UV data, we find that ten out of 13 LINERs in our sample show evidence of long term variability in at least one energy band. Hence, variability is very common in LINERs.

We find that X-ray variations are due to changes in the continuum of the AGN. These results agree well with the expected variations according to their BH masses and accretion rates.  In this sense, LINERs would be in the same plane than more powerful AGN and XRB. However, we find an anticorrelation between the slope of the power law, $\Gamma$, and the Eddington ratio, that may suggest a different accretion mechanism in LINERs, more similar to the hard state of XRB.

On the other hand, the result that some type 2 LINERs eventually vary at UV frequencies may suggest that a naked AGN can be observed at these wavelenghts, what could be explained in the scenario where the torus dissapears at low luminosities.

\begin{acknowledgements}

          We thank the anonymous referee for his/her helpful comments that helped to improve the paper, and the AGN group at the IAA for helpful comments during this work. This work was financed by MINECO grant AYA 2010-15169, Junta de Andaluc\'{i}a TIC114 and Proyecto de Excelencia de la Junta de Andaluc\'{i}a P08-TIC-03531. LHG acknowledges financial support from the Ministerio de Econom\'{i}a y Competitividad through the Spanish grant FPI BES-2011-043319. OGM thanks Spanish MINECO through a Juan de la Cierva Fellowship. This research made use of data obtained from the \emph{Chandra} Data Archive provided by the \emph{Chandra} X-ray Center (CXC). This research made use of data obtained from the \emph{XMM}-Newton Data Archive provided by the \emph{XMM}-Newton Science Archive (XSA). This research made use of the NASA/IPAC extragalactic database (NED), which is operated by the Jet Propulsion Laboratory under contract with the National Aeronautics and Space Administration. We acknowledge the usage of the HyperLeda database (http://leda.univ-lyon1.fr).

\end{acknowledgements}

\bibliographystyle{aa}
\bibliography{000referencias}

\begin{thebibliography}{90}
\expandafter\ifx\csname natexlab\endcsname\relax\def\natexlab#1{#1}\fi

\bibitem[{{Antonucci}(1993)}]{antonucci1993}
{Antonucci}, R. 1993, \araa, 31, 473

\bibitem[{{Birkinshaw} \& {Davies}(1985)}]{birkinshawdavies1985}
{Birkinshaw}, M. \& {Davies}, R.~L. 1985, \apj, 291, 32

\bibitem[{{Brassington} {et~al.}(2009){Brassington}, {Fabbiano}, {Kim},
  {Zezas}, {Zepf}, {Kundu}, {Angelini}, {Davies}, {Gallagher}, {Kalogera},
  {Fragos}, {King}, {Pellegrini}, \& {Trinchieri}}]{brassington2009}
{Brassington}, N.~J., {Fabbiano}, G., {Kim}, D.-W., {et~al.} 2009, \apjs, 181,
  605

\bibitem[{{Brenneman} {et~al.}(2009){Brenneman}, {Weaver}, {Kadler}, {Tueller},
  {Marscher}, {Ros}, {Zensus}, {Kovalev}, {Aller}, {Aller}, {Irwin}, {Kerp}, \&
  {Kaufmann}}]{brenneman2009}
{Brenneman}, L.~W., {Weaver}, K.~A., {Kadler}, M., {et~al.} 2009, \apj, 698,
  528

\bibitem[{{Brightman} \& {Nandra}(2011)}]{brightman2011}
{Brightman}, M. \& {Nandra}, K. 2011, \mnras, 413, 1206

\bibitem[{{Cappellari} {et~al.}(1999){Cappellari}, {Renzini}, {Greggio}, {di
  Serego Alighieri}, {Buson}, {Burstein}, \& {Bertola}}]{cappellari1999}
{Cappellari}, M., {Renzini}, A., {Greggio}, L., {et~al.} 1999, \apj, 519, 117

\bibitem[{{Cardullo} {et~al.}(2008){Cardullo}, {Corsini}, {Beifiori},
  {Pizzella}, \& {Buson}}]{cardullo2008}
{Cardullo}, A., {Corsini}, E.~M., {Beifiori}, A., {Pizzella}, A., \& {Buson},
  L.~M. 2008, in Astronomical Society of the Pacific Conference Series, Vol.
  396, Formation and Evolution of Galaxy Disks, ed. {J.~G.~Funes \&
  E.~M.~Corsini}, 53

\bibitem[{{Carollo} {et~al.}(1997){Carollo}, {Franx}, {Illingworth}, \&
  {Forbes}}]{carollo1997}
{Carollo}, C.~M., {Franx}, M., {Illingworth}, G.~D., \& {Forbes}, D.~A. 1997,
  \apj, 481, 710

\bibitem[{{de Vaucouleurs}(1975)}]{deVaucouleurs1975}
{de Vaucouleurs}, G. 1975, Social Studies of Science, 9, 557

\bibitem[{{Degenaar} {et~al.}(2014){Degenaar}, {Maitra}, {Cackett}, {Reynolds},
  {Miller}, {Reis}, {King}, {Gultekin}, {Bailyn}, {Buxton}, {MacDonald},
  {Fabian}, {Fox}, \& {Rykoff}}]{degenaar2014}
{Degenaar}, N., {Maitra}, D., {Cackett}, E.~M., {et~al.} 2014, ArXiv e-prints

\bibitem[{{Dickey} \& {Lockman}(1990)}]{dickeylockman1990}
{Dickey}, J.~M. \& {Lockman}, F.~J. 1990, \araa, 28, 215

\bibitem[{{D'Onofrio} {et~al.}(2012){D'Onofrio}, {Marziani}, \&
  {Sulentic}}]{donofriomarzianisulentic2012}
{D'Onofrio}, M., {Marziani}, P., \& {Sulentic}, J.~W. 2012, {Fifty Years of
  Quasars: Current Impressions and Future Perspectives}, ed. M.~{D'Onofrio},
  P.~{Marziani}, \& J.~W. {Sulentic}, 549

\bibitem[{{Dudik} {et~al.}(2005){Dudik}, {Satyapal}, {Gliozzi}, \&
  {Sambruna}}]{dudik2005}
{Dudik}, R.~P., {Satyapal}, S., {Gliozzi}, M., \& {Sambruna}, R.~M. 2005, \apj,
  620, 113

\bibitem[{{Elitzur} \& {Ho}(2009)}]{elitzur2009}
{Elitzur}, M. \& {Ho}, L.~C. 2009, \apjl, 701, L91

\bibitem[{{Elitzur} \& {Shlosman}(2006)}]{elitzur2006}
{Elitzur}, M. \& {Shlosman}, I. 2006, \apjl, 648, L101

\bibitem[{{Emmanoulopoulos} {et~al.}(2012){Emmanoulopoulos}, {Papadakis},
  {McHardy}, {Ar{\'e}valo}, {Calvelo}, \& {Uttley}}]{emmanoulopoulos2012}
{Emmanoulopoulos}, D., {Papadakis}, I.~E., {McHardy}, I.~M., {et~al.} 2012,
  \mnras, 424, 1327

\bibitem[{{Eracleous} {et~al.}(2010{\natexlab{a}}){Eracleous}, {Hwang}, \&
  {Flohic}}]{eracleous2010b}
{Eracleous}, M., {Hwang}, J.~A., \& {Flohic}, H.~M.~L.~G. 2010{\natexlab{a}},
  \apj, 711, 796

\bibitem[{{Eracleous} {et~al.}(2010{\natexlab{b}}){Eracleous}, {Hwang}, \&
  {Flohic}}]{eracleous2010}
{Eracleous}, M., {Hwang}, J.~A., \& {Flohic}, H.~M.~L.~G. 2010{\natexlab{b}},
  \apjs, 187, 135

\bibitem[{{Eracleous} {et~al.}(2002){Eracleous}, {Shields}, {Chartas}, \&
  {Moran}}]{eracleous2002}
{Eracleous}, M., {Shields}, J.~C., {Chartas}, G., \& {Moran}, E.~C. 2002, \apj,
  565, 108

\bibitem[{{Ferrarese} {et~al.}(1996){Ferrarese}, {Ford}, \&
  {Jaffe}}]{ferrarese1996}
{Ferrarese}, L., {Ford}, H.~C., \& {Jaffe}, W. 1996, \apj, 470, 444

\bibitem[{{Filho} {et~al.}(2004){Filho}, {Fraternali}, {Markoff}, {Nagar},
  {Barthel}, {Ho}, \& {Yuan}}]{filho2004}
{Filho}, M.~E., {Fraternali}, F., {Markoff}, S., {et~al.} 2004, \aap, 418, 429

\bibitem[{{Garcia}(1993)}]{garcia1993}
{Garcia}, A.~M. 1993, \aaps, 100, 47

\bibitem[{{Garmire} {et~al.}(2003){Garmire}, {Bautz}, {Ford}, {Nousek}, \&
  {Ricker}}]{garmire2003}
{Garmire}, G.~P., {Bautz}, M.~W., {Ford}, P.~G., {Nousek}, J.~A., \& {Ricker},
  J. G.~R. 2003, in Society of Photo-Optical Instrumentation Engineers (SPIE)
  Conference Series, Vol. 4851, Society of Photo-Optical Instrumentation
  Engineers (SPIE) Conference Series, ed. J.~E. {Truemper} \& H.~D.
  {Tananbaum}, 28--44

\bibitem[{{Giacintucci} {et~al.}(2011){Giacintucci}, {O'Sullivan}, {Vrtilek},
  {David}, {Raychaudhury}, {Venturi}, {Athreya}, {Clarke}, {Murgia},
  {Mazzotta}, {Gitti}, {Ponman}, {Ishwara-Chandra}, {Jones}, \&
  {Forman}}]{giacintucci2011}
{Giacintucci}, S., {O'Sullivan}, E., {Vrtilek}, J., {et~al.} 2011, \apj, 732,
  95

\bibitem[{{Giroletti} {et~al.}(2005){Giroletti}, {Taylor}, \&
  {Giovannini}}]{giroletti2005}
{Giroletti}, M., {Taylor}, G.~B., \& {Giovannini}, G. 2005, \apj, 622, 178

\bibitem[{{Gonzalez-Martin} {et~al.}(2014){Gonzalez-Martin}, {Diaz-Gonzalez},
  {Acosta-Pulido}, {Masegosa}, {Papadakis}, {Rodriguez-Espinosa}, {Marquez}, \&
  {Hernandez-Garcia}}]{omaira2014}
{Gonzalez-Martin}, O., {Diaz-Gonzalez}, D., {Acosta-Pulido}, J.~A., {et~al.}
  2014, ArXiv e-prints

\bibitem[{{Gonz{\'a}lez-Mart{\'i}n}
  {et~al.}(2009{\natexlab{a}}){Gonz{\'a}lez-Mart{\'i}n}, {Masegosa},
  {M{\'a}rquez}, \& {Guainazzi}}]{omaira2009b}
{Gonz{\'a}lez-Mart{\'i}n}, O., {Masegosa}, J., {M{\'a}rquez}, I., \&
  {Guainazzi}, M. 2009{\natexlab{a}}, \apj, 704, 1570

\bibitem[{{Gonz{\'a}lez-Mart{\'i}n}
  {et~al.}(2009{\natexlab{b}}){Gonz{\'a}lez-Mart{\'i}n}, {Masegosa},
  {M{\'a}rquez}, {Guainazzi}, \& {Jim{\'e}nez-Bail{\'o}n}}]{omaira2009a}
{Gonz{\'a}lez-Mart{\'i}n}, O., {Masegosa}, J., {M{\'a}rquez}, I., {Guainazzi},
  M., \& {Jim{\'e}nez-Bail{\'o}n}, E. 2009{\natexlab{b}}, \aap, 506, 1107

\bibitem[{{Gonz{\'a}lez-Mart{\'i}n}
  {et~al.}(2011{\natexlab{a}}){Gonz{\'a}lez-Mart{\'i}n}, {Papadakis}, {Braito},
  {Masegosa}, {M{\'a}rquez}, {Mateos}, {Acosta-Pulido}, {Mart{\'i}nez},
  {Ebrero}, {Esquej}, {O'Brien}, {Tueller}, {Warwick}, \&
  {Watson}}]{omaira2011b}
{Gonz{\'a}lez-Mart{\'i}n}, O., {Papadakis}, I., {Braito}, V., {et~al.}
  2011{\natexlab{a}}, \aap, 527, A142

\bibitem[{{Gonz{\'a}lez-Mart{\'i}n}
  {et~al.}(2011{\natexlab{b}}){Gonz{\'a}lez-Mart{\'i}n}, {Papadakis}, {Reig},
  \& {Zezas}}]{omaira2011a}
{Gonz{\'a}lez-Mart{\'i}n}, O., {Papadakis}, I., {Reig}, P., \& {Zezas}, A.
  2011{\natexlab{b}}, \aap, 526, A132

\bibitem[{{Gonz{\'a}lez-Mart{\'i}n} \& {Vaughan}(2012)}]{omairavaughan2012}
{Gonz{\'a}lez-Mart{\'i}n}, O. \& {Vaughan}, S. 2012, \aap, 544, A80

\bibitem[{{Gu} \& {Cao}(2009)}]{gucao2009}
{Gu}, M. \& {Cao}, X. 2009, \mnras, 399, 349

\bibitem[{{Guainazzi} {et~al.}(2012){Guainazzi}, {La Parola}, {Miniutti},
  {Segreto}, \& {Longinotti}}]{guainazzi2012}
{Guainazzi}, M., {La Parola}, V., {Miniutti}, G., {Segreto}, A., \&
  {Longinotti}, A.~L. 2012, \aap, 547, A31

\bibitem[{{Guainazzi} {et~al.}(2000){Guainazzi}, {Oosterbroek}, {Antonelli}, \&
  {Matt}}]{guainazzi2000}
{Guainazzi}, M., {Oosterbroek}, T., {Antonelli}, L.~A., \& {Matt}, G. 2000,
  \aap, 364, L80

\bibitem[{{Haardt} \& {Maraschi}(1991)}]{haardt1991}
{Haardt}, F. \& {Maraschi}, L. 1991, \apjl, 380, L51

\bibitem[{{Harris} {et~al.}(2003){Harris}, {Biretta}, {Junor}, {Perlman},
  {Sparks}, \& {Wilson}}]{harris2003}
{Harris}, D.~E., {Biretta}, J.~A., {Junor}, W., {et~al.} 2003, \apjl, 586, L41

\bibitem[{{Harris} {et~al.}(2006){Harris}, {Cheung}, {Biretta}, {Sparks},
  {Junor}, {Perlman}, \& {Wilson}}]{harris2006}
{Harris}, D.~E., {Cheung}, C.~C., {Biretta}, J.~A., {et~al.} 2006, \apj, 640,
  211

\bibitem[{{Harris} {et~al.}(2009){Harris}, {Cheung}, {Stawarz}, {Biretta}, \&
  {Perlman}}]{harris2009}
{Harris}, D.~E., {Cheung}, C.~C., {Stawarz}, {\L}., {Biretta}, J.~A., \&
  {Perlman}, E.~S. 2009, \apj, 699, 305

\bibitem[{{Harris} {et~al.}(2011){Harris}, {Massaro}, {Cheung}, {Horns},
  {Raue}, {Stawarz}, {Wagner}, {Colin}, {Mazin}, {Wagner}, {Beilicke},
  {LeBohec}, {Hui}, \& {Mukherjee}}]{harris2011}
{Harris}, D.~E., {Massaro}, F., {Cheung}, C.~C., {et~al.} 2011, \apj, 743, 177

\bibitem[{{Heckman}(1980)}]{heckman1980}
{Heckman}, T.~M. 1980, \aap, 87, 152

\bibitem[{{Hern{\'a}ndez-Garc{\'i}a} {et~al.}(2013){Hern{\'a}ndez-Garc{\'i}a},
  {Gonz{\'a}lez-Mart{\'i}n}, {M{\'a}rquez}, \& {Masegosa}}]{lore2013}
{Hern{\'a}ndez-Garc{\'i}a}, L., {Gonz{\'a}lez-Mart{\'i}n}, O., {M{\'a}rquez},
  I., \& {Masegosa}, J. 2013, \aap, 556, A47

\bibitem[{{Ho}(2008)}]{ho2008}
{Ho}, L.~C. 2008, \araa, 46, 475

\bibitem[{{Ho} {et~al.}(1997){Ho}, {Filippenko}, {Sargent}, \& {Peng}}]{ho1997}
{Ho}, L.~C., {Filippenko}, A.~V., {Sargent}, W.~L.~W., \& {Peng}, C.~Y. 1997,
  \apjs, 112, 391

\bibitem[{{Ho} \& {Ulvestad}(2001)}]{houlvestad2001}
{Ho}, L.~C. \& {Ulvestad}, J.~S. 2001, \apjs, 133, 77

\bibitem[{{Ishwara-Chandra} \& {Saikia}(1999)}]{ishwara1999}
{Ishwara-Chandra}, C.~H. \& {Saikia}, D.~J. 1999, \mnras, 309, 100

\bibitem[{{Kalberla} {et~al.}(2005){Kalberla}, {Burton}, {Hartmann}, {Arnal},
  {Bajaja}, {Morras}, \& {P{\"o}ppel}}]{kalberla2005}
{Kalberla}, P.~M.~W., {Burton}, W.~B., {Hartmann}, D., {et~al.} 2005, \aap,
  440, 775

\bibitem[{{Krips} {et~al.}(2007){Krips}, {Eckart}, {Krichbaum}, {Pott}, {Leon},
  {Neri}, {Garc{\'i}a-Burillo}, {Combes}, {Boone}, {Baker}, {Tacconi},
  {Schinnerer}, \& {Hunt}}]{krips2007}
{Krips}, M., {Eckart}, A., {Krichbaum}, T.~P., {et~al.} 2007, \aap, 464, 553

\bibitem[{{Lawrence} {et~al.}(1987){Lawrence}, {Watson}, {Pounds}, \&
  {Elvis}}]{lawrence1987}
{Lawrence}, A., {Watson}, M.~G., {Pounds}, K.~A., \& {Elvis}, M. 1987, \nat,
  325, 694

\bibitem[{{Maiolino} {et~al.}(1998){Maiolino}, {Salvati}, {Bassani}, {Dadina},
  {della Ceca}, {Matt}, {Risaliti}, \& {Zamorani}}]{maiolino1998}
{Maiolino}, R., {Salvati}, M., {Bassani}, L., {et~al.} 1998, \aap, 338, 781

\bibitem[{{Maoz} {et~al.}(2005){Maoz}, {Nagar}, {Falcke}, \&
  {Wilson}}]{maoz2005}
{Maoz}, D., {Nagar}, N.~M., {Falcke}, H., \& {Wilson}, A.~S. 2005, \apj, 625,
  699

\bibitem[{{Masegosa} {et~al.}(2011){Masegosa}, {M{\'a}rquez}, {Ramirez}, \&
  {Gonz{\'a}lez-Mart{\'i}n}}]{masegosa2011}
{Masegosa}, J., {M{\'a}rquez}, I., {Ramirez}, A., \& {Gonz{\'a}lez-Mart{\'i}n},
  O. 2011, \aap, 527, A23

\bibitem[{{McHardy}(2010)}]{mchardy2009}
{McHardy}, I. 2010, in Lecture Notes in Physics, Berlin Springer Verlag, Vol.
  794, Lecture Notes in Physics, Berlin Springer Verlag, ed. T.~{Belloni}, 203

\bibitem[{{McHardy} {et~al.}(2006){McHardy}, {Koerding}, {Knigge}, {Uttley}, \&
  {Fender}}]{mchardy2006}
{McHardy}, I.~M., {Koerding}, E., {Knigge}, C., {Uttley}, P., \& {Fender},
  R.~P. 2006, \nat, 444, 730

\bibitem[{{Nagar} {et~al.}(2005){Nagar}, {Falcke}, \& {Wilson}}]{nagar2005}
{Nagar}, N.~M., {Falcke}, H., \& {Wilson}, A.~S. 2005, \aap, 435, 521

\bibitem[{{Nagar} {et~al.}(2002){Nagar}, {Falcke}, {Wilson}, \&
  {Ulvestad}}]{nagar2002}
{Nagar}, N.~M., {Falcke}, H., {Wilson}, A.~S., \& {Ulvestad}, J.~S. 2002, \aap,
  392, 53

\bibitem[{{Nandra} {et~al.}(1997){Nandra}, {George}, {Mushotzky}, {Turner}, \&
  {Yaqoob}}]{nandra1997}
{Nandra}, K., {George}, I.~M., {Mushotzky}, R.~F., {Turner}, T.~J., \&
  {Yaqoob}, T. 1997, \apj, 476, 70

\bibitem[{{Narayan} \& {Yi}(1994)}]{narayanyi1994}
{Narayan}, R. \& {Yi}, I. 1994, \apjl, 428, L13

\bibitem[{{O'Sullivan} {et~al.}(2005){O'Sullivan}, {Vrtilek}, \&
  {Kempner}}]{osullivan2005}
{O'Sullivan}, E., {Vrtilek}, J.~M., \& {Kempner}, J.~C. 2005, \apjl, 624, L77

\bibitem[{{Pellegrini} {et~al.}(2002){Pellegrini}, {Fabbiano}, {Fiore},
  {Trinchieri}, \& {Antonelli}}]{pellegrini2002}
{Pellegrini}, S., {Fabbiano}, G., {Fiore}, F., {Trinchieri}, G., \&
  {Antonelli}, A. 2002, \aap, 383, 1

\bibitem[{{Pellegrini} {et~al.}(2012){Pellegrini}, {Wang}, {Fabbiano}, {Kim},
  {Brassington}, {Gallagher}, {Trinchieri}, \& {Zezas}}]{pellegrini2012}
{Pellegrini}, S., {Wang}, J., {Fabbiano}, G., {et~al.} 2012, \apj, 758, 94

\bibitem[{{Peterson}(1997)}]{bradley1997}
{Peterson}, B.~M. 1997, {An Introduction to Active Galactic Nuclei}, ed.
  {Peterson, B.~M.}

\bibitem[{{Pian} {et~al.}(2010){Pian}, {Romano}, {Maoz}, {Cucchiara}, {Pagani},
  \& {Parola}}]{pian2010}
{Pian}, E., {Romano}, P., {Maoz}, D., {et~al.} 2010, \mnras, 401, 677

\bibitem[{{Pogge} {et~al.}(2000){Pogge}, {Maoz}, {Ho}, \&
  {Eracleous}}]{pogge2000}
{Pogge}, R.~W., {Maoz}, D., {Ho}, L.~C., \& {Eracleous}, M. 2000, \apj, 532,
  323

\bibitem[{{Puccetti} {et~al.}(2007){Puccetti}, {Fiore}, {Risaliti}, {Capalbi},
  {Elvis}, \& {Nicastro}}]{puccetti2007}
{Puccetti}, S., {Fiore}, F., {Risaliti}, G., {et~al.} 2007, \mnras, 377, 607

\bibitem[{{Qiao} \& {Liu}(2013)}]{qiao2013a}
{Qiao}, E. \& {Liu}, B.~F. 2013, \apj, 764, 2

\bibitem[{{Qiao} {et~al.}(2013){Qiao}, {Liu}, {Panessa}, \& {Liu}}]{qiao2013b}
{Qiao}, E., {Liu}, B.~F., {Panessa}, F., \& {Liu}, J.~Y. 2013, \apj, 777, 102

\bibitem[{{Quataert}(2004)}]{quataert2004}
{Quataert}, E. 2004, in Astronomical Society of the Pacific Conference Series,
  Vol. 311, AGN Physics with the Sloan Digital Sky Survey, ed. G.~T. {Richards}
  \& P.~B. {Hall}, 131

\bibitem[{{Randall} {et~al.}(2011){Randall}, {Forman}, {Giacintucci}, {Nulsen},
  {Sun}, {Jones}, {Churazov}, {David}, {Kraft}, {Donahue}, {Blanton},
  {Simionescu}, \& {Werner}}]{randall2011}
{Randall}, S.~W., {Forman}, W.~R., {Giacintucci}, S., {et~al.} 2011, \apj, 726,
  86

\bibitem[{{Remillard} \& {McClintock}(2006)}]{remillard2006}
{Remillard}, R.~A. \& {McClintock}, J.~E. 2006, \araa, 44, 49

\bibitem[{{Risaliti} {et~al.}(2007){Risaliti}, {Elvis}, {Fabbiano}, {Baldi},
  {Zezas}, \& {Salvati}}]{risaliti2007}
{Risaliti}, G., {Elvis}, M., {Fabbiano}, G., {et~al.} 2007, \apjl, 659, L111

\bibitem[{{Risaliti} {et~al.}(2011){Risaliti}, {Nardini}, {Salvati}, {Elvis},
  {Fabbiano}, {Maiolino}, {Pietrini}, \& {Torricelli-Ciamponi}}]{risaliti2011}
{Risaliti}, G., {Nardini}, E., {Salvati}, M., {et~al.} 2011, \mnras, 410, 1027

\bibitem[{{Rubin} {et~al.}(1979){Rubin}, {Ford}, \& {Roberts}}]{rubin1979}
{Rubin}, V.~C., {Ford}, W.~K.~J., \& {Roberts}, M.~S. 1979, \apj, 230, 35

\bibitem[{{Sambruna} {et~al.}(2003){Sambruna}, {Gliozzi}, {Eracleous},
  {Brandt}, \& {Mushotzky}}]{sambruna2003}
{Sambruna}, R.~M., {Gliozzi}, M., {Eracleous}, M., {Brandt}, W.~N., \&
  {Mushotzky}, R. 2003, \apjl, 586, L37

\bibitem[{{Sanfrutos} {et~al.}(2013){Sanfrutos}, {Miniutti},
  {Ag{\'i}s-Gonz{\'a}lez}, {Fabian}, {Miller}, {Panessa}, \&
  {Zoghbi}}]{sanfrutos2013}
{Sanfrutos}, M., {Miniutti}, G., {Ag{\'i}s-Gonz{\'a}lez}, B., {et~al.} 2013,
  \mnras, 436, 1588

\bibitem[{{Satyapal} {et~al.}(2005){Satyapal}, {Dudik}, {O'Halloran}, \&
  {Gliozzi}}]{satyapal2005}
{Satyapal}, S., {Dudik}, R.~P., {O'Halloran}, B., \& {Gliozzi}, M. 2005, \apj,
  633, 86

\bibitem[{{Satyapal} {et~al.}(2004){Satyapal}, {Sambruna}, \&
  {Dudik}}]{satyapal2004}
{Satyapal}, S., {Sambruna}, R.~M., \& {Dudik}, R.~P. 2004, \aap, 414, 825

\bibitem[{{Str{\"u}der} {et~al.}(2001){Str{\"u}der}, {Briel}, {Dennerl},
  {Hartmann}, {Kendziorra}, {Meidinger}, {Pfeffermann}, {Reppin}, {Aschenbach},
  {Bornemann}, {Br{\"a}uninger}, {Burkert}, {Elender}, {Freyberg}, {Haberl},
  {Hartner}, {Heuschmann}, {Hippmann}, {Kastelic}, {Kemmer}, {Kettenring},
  {Kink}, {Krause}, {M{\"u}ller}, {Oppitz}, {Pietsch}, {Popp}, {Predehl},
  {Read}, {Stephan}, {St{\"o}tter}, {Tr{\"u}mper}, {Holl}, {Kemmer}, {Soltau},
  {St{\"o}tter}, {Weber}, {Weichert}, {von Zanthier}, {Carathanassis}, {Lutz},
  {Richter}, {Solc}, {B{\"o}ttcher}, {Kuster}, {Staubert}, {Abbey}, {Holland},
  {Turner}, {Balasini}, {Bignami}, {La Palombara}, {Villa}, {Buttler},
  {Gianini}, {Lain{\'e}}, {Lumb}, \& {Dhez}}]{struder2001}
{Str{\"u}der}, L., {Briel}, U., {Dennerl}, K., {et~al.} 2001, \aap, 365, L18

\bibitem[{{Terashima} \& {Wilson}(2003)}]{terashimawilson2003}
{Terashima}, Y. \& {Wilson}, A.~S. 2003, \apj, 583, 145

\bibitem[{{Terashima} \& {Wilson}(2004)}]{terashimawilson2004}
{Terashima}, Y. \& {Wilson}, A.~S. 2004, \apj, 601, 735

\bibitem[{{Tremaine} {et~al.}(2002){Tremaine}, {Gebhardt}, {Bender}, {Bower},
  {Dressler}, {Faber}, {Filippenko}, {Green}, {Grillmair}, {Ho}, {Kormendy},
  {Lauer}, {Magorrian}, {Pinkney}, \& {Richstone}}]{tremaine2002}
{Tremaine}, S., {Gebhardt}, K., {Bender}, R., {et~al.} 2002, \apj, 574, 740

\bibitem[{{Turner} {et~al.}(1997){Turner}, {George}, {Nandra}, \&
  {Mushotzky}}]{turner1997}
{Turner}, T.~J., {George}, I.~M., {Nandra}, K., \& {Mushotzky}, R.~F. 1997,
  \apjs, 113, 23

\bibitem[{{Urry} \& {Padovani}(1995)}]{urrypadovani1995}
{Urry}, C.~M. \& {Padovani}, P. 1995, \pasp, 107, 803

\bibitem[{{Vaughan} {et~al.}(2003){Vaughan}, {Edelson}, {Warwick}, \&
  {Uttley}}]{vaughan2003}
{Vaughan}, S., {Edelson}, R., {Warwick}, R.~S., \& {Uttley}, P. 2003, \mnras,
  345, 1271

\bibitem[{{Venturi} {et~al.}(1993){Venturi}, {Giovannini}, {Feretti},
  {Comoretto}, \& {Wehrle}}]{venturi1993}
{Venturi}, T., {Giovannini}, G., {Feretti}, L., {Comoretto}, G., \& {Wehrle},
  A.~E. 1993, \apj, 408, 81

\bibitem[{{Vermeulen} {et~al.}(2003){Vermeulen}, {Ros}, {Kellermann}, {Cohen},
  {Zensus}, \& {van Langevelde}}]{vermeulen2003}
{Vermeulen}, R.~C., {Ros}, E., {Kellermann}, K.~I., {et~al.} 2003, \aap, 401,
  113

\bibitem[{{Vrtilek} {et~al.}(2013){Vrtilek}, {O'Sullivan}, {David},
  {Kolokythas}, {Giacintucci}, {Raychaudhury}, \& {Ponman}}]{vrtilek2013}
{Vrtilek}, J.~M., {O'Sullivan}, E., {David}, L.~P., {et~al.} 2013, in AAS/High
  Energy Astrophysics Division, Vol.~13, AAS/High Energy Astrophysics Division,
  \#116.06

\bibitem[{{Xu} {et~al.}(2000){Xu}, {Baum}, {O'Dea}, {Wrobel}, \&
  {Condon}}]{xu2000}
{Xu}, C., {Baum}, S.~A., {O'Dea}, C.~P., {Wrobel}, J.~M., \& {Condon}, J.~J.
  2000, \aj, 120, 2950

\bibitem[{{Younes} {et~al.}(2010){Younes}, {Porquet}, {Sabra}, {Grosso},
  {Reeves}, \& {Allen}}]{younes2010}
{Younes}, G., {Porquet}, D., {Sabra}, B., {et~al.} 2010, \aap, 517, A33

\bibitem[{{Younes} {et~al.}(2011){Younes}, {Porquet}, {Sabra}, \&
  {Reeves}}]{younes2011}
{Younes}, G., {Porquet}, D., {Sabra}, B., \& {Reeves}, J.~N. 2011, \aap, 530,
  A149

\bibitem[{{Younes} {et~al.}(2012){Younes}, {Porquet}, {Sabra}, {Reeves}, \&
  {Grosso}}]{younes2012}
{Younes}, G., {Porquet}, D., {Sabra}, B., {Reeves}, J.~N., \& {Grosso}, N.
  2012, \aap, 539, A104

\end{thebibliography}

\newpage

\appendix

\section{Tables}

\onecolumn

\tiny
\renewcommand{\arraystretch}{1.4}
\begin{longtable}{lccccccccc}
\caption[Observational details]{\label{obs} Observational details.} \\   \hline \hline
 Name & Instrument & ObsID  & Date & R  & Net Exptime & Counts & HR &  log($L_{UV}$) & Filter \\
 &     &           &                    & ($\arcsec$)  & (ksec) & & & (erg/s) &  \\
 (1) & (2) & (3) & (4) & (5) & (6) & (7) & (8) & (9) & (10) \\
\hline 
\endfirsthead
\caption{(Cont.)} \\
\hline \hline
 Name & Instrument & ObsID  & Date & R  & Net Exptime & Counts & HR &  log($L_{UV}$) & Filter \\
 &     &           &                    & ($\arcsec$)  & (ksec) & & & (erg/s) &  \\
 (1) & (2) & (3) & (4) & (5) & (6) & (7) & (8) & (9) & (10) \\
 \hline  
\endhead
%\multicolumn{11}{r}{continued on next page\ldots}\\
\endfoot
\\
\endlastfoot
 NGC\,315 & \emph{XMM}--Newton$^c$ & 0305290201  & 2005-07-02 & 25  & 14 & 5799 & -0.55$^+_-$0.01 & 41.75 $^+_-$    0.05 &    UVM2  \\
        & \emph{Chandra} & 855 &  2000-10-08 & 3 &  5 & 501 & -0.27$^+_-$0.13 & -  \\
          & \emph{Chandra}$^c$ & 4156 & 2003-02-22 & 3  & 55 & 5550 & -0.23$^+_-$0.05 & -  \\
 \hline
 NGC\,1052 & \emph{XMM}--Newton$^c$ &  0093630101  & 2001-08-15 & 25 &  11.2 & 5818 & 0.28 $^+_-$ 0.06 & -  \\
  & \emph{XMM}--Newton &  0306230101  & 2006-01-12 & 25 &  44.8 & 25565 & 0.34 $^+_-$ 0.03 & 41.04 $^+_-$ 0.01  & UVM2 \\
   &                    &             &            &     &              &               &    &  40.94 $^+_-$ 0.03 & UVW2 \\
  & \emph{XMM}--Newton &   0553300301   & 2009-01-14 & 25 &  42.4 & 27367 & 0.39 $^+_-$ 0.02 &  41.14 $^+_-$ 0.01 & UVM2 \\
     &                    &             &            &     &              &               &    & 40.95 $^+_-$ 0.03 & UVW2 \\
\end{longtable}

\tiny
\renewcommand{\arraystretch}{1.4}
\begin{longtable}{lccccccccc}
\caption[Spectral fittings]{\label{bestfit}  Final compilation of the best-fit models for the sample, including the individual best-fit model for each observation, and the simultaneous best-fit model with the varying parameters. }  \\  \hline \hline
Analysis & ObsID & Model & $N_{H1}$ & $N_{H2}$ & kT & $\Gamma$ & $Norm_1$ & $Norm_2$ & $\chi^2/d.o.f$  \\
  & & &  &  & keV & & ($10^{-4}$) & ($10^{-4}$) & F-test \\
(1) & (2) & (3) & (4) & (5) & (6) & (7) & (8) & (9) & (10)  \\
\hline 
\endfirsthead
\caption[]{(Cont.)} \\
\hline \hline
Analysis & ObsID & Model & $N_{H1}$ & $N_{H2}$ & kT & $\Gamma$ & $Norm_1$ & $Norm_2$ & $\chi^2/d.o.f$  \\
  & & &  &  & keV & & ($10^{-4}$) & ($10^{-4}$) & F-test \\
(1) & (2) & (3) & (4) & (5) & (6) & (7) & (8) & (9) & (10)  \\
\hline  
\endhead
%\multicolumn{11}{r}{continued on next page\ldots}\\
\endfoot
\\
\endlastfoot
\multicolumn{10}{c}{NGC\,315} \\ 
\hline
Ind & 0305290201 &   MEPL & - & 0.17$ _{ 0.11 }^{ 0.25}$ & 0.56$ _{ 0.54 }^{ 0.58 }$ & 1.59$ _{ 1.50 }^{ 1.69 }$ & 1.20$ _{ 1.11 }^{ 1.29 }$ & 1.58$ _{ 1.41 }^{ 1.77 }$ & 197.97/198 \\
Ind & 855 &   MEPL & - & 0.35$ _{ 0.00 }^{ 0.87 }$ & 0.33$ _{ 0.17 }^{ 0.55 }$ & 1.53$ _{ 1.10 }^{ 2.01 }$ & 0.52$ _{ 0.25 }^{ 0.87 }$ & 1.66$ _{ 1.01 }^{ 3.27 }$ & 15.16/17 \\
Ind (3$\arcsec$) & 4156* &   MEPL & 0.00$ _{ 0.00 }^{ 0.02 }$ & 0.77$ _{ 0.62 }^{ 0.92 }$ & 0.52$ _{ 0.49 }^{ 0.55 }$ & 1.61$ _{ 1.49 }^{ 1.76 }$ & 0.40$ _{ 0.37 }^{ 0.43 }$ & 2.04$ _{ 1.71 }^{ 2.44 }$ & 190.43/172 \\
Ind (25$\arcsec$) & 4156 & MEPL & - & 0.16$ _{ 0.12 }^{ 0.23 }$ & 0.56$ _{ 0.55 }^{ 0.57 }$ & 1.30$ _{ 1.22 }^{ 1.38 }$ & 1.17$ _{ 1.11 }^{ 1.25 }$ & 1.51$ _{ 1.39 }^{ 1.67 }$ & 258.47/206 \\
\hline
\vspace*{0.15cm}
 SMF0 &  All &  MEPL &  - &  0.75$ _{ 0.61 }^{0.90 }$ &  0.52$ _{ 0.49 }^{ 0.55 }$ &  1.62$ _{ 1.49 }^{ 1.75 }$ &  0.40$ _{ 0.39 }^{ 0.43 }$ &  2.02$ _{ 1.70 }^{ 2.40 }$ &  209.89/195 \\
\multicolumn{10}{c}{NGC\,1052} \\ 
\hline
Ind & 093630101 & ME2PL & - & 13.77$^{15.78}_{11.95}$ & 0.62$^{0.69}_{0.51}$ & 1.24$^{1.35}_{1.11}$ & 1.05$^{1.13}_{0.97}$ & 6.09$^{4.55}_{4.79}$ & 269.16/234  \\
\end{longtable}

\begin{longtable}{ccccccc}
\caption[]{\label{lumincorr} X-ray luminosities.}  \\  \hline \hline
\multicolumn{7}{c}{ \hspace*{5.7cm} Individual \hspace*{3.5cm} Simultaneous } \\  \cmidrule(r){4-5} \cmidrule{6-7}
Name & Satellite & ObsID & log(L(0.5-2 keV))  & log(L(2-10 keV)) & log(L(0.5-2 keV))  & log(L(2-10 keV)) \\ 
(1) & (2) & (3) & (4) & (5) & (6) & (7)  \\ 
\hline 
\endfirsthead
\caption[]{ (Cont.)} \\
\hline \hline
\multicolumn{7}{c}{ \hspace*{5.7cm} Individual \hspace*{3.5cm} Simultaneous } \\ \cmidrule(r){4-5} \cmidrule{6-7}
Name & Satellite & ObsID & log(L(0.5-2 keV))  & log(L(2-10 keV)) & log(L(0.5-2 keV))  & log(L(2-10 keV)) \\ 
(1) & (2) & (3) & (4) & (5) & (6) & (7)  \\ 
\hline 
\endhead
%\multicolumn{11}{r}{continued on next page\ldots}\\
\endfoot
\\
\endlastfoot
NGC\,315 & \emph{XMM}--Newton & 0305290201 & 41.46$_{ 41.44 }^{ 41.47}$ &   41.55$_{  41.52 }^{ 41.57 }$ &  \\
         & \emph{Chandra} & 855 & 41.31$_{ 41.15}^{ 41.43}$ &   41.55$_{ 41.47 }^{ 41.63}$ &  41.38$_{ 41.37}^{ 41.40}$ &   41.58$_{ 41.55}^{ 41.61}$ \\
         & \emph{Chandra} (3$\arcsec$) & 4156 & 41.38$_{ 41.37 }^{ 41.39}$  &   41.58$_{ 41.56}^{ 41.61}$ & 41.38$_{ 41.37}^{ 41.40}$ &   41.58$_{ 41.55}^{ 41.61}$ \\
         & \emph{Chandra} (25$\arcsec$) & 4156 & 41.44$_{ 41.44}^{41.45}$ &   41.68$_{ 41.66}^{ 41.700}$ & \\
\hline
NGC\,1052 & \emph{XMM}--Newton &  093630101 & 40.96    $_{40.94  }^{  40.97}$ &  41.46    $_{41.45  }^{ 41.47}$ & 40.99    $_{40.99  }^{  41.00}$ & 41.47	$_{41.46  }^{  41.49}$  \\
& \emph{XMM}--Newton &  306230101 & 40.96    $_{40.95  }^{  40.97}$ &  41.48    $_{41.48  }^{ 41.49}$ & 41.01    $_{41.00  }^{  41.01}$ & 41.49    $_{41.48  }^{  41.49}$ \\
& \emph{XMM}--Newton &  553300301 & 41.10    $_{41.10  }^{  41.11}$ &  41.56    $_{41.55  }^{ 41.56}$ & 41.06    $_{41.06  }^{  41.07}$ & 41.55    $_{41.54  }^{  41.55}$ \\
& \emph{XMM}--Newton & 553300401  & 41.12    $_{41.11  }^{  41.13}$ &  41.55    $_{41.55  }^{ 41.56}$ & 41.09    $_{41.08  }^{  41.09}$ & 41.57    $_{41.57  }^{  41.58}$ \\
\end{longtable}

\begin{longtable}{lcccccccccc}
\caption[]{\label{annulus} Results for the best fit of the annular region (ring) in \emph{Chandra} data, and the best fit obtained for the nucleus of \emph{XMM}--Newton data when the contribution from the annular region was removed. }  \\  \hline \hline
Name (obsID) & Region & Model & $N_{H1}$ & $N_{H2}$ & kT & $\Gamma$ & $\chi^2_r$ & $log(L_{soft})$ & $log(L_{hard})$ & Cont. \\
 & & & ($10^{22} cm^{-2}$) & ($10^{22} cm^{-2}$) & (keV) & & & (0.5-2 keV) & (2-10 keV) & \%  \\
(1) & (2) & (3)          & (4) & (5) & (6) & (7) & (8) & (9) & (10) & (11) \\ 
\hline 
\endfirsthead
\caption[]{(Cont.)} \\
\hline \hline
Name (obsID) & Region & Model & $N_{H1}$ & $N_{H2}$ & kT & $\Gamma$ & $\chi^2_r$ & $log(L_{soft})$ & $log(L_{hard})$ & Cont. \\
 & & & ($10^{22} cm^{-2}$) & ($10^{22} cm^{-2}$) & (keV) & & & (0.5-2 keV) & (2-10 keV) & \%  \\
(1) & (2) & (3)          & (4) & (5) & (6) & (7) & (8) & (9) & (10) & (11) \\ \hline  
\hline 
\endhead
%\multicolumn{11}{r}{continued on next page\ldots}\\
\endfoot
\\
\endlastfoot
NGC\,315 (4156) & Ring* & MEPL & - & - & 0.59$_{ 0.57}^{ 0.61}$ & 1.47$_{ 1.31}^{ 1.63}$ & 1.49 & 41.13 & 40.88 & 3 \\
NGC\,315 (0305290201) & Nucleus** & MEPL & - & 0.77$_{0.47}^{1.08}$ & 0.53$_{ 0.45}^{ 0.59}$ & 1.96$_{ 1.74}^{ 2.20}$ & 0.97 & 41.41 & 41.43 & - \\ \hline
NGC\,1052 (5910) & Ring* & ME2PL & - & 26.10$ _{ 16.93 }^{ 38.14 }$ & 0.31$ _{ 0.28 }^{ 0.39 }$ & 1.98$ _{ 1.71 }^{ 2.23 }$ & 1.47 & 40.185 & 40.563 & 10  \\
NGC\,1052 (093630101) & Nucleus** & ME2PL & - & 7.54$ _{ 5.20 }^{ 9.98 }$ & 0.78$ _{ 0.68 }^{ 0.89 }$ & 1.69$ _{ 1.50 }^{ 1.86 }$ & 1.13 & 39.875 & 41.235 & - \\ \hline
NGC\,2787 (4689) & Ring* & PL & 0.00$ _{ 0.00 }^{ 0.05 }$ & - & - & 1.59$_{ 1.43}^{ 1.92 }$  & 1.05 & 38.82 & 39.23 & 53 \\
NGC\,2787 (0200250101) & Nucleus** & PL & - & - & - & 1.65$_{ 1.54}^{ 1.77 }$  & 1.49 & 39.12 & 39.47 & - \\ \hline
NGC\,2841 (6096) & Ring* & PL & 0.02$ _{ 0.00 }^{ 0.11 }$ & - & - & 1.96$_{ 1.72}^{ 2.45 }$  & 1.09 & 39.54 & 39.60 & 60 \\
NGC\,2841 (0201440101) & Nucleus** & MEPL & 0.43$^{0.51}_{0.00}$ & - & 0.31$^{0.60}_{0.19}$ & 2.19$^{2.34}_{1.93}$ & 1.41 & 39.73 & 39.77 & - \\
\end{longtable}

\begin{longtable}{lccccccccr}
\caption[]{\label{simultanillo} Simultaneous fittings taking into account the contribution from the annular region given in Table \ref{annulus}.}  \\  \hline \hline
 ObsID & $N_{H1}$ & $N_{H2}$ & kT & $\Gamma$ & $Norm_1$ & $Norm_2$ & $\chi^2/d.o.f$ & $log(L_{soft})$ & $log(L_{hard})$ \\
  & ($10^{22} cm^{-2}$) & ($10^{22} cm^{-2}$) & (keV) & & ($10^{-4}$) & ($10^{-4}$) & & (0.5-2 keV) & (2-10 keV)  \\
(1) & (2) & (3) & (4) & (5) & (6) & (7) & (8) & (9) & (10)  \\
\hline 
\endfirsthead
\caption[]{\textbf{(Cont.)}} \\
\hline \hline
 ObsID & $N_{H1}$ & $N_{H2}$ & kT & $\Gamma$ & $Norm_1$ & $Norm_2$ & $\chi^2/d.o.f$ & $log(L_{soft})$ & $log(L_{hard})$ \\
  & ($10^{22} cm^{-2}$) & ($10^{22} cm^{-2}$) & (keV) & & ($10^{-4}$) & ($10^{-4}$) & & (0.5-2 keV) & (2-10 keV)  \\
(1) & (2) & (3) & (4) & (5) & (6) & (7) & (8) & (9) & (10)  \\ 
\hline 
\endhead
%\multicolumn{11}{r}{continued on next page\ldots}\\
\endfoot
\\
\endlastfoot
\multicolumn{10}{c}{NGC\,315} \\  \hline
0305290201 & - & 0.86$_{0.73}^{1.00}$ & 0.52$_{0.50}^{0.55}$ & 1.85$_{1.74}^{1.96}$ & 0.42$_{0.39}^{0.44}$ & 2.38$_{2.05}^{2.77}$ & 467.58/379 & 41.44$_{41.43}^{41.55}$ & 41.54$_{41.52}^{41.56}$ \\
4156 & & & & & & & & 41.44$_{41.43}^{41.55}$ & 41.51$_{41.49}^{41.53}$ \\

\multicolumn{10}{c}{NGC\,1052} \\  \hline
0093630101 & - & 6.11$_{5.07}^{7.18}$ & 0.68$_{0.63}^{0.73}$ & 1.58$_{1.47}^{1.70}$ & 0.48$_{0.44}^{0.53}$ & 1.28$_{0.95}^{1.69}$ & 577.98/477 & 40.35$_{40.34}^{40.36}$ & 41.33$_{41.32}^{41.35}$ \\
5910 & & & & & & & & 40.35$_{40.34}^{40.36}$ & 41.27$_{41.26}^{41.28}$ \\

\multicolumn{10}{c}{NGC\,4261} \\  \hline
0502120101 & 0.03$_{0.00}^{0.08}$ & 7.64$_{6.58}^{8.74}$ & 0.60$_{0.59}^{0.61}$ & 1.55$_{1.36}^{1.75}$ & 0.22$_{0.19}^{0.26}$ & 1.53$_{1.04}^{2.25}$ & 972.83/627 & 40.84$_{40.84}^{40.84}$ & 41.03$_{41.01}^{41.05}$ \\
\end{longtable}

\begin{longtable}{cccccccc}
\caption[]{\label{estcurvas} Statistics of the light curves.}  \\  \hline \hline
Name &  ObsID & Energy & $\chi^2/d.o.f$  & Prob.(\%) & $\sigma_{NXS}^2$  &  $<\sigma^2_{NXS}>$ \\
(1) & (2) & (3) & (4) & (5) & (6) & (7)   \\ 
\hline 
\endfirsthead
\caption[]{(Cont.)} \\
\hline \hline
Name &  ObsID & Energy & $\chi^2/d.o.f$  & Prob.(\%) & $\sigma_{NXS}^2$  &  $<\sigma^2_{NXS}>$  \\
(1) & (2) & (3) & (4) & (5) & (6) & (7)  \\ 
\hline 
\endhead
%\multicolumn{11}{r}{continued on next page\ldots}\\
\endfoot
\\
\endlastfoot
NGC\,315 & 4156        & 0.5-10     & 50.2/55 & 34 &  $<$0.0046 \\
         &             & 0.5-2      &  &  &  $<$0.0086 \\
         &             & 2-10       &  &  & 0.0096  $^+_-$	 0.0059 \\
NGC\,1052 & 0306230101  & 0.5-10    & 57.1/49 & 80 &  $<$0.0012 \\
         &             & 0.5-2      &  &  & $<$0.0039 \\
         &             & 2-10       &  &  &  $<$0.0018 \\
         & 0553300301  & 0.5-10     & 49.5/46 & 66 & $<$0.0012 \\
         &             & 0.5-2      &  &  & $<$0.0037 \\
\end{longtable}

\newpage

\twocolumn

\normalsize

\section{\label{indivnotes} Notes and comparisons with previous results for individual objects}

\subsection{NGC\,315}

NGC\,315 is a radiogalaxy located in the Zwicky cluster 0107.5+3212.
It was classified optically as a type 1.9 LINER by \cite{ho1997}, and 
as an AGN candidate at X-ray frequencies \citep{omaira2009a}.

At radio frequencies (\emph{VLBI} and \emph{VLA}) the galaxy shows an asymmetric morphology, with a compact nuclear emission and a one-sided
jet \citep{venturi1993}. The jet can also be observed in X-rays (see Appendix \ref{Ximages}). Using \emph{VLA} data,
\cite{ishwara1999} did not find significant
variability over a timescale of $\sim$ 12 years.

In X-rays, it was observed twice with 
\emph{Chandra} in 2000 and 2003 and once with \emph{XMM}--Newton in 2005.
\cite{younes2011} found variations in $\Gamma$ (from 1.5$^+_-$0.1
to 2.1$^{+0.1}_{-0.2}$), and a decreasing in the hard luminosity of 53\% between 
2003 and 2005. 
They included the emission of the jet in \emph{XMM}-Newton data to derive the nuclear spectral parameters. With the same data set, we obtained very similar individual spectral fittings and luminosities (see Table \ref{bestfit} and \ref{lumincorr} for \emph{Chandra} data and Table \ref{annulus} for the nuclear region in \emph{XMM}-Newton data). However, we do not find spectral variations, since SMF0 was used both for \emph{Chandra} data and when comparing \emph{Chandra} and \emph{XMM}-Newton. The difference found with the results reported by \cite{younes2011} might be due to the different errors.
 A large set of X-ray observations would be desirable
to obtain conclusive results. 
%With the same data set, we found very similar individual
%spectral fittings and luminosities (Tables \ref{bestfit} and
%\ref{lumincorr}). The only difference is $\Gamma$
%for \emph{XMM}--Newton data, where we found a flatter value ($1.59^{+0.10}_{-0.09}$). 
%We did not find any
%variations neither in the spectral parameters nor in the flux, most probably because \cite{younes2011} included the emission of the jet, what explains that they get a stepper $\Gamma$.

\emph{XMM}--Newton data were used to study short term variations.
From its PSD analysis, \cite{omairavaughan2012} did not find them in any
of the energy bands (soft, hard, total). From the light curve in the
0.5--10 keV energy band, \cite{younes2011} reported no
variations. We found
$\sigma^2_{NXS} > 0$ at 1.6$\sigma$ confidence level in the 2--10 keV
energy band, consistent with no variability.

At UV frequencies, \cite{younes2012} derived the luminosities from
the OM onboard \emph{XMM}--Newton with UVW2 and UVM2 filters, that
agree with our results. Variability can not be studied since OM data
are only  available at one epoch.

\subsection{NGC\,1052}

It is the brightest elliptical galaxy in the Cetus~I group. Previously
classed as a LINER in the pioneering work by \cite{heckman1980}, it was classified optically as a type 1.9 LINER \citep{ho1997}, and as an AGN candidate at X-ray frequencies
\citep{omaira2009a}. \emph{VLA} data show a core-dominated and a two-sided jet structure at
radio frequencies \citep{vermeulen2003}.

NGC\,1052 was observed twice with \emph{Chandra} and five times
with \emph{XMM}--Newton.  
Long term variability studies are not found in the literature. 
We find variations due to the nuclear power, $Norm_2$ (49\%) and the column density, $N_{H2}$ (31\%), both at hard energies, in an eight years period.

\cite{omairavaughan2012} studied short
term variations from the PSD with \emph{XMM}--Newton data with null results in any of the
energy bands. We analysed \emph{Chandra} and \emph{XMM}--Newton
light curves and variations were not found.
Short term variations were previously studied with other instruments;
\cite{guainazzi2000} studied \emph{BeppoSAX} data and did not
find short term variations. The most recent observation in X-rays
reported so far is a 100 ks observation taken with \emph{Suzaku} in
2007, the derived spectral characteristics reported by
\cite{brenneman2009} appear to be similar to those from
\emph{XMM}--Newton, which are compatible with the values in
\cite{omaira2009a}, \cite{brightman2011}, and this paper (intrinsic
luminosity of log(L(2-10 keV)) $\sim$ 41.5), and no
variations along the observation.

In the UV range, \cite{maoz2005} studied this galaxy with
\emph{HST} ACS and found a factor of 2 decrease in the flux of the source between the 1997 data reported by \cite{pogge2000} and
their 2002 dataset. We found a factor of
1.3 UV flux variations using \emph{XMM}-OM data in a seven months period.

\subsection{NGC\,1961}

NGC\,1961 is one of the most massive spiral galaxies known
\citep{rubin1979}. It was classified as a type 2 LINER by
\cite{ho1997}. \emph{MERLIN} and \emph{EVN} data
show a core plus two sided jet structure for this source at
radio frequencies \citep{krips2007}, what makes it a suitable AGN candidate.

This galaxy was observed once with \emph{Chandra} in 2010, and
twice with \emph{XMM}--Newton in 2011. X-ray variability from these
data was not studied before. We did not find variations in one month period.

No information in the UV is found for this object in the literature.

\subsection{NGC\,2681}

The nucleus of this galaxy was classified optically as a type 1.9 LINER
\citep{ho1997}. Classified as an AGN and as a
\emph{Compton}-thick candidate in X-rays \citep{omaira2009a,omaira2009b}, a
nuclear counterpart at radio frequencies has not been detected
\citep{nagar2005}.

The source was observed twice with \emph{Chandra} in January and
May 2001.
\cite{younes2011} did not find short time scale variations from the 
analysis of the
light curves neither long term variations from the spectral analysis.  
These results agree with our variability analysis.

At UV frequencies, no variations were found \citep{cappellari1999}.

\subsection{NGC\,2787}

The nucleus of NGC\,2787 is surrounded by diffuse emission extending up to
$\sim$ 30$\arcsec$ \citep{terashimawilson2003}. It was optically classified as a type 1.9 LINER \citep{ho1997}, and as an
AGN candidate at X-ray frequencies \citep{omaira2009a}.

\cite{nagar2005} detected a radio core with \emph{VLA}, while
evidence of a jet structure has not been found in the literature. Flux
variations were obtained at 2 and 3.6 cm on timescales of months
\citep{nagar2002}.

In X-rays, this galaxy was observed twice with \emph{Chandra} in
2000 (snapshot) and 2004, and once with \emph{XMM}--Newton in 2004.
\cite{younes2011}
found this to be a non-variable object at long timescales after
correcting \emph{XMM}--Newton data from contamination of X-ray sources. Due to
the high contamination from the extranuclear emission in
\emph{XMM}--Newton data, we have not performed a simultaneous fit for
this object.  

From one \emph{Chandra} light curve, \cite{younes2011} calculated an upper limit of $\sigma_{NXS}^2$ with which our value agrees.

No UV data are found in the literature.

\subsection{NGC\,2841}

\cite{ho1997} optically
classified NGC\,2841 as a type 2 LINER. It was classified as an AGN
candidate at X-ray frequencies by \cite{omaira2009a}. This galaxy shows some
X-ray sources in the surroundings \citep{omaira2009a}.
A core structure was found with
\emph{VLA} by \cite{nagar2005}, without evidence of any jet structure.

NGC\,2841 was observed twice with \emph{Chandra} in 1999
(snapshot) and 2004, and once with \emph{XMM}--Newton in 2004. We did
not use in the analysis the snapshot \emph{Chandra} data because it does not have
enough count rate for the spectral analysis. Moreover, 
since the extranuclear emission in \emph{Chandra} data
contributed with 60\% in the 0.5--10.0 keV energy band, we can not analyze the
spectral variations in this source. 
No information on variability is reported in the literature for this source.

\subsection{NGC\,3226}

NGC\,3226 is a dwarf elliptical galaxy that is strongly interacting
with the type 1.5 Seyfert NGC\,3227, located at 2' in
projected distance \citep[see Figure C.19
  in][]{omaira2009a}. NGC\,3226 was optically classified by \cite{ho1997}
as a type 1.9 LINER, and as an AGN candidate at X-ray frequencies by \cite{omaira2009a}. A
compact source is detected with \emph{VLA} \citep{nagar2005}, without
evidence of any jet structure.

This galaxy was observed twice with \emph{Chandra} in 1999 and
2001 and four times with \emph{XMM}--Newton from 2000 to 2006. 
The possible contamination of NGC\,3227 avoids the analysis of long term variations.
We refer the reader to HG13 for details on this subject.

We analysed one \emph{Chandra} light curve and we obtain $\sigma_{NXS}^2 > 0$ below 2$\sigma$, consistent with no short term variations. 

UV variations are not found in the literature. We found
11\% variations in the UVW1 filter from OM data.

\subsection{NGC\,3608}

NGC\,3608 is a member of the Leo II group, which forms a non
interacting pair with NGC\,3607. It was optically classified as a
type 2 LINER \citep{ho1997}. No hard nuclear point source was
detected in \emph{Chandra} images \citep{omaira2009a}, thus it was classified in X-rays as a non-AGN candidate, and also 
it appears to be a \emph{Compton}-thick
candidate \citep{omaira2009b}. A compact nuclear source at radio frequencies has not been detected \citep{nagar2005}.

This galaxy was observed once with \emph{Chandra} and twice with
\emph{XMM}--Newton in 2000 and 2012.  Variability studies are not found at any frequency in the literature. We did not find variations
in the 12 years period analyzed.

\subsection{NGC\,3718}

\begin{figure}[H]
\centering
\includegraphics[width=0.49\textwidth]{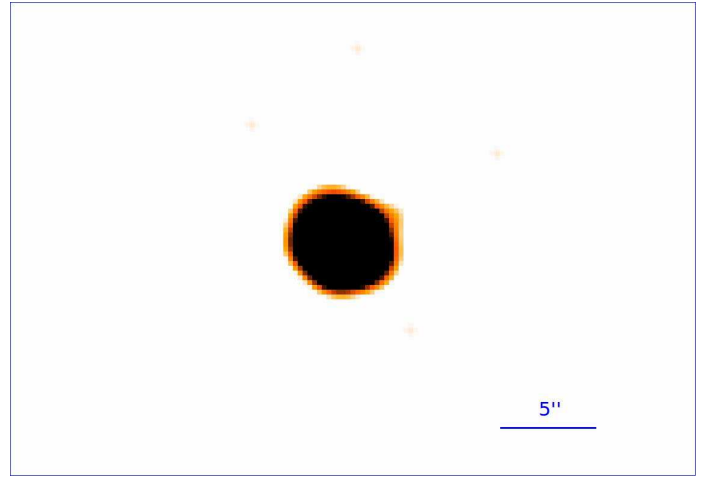}
\caption{\emph{Chandra} image in the 4.5--8.0 keV energy band of NGC\,3718, where a
  point like source can be distingueshed.}
\label{3718_458}
\end{figure}

NGC\,3718 has a distorted gas and a dusty disc, maybe caused by the
interaction with a close companion \citep{krips2007}. It was optically classified as a type 1.9 LINER \citep{ho1997}. It shows a
point like source in the 4.5--8.0 keV energy band (see Fig. \ref{3718_458}), and 
therefore we can classify it as an AGN candidate following \cite{omaira2009a}.

At radio frequencies, NGC\,3718 was observed with the \emph{VLA}
by \cite{nagar2005}, and with \emph{MERLIN} at 18 cm by \cite{krips2007},
where it shows a core and a compact jet. \cite{nagar2002}
reported radio variability at 2 cm with \emph{VLA} data, although the
result ``is not totally reliable".

In X-rays, this galaxy was observed once with \emph{Chandra} in
1999 and twice with \emph{XMM}--Newton in 2004.  \cite{younes2011}
studied all the available data for this object and report it 
as variable. When jointly fit
\emph{Chandra} and \emph{XMM}--Newton data, we found spectral
variations in $Norm_2$ (37\%). 

\cite{younes2011} did not find short term variations from the
analysis of the lightcurves. We did not analyse short term
variations because the lenght of the observations are $<$ 30 ksec.

At UV frequencies, \cite{younes2012} studied this galaxy with
\emph{XMM}--Newton but the nucleus was not detected, so they
estimated upper limits for the flux in one epoch.

\subsection{NGC\,4261}

\cite{ho1997} optically classified this galaxy as a type 2
LINER. \cite{omaira2009a} classified it as an AGN candidate at X-ray frequencies.
NGC\,4261 contains a pair of symmetric kpc-scale jets
\citep{birkinshawdavies1985} and a nuclear disc of dust roughly
perpendicular to the radio jet \citep{ferrarese1996}.

It was observed twice with \emph{Chandra}, in 2000 and 2008, and
with \emph{XMM}--Newton in another three epochs from 2001 to 2007.
Long term variability studies are not found in the literature.
We did not find variations in six years period.

\cite{sambruna2003} found variations of 3-5 ksec in the 2--10 keV
and 0.3--8.0 keV energy bands in the light curve from 2001, and
argued in favour of these variations being more related to the inner
X-ray jet than to an advection dominated accretion flow (ADAF), since
the expected timescale for the light crossing time of an ADAF was
$\sim$ 2 orders of magnitude longer than the observed variability
timescale. 
In HG13 we analysed the same observation in the 0.5--10
keV band, and reported it as non variable. However, we notice that 
$\sigma^2_{NXS} = 0.109 ^+_- 0.099$
at 1$\sigma$ confidence level.
In the present paper we did not analyse this light curve
since the net exposure time is shorter than 30 ksec.
Other light curves were studied. \cite{omairavaughan2012} did not
find short term variations from the PSD analysis of \emph{XMM}--Newton data. In the present
study we analysed two \emph{Chandra} observations and we can not
confirm rapid variations in this source, since upper limits for the
$\sigma^2_{NXS}$ were obtained in both cases.

No information from the UV is found in the literature.
We found 10\% (33\%) variation in the UVW1(UVW2) filter.

\subsection{NGC\,4278}

The north-northwest side of NGC\,4278 is heavily obscured by
large-scale dust-lanes, whose distribution show several dense
knots interconnected by filaments \citep{carollo1997}.
It is an elliptical galaxy with a relatively weak, broad $H_{\alpha}$
line, that made \cite{ho1997} classify it optically as a type 1.9 LINER.
It was
classified at X-ray frequencies as an AGN candidate \citep{omaira2009a}.

A two-sided jet is observed at radio frequencies wih \emph{VLBA}
and \emph{VLA} \citep{giroletti2005}. \cite{nagar2002} reported
radio variability at 2 and 3.6 cm with \emph{VLA} data. However, these
results ``are not totally reliable".

In X-rays this galaxy was observed in nine occasions with
\emph{Chandra} from 2000 to 2010 and once with \emph{XMM}--Newton in
2004. \cite{brassington2009} used six \emph{Chandra} observations
and found 97 variable sources within NGC\,4278, in a 4' elliptical area
centered on the nucleus, none of them within the
aperture we used for the nuclear extraction.
\cite{pellegrini2012} studied \emph{Chandra} observations of
NGC\,4278 and found an X-ray luminosity decrease by a factor of $\sim$
18 between 2005 and 2010. \cite{younes2010} detected a factor of $\sim$ 3 flux increase on
a timescale of a few months and a factor of 5 variation between the faintest and
the brightest observations (separated by about three years). We used
three of these observations (others were affected by pileup or did not met the minimum count number), our spectral fittings being 
in good
agreement with theirs, although we found weaker variations in
luminosities. 

Whereas the different \emph{Chandra} observations did
not show short timescale variability, during the \emph{XMM}--Newton
observation \cite{younes2010} found a 10\% flux increase in few
hours. With the same dataset, HG13 obtained a 3\%
variation in the same time range, the difference being most probably
due to the different apertures used for the analysis (10$\arcsec$ vs 25$\arcsec$).

In the UV, \cite{cardullo2008} found that the luminosity
increased a factor of 1.6 in about six months using data from
\emph{HST} WFPC2/F218W.

\subsection{NGC\,4374}

NGC\,4374 is one of the brightest giant elliptical galaxies in the
center of the Virgo cluster. Optically classified as a type type 2
LINER \citep{ho1997}, at X-ray frequencies it is a \emph{Compton}-thick AGN
candidate \citep{omaira2009b,omaira2009a}.

At radio frecuencies, it shows a core-jet structure, with two-sided
jets emerging from its compact core \citep{xu2000}. \cite{nagar2002}
reported flux variations at 3.6 cm with \emph{VLA}, and
variations at 2 cm which ``are not fully reliable".

This galaxy was observed four times with \emph{Chandra}, twice in
2000 (ObsID 401 is a snapshot) and twice in 2005, and once with
\emph{XMM}--Newton in 2011.  No information about variability in X-ray or UV
is found in the literature. Here we report large 
variations at hard energies (73\% in $Norm_2$).

We analysed two \emph{Chandra} light curves, one of them with $\sigma_{NXS}^2 > 0$ below 2$\sigma$ confidence level, which is compatible with no variations.

\subsection{NGC\,4494}

NGC\,4494 is a elliptical galaxy located in the Coma I cloud. It was optically classified as a type 2 LINER \citep{ho1997}, and at
X-rays as an AGN candidate \citep{omaira2009a}. The nucleus of this
galaxy was not detected in radio with \emph{VLA} data
\citep{nagar2005}.

This galaxy was observed twice with \emph{Chandra} in 1999
(snapshot) and 2001, and once with \emph{XMM}--Newton in 2001.
Variability analyses are not 
found in the literature. We report the source as variable at X-ray frequencies.

\subsection{NGC\,4636}

NGC\,4636 was optically classified as a type 1.9 LINER by \cite{ho1997}. At 
X-rays it 
does not show emission at hard energies and therefore it was classified as a
non-AGN candidate \citep{omaira2009a}. It was also classified as a
\emph{Compton}-thick candidate \citep{omaira2009b}.

At radio frequencies, it shows a compact core with \emph{VLA} data \citep{nagar2005}. Recently, \cite{giacintucci2011} found bright jets at radio
frequencies. No variations were found at 2 cm with \emph{VLA}
data \citep{nagar2002}.

This galaxy was observed four times with \emph{Chandra} data
between 1999 and 2003, and three times with \emph{XMM}--Newton between
2000 and 2001.  \cite{osullivan2005} studied the X-ray morphology
of the galaxy and suggest that it can be the result of a past
AGN that is actually quiescent. Long term variations were not found in the present analysis.

\cite{omairavaughan2012} did not
find short term variations from the analysis of \emph{XMM}--Newton
light curves. From one \emph{XMM}--Newton light curve, we found
$\sigma^2_{NXS} > 0$ at 1.4$\sigma$ confidence level.  We did not find long
term variations.

No UV variability studies are found in the literature. Our analysis let us conclude that it is
variable at UV frequencies.

\subsection{NGC\,4736}

NGC\,4736 is a Sab spiral galaxy, member of the Canes Venatici I (CVn
I) Cloud \citep{deVaucouleurs1975}. Optically classified as a
type 2 LINER \citep{ho1997}, it is an AGN candidate at X-ray frequencies
\citep{omaira2009a}. \cite{nagar2005} reported an unresolved
nuclear source at its nucleus, using 0.15$\arcsec$ resolution
\emph{VLA} data, without evidence of any jet structure.

This galaxy was observed three times with \emph{Chandra} between
2000 and 2008 and three times with \emph{XMM}--Newton between 2002 and
2006.  
No long term variability information is found in the literature.
In the present work we did not find any variation in four years period.

It harbors a plethora of discrete X-ray sources in and around
its nucleus (see Appendix \ref{Ximages}). \cite{eracleous2002} studied \emph{Chandra} data
from 2000. They found a very dense cluster of ten discrete
sources in the innermost 400 $\times$ 400 pc of the galaxy. They
studied the brightest four sources (namely X-1 to X-4), and found
that spectra are well described by a single power law with photon
indices in the range 1.1--1.8, and 2--10 keV luminosities between $4-9
\times 10^{39}$ erg $s^{-1}$. They also studied short timescale
variability from the analysis of the light curves. They estimated
the normalised excess variance ($\sigma^2 = 0.06 ^+_-0.04$) of the
nucleus of NGC\,4736 (X-2), and reported it as variable. The
other sources also showed short-term variations (see Table 5 in
\citealt{eracleous2002}). They argued that there is no evidence for the
presence of an AGN, and concluded that this LINER spectrum could be the result either from a
current or recent starburst or from an AGN.
However, they noted that X-2 is the only source
with an UV counterpart, detected by \emph{HST}. \cite{omaira2009a} assigned X-2 to the nucleus of the galaxy, since it coincides
with the 2MASS near-IR nucleus within 0.82$\arcsec$.

By studying \emph{BeppoSAX} and \emph{ROSAT} data,
\cite{pellegrini2002} excluded variations of the 2--10 keV flux
larger than $\sim $50\% on time scales of the order of one day.
Comparing data from both instruments, they did not find variations
between 1995 and 2000. They concluded that the X-ray emission is due to
a recent starburst in NGC 4736. However, they mentioned that an
extremely low luminosity AGN could still be present, though, due to
the presence of a compact non-thermal radio source coincident with an
X-ray faint central point source.

\cite{omairavaughan2012} studied the PSD of the
\emph{XMM}--Newton data from 2006 and found no short term
variations.

We analysed \emph{Chandra} and \emph{XMM}--Newton light
curves. Variations were found but in all cases below 2$\sigma$
confidence level, in agreement with
\cite{eracleous2002}.

At UV frequencies, \cite{maoz2005} found long term variations
between 1993 and 2003, the nucleus being 2.5 times brighter in
2003. From
the OM data, we found variations of 66\% in the U filter between 2002
and 2006.

\subsection{NGC\,5195}

\begin{figure}[H]
\centering
\includegraphics[width=0.49\textwidth]{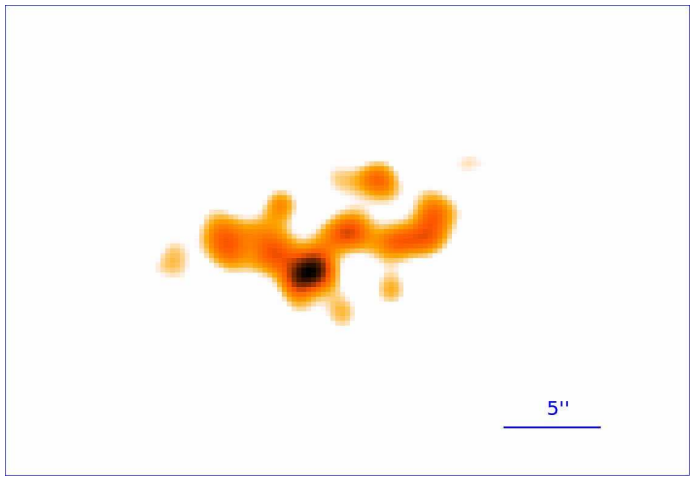}
\caption{\emph{Chandra} image of NGC\,5195 in the 4.5--8.0 keV energy band, where a
  point like source can be distingueshed.}
\label{5195_458}
\end{figure}

NGC\,5195 is tidally interacting with a companion SB0 galaxy NGC\,5194
(M51). It was optically classified as a type 2 LINER by
\cite{ho1997}. It shows a
point like source in the 4.5--8.0 keV energy band (see Fig. \ref{5195_458}), and 
therefore we can classify it as an AGN candidate following \cite{omaira2009a}. A radio
counterpart was found by \cite{houlvestad2001} with \emph{VLA}
data at 6 and 20 cm, without any jet indications.

This source was observed eight times with \emph{Chandra} between
2000 and 2012, and five times with \emph{XMM}--Newton between 2003 and
2011. \cite{terashimawilson2004}
studied \emph{Chandra} data from 2000 and 2001 and found neither long
term, nor short term variability in the full, soft, nor hard energy
bands.
Since the \emph{Chandra} observations from 2000 and 2001 were
rejected from our sample because of the low number counts, we can not
compare our spectral fittings with theirs. However, our estimation 
of the
luminosity in \emph{Chandra} data agrees with their results. We
found this object to be variable on long timescales, whereas
short term variations were not detected.

At UV frequencies, no references are found in the literature.
We found variations in the UVW1 filter.

\subsection{NGC\,5813}

NGC\,5813 is one of the galaxies in the group catalogue by
\citet{deVaucouleurs1975}, with NGC\,5846 being the brightest member
of the group. It was classified as a type 2 LINER by
\cite{ho1997}. The X-ray morphology is extremely diffuse, with very
extended emission at softer energies and without emission above 4
keV, what made \cite{omaira2009a} classify it as a non-AGN
candidate. It was also classified as \emph{Compton}-thick candidate
\citep{omaira2009b}. At radio frequencies, it shows a compact core \citep{nagar2005} and a jet-like structure
\citep{randall2011}.

This source was observed nine times with \emph{Chandra} between
2005 and 2011, and three times with \emph{XMM}--Newton between 2005
and 2009.  Variability studies at X-ray and UV frequencies are not reported in the literature.
We did not find either long term or short term variations in X-rays. UV variations were found in the UVW1 filter.

\subsection{NGC\,5982}

NGC\,5982 is the brightest galaxy in the LGG 402 group, which is composed
by four members \citep{garcia1993}. Recently, \cite{vrtilek2013} using
\emph{GMRT} 610 MHz observations found a compact radio core in the
position of the source, indicating an AGN-like object; jets were not detected.

This galaxy was observed twice with \emph{XMM}-Newton in 2011 and
2012. Variability
studies are not reported in the literature.
We found variations in the nuclear power (50\%) in a one year period, whereas UV variations were not found.

\section{\label{Ximages}}

In this appendix we present the images from \emph{Chandra} (left) and \emph{XMM}-Newton (right) that were used to compare the spectra from these two instruments in the 0.5-10 keV band. Big circles represent \emph{XMM}--Newton apertures. Small circles in the figures to the left represent the nuclear extraction aperture used with \emph{Chandra} observations (see Table \ref{obs}). In all cases, the gray levels extend from twice the value of the background dispersion to the maximum value at the center of each galaxy.

\onecolumn

\begin{figure}[H]
\caption{ \label{images} Images for \emph{Chandra} data (left) and \emph{XMM}-Newton data (right) for the sources in the 0.5-10 keV band. Big circles represent \emph{XMM}--Newton datas apertures. Small circles in the figures to the left represent the nuclear extraction aperture used with \emph{Chandra} observations (see Table \ref{obs}). }
\centering
\includegraphics[width=\textwidth]{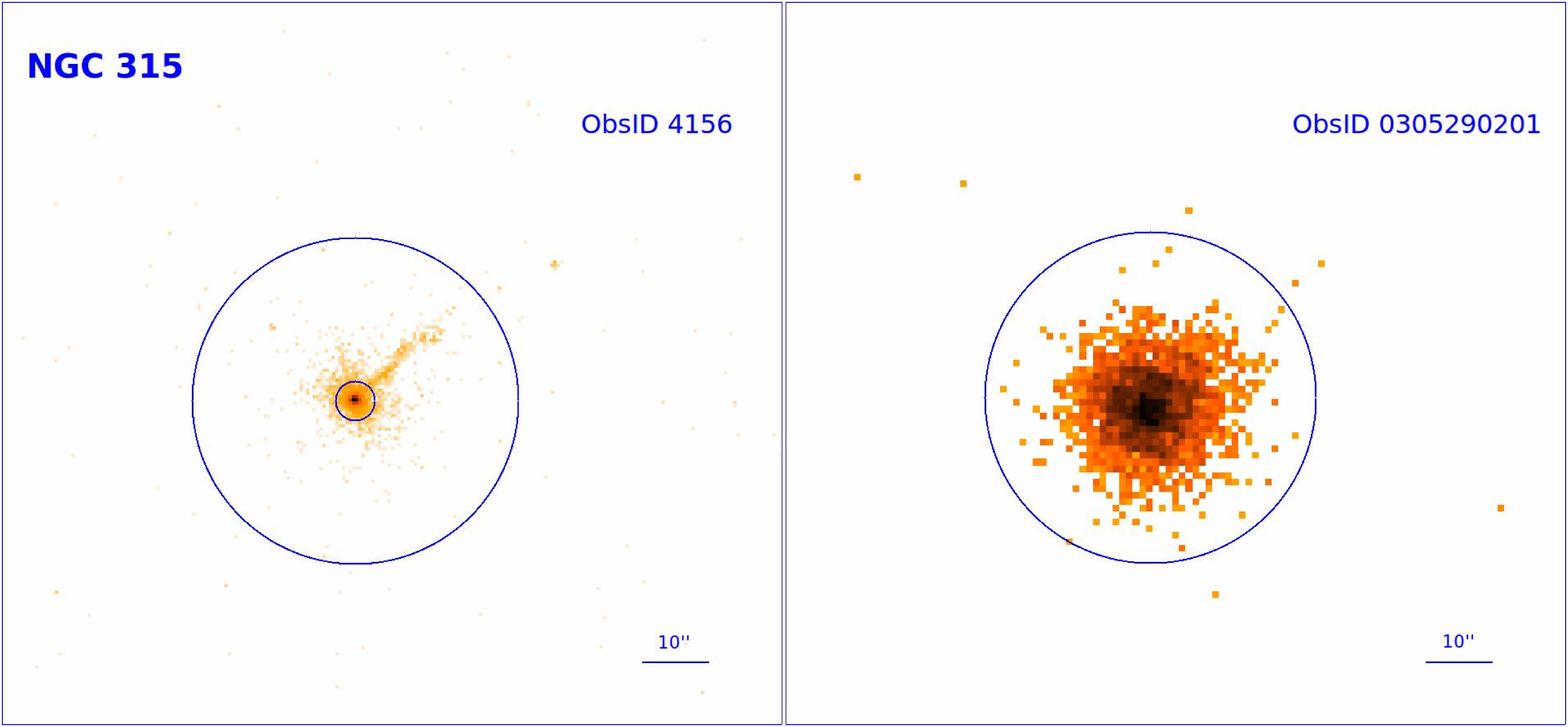}

\includegraphics[width=\textwidth]{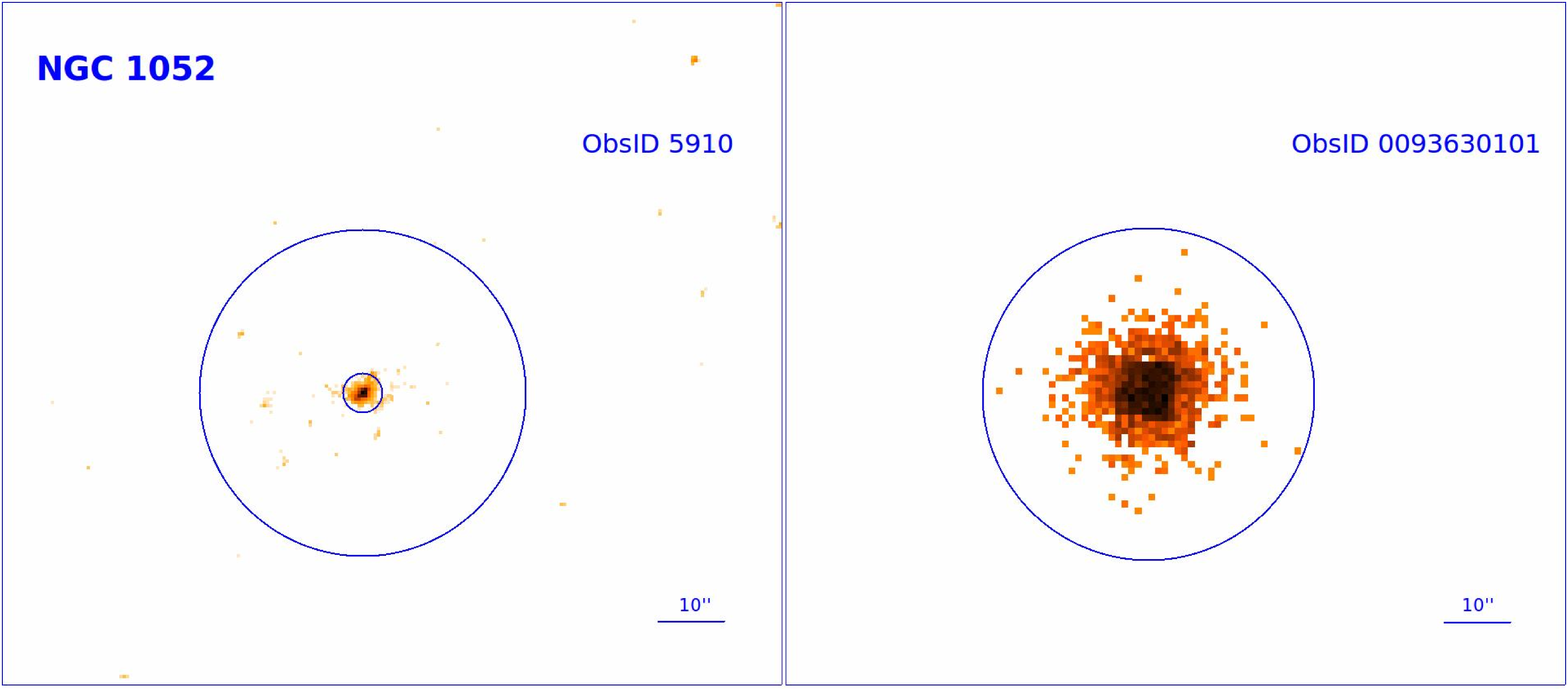}

\includegraphics[width=\textwidth]{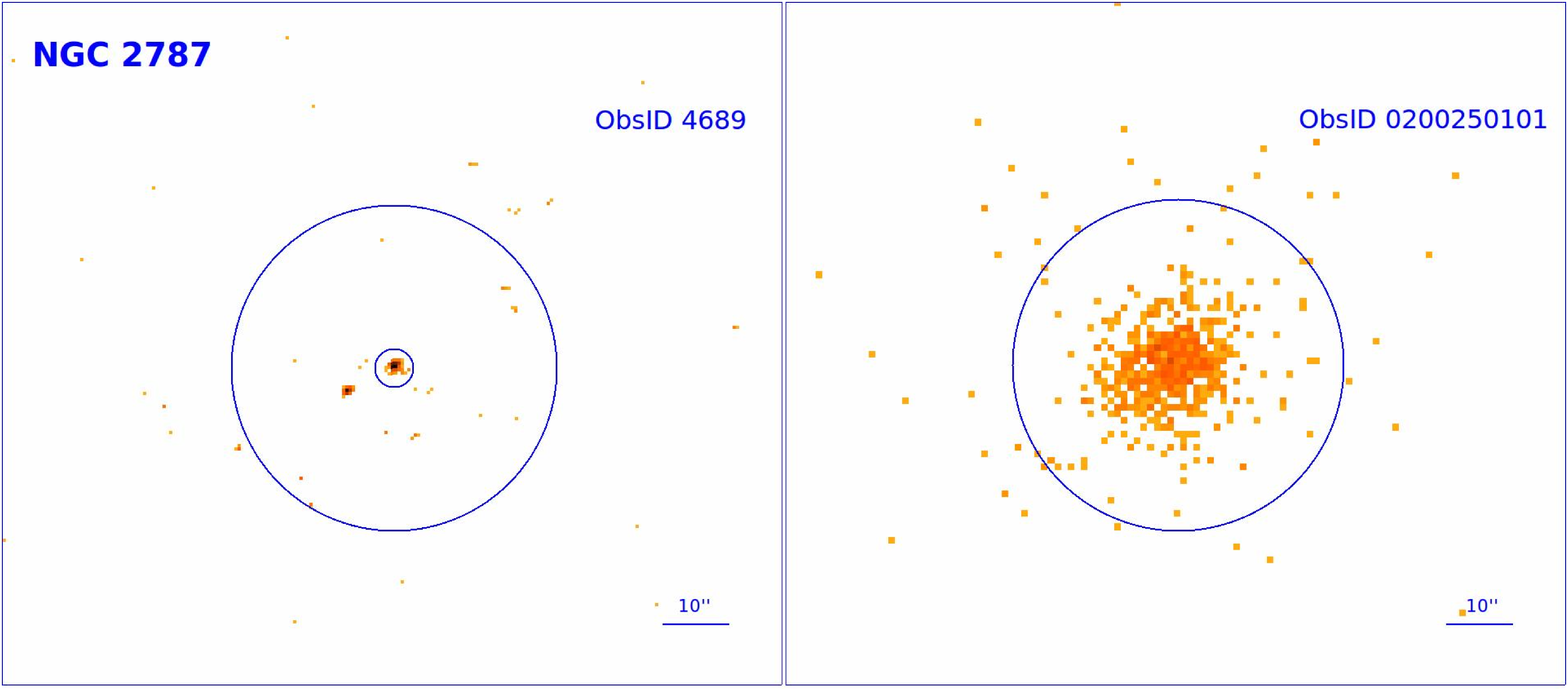}

\end{figure}

\begin{figure}[H]
\centering
\includegraphics[width=\textwidth]{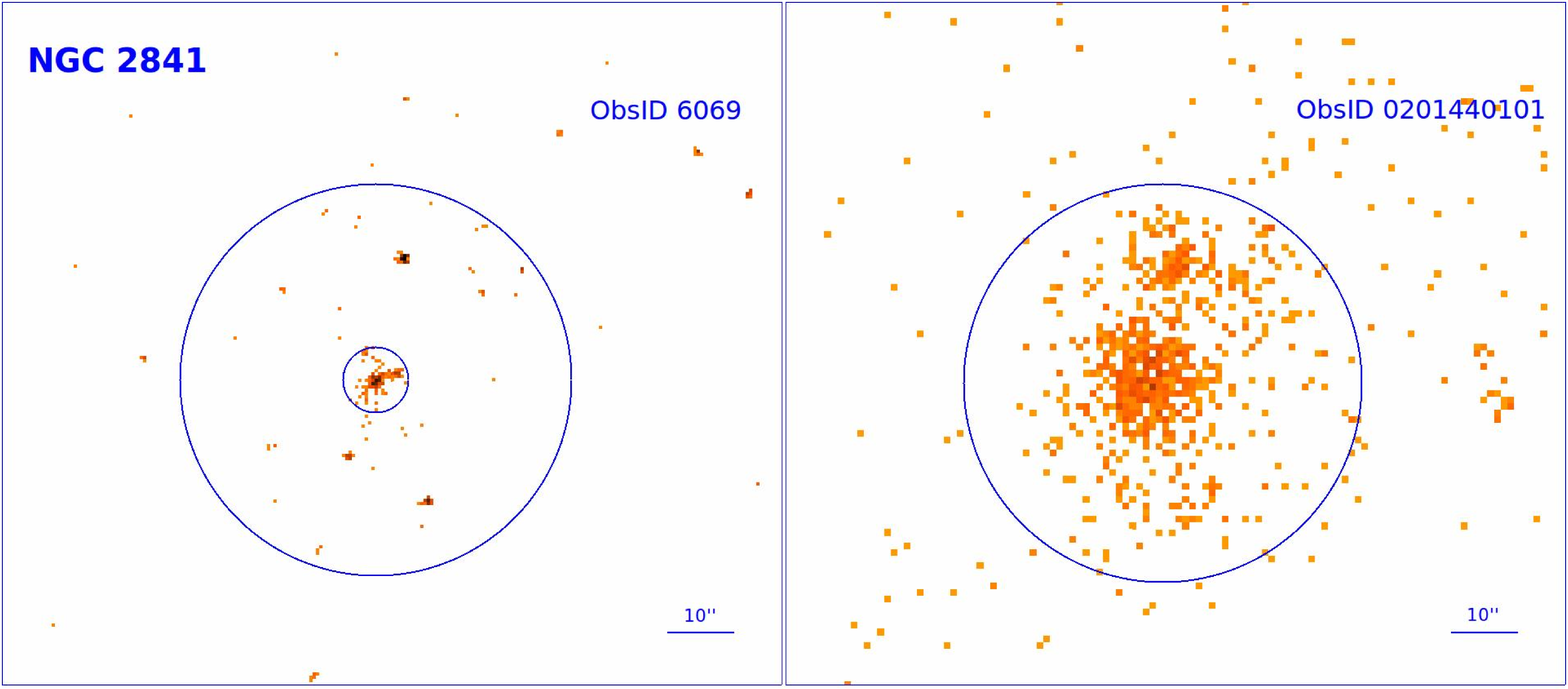}

\includegraphics[width=\textwidth]{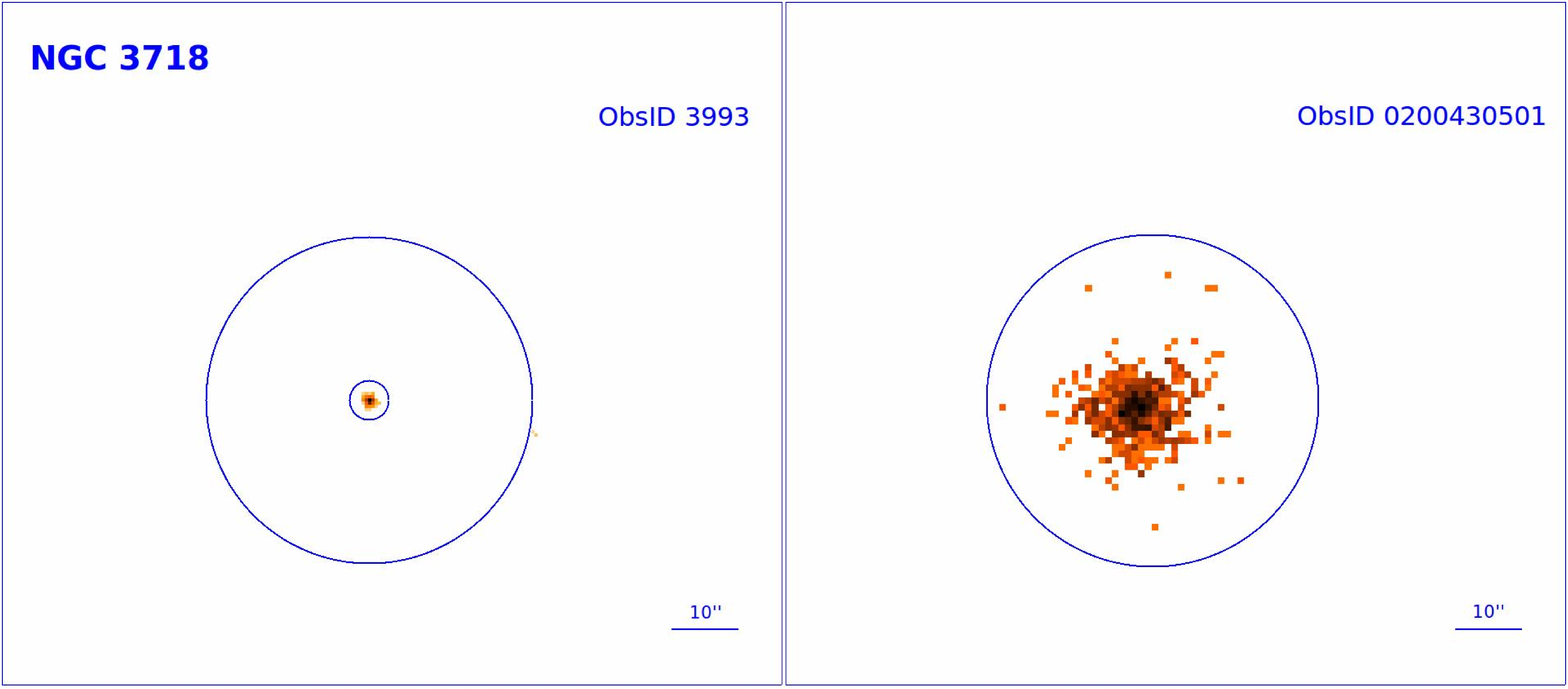}

\includegraphics[width=\textwidth]{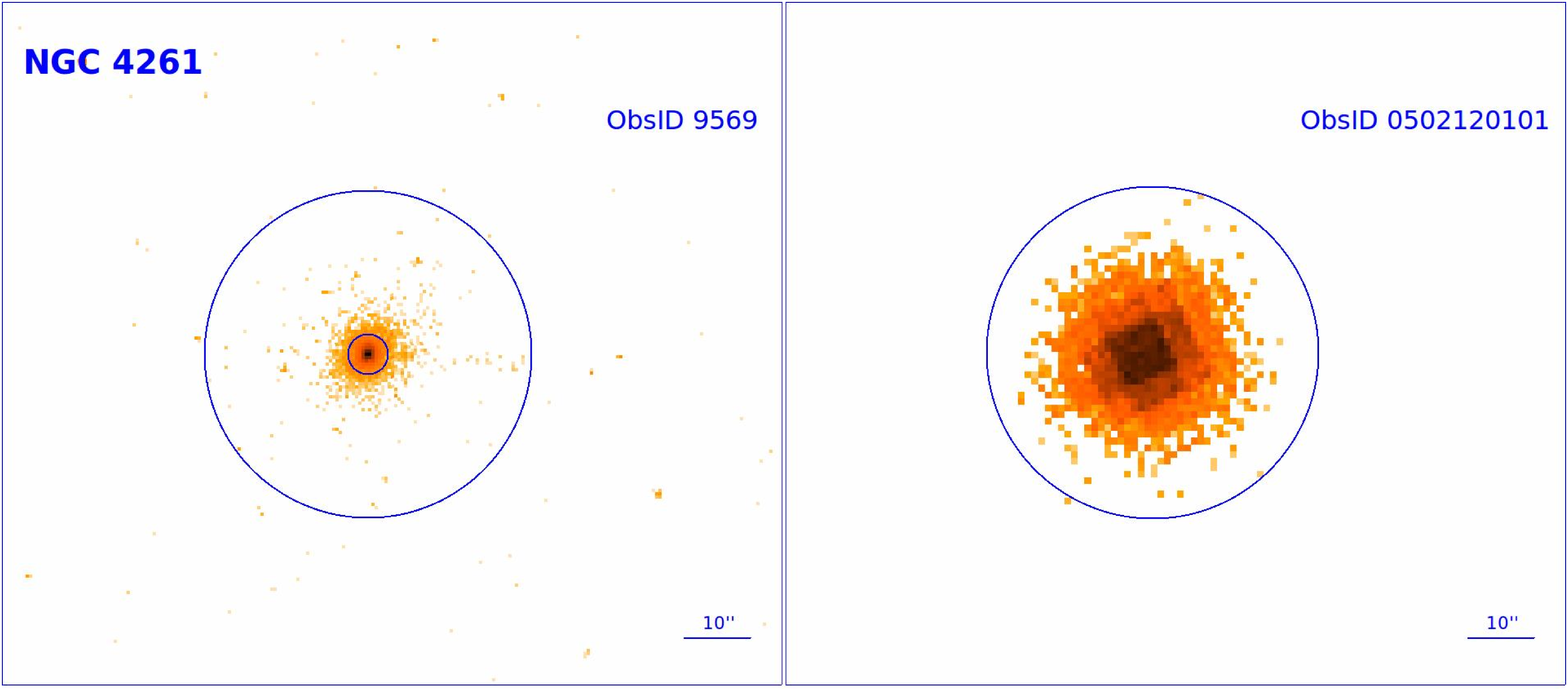}

\end{figure}

\begin{figure}[H]
\centering
\includegraphics[width=\textwidth]{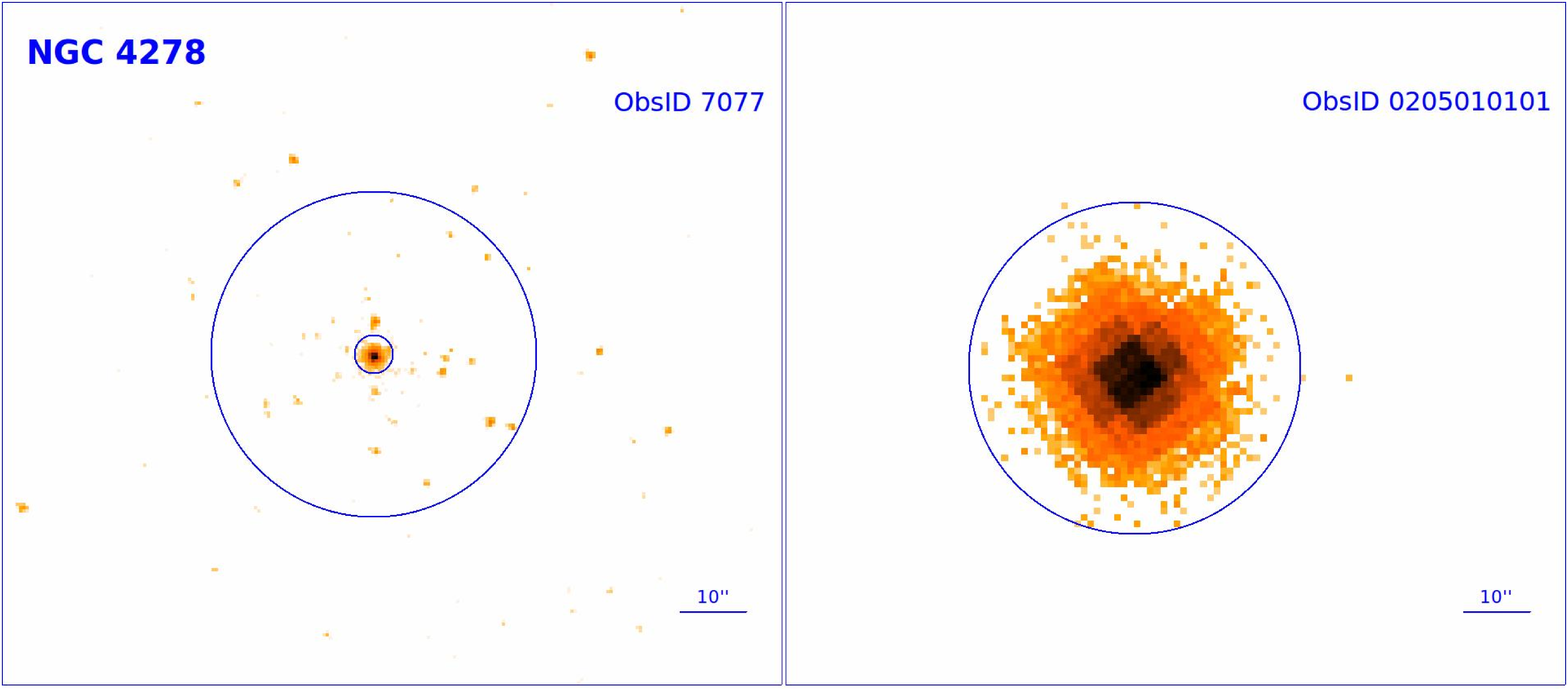}

\includegraphics[width=\textwidth]{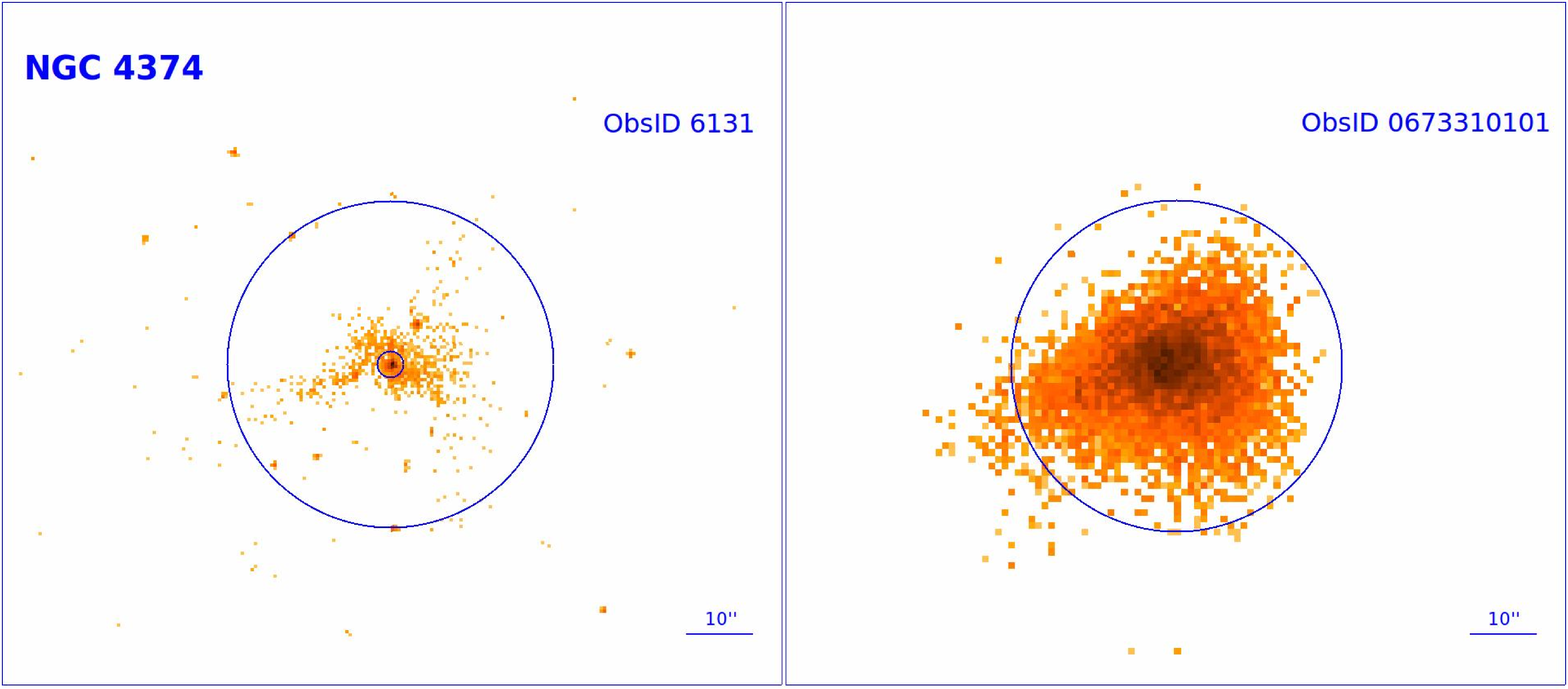}

\includegraphics[width=\textwidth]{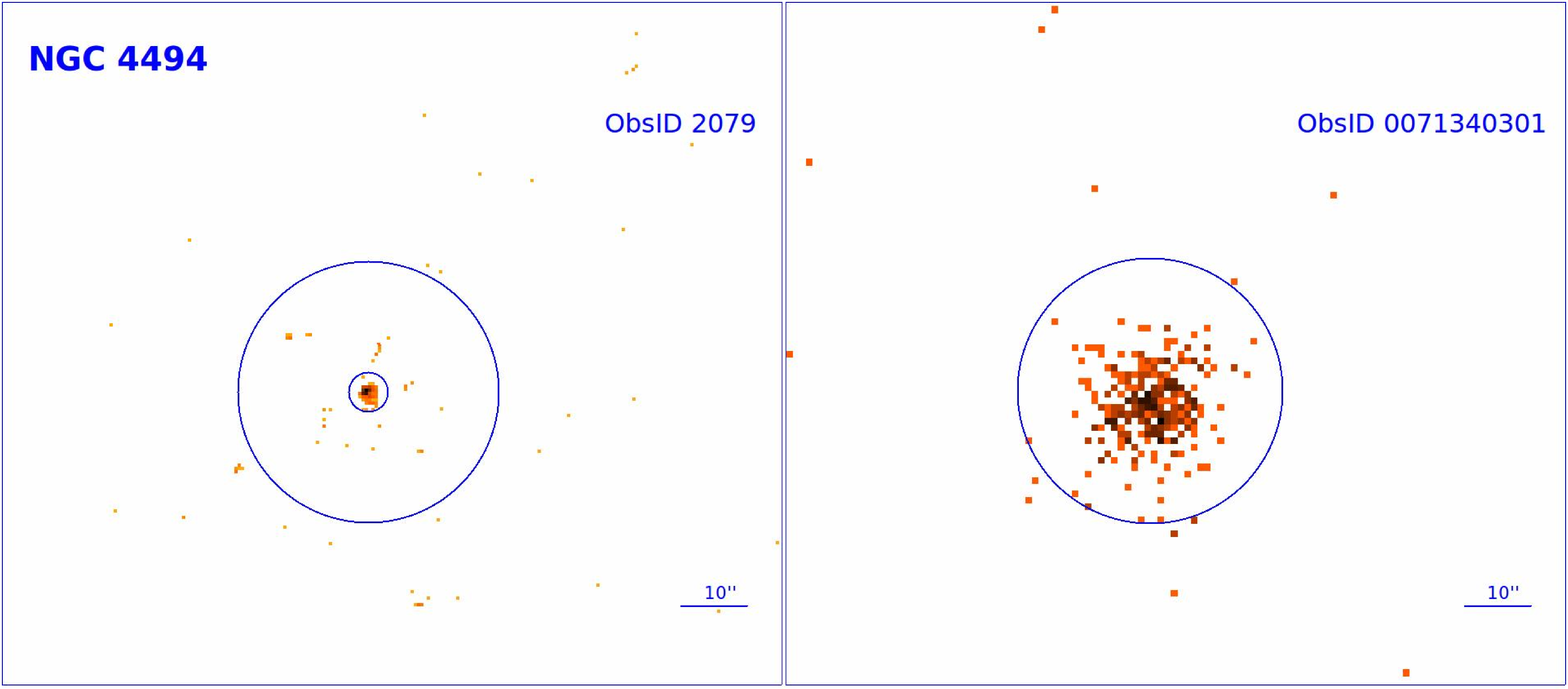}

\end{figure}

\begin{figure}[H]
\centering
\includegraphics[width=\textwidth]{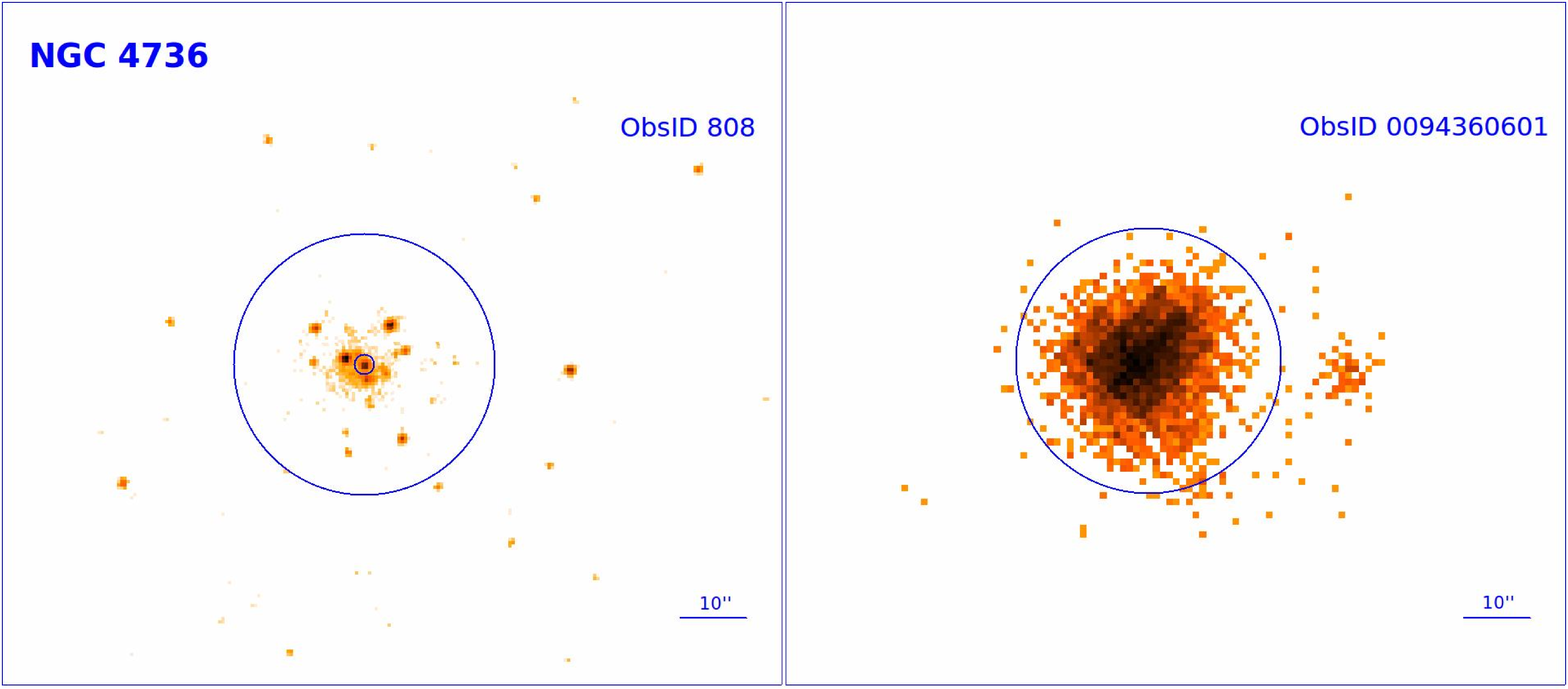}

\includegraphics[width=\textwidth]{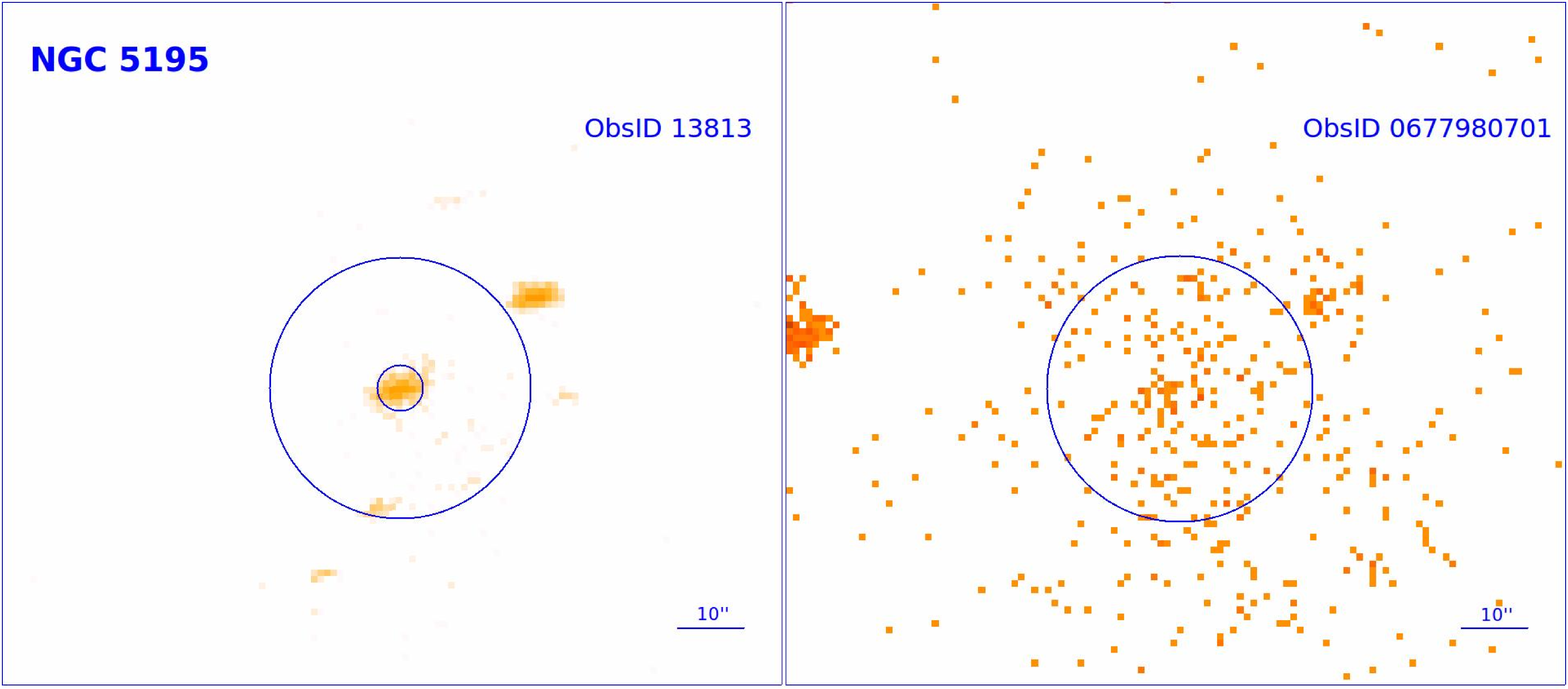}

\includegraphics[width=\textwidth]{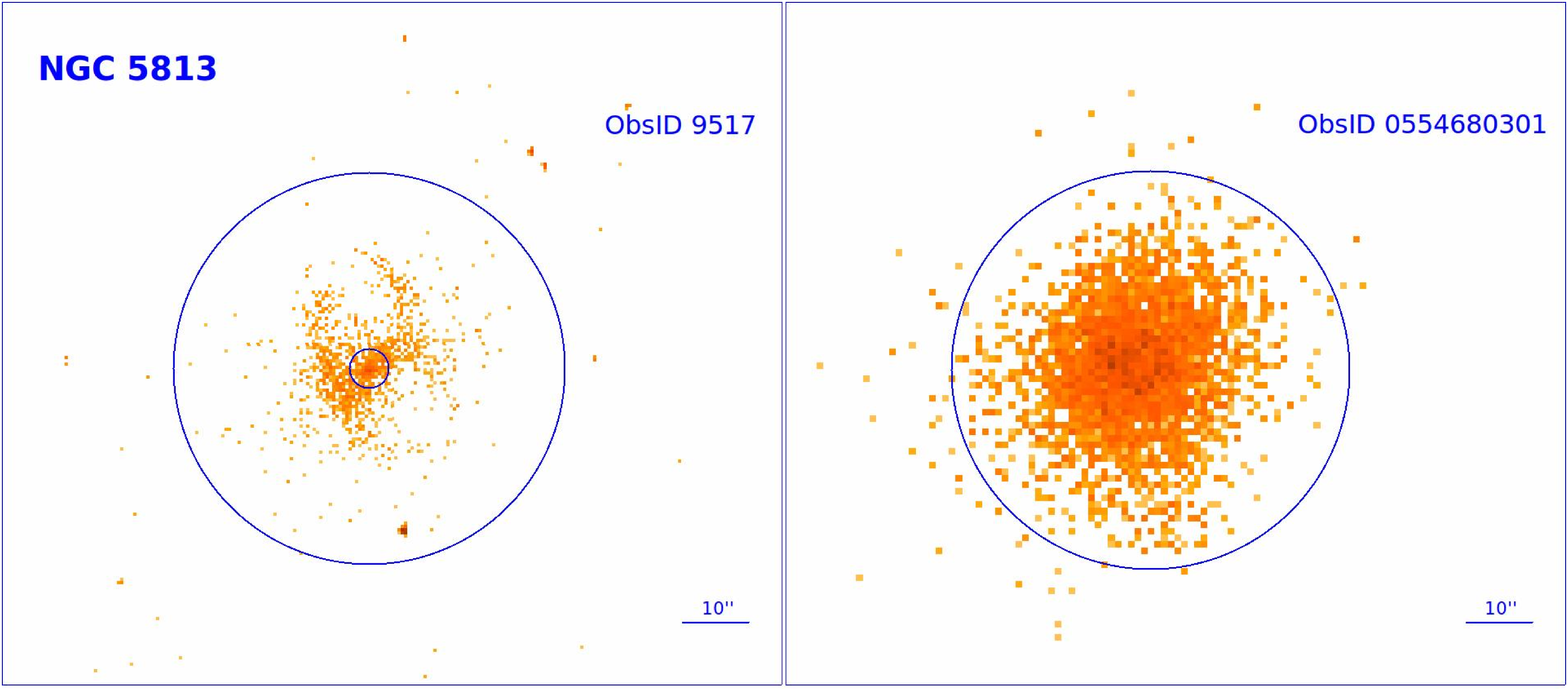}
\end{figure}

\section{\label{lightcurves} Light curves}

In this appendix the plots corresponding to the light curves are provided. Three plots per observation are presented, corresponding to soft (left), hard (middle), and total (right) energy bands. Each light curve has a minimum of 30 ksec (i.e., 8 hours) exposure time, whereas long light curves are divided in segments of 40 ksec (i.e., 11 hours). Each segment is enumerated in the title of the light curve. Count rates versus time continuous are represented. The solid line represent the mean value and dashed lines $^+_-$1$\sigma$ from the average.

\begin{figure}[H]
\centering
\subfloat{\includegraphics[width=0.30\textwidth]{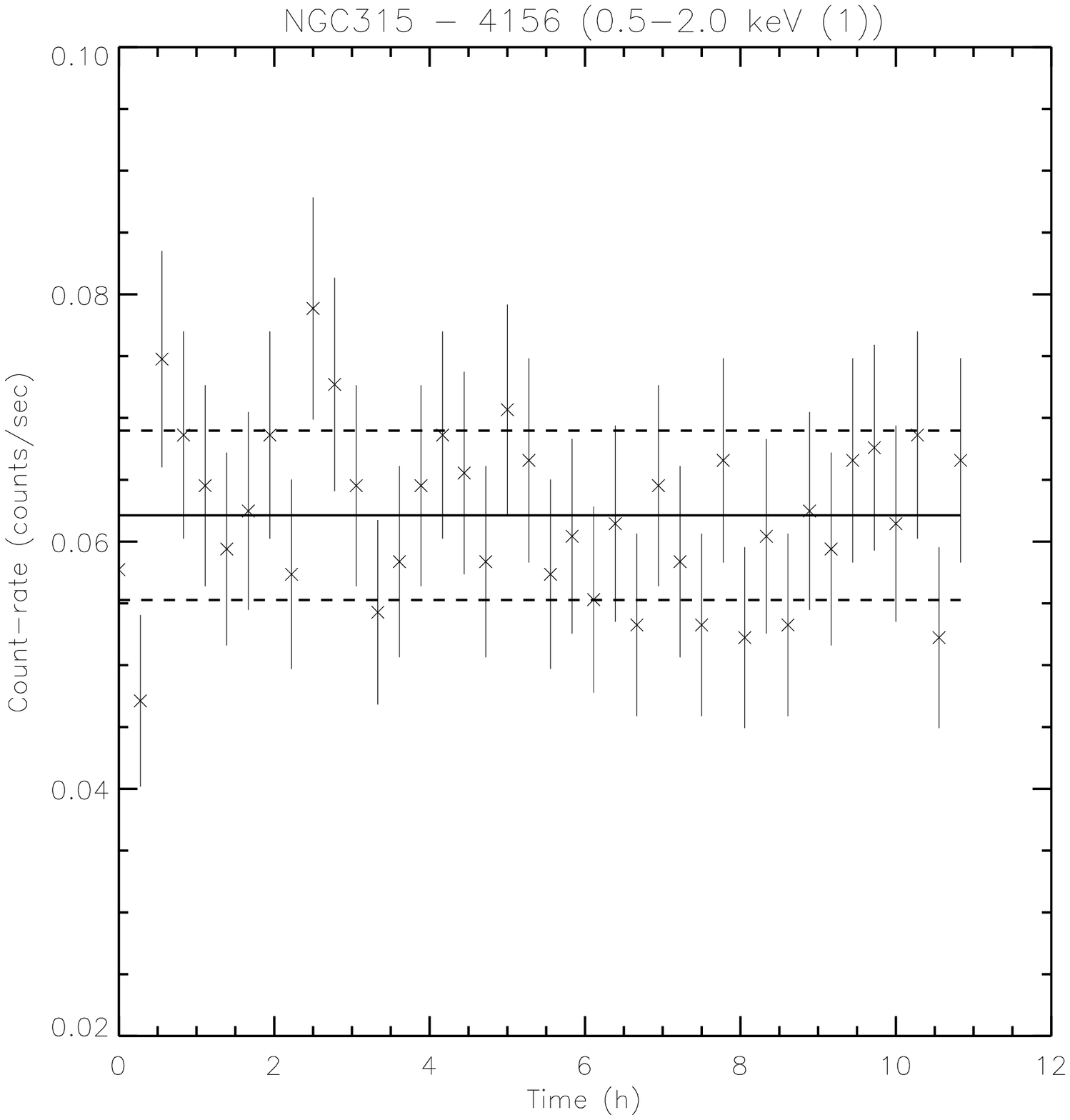}}
\subfloat{\includegraphics[width=0.30\textwidth]{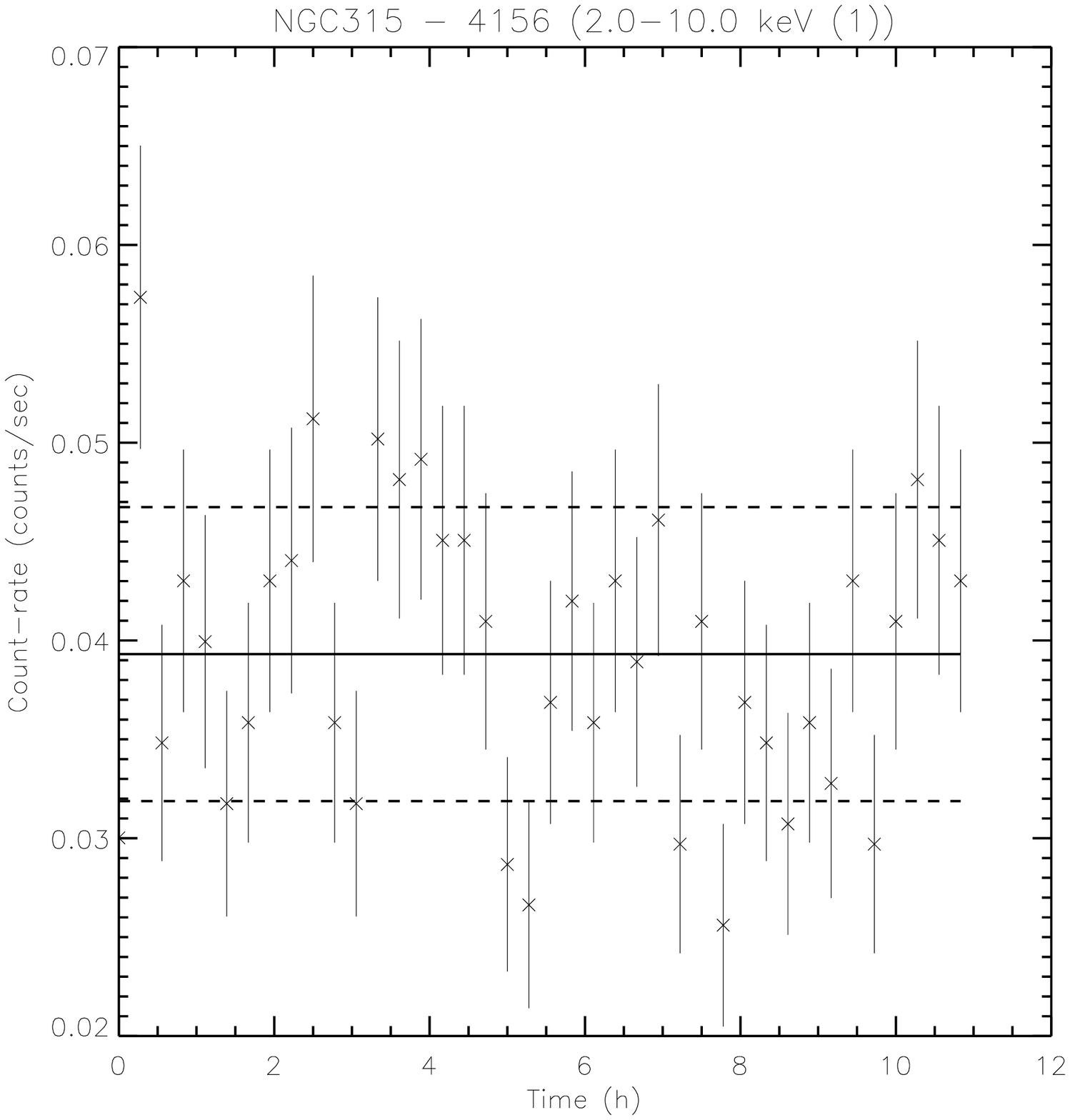}}
\subfloat{\includegraphics[width=0.30\textwidth]{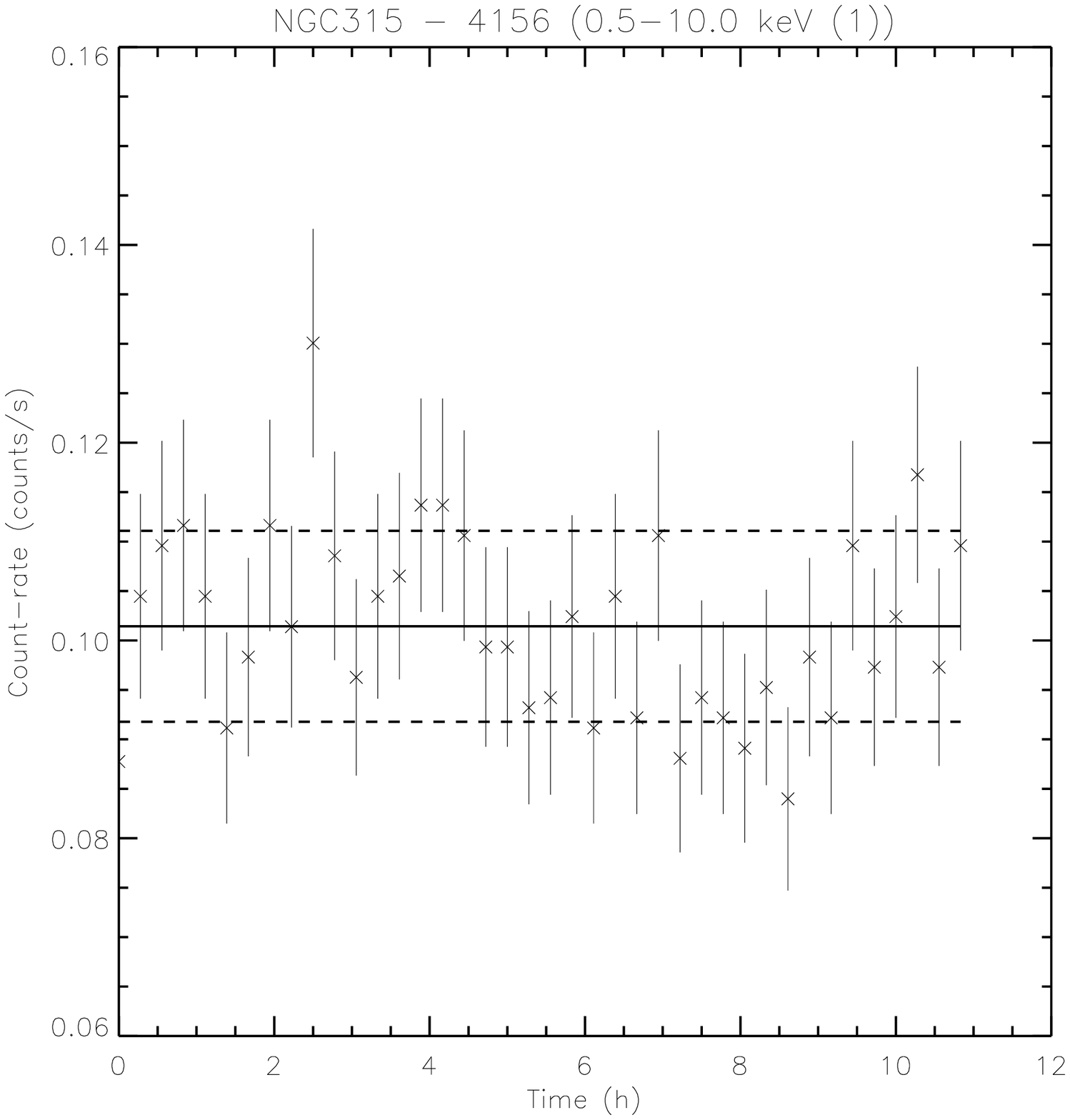}}
\caption{Light curves of NGC\,315 from \emph{Chandra} data.}
\label{l315}
\end{figure}

\begin{figure}[H]
\centering
\subfloat{\includegraphics[width=0.30\textwidth]{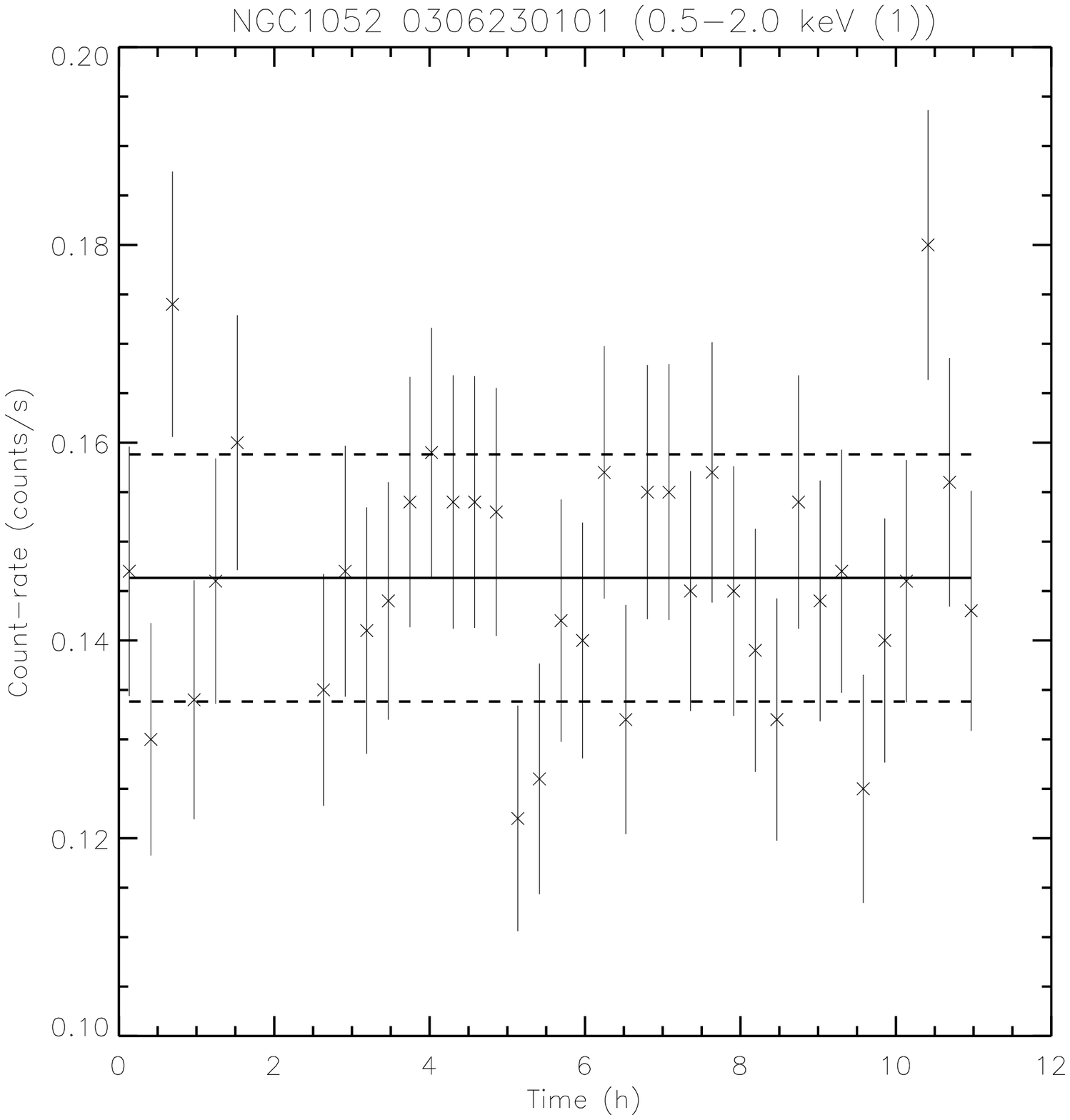}}
\subfloat{\includegraphics[width=0.30\textwidth]{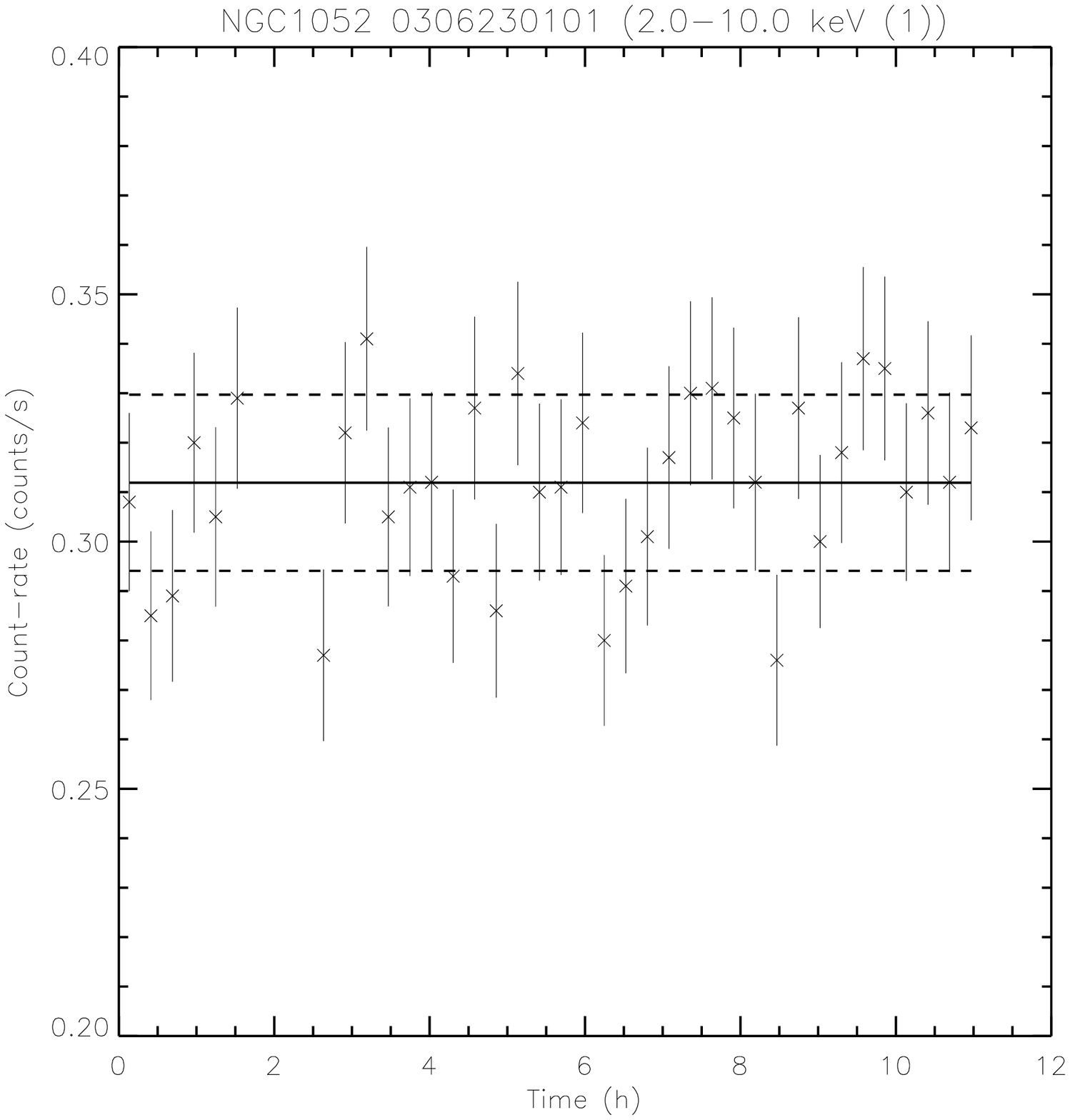}}
\subfloat{\includegraphics[width=0.30\textwidth]{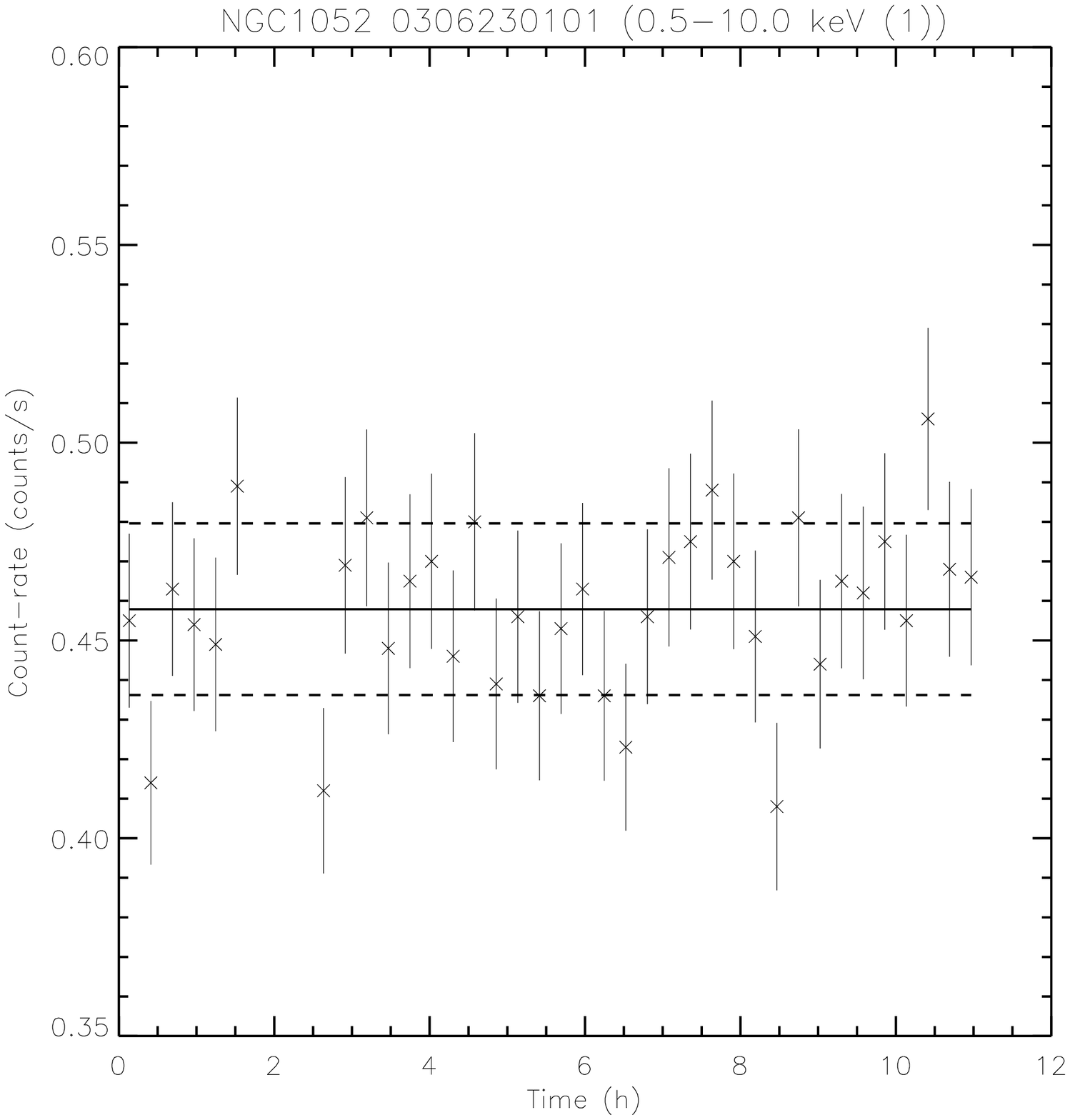}}

\subfloat{\includegraphics[width=0.30\textwidth]{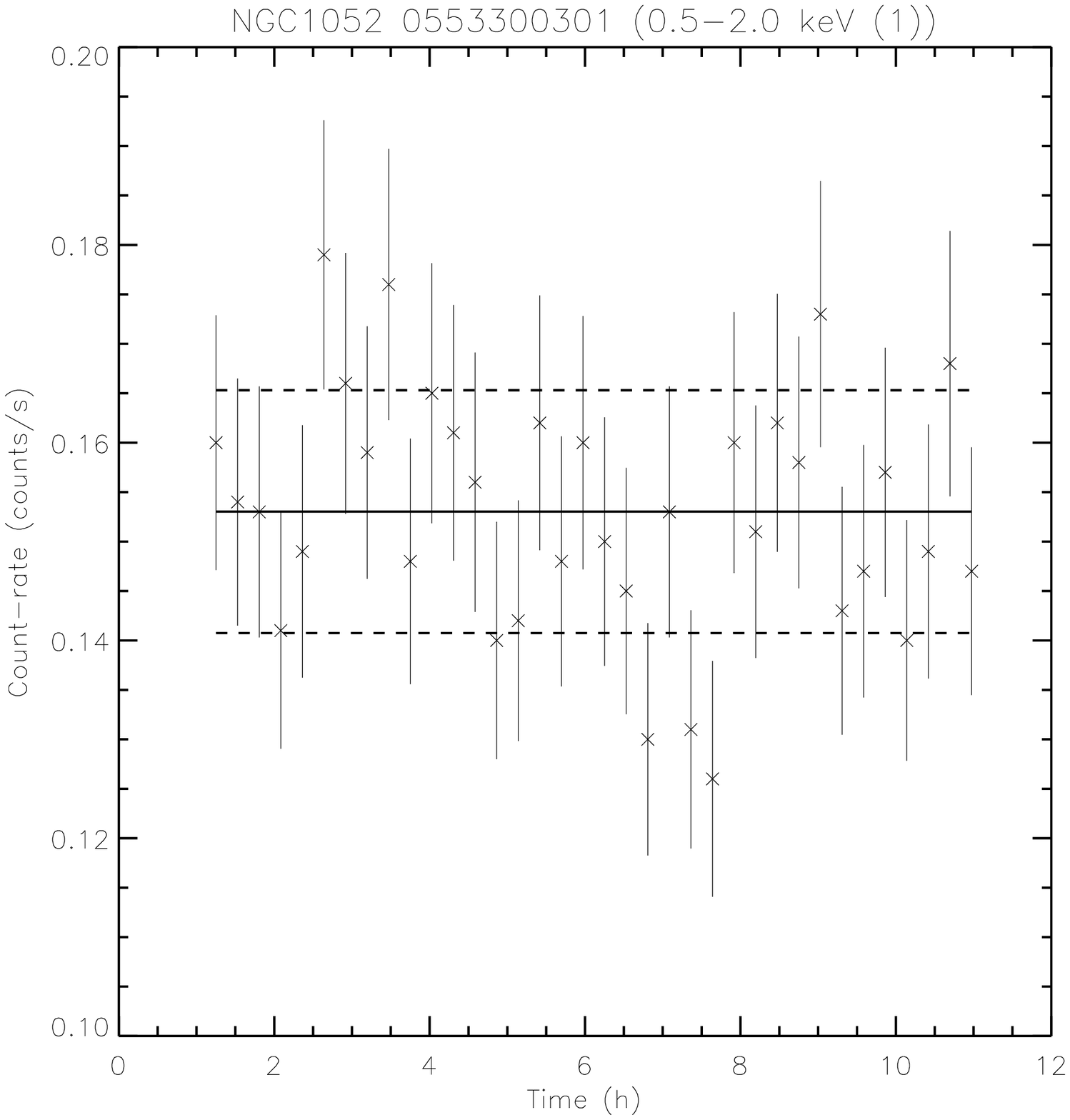}}
\subfloat{\includegraphics[width=0.30\textwidth]{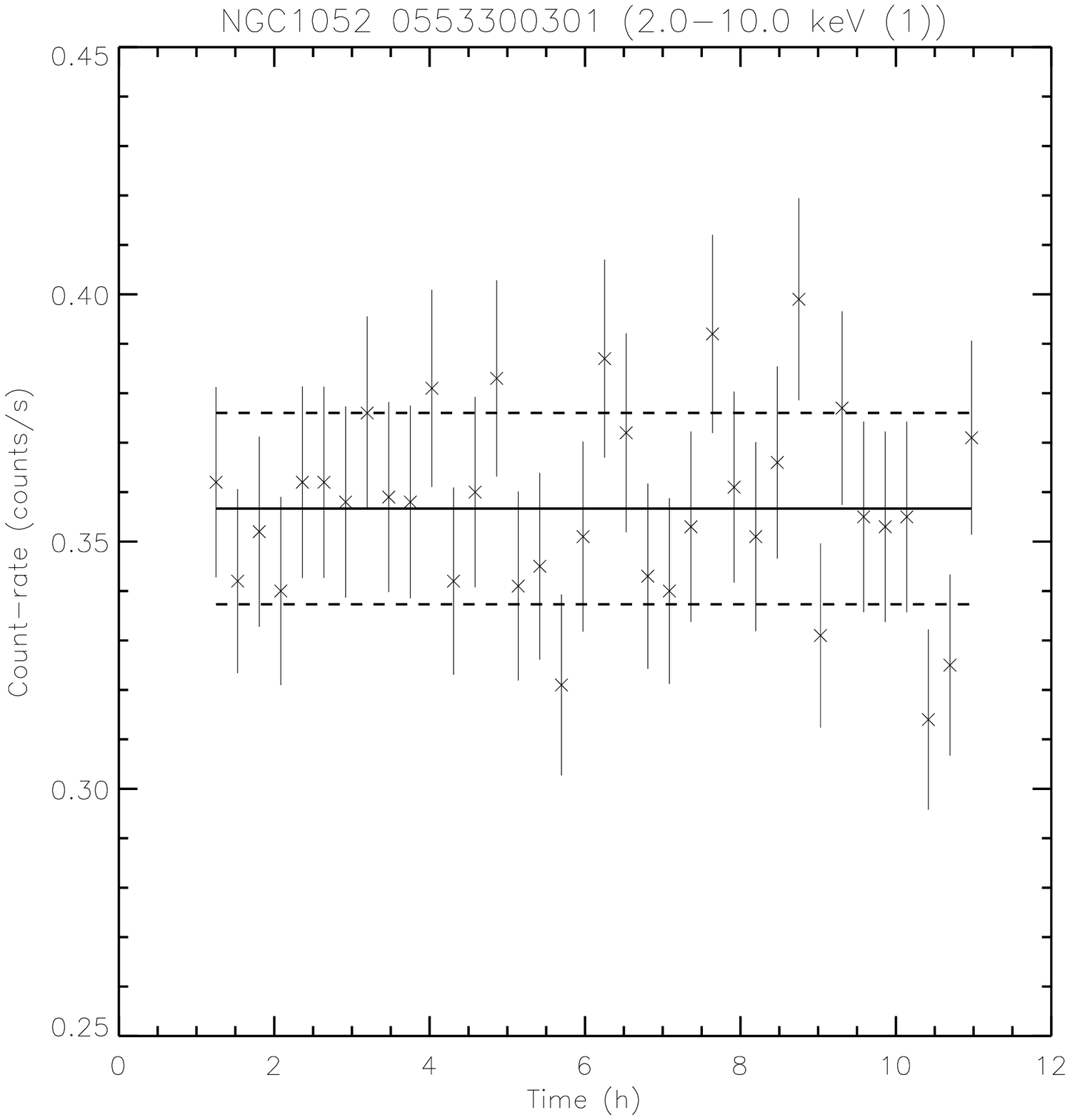}}
\subfloat{\includegraphics[width=0.30\textwidth]{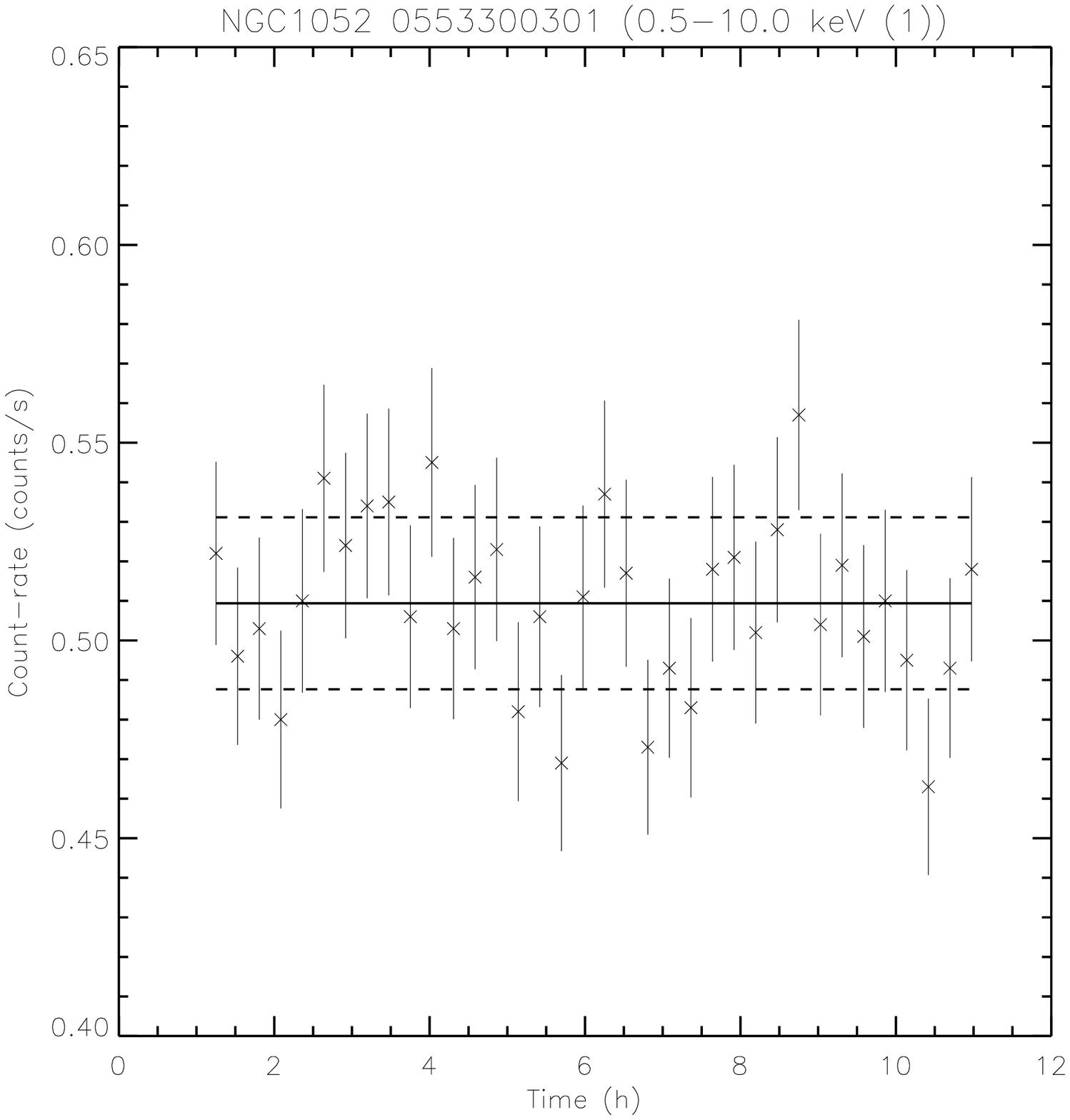}}

\subfloat{\includegraphics[width=0.30\textwidth]{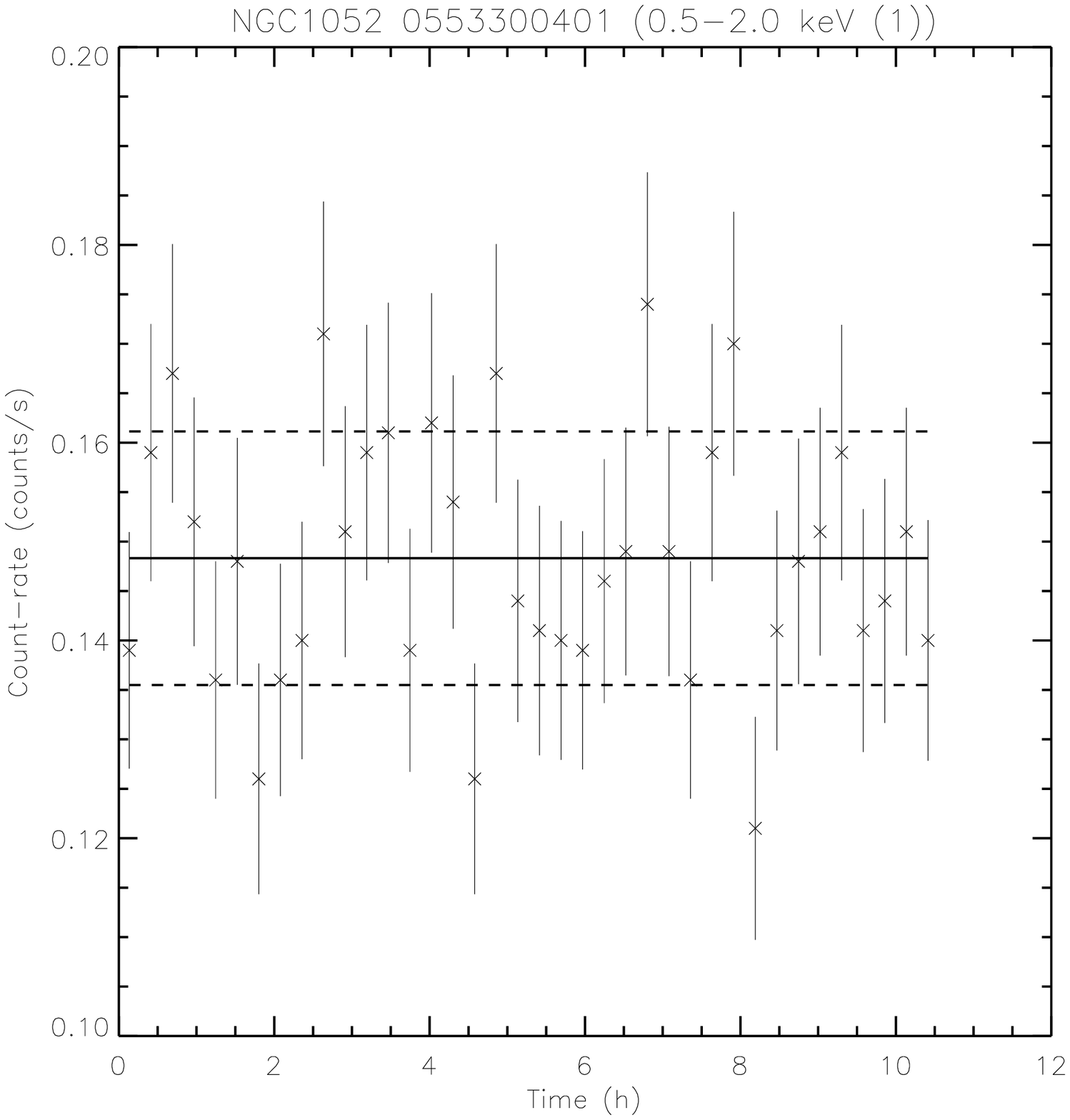}}
\subfloat{\includegraphics[width=0.30\textwidth]{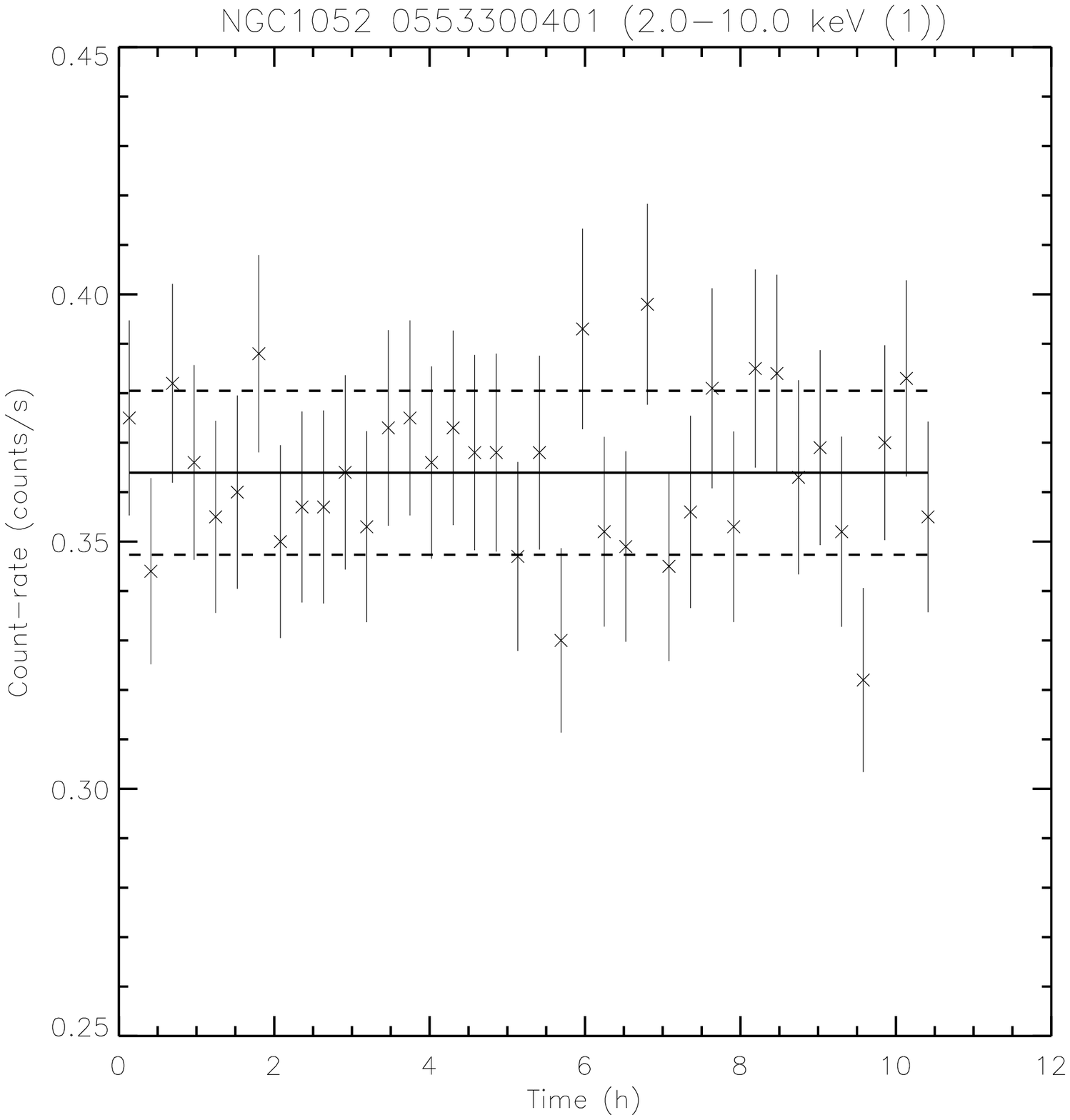}}
\subfloat{\includegraphics[width=0.30\textwidth]{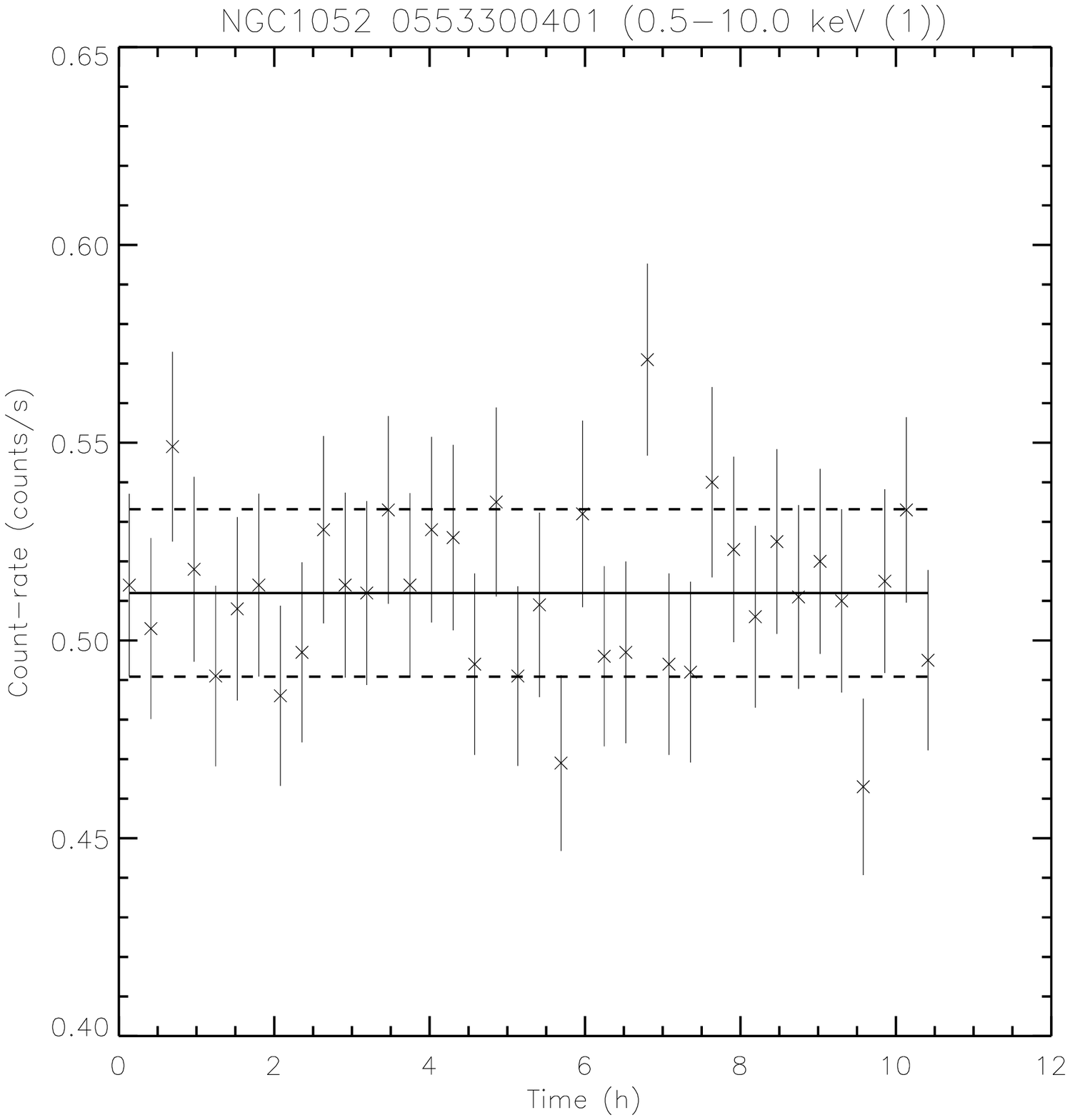}}
\caption{Light curves of NGC\,1052 from \emph{XMM}--Newton data.}
\label{l1052}
\end{figure}

\begin{figure}[H]
\centering
\subfloat{\includegraphics[width=0.30\textwidth]{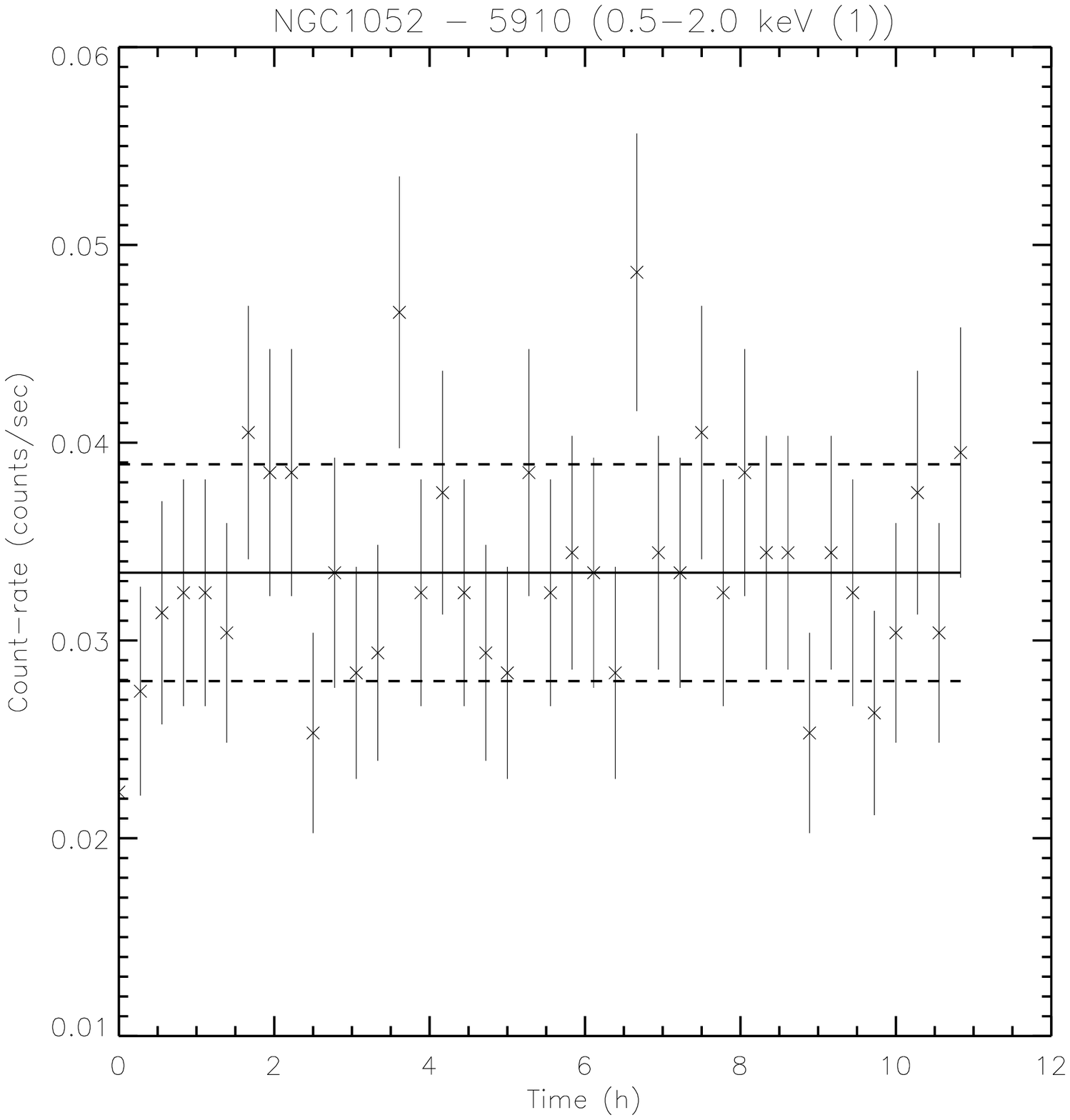}}
\subfloat{\includegraphics[width=0.30\textwidth]{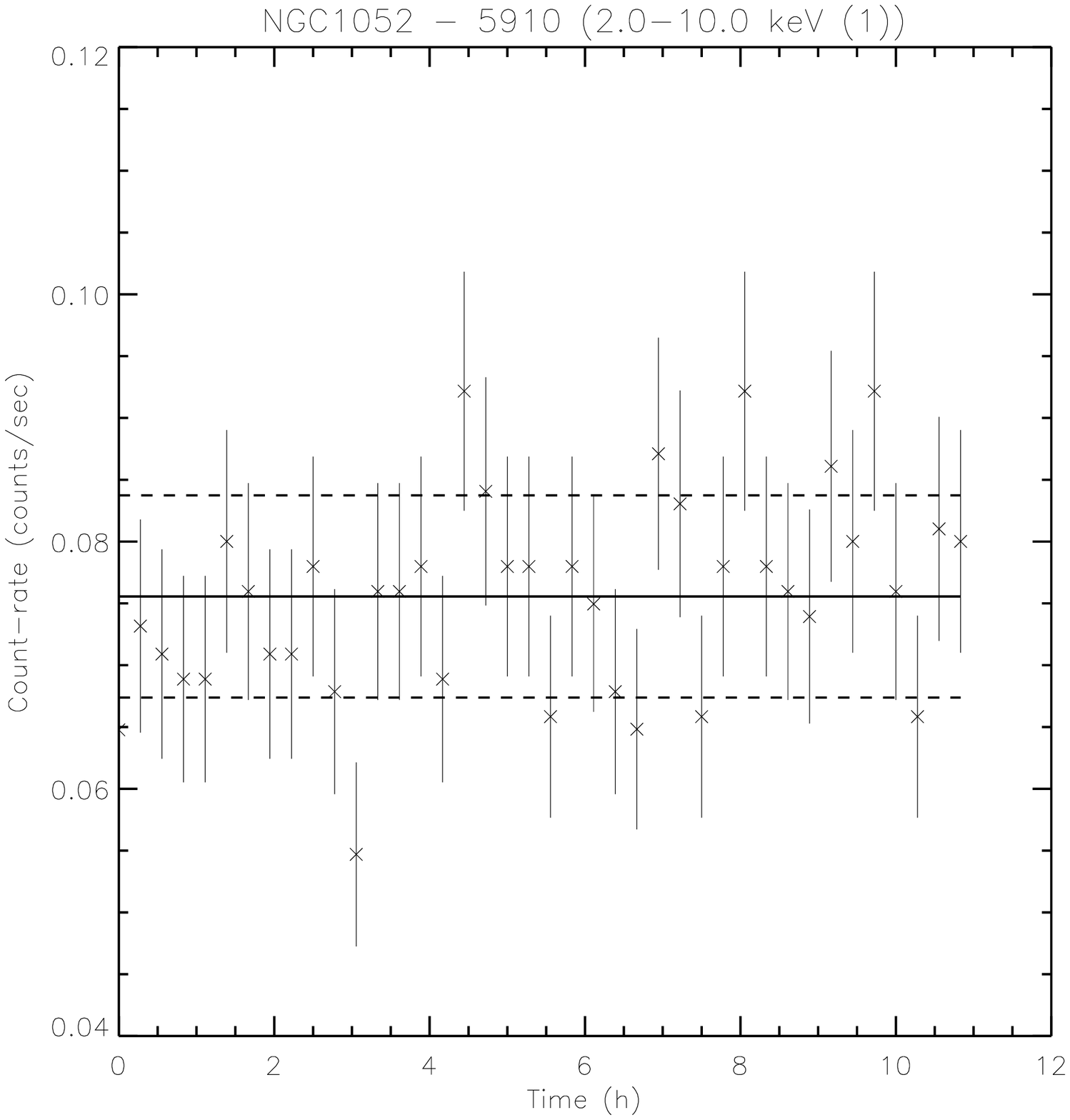}}
\subfloat{\includegraphics[width=0.30\textwidth]{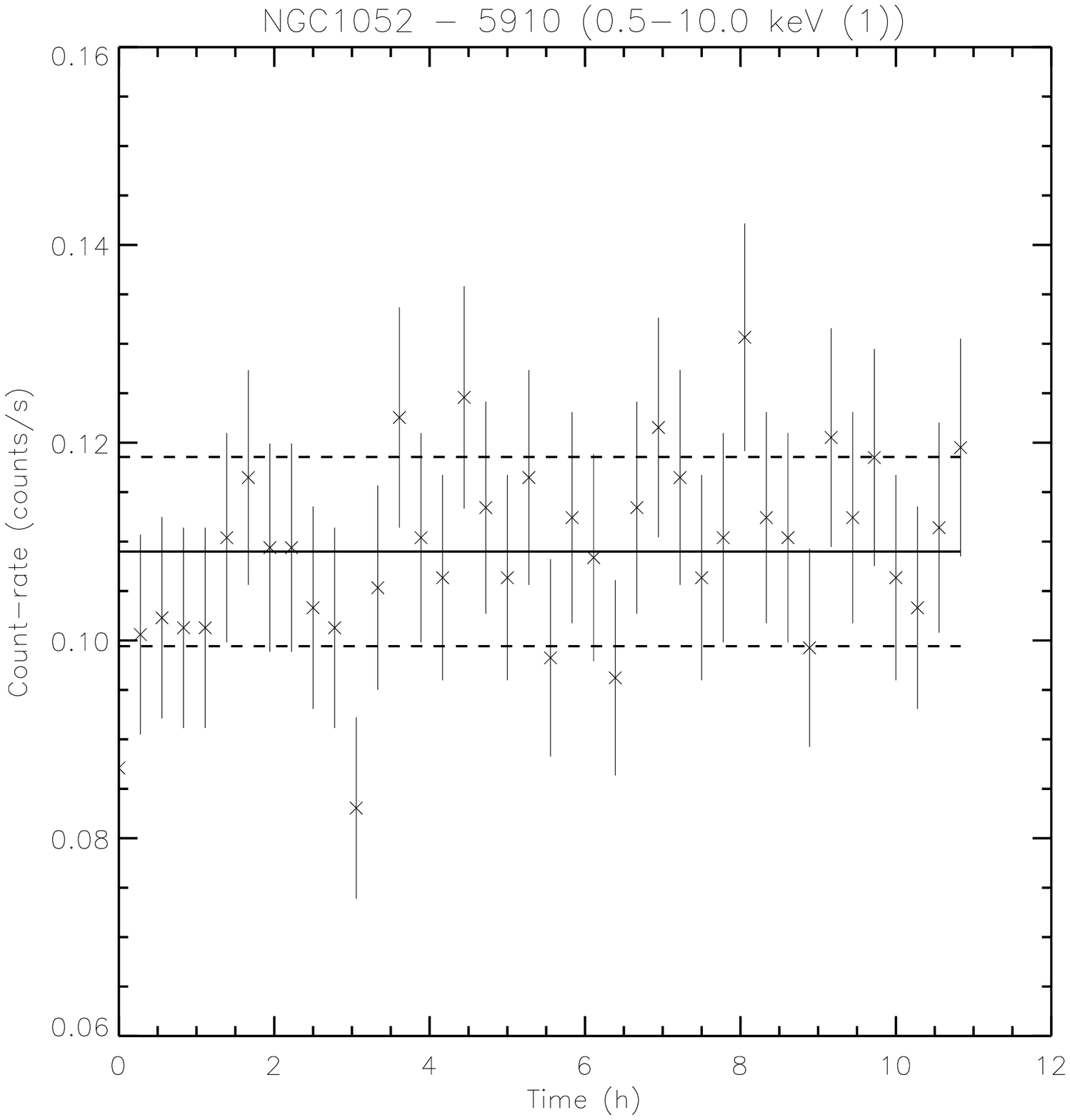}}
\caption{Light curves of NGC\,1052 from \emph{Chandra} data.}
\end{figure}

\begin{figure}[H]
\centering
\subfloat{\includegraphics[width=0.30\textwidth]{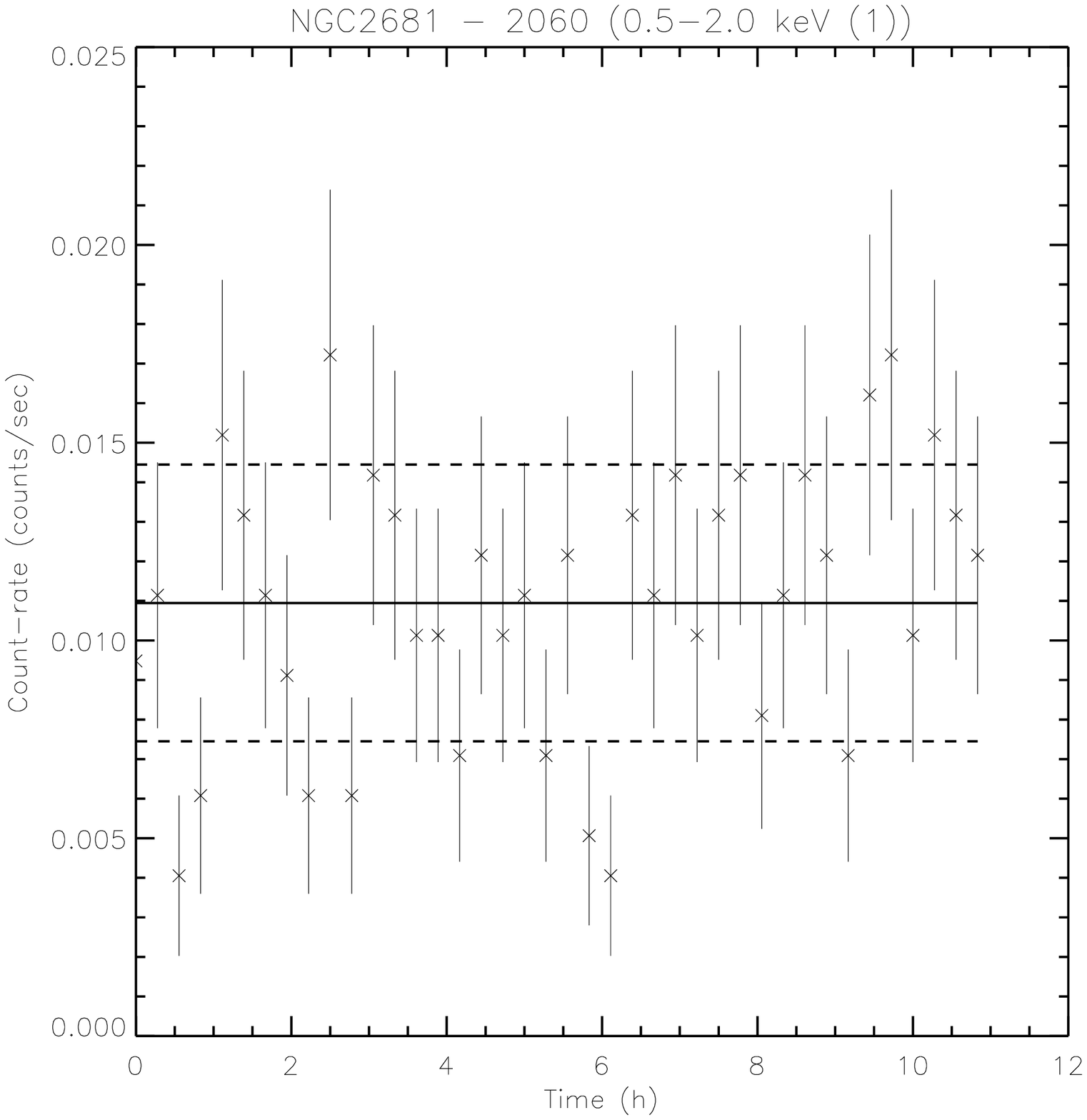}}
\subfloat{\includegraphics[width=0.30\textwidth]{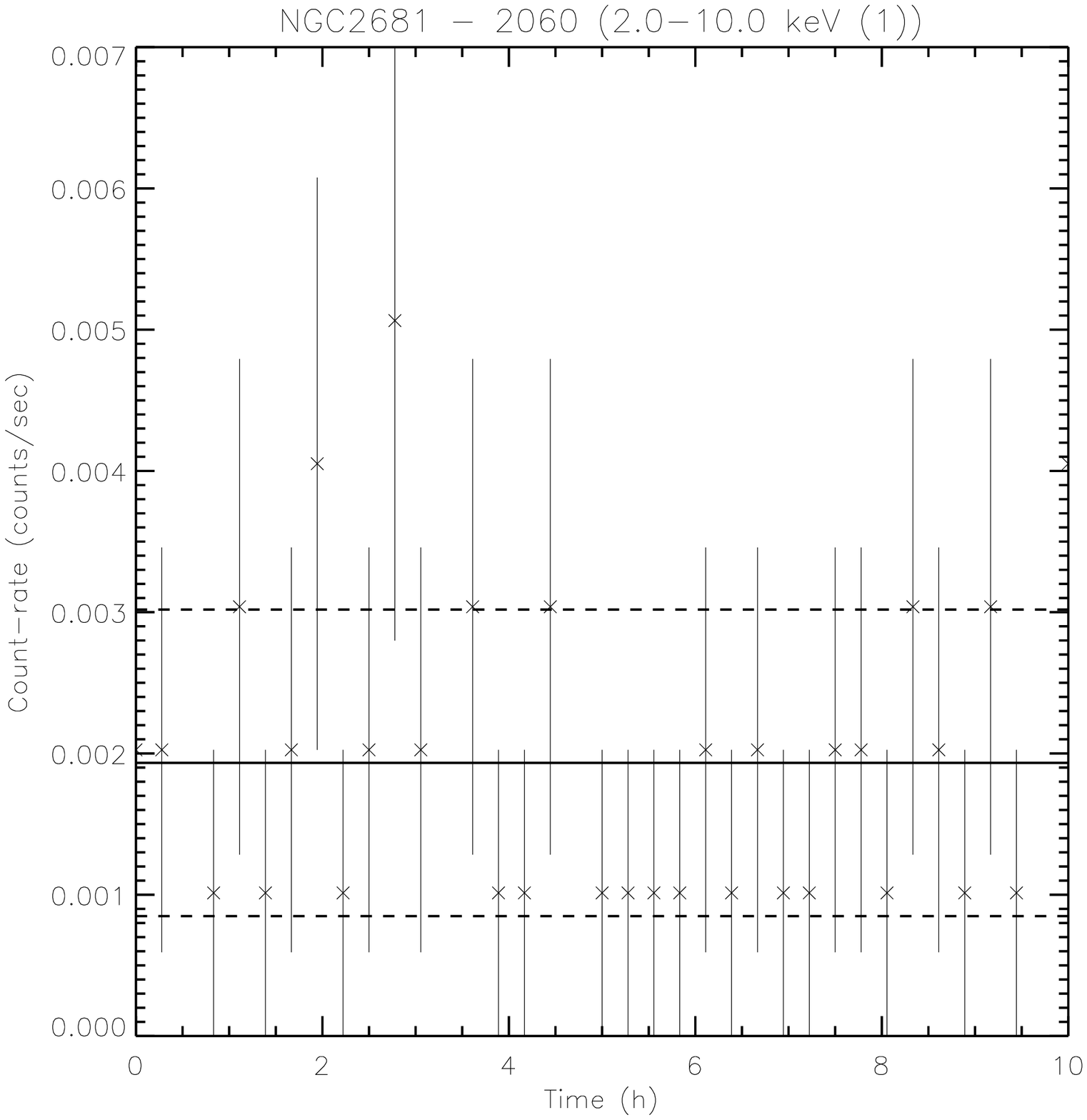}}
\subfloat{\includegraphics[width=0.30\textwidth]{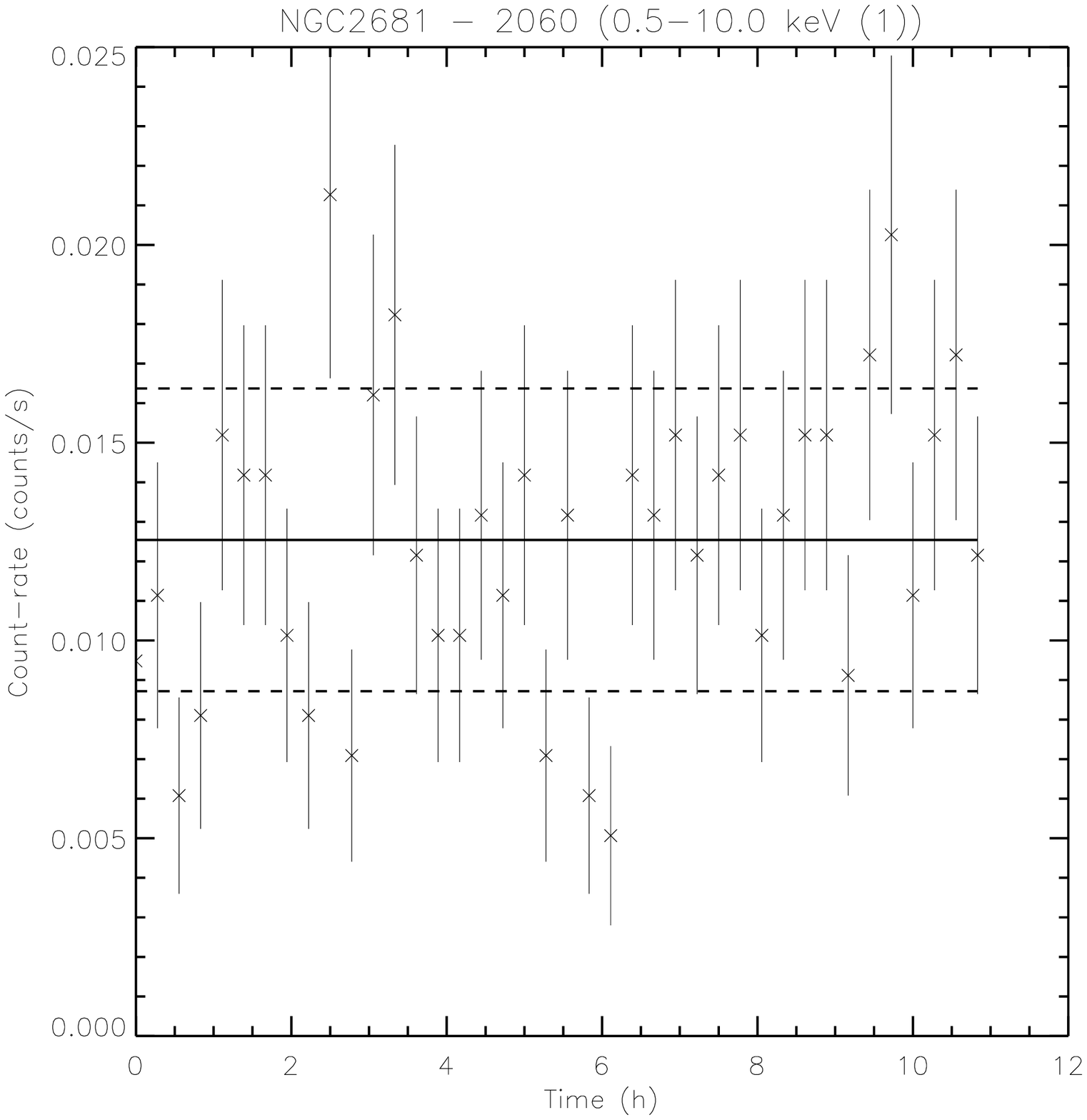}}

\subfloat{\includegraphics[width=0.30\textwidth]{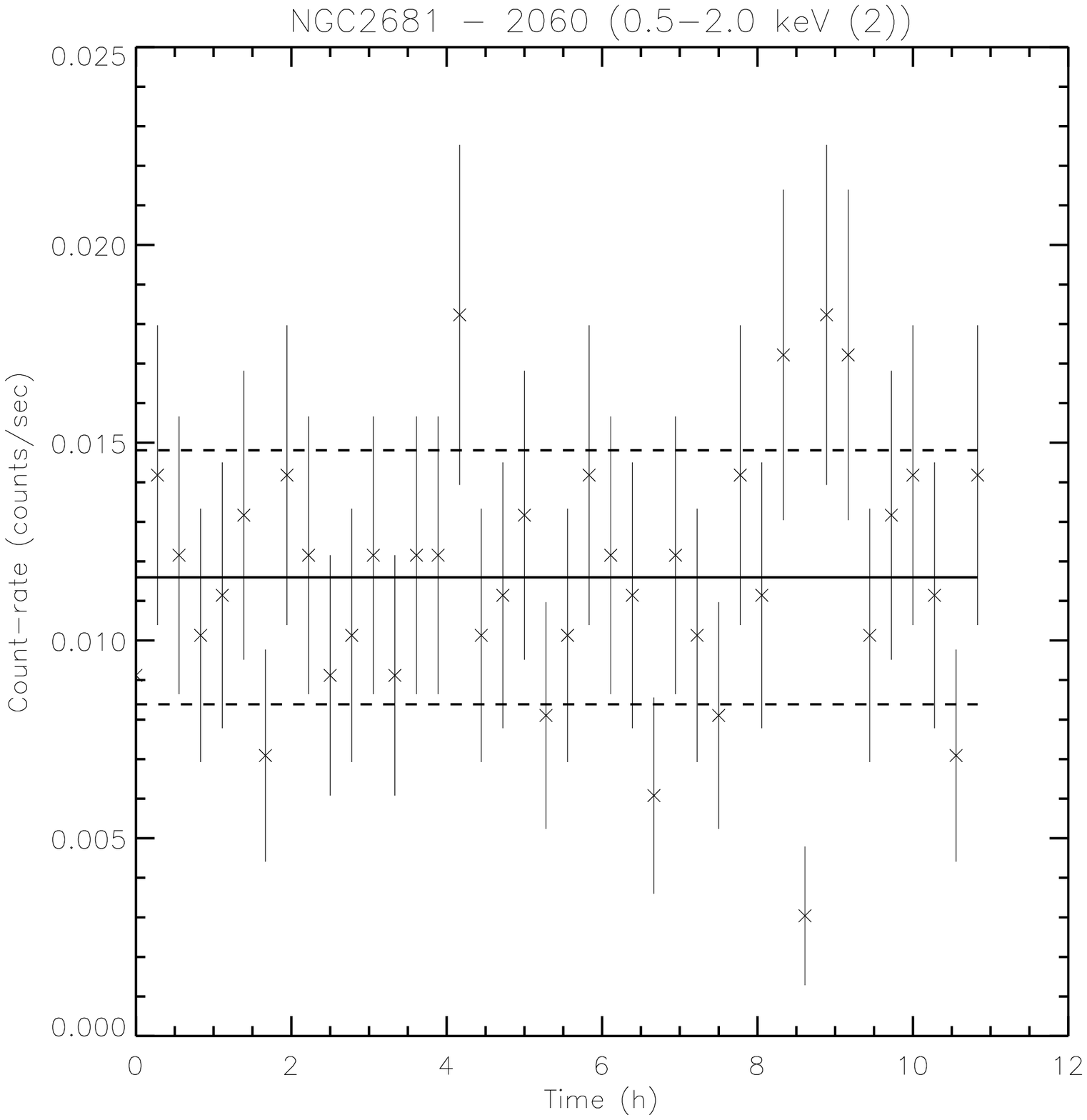}}
\subfloat{\includegraphics[width=0.30\textwidth]{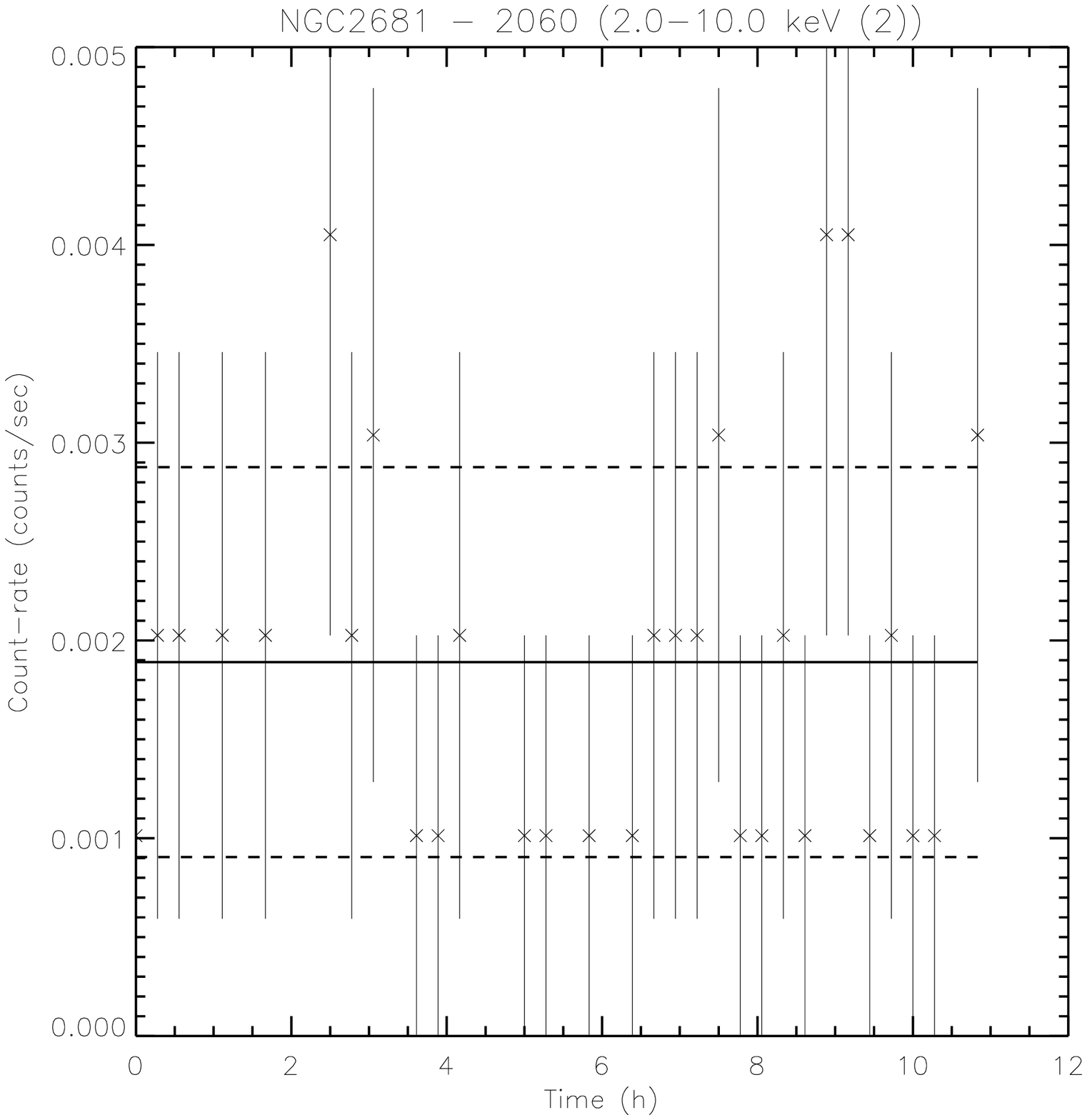}}
\subfloat{\includegraphics[width=0.30\textwidth]{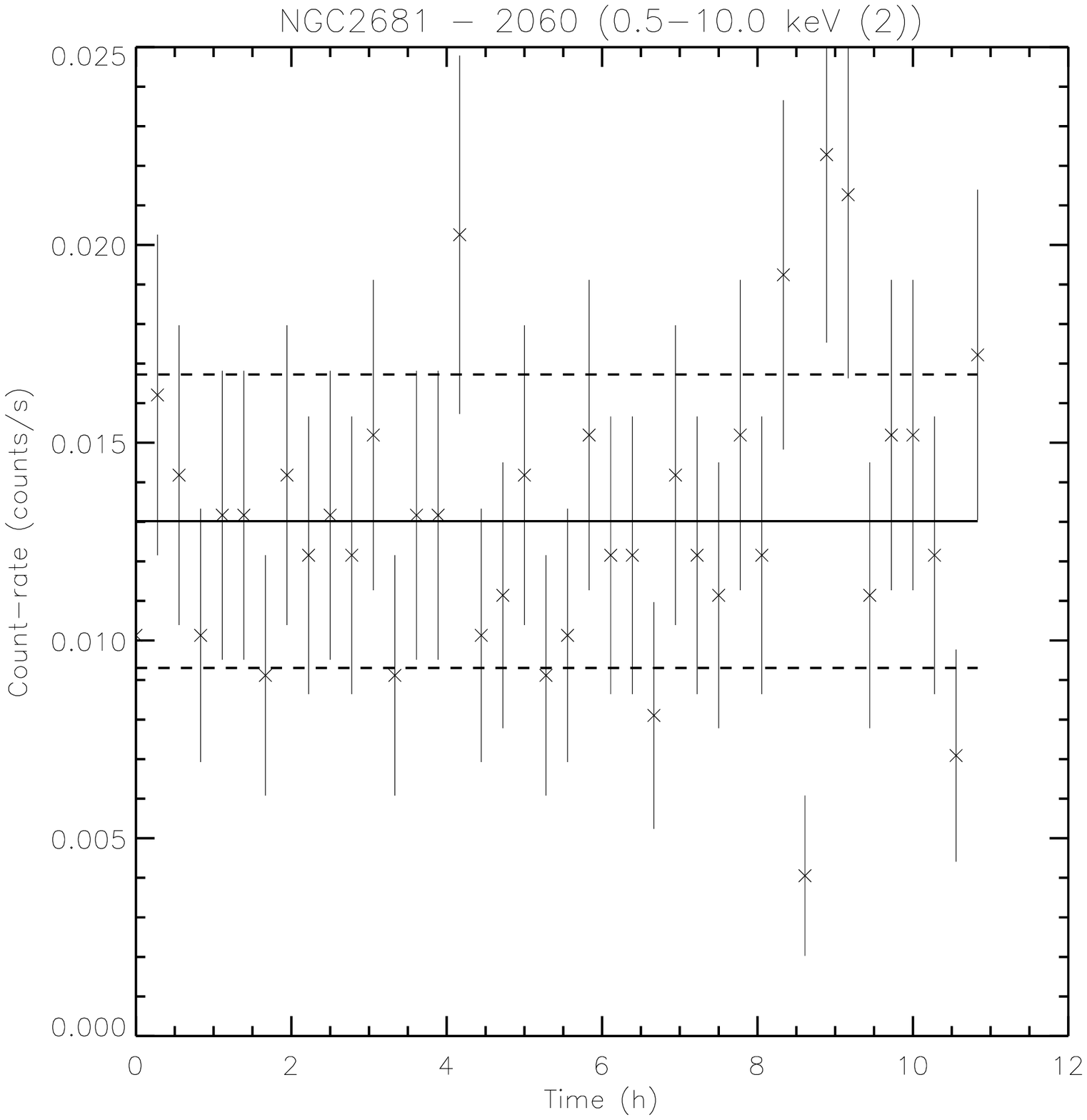}}

\subfloat{\includegraphics[width=0.30\textwidth]{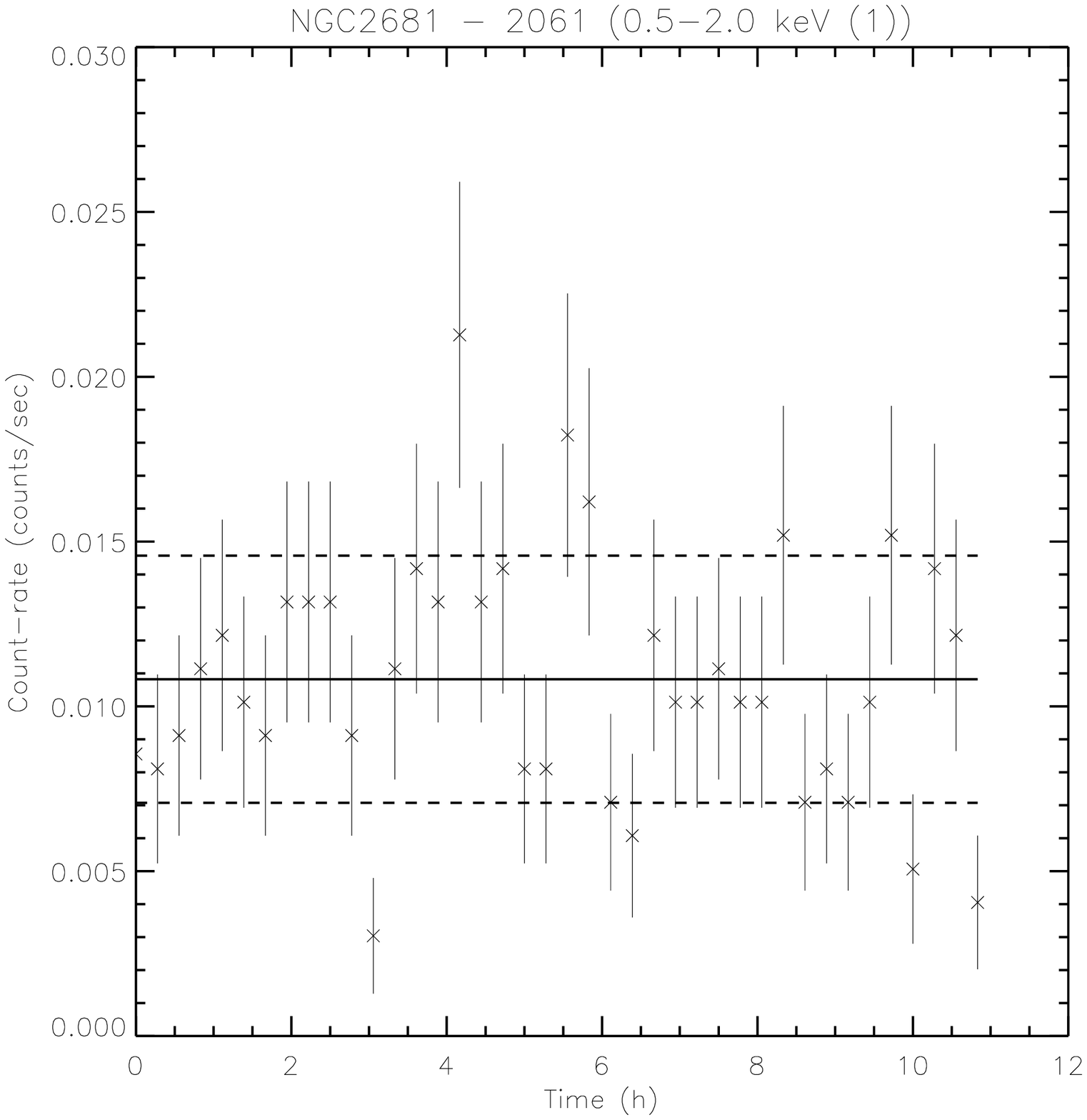}}
\subfloat{\includegraphics[width=0.30\textwidth]{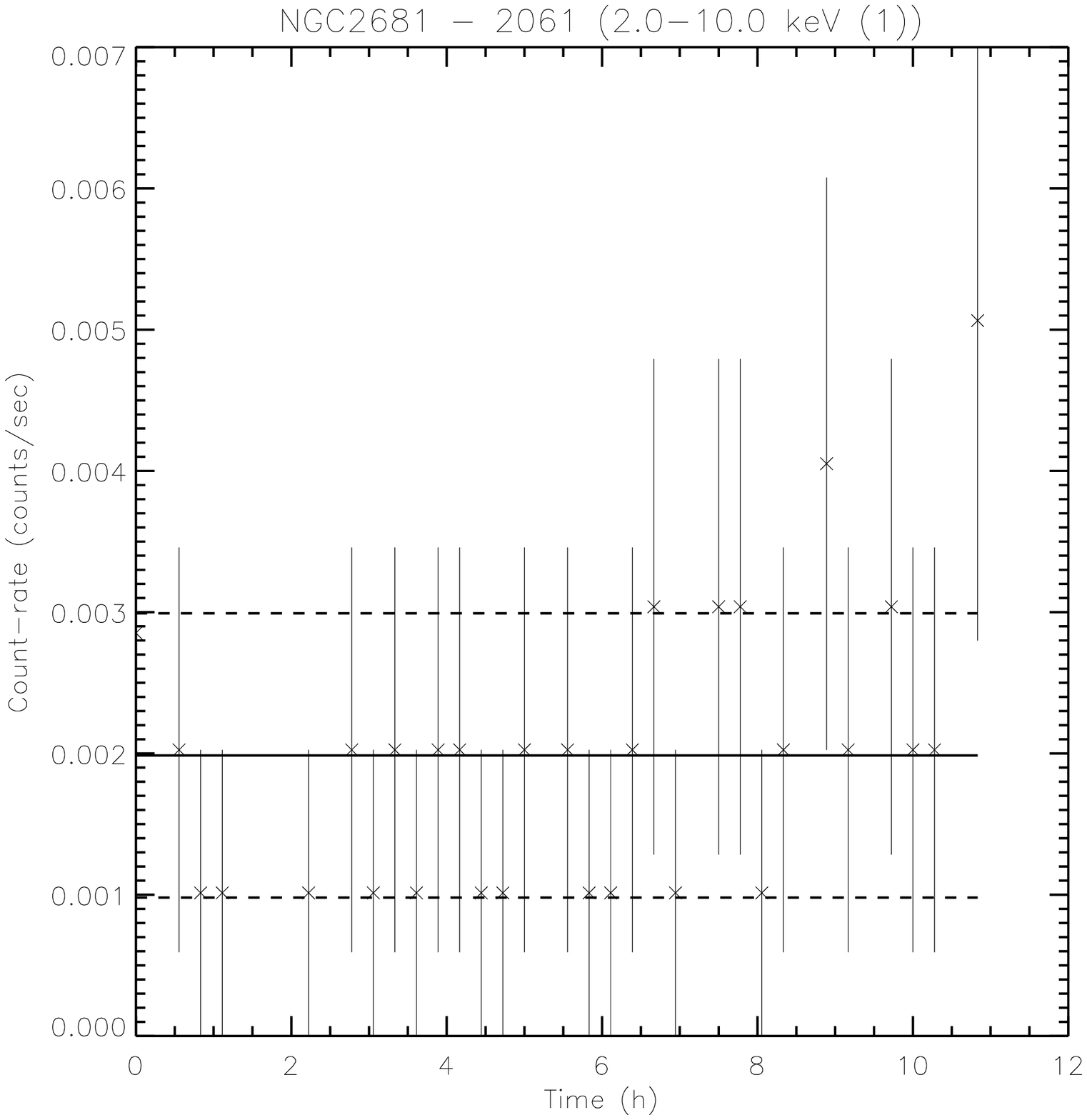}}
\subfloat{\includegraphics[width=0.30\textwidth]{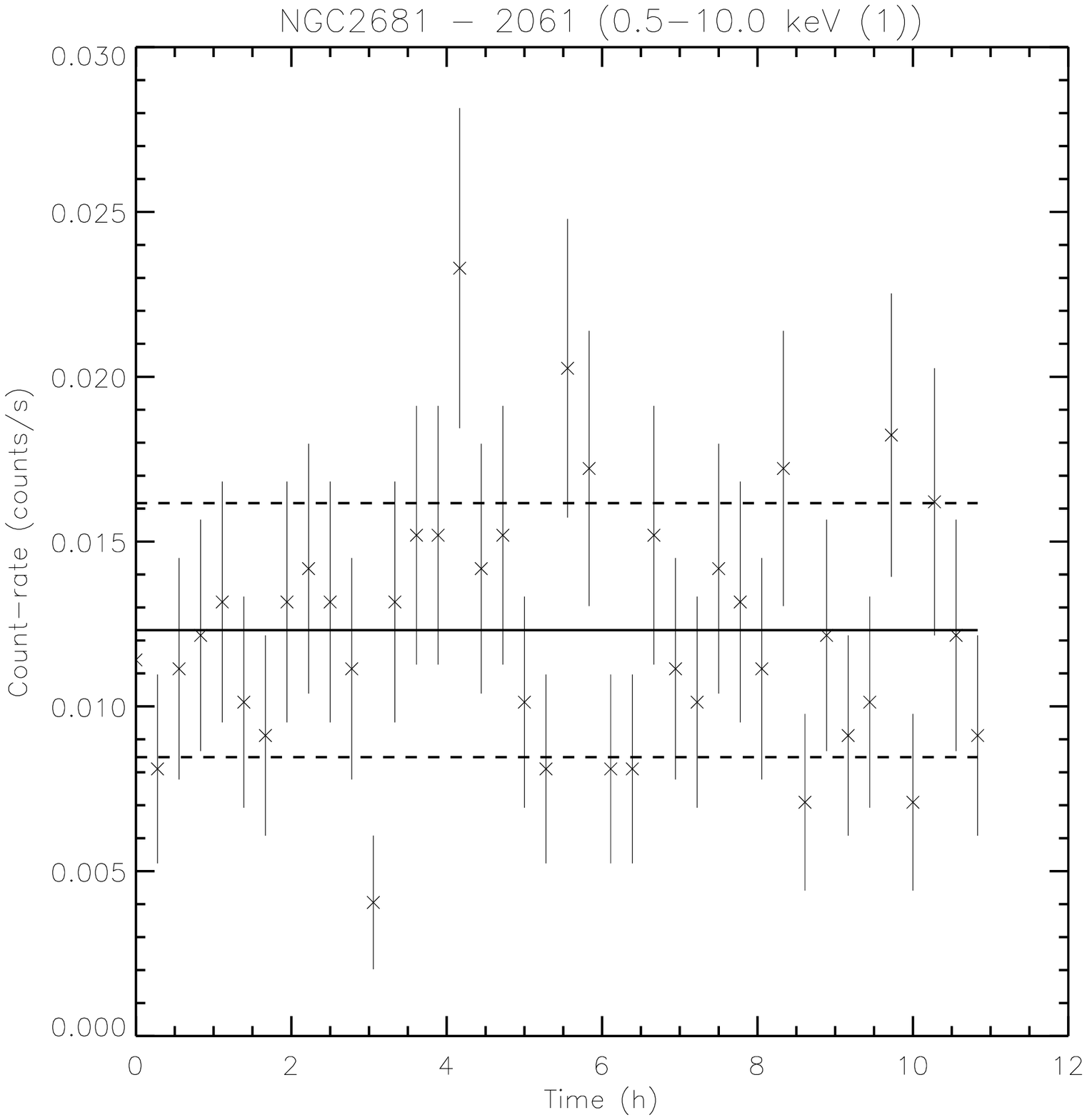}}
\caption{Light curves of NGC\,2681 from \emph{Chandra} data. Note that ObsID. 2060 is divided in two segments.}
\label{l2681}
\end{figure}

\begin{figure}[H]
\centering
\subfloat{\includegraphics[width=0.30\textwidth]{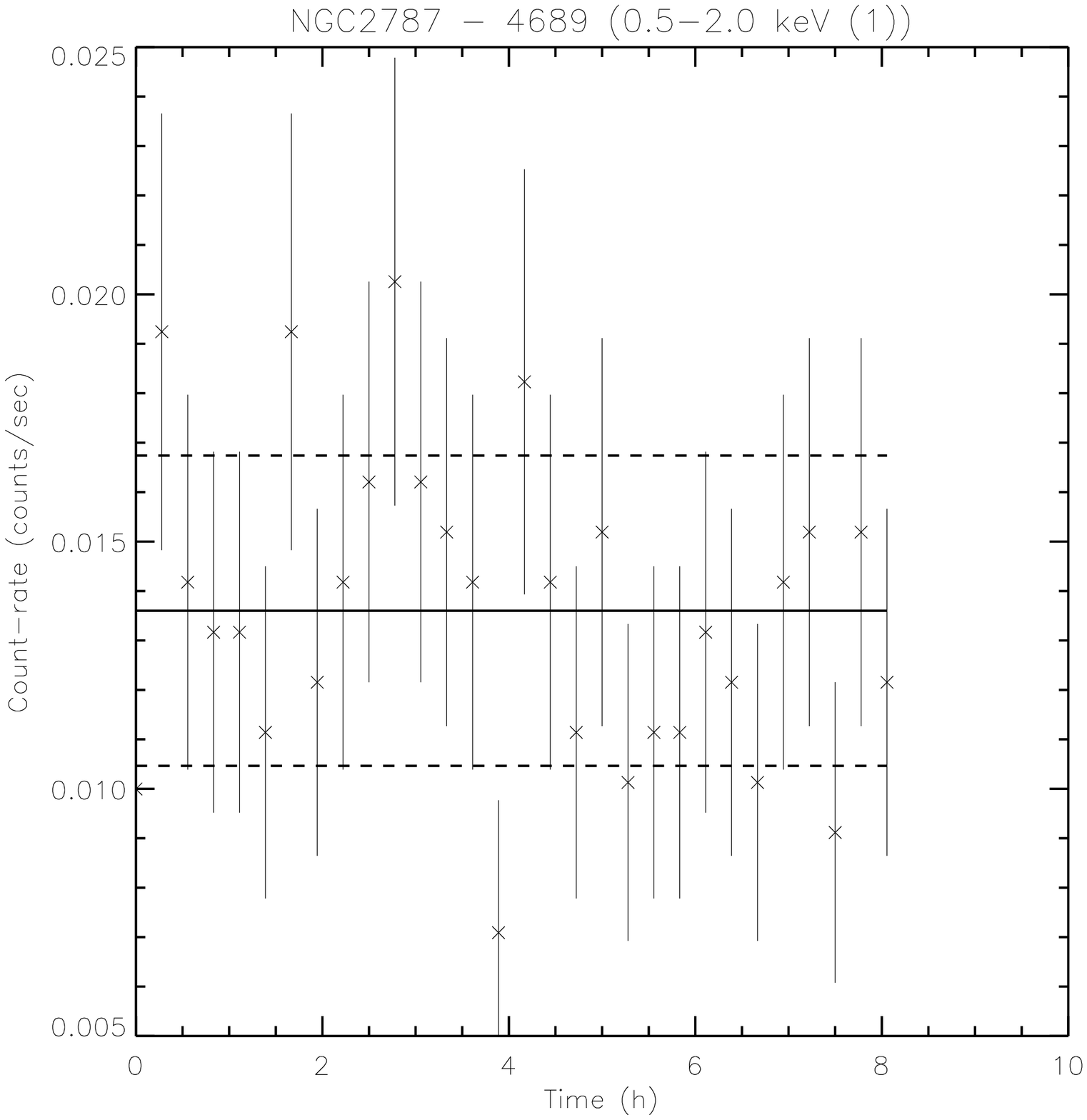}}
\subfloat{\includegraphics[width=0.30\textwidth]{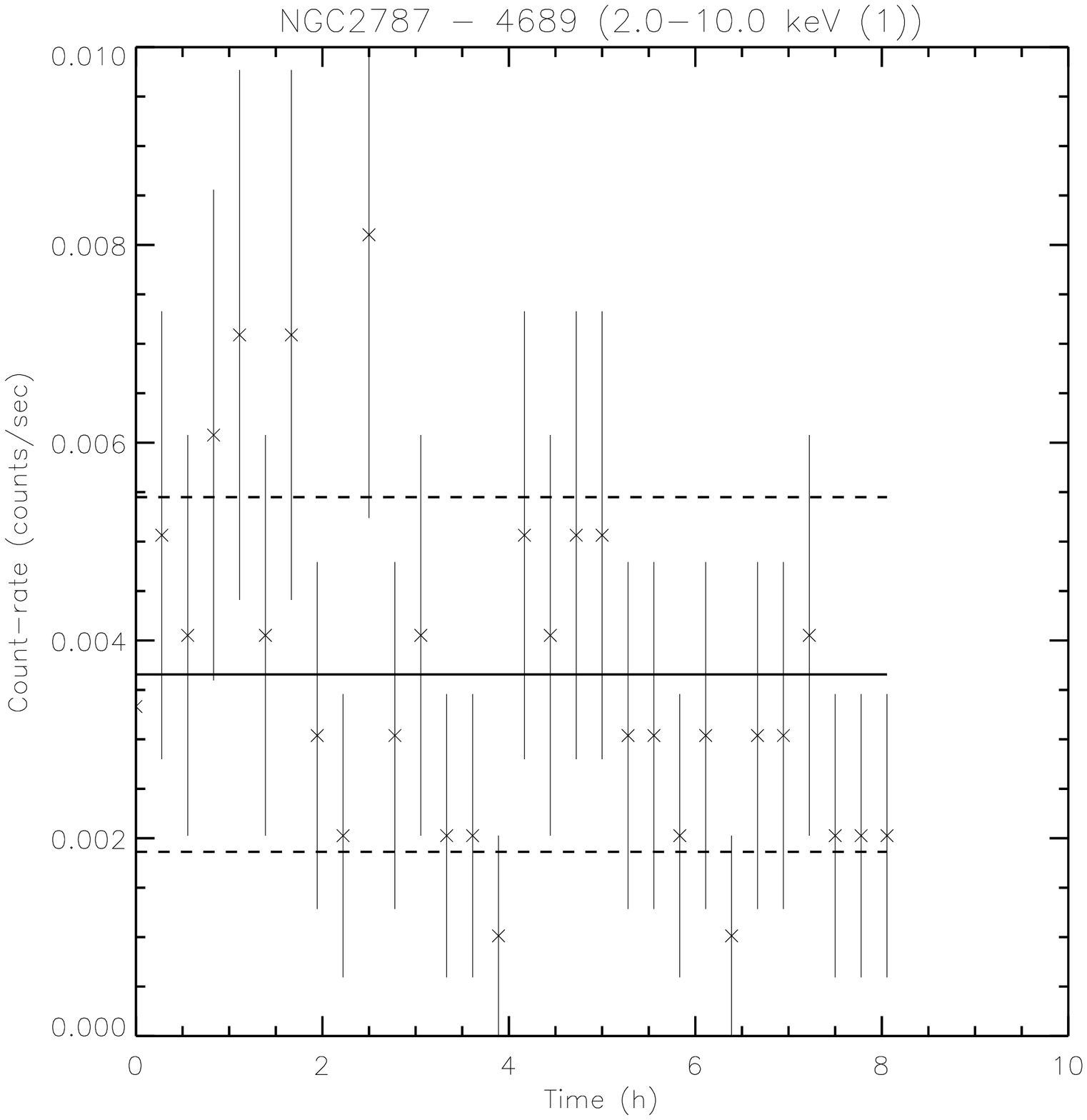}}
\subfloat{\includegraphics[width=0.30\textwidth]{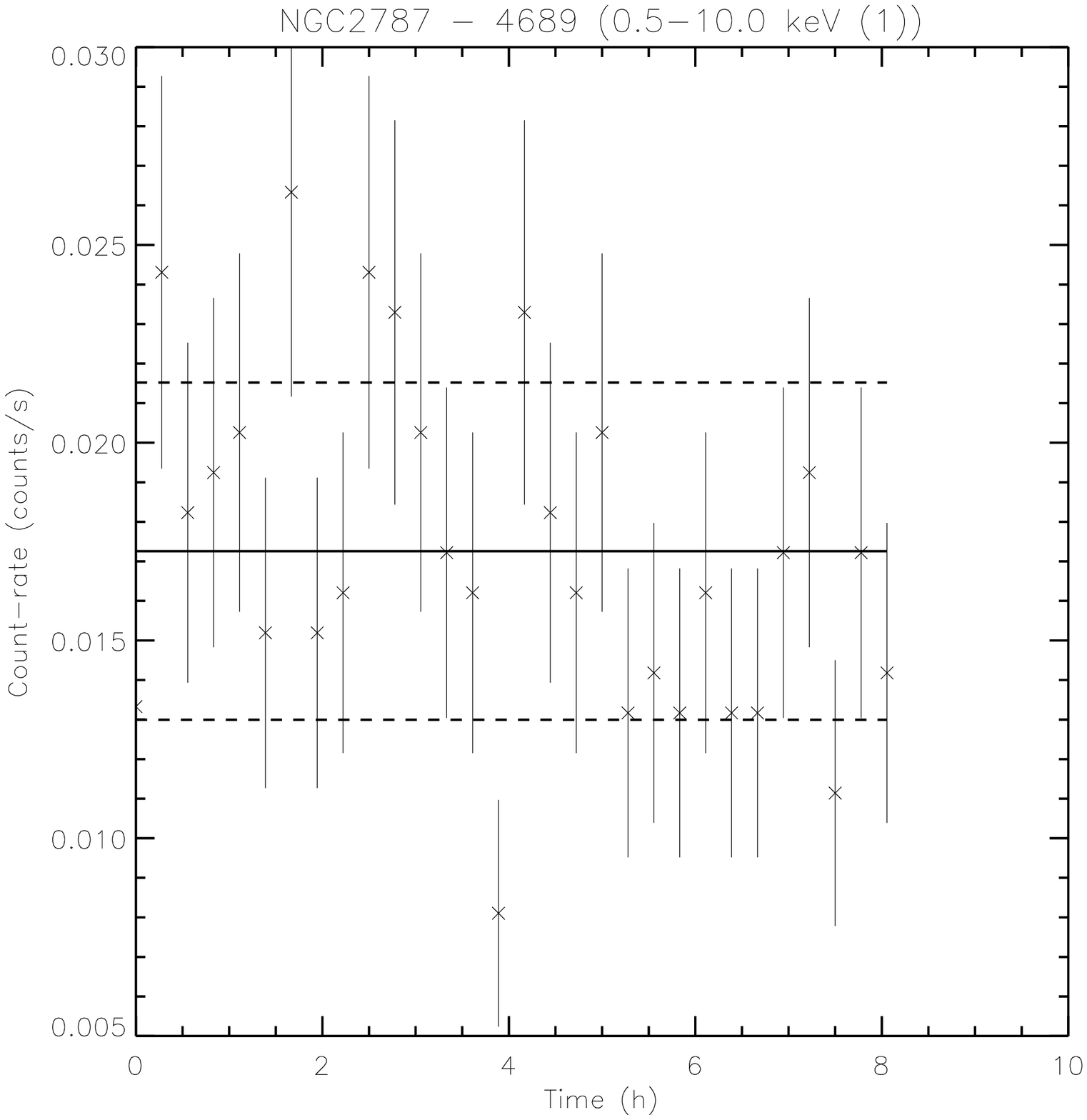}}
\caption{Light curves of NGC\,2787 from \emph{Chandra} data.}
\label{l2787}
\end{figure}

\begin{figure}[H]
\centering
\subfloat{\includegraphics[width=0.30\textwidth]{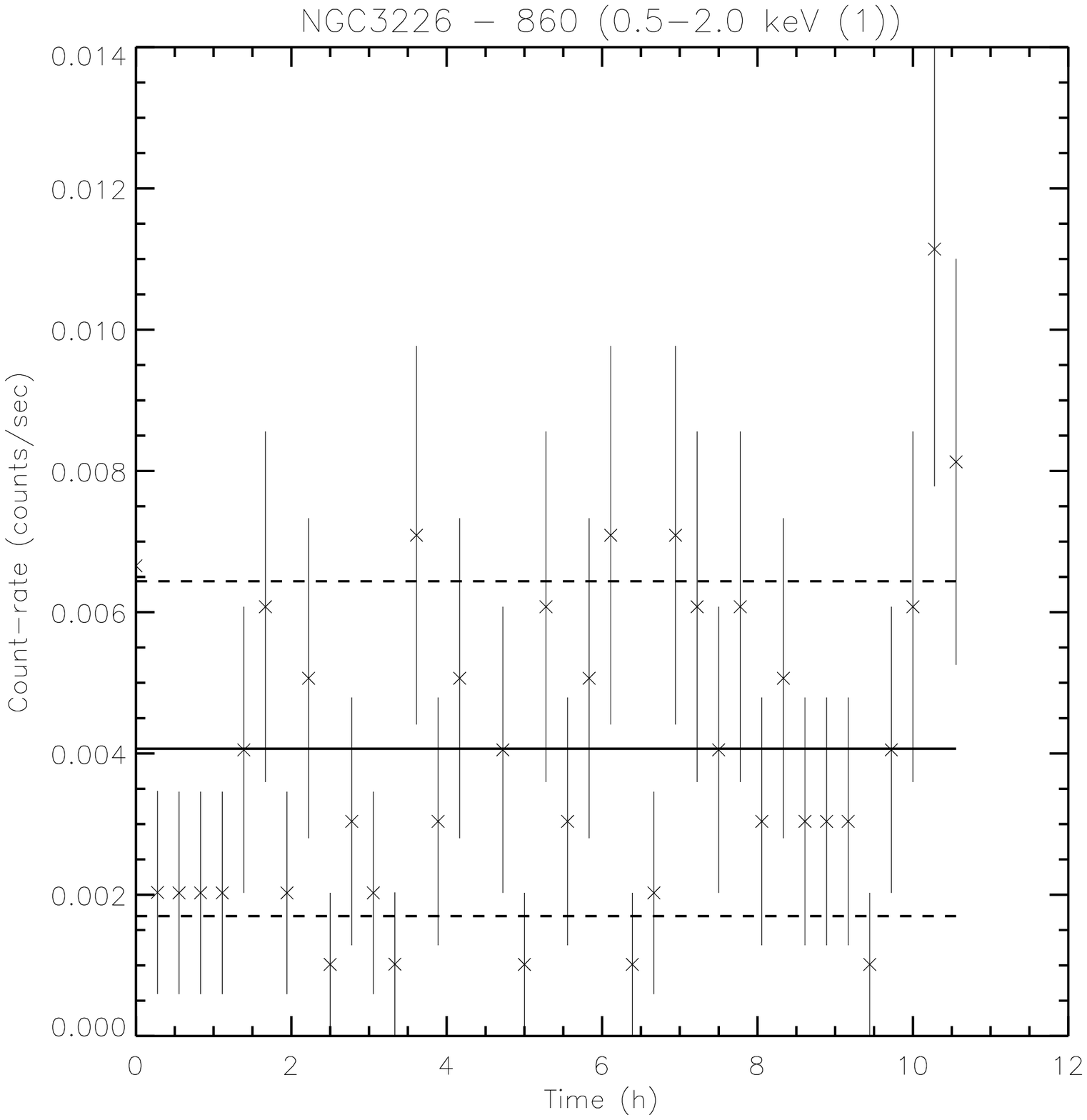}}
\subfloat{\includegraphics[width=0.30\textwidth]{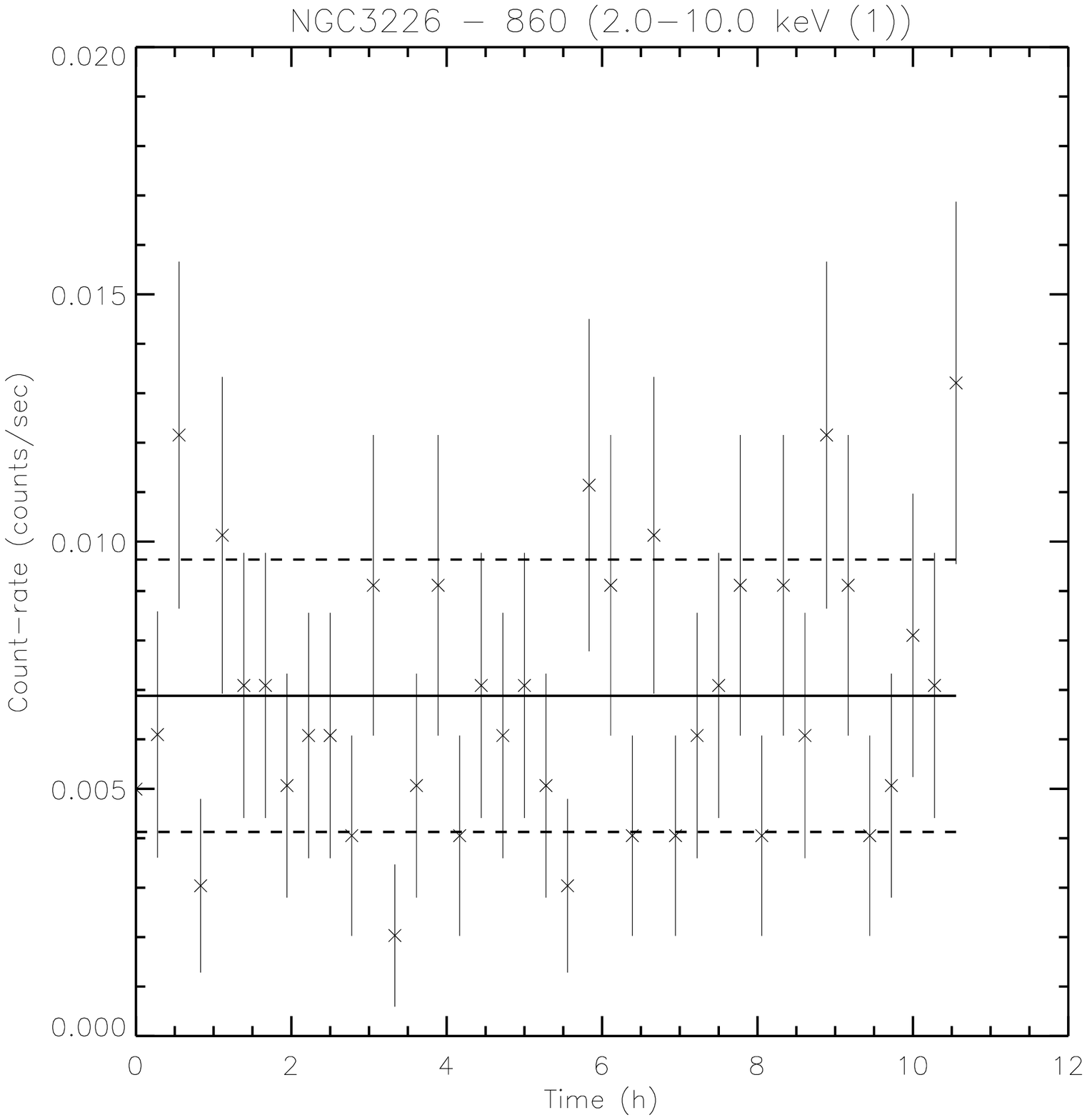}}
\subfloat{\includegraphics[width=0.30\textwidth]{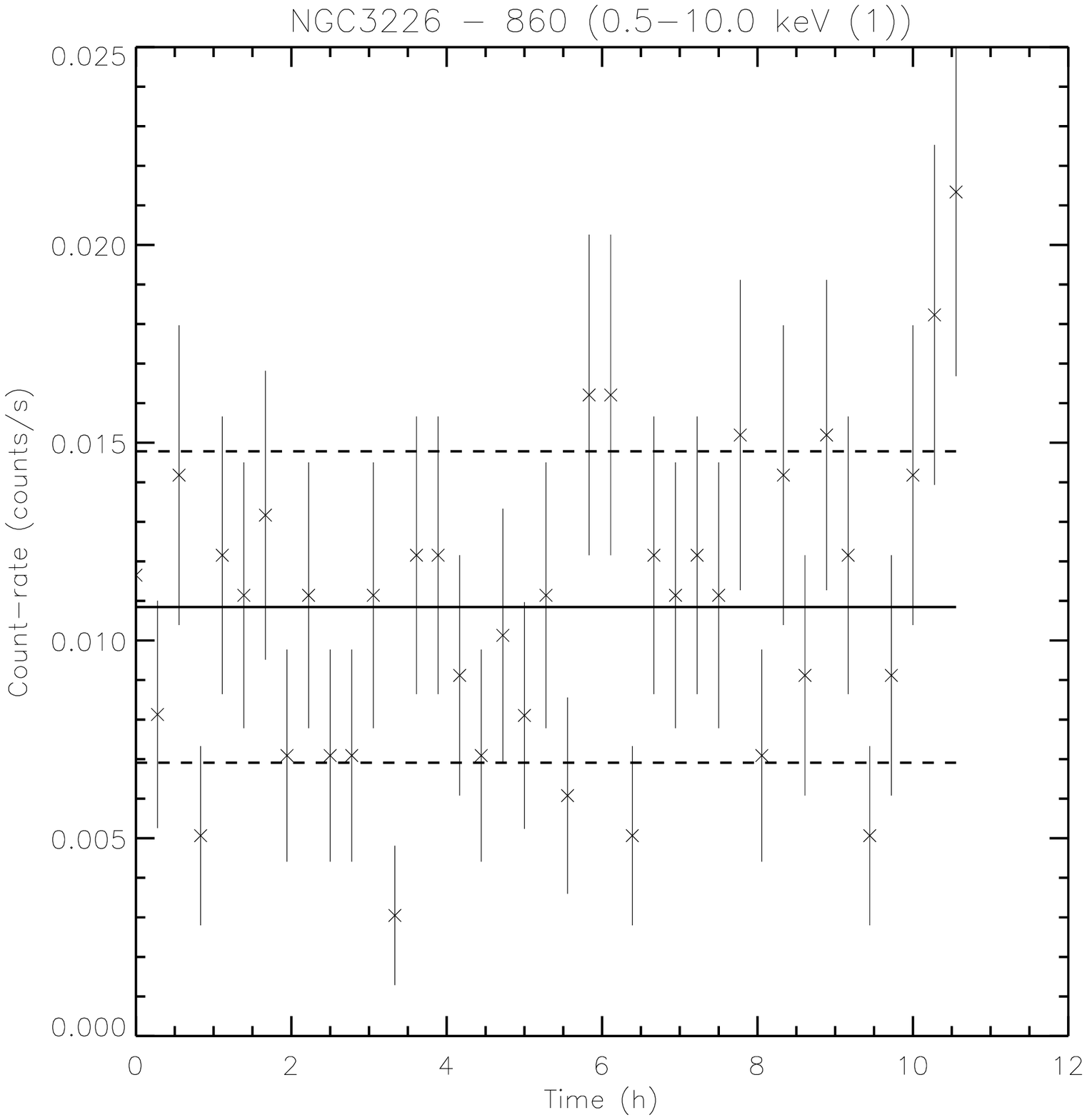}}
\caption{Light curves of NGC\,3226 from \emph{Chandra} data.}
\label{l3226}
\end{figure}

\begin{figure}[H]
\centering
\subfloat{\includegraphics[width=0.30\textwidth]{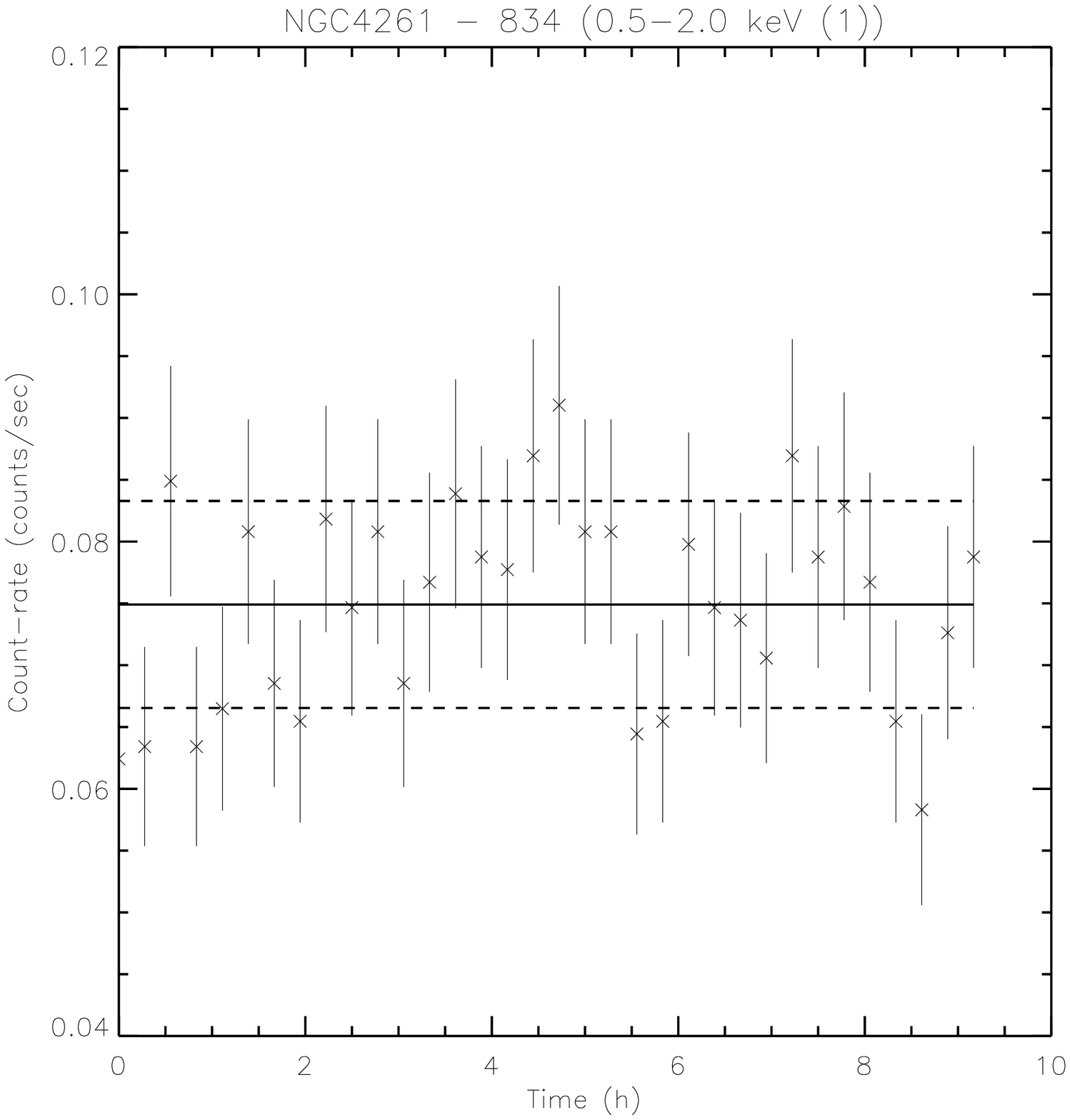}}
\subfloat{\includegraphics[width=0.30\textwidth]{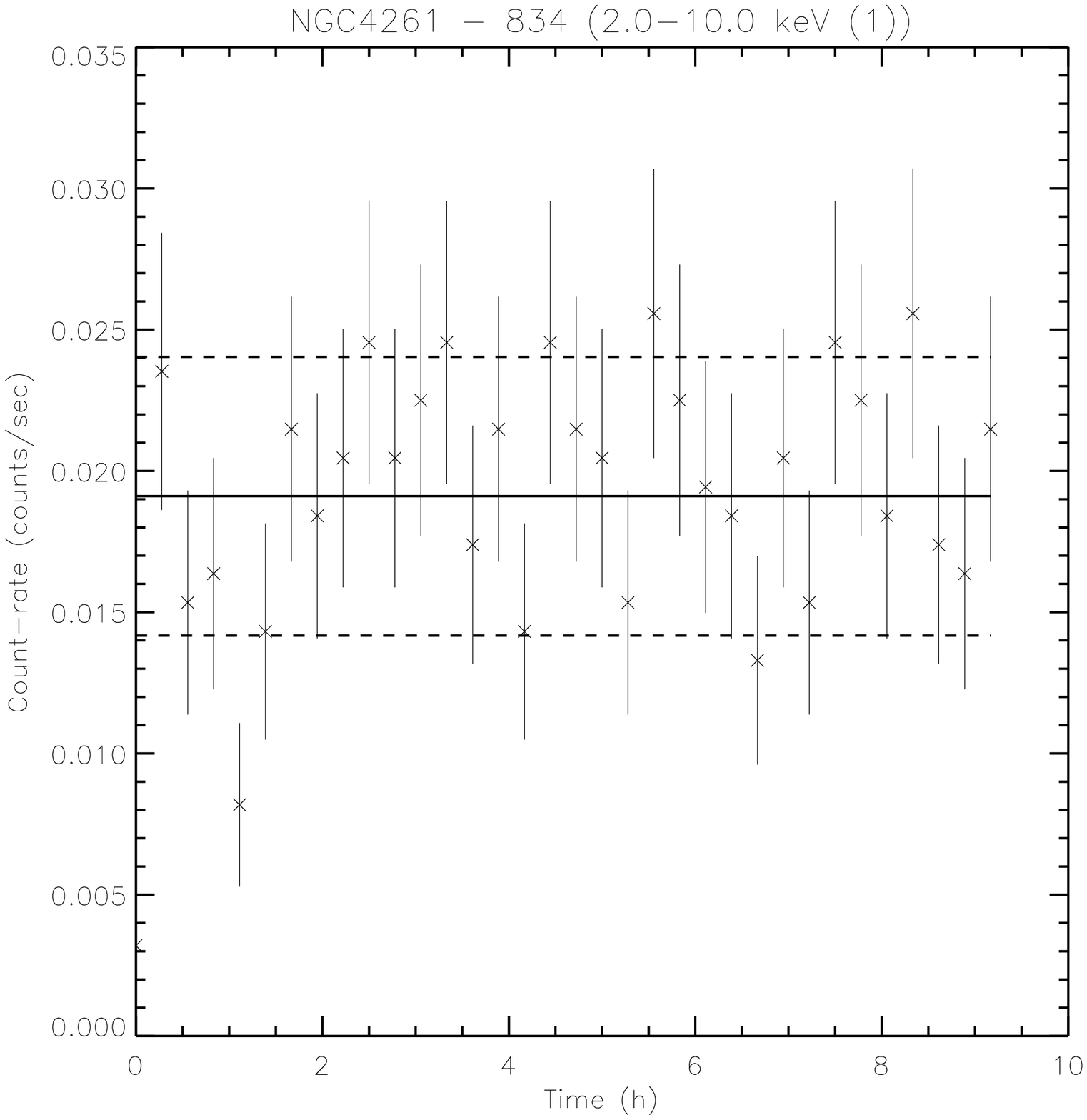}}
\subfloat{\includegraphics[width=0.30\textwidth]{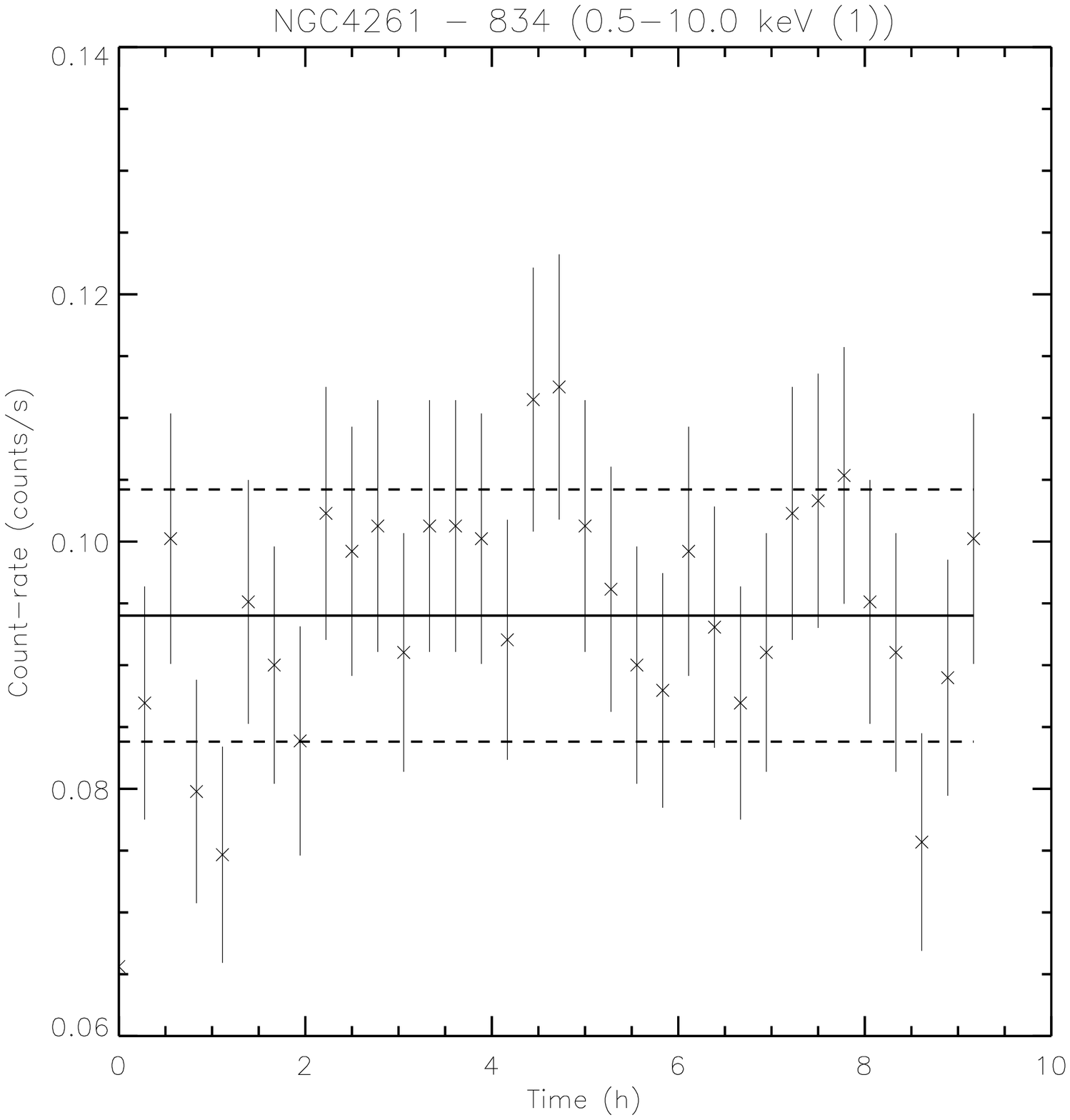}}

\subfloat{\includegraphics[width=0.30\textwidth]{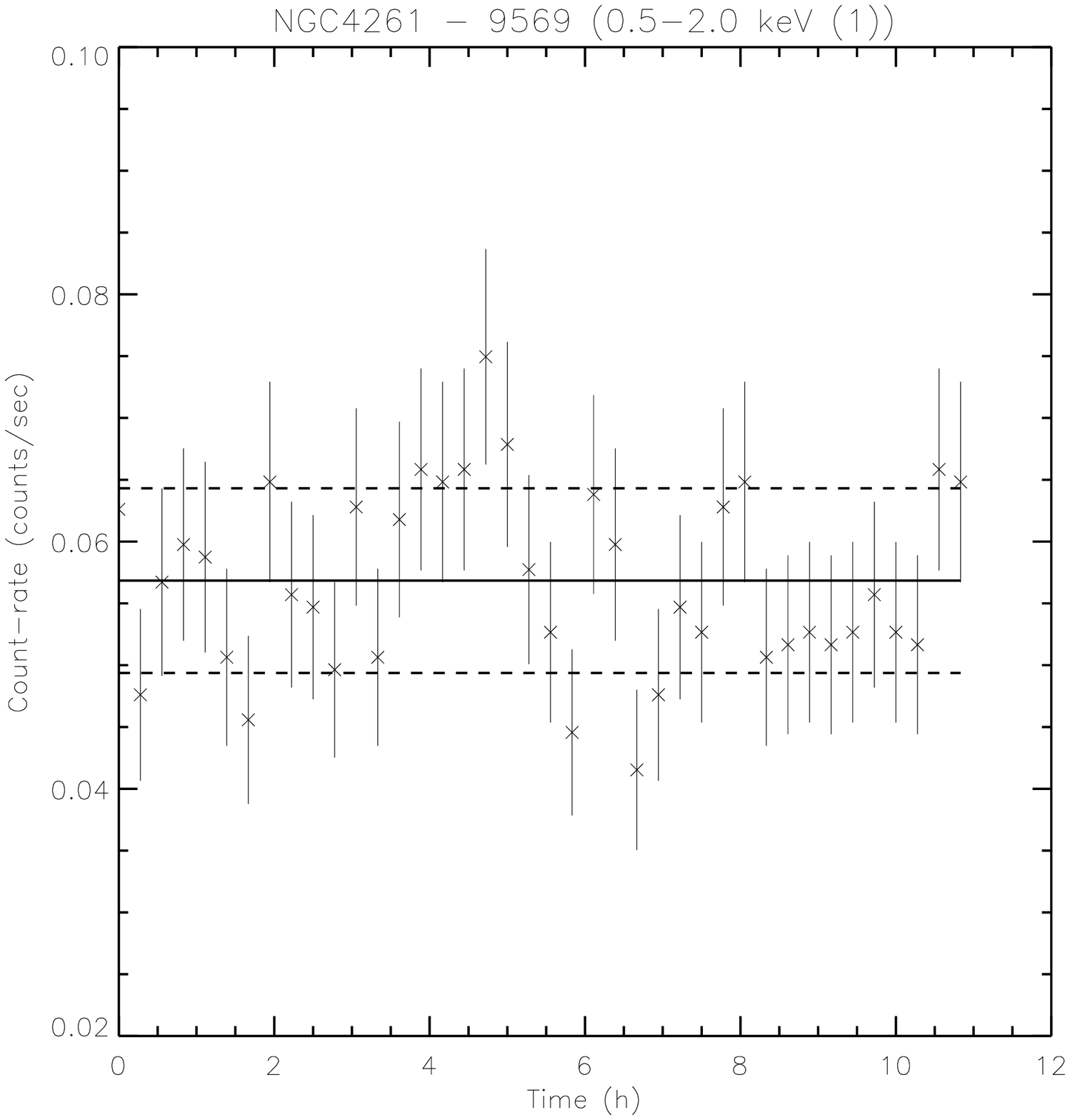}}
\subfloat{\includegraphics[width=0.30\textwidth]{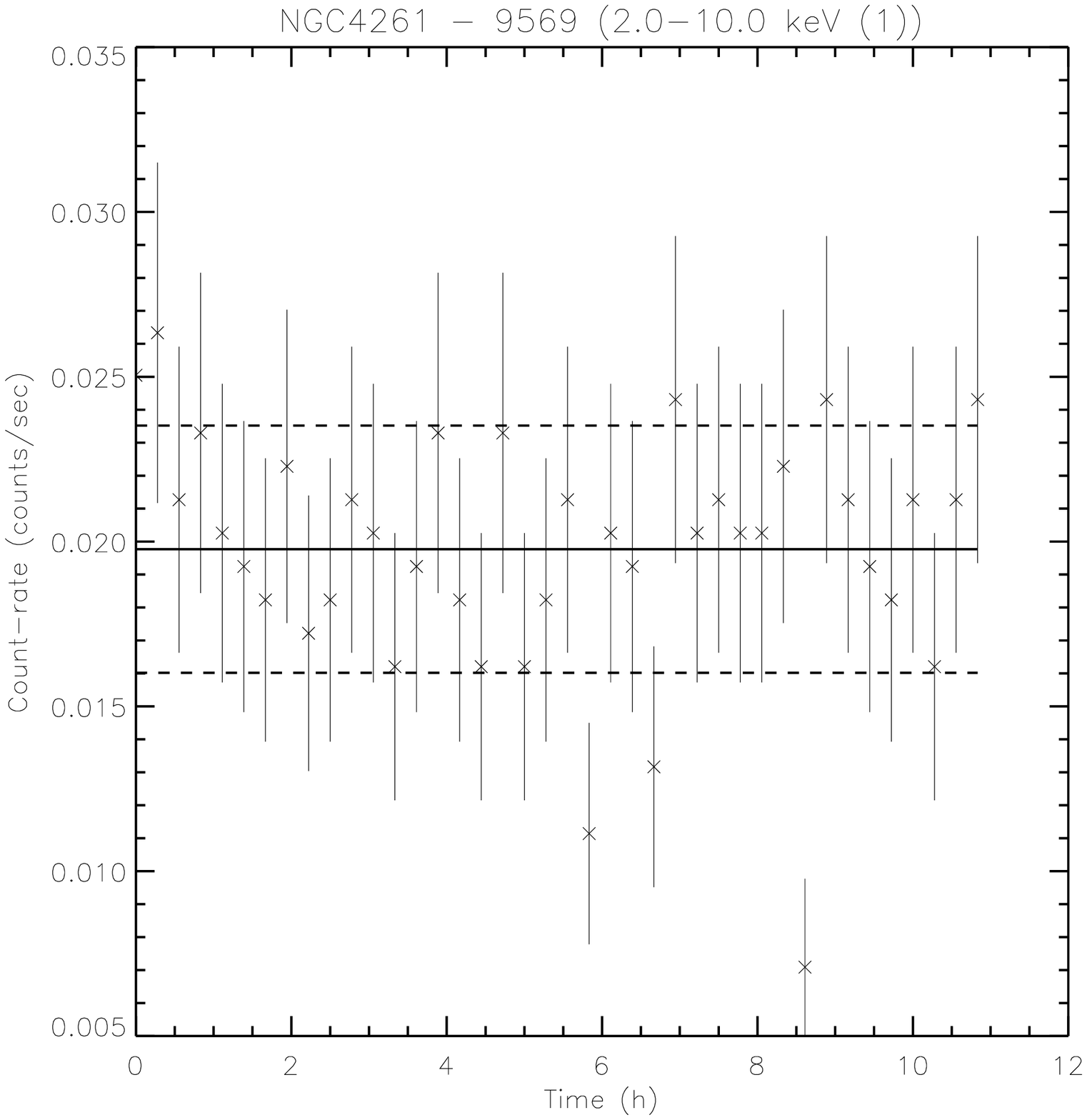}}
\subfloat{\includegraphics[width=0.30\textwidth]{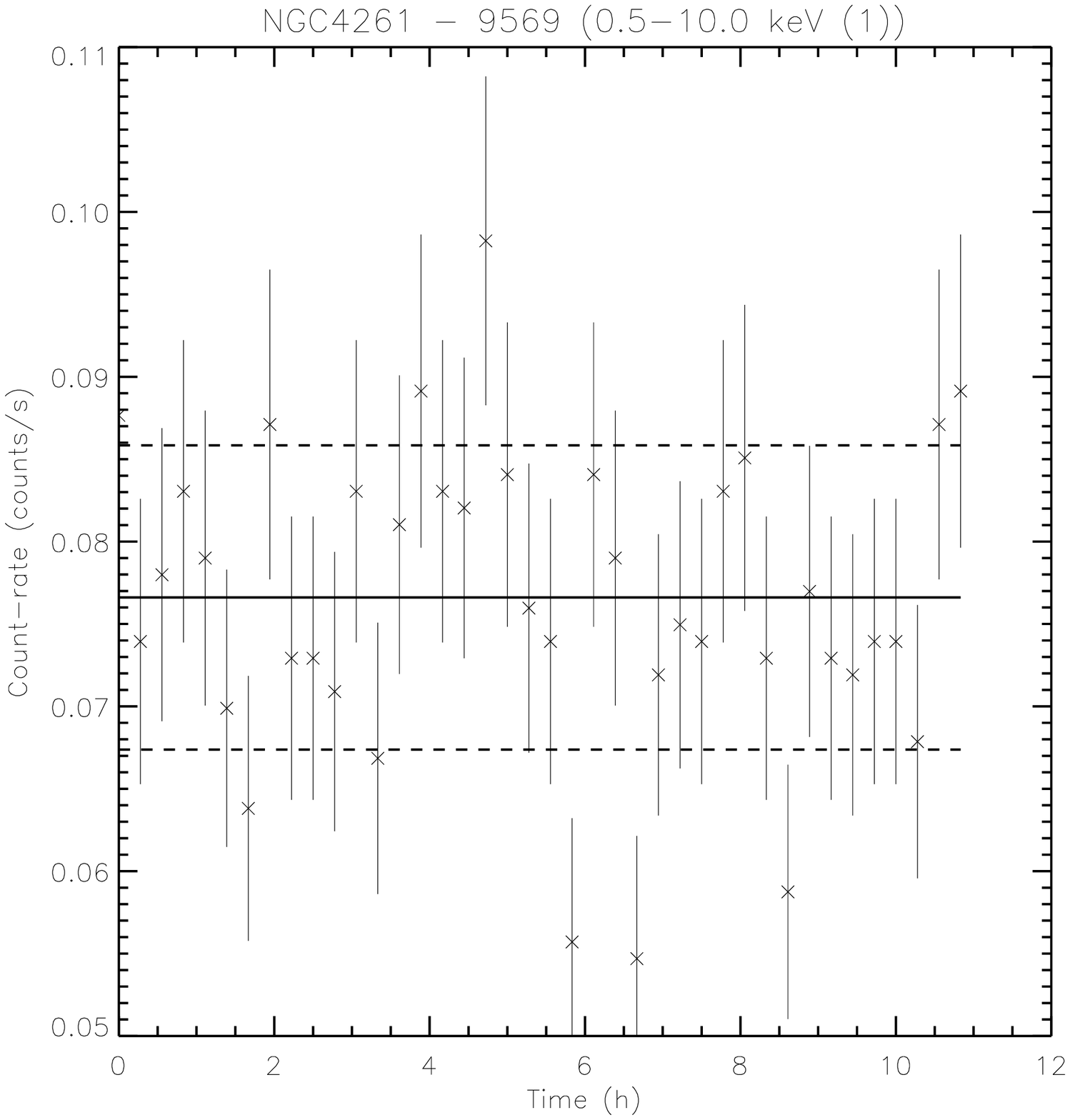}}

\subfloat{\includegraphics[width=0.30\textwidth]{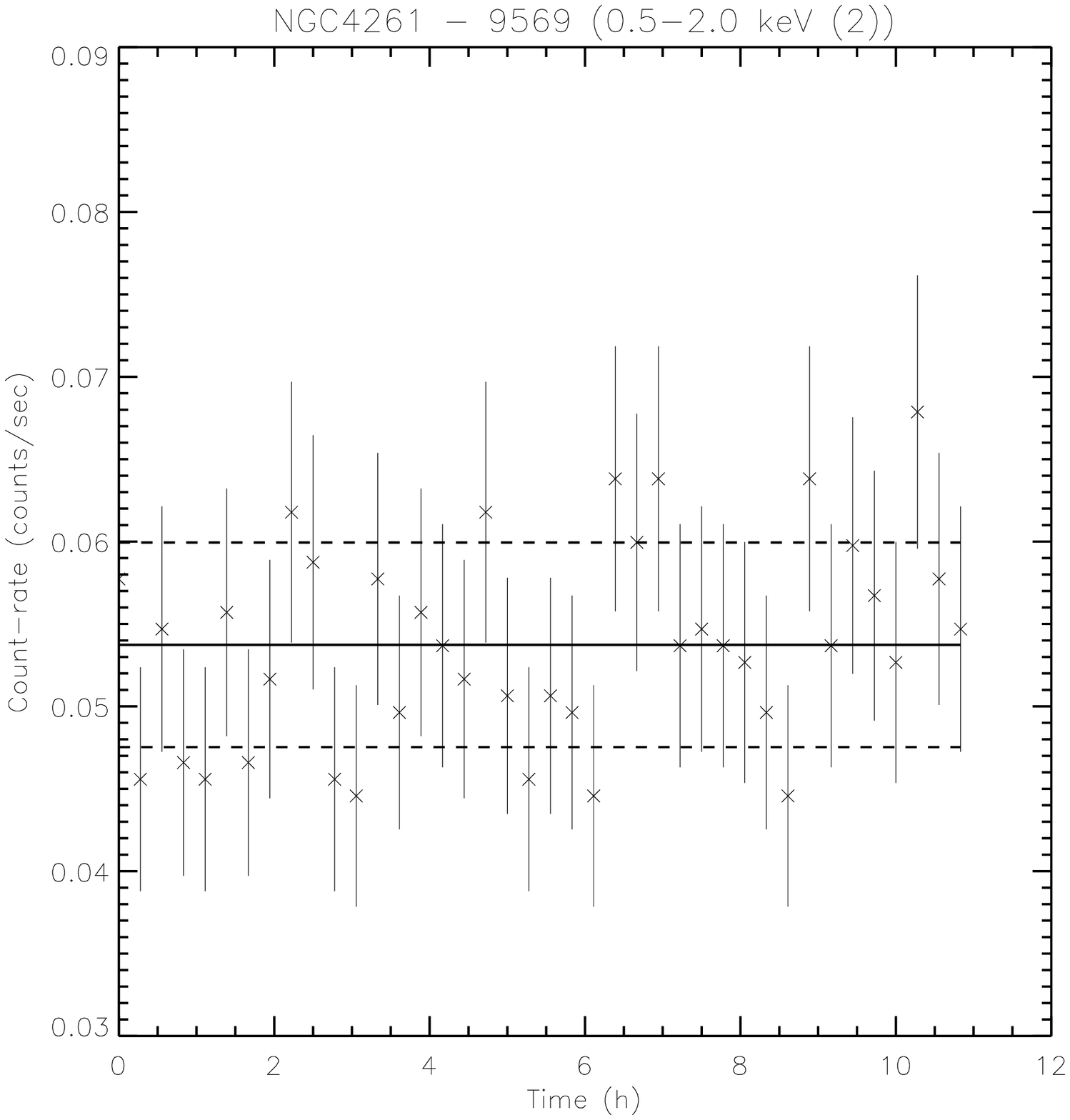}}
\subfloat{\includegraphics[width=0.30\textwidth]{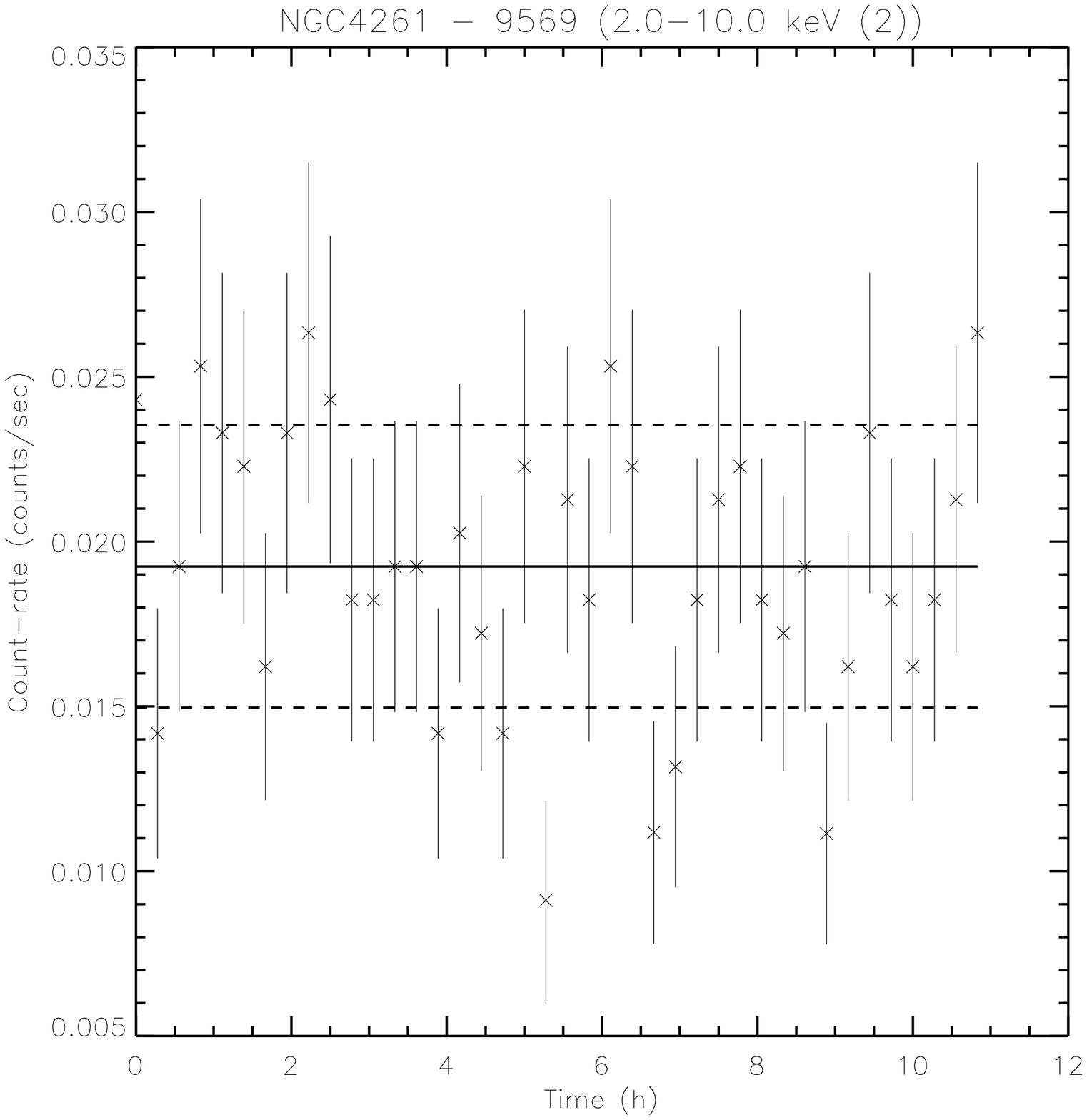}}
\subfloat{\includegraphics[width=0.30\textwidth]{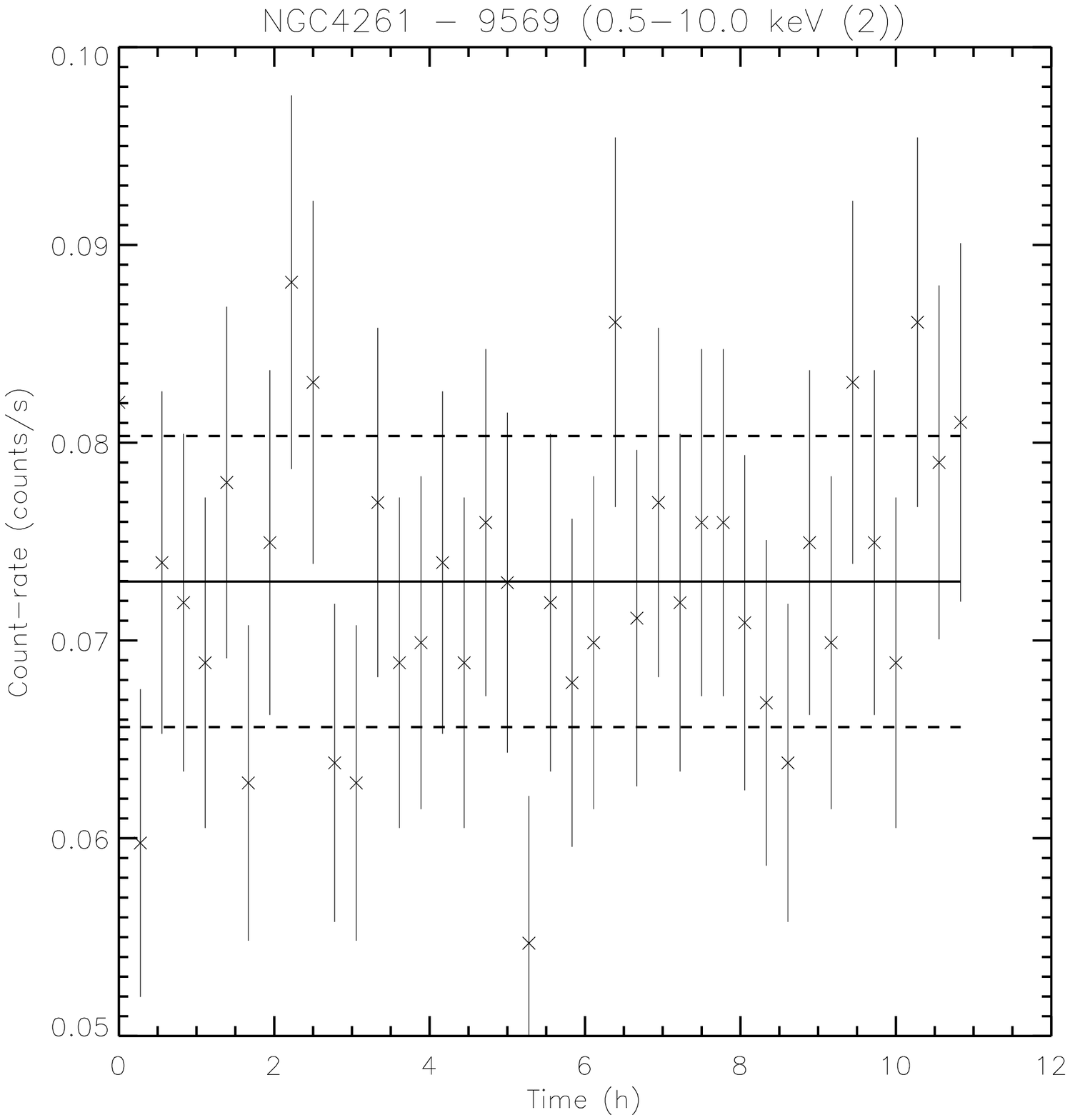}}
\caption{Light curves of NGC\,4261 from \emph{Chandra} data. Note that ObsID. 9569 is divided in two segments.}
\label{l4261}
\end{figure}

\begin{figure}[H]
\centering
\subfloat{\includegraphics[width=0.30\textwidth]{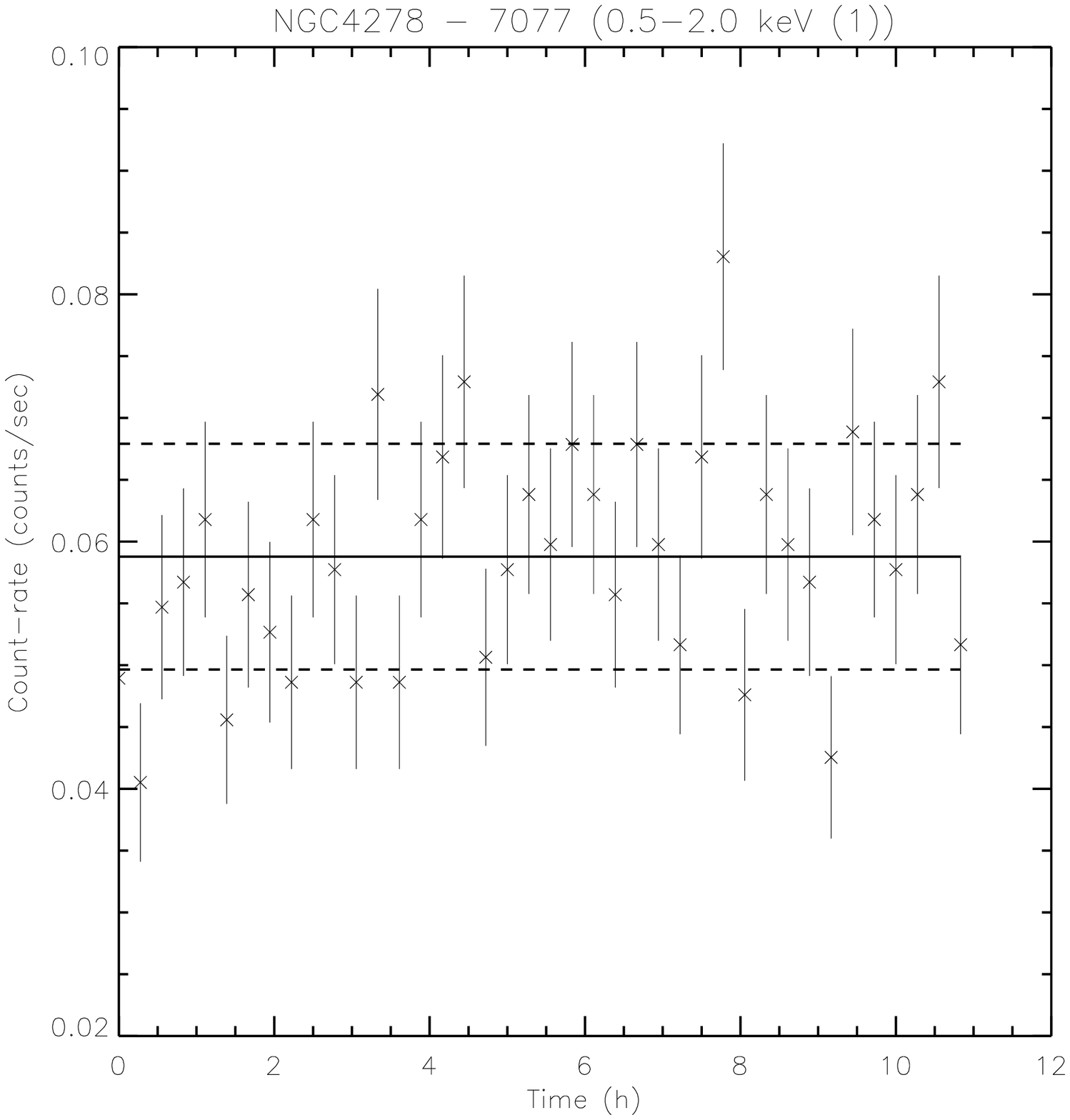}}
\subfloat{\includegraphics[width=0.30\textwidth]{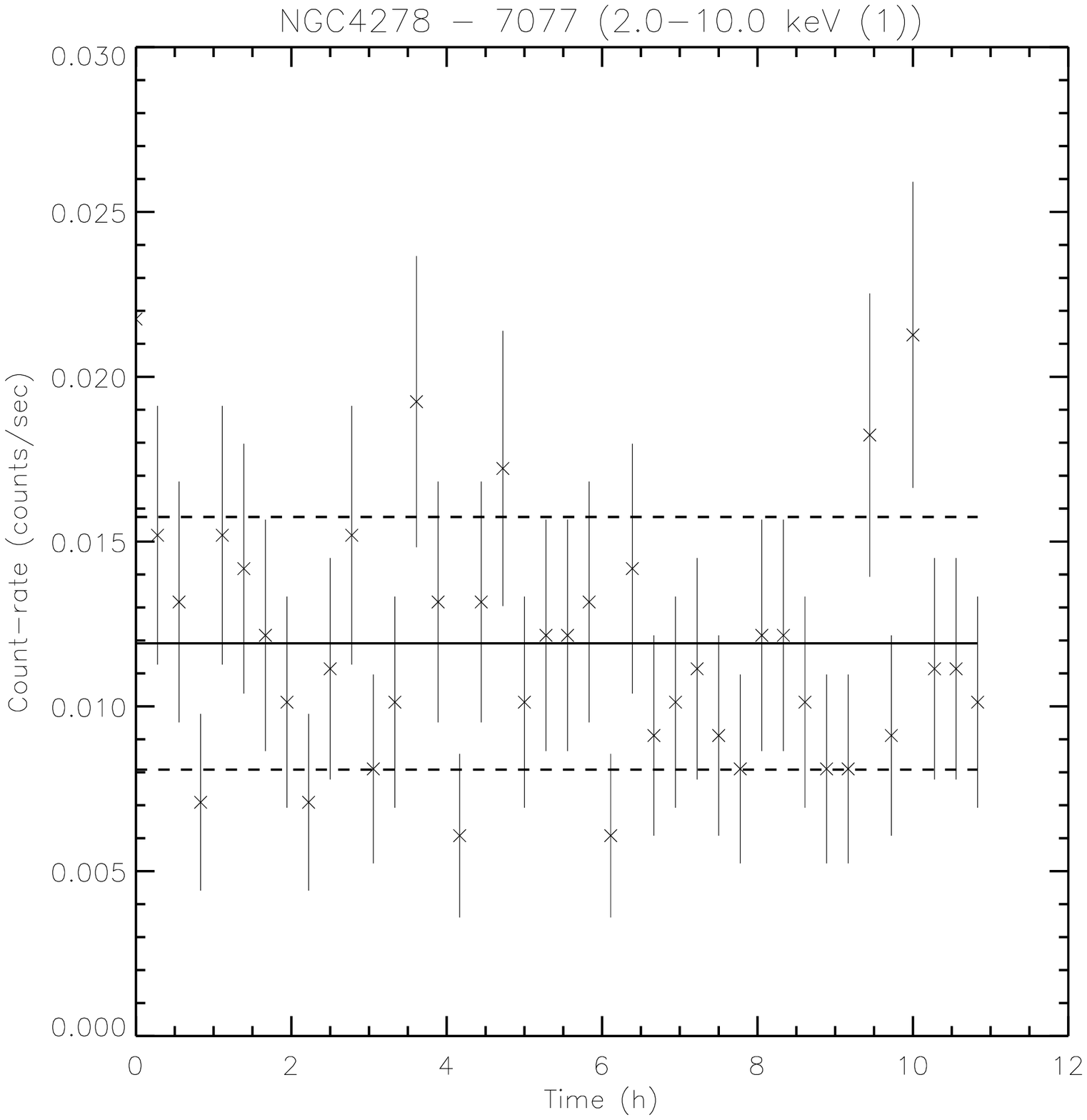}}
\subfloat{\includegraphics[width=0.30\textwidth]{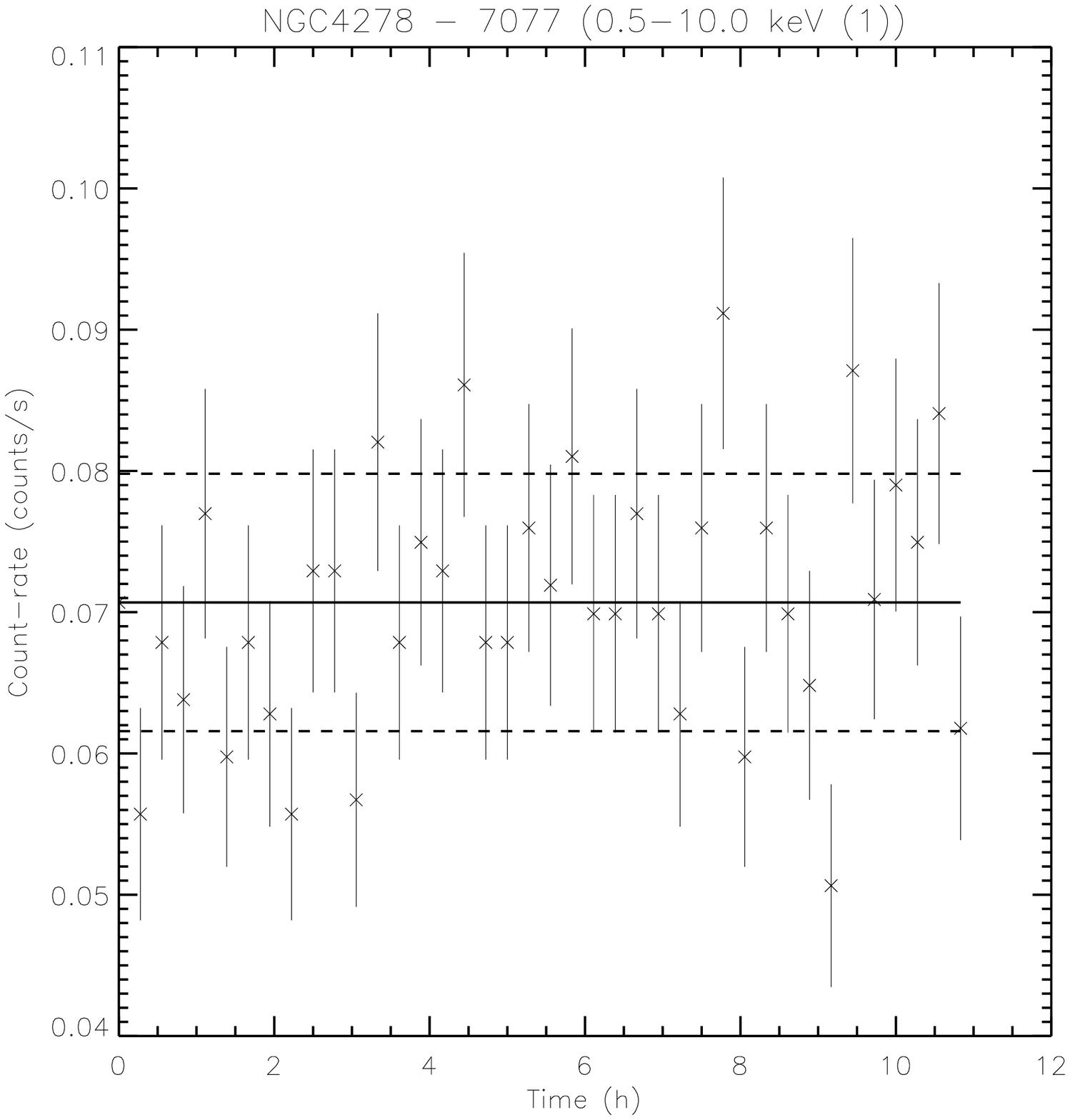}}

\subfloat{\includegraphics[width=0.30\textwidth]{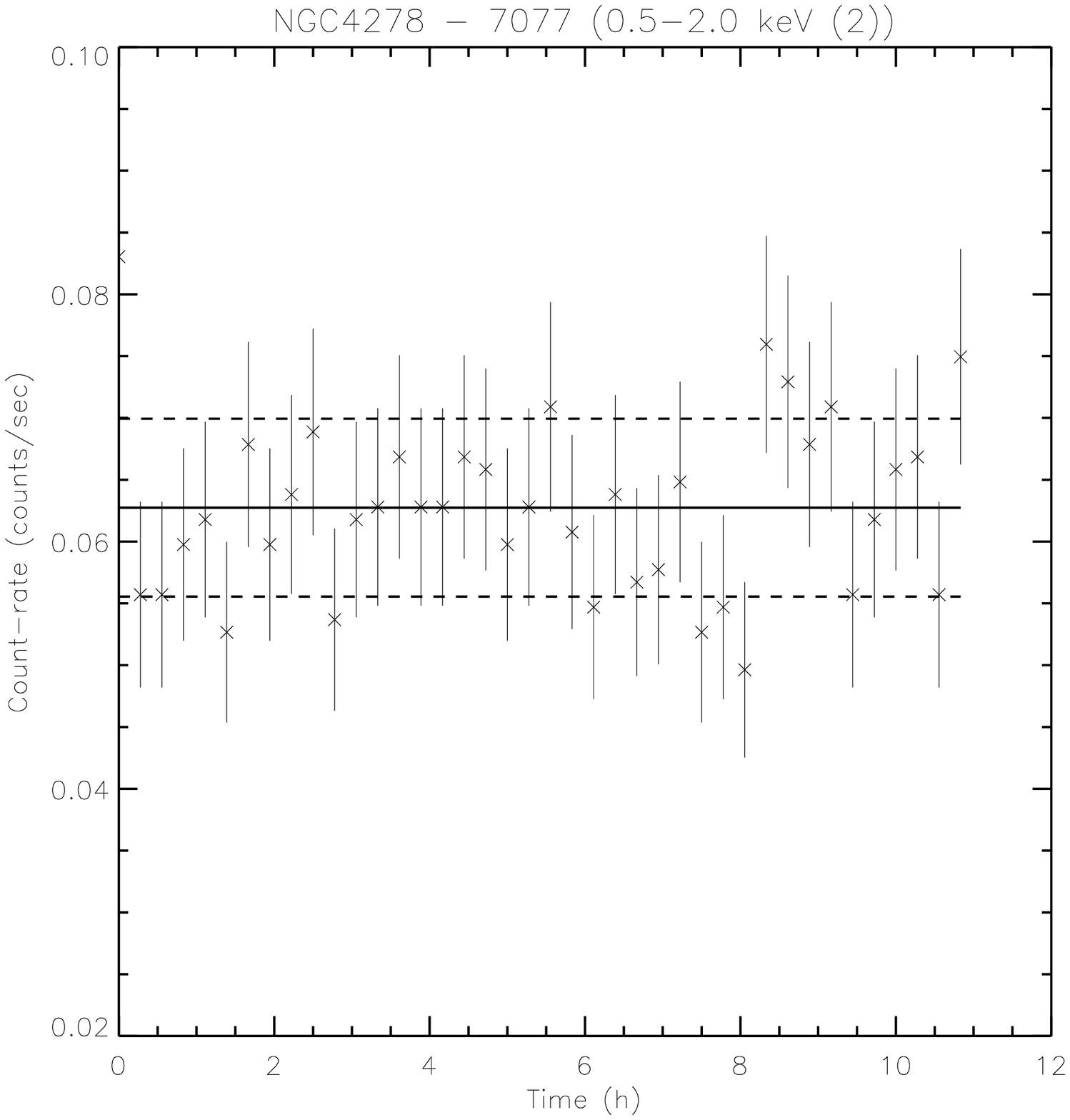}}
\subfloat{\includegraphics[width=0.30\textwidth]{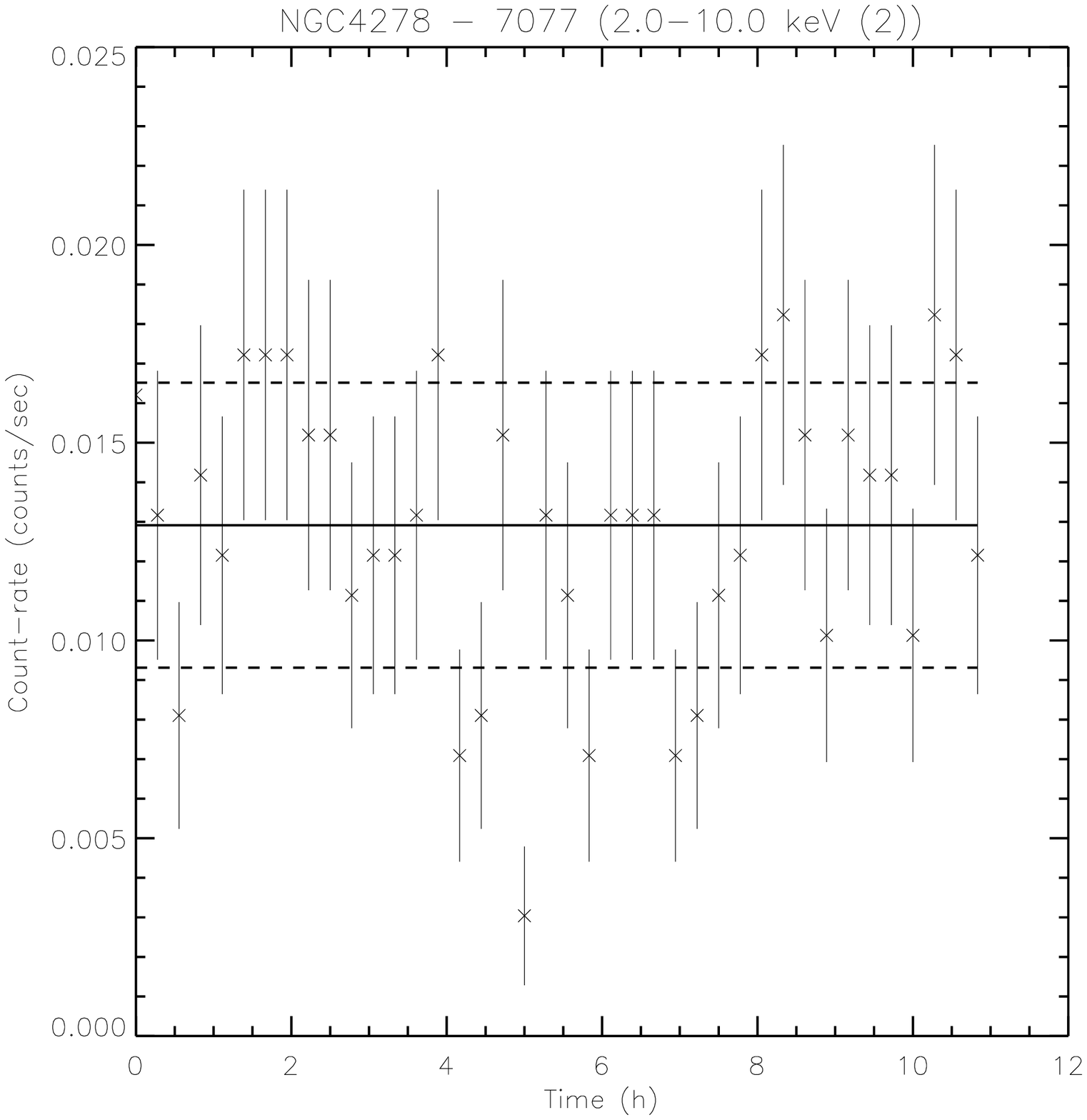}}
\subfloat{\includegraphics[width=0.30\textwidth]{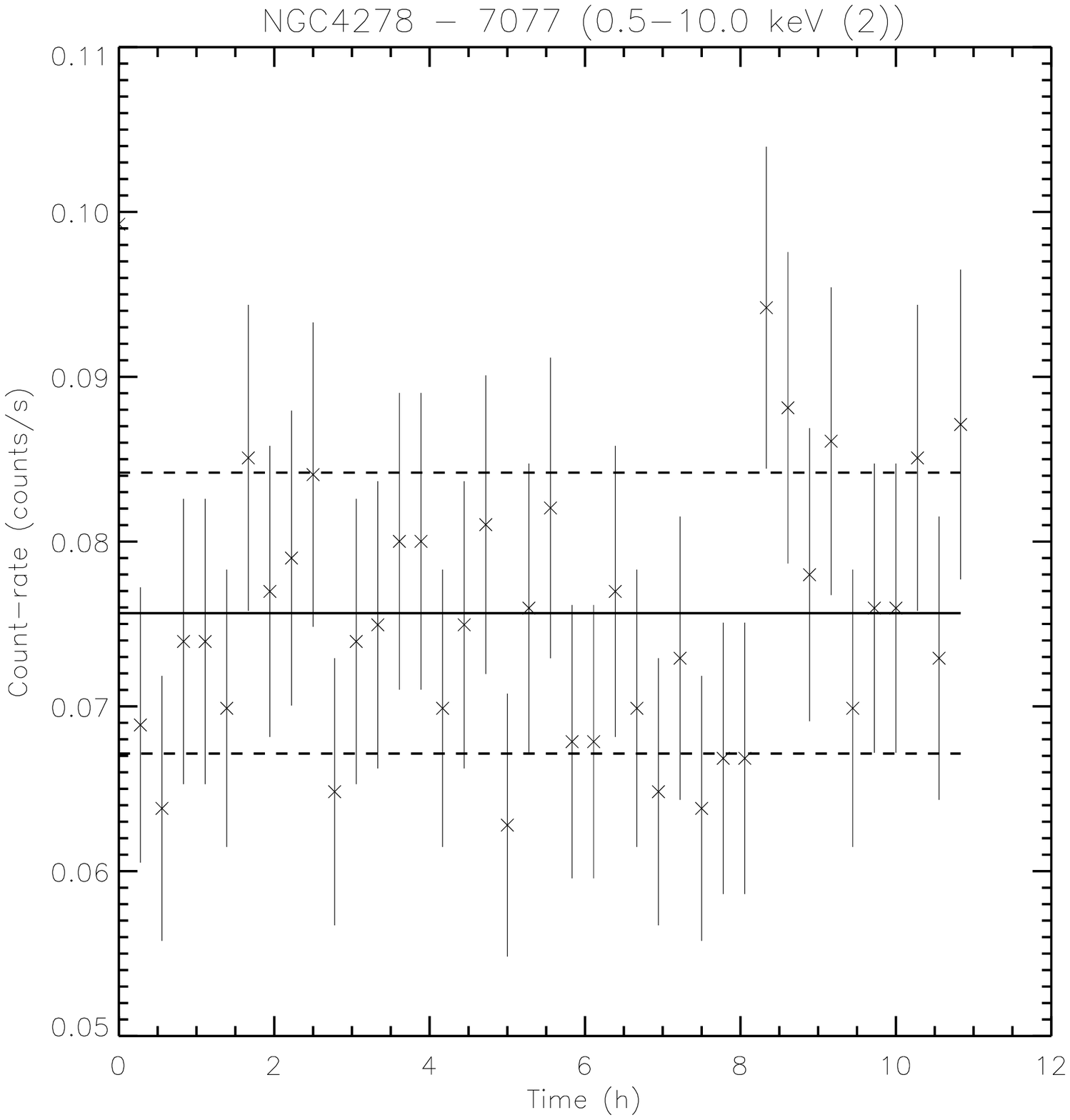}}

\subfloat{\includegraphics[width=0.30\textwidth]{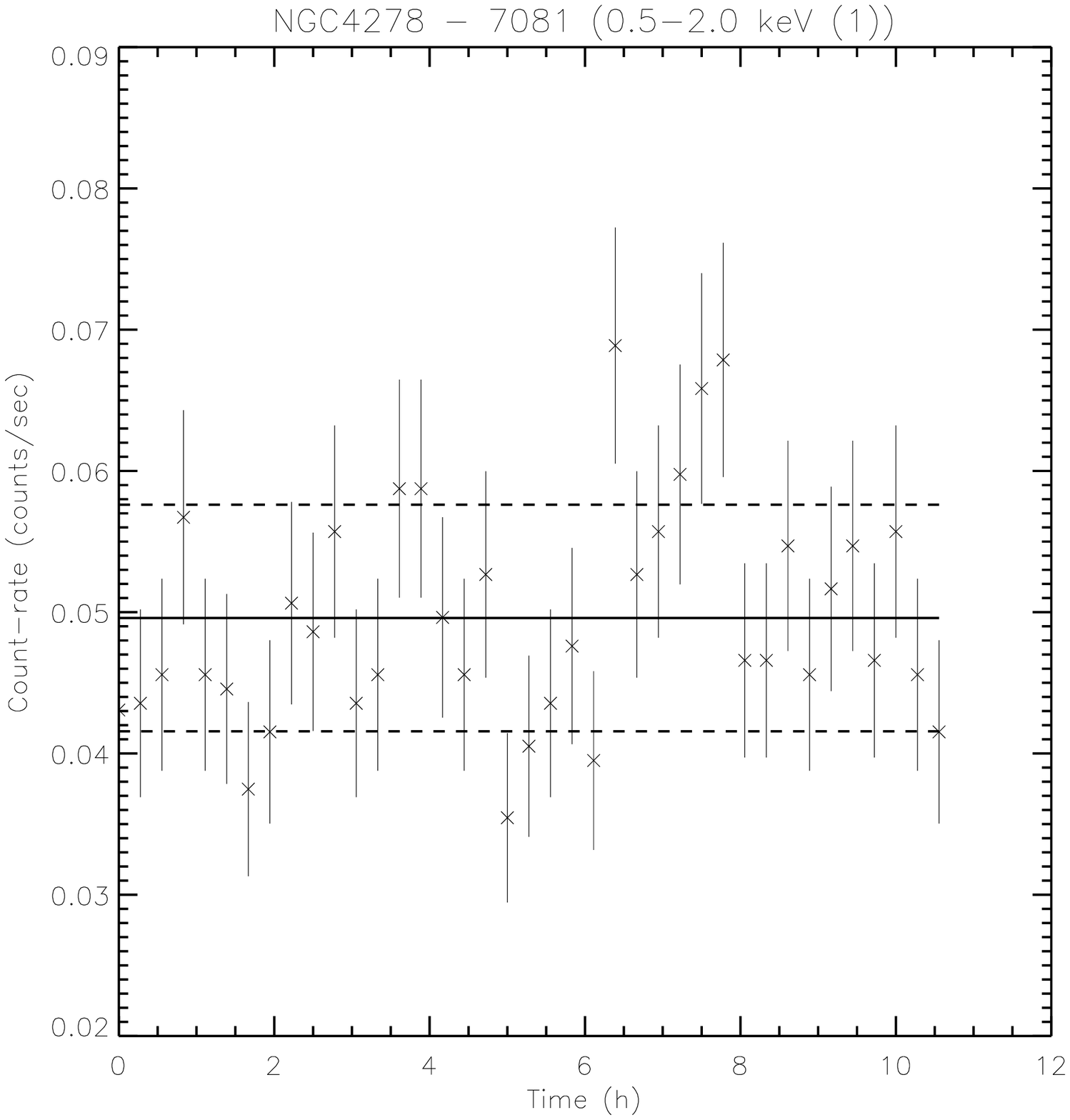}}
\subfloat{\includegraphics[width=0.30\textwidth]{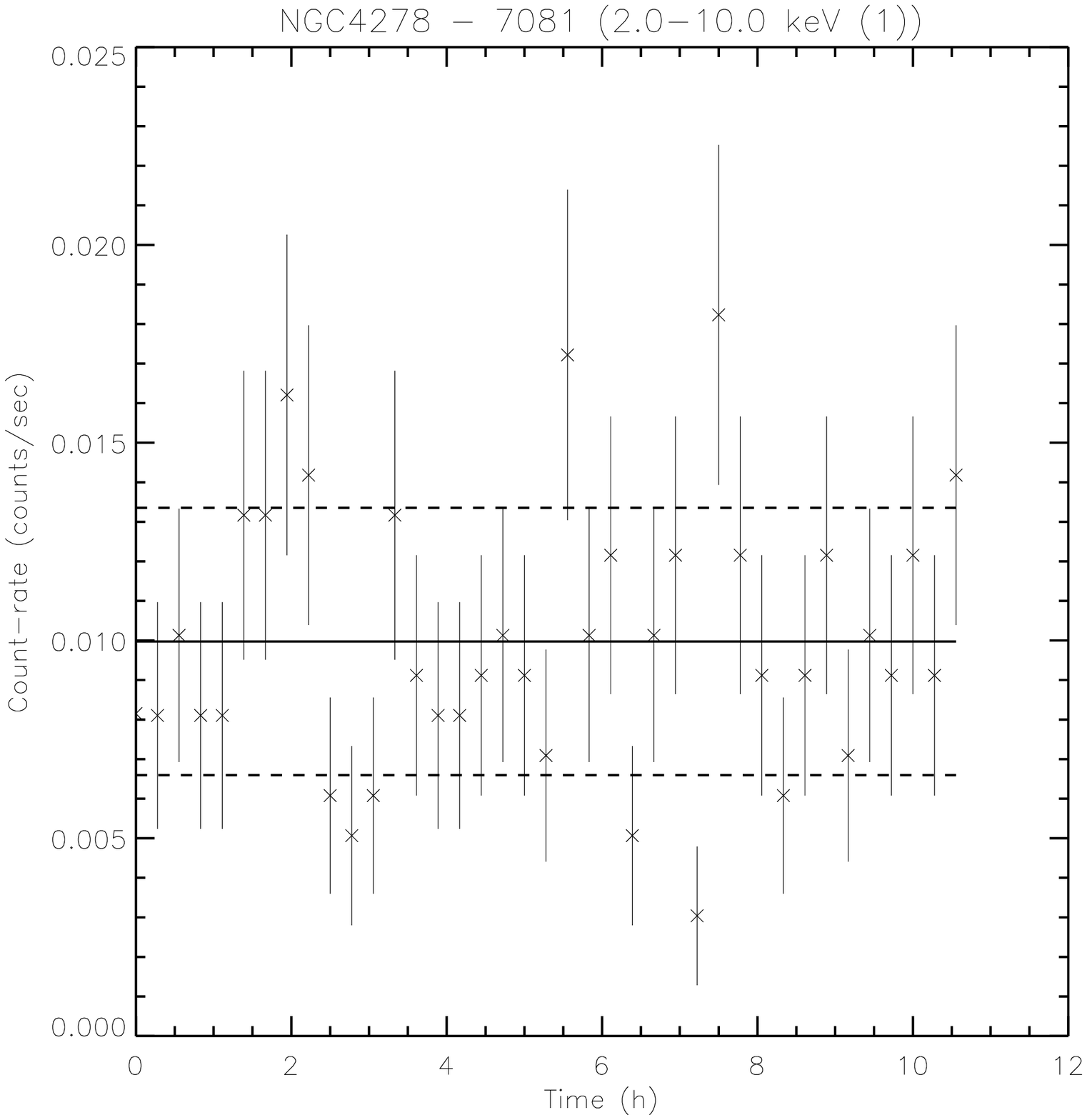}}
\subfloat{\includegraphics[width=0.30\textwidth]{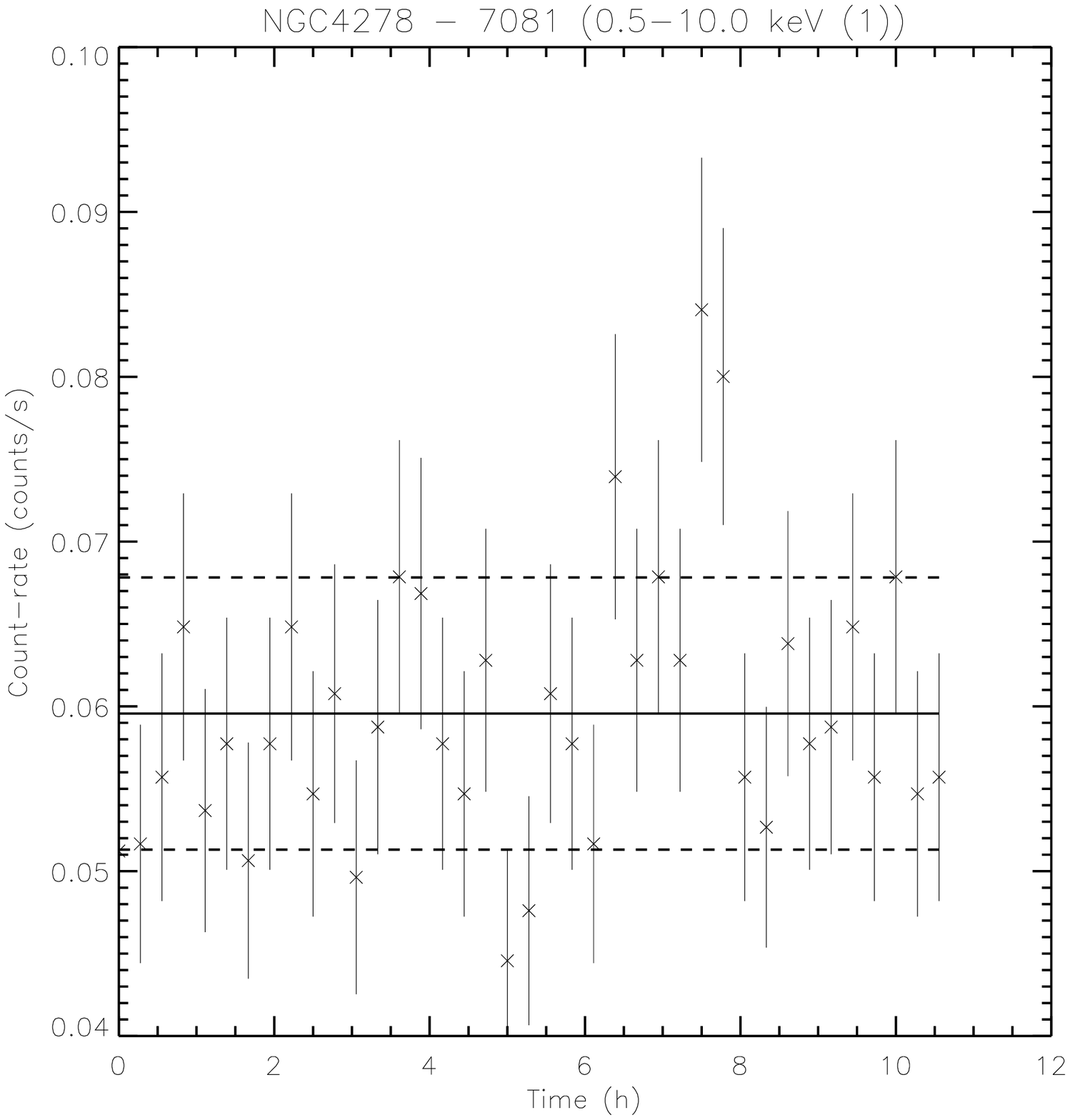}}

\subfloat{\includegraphics[width=0.30\textwidth]{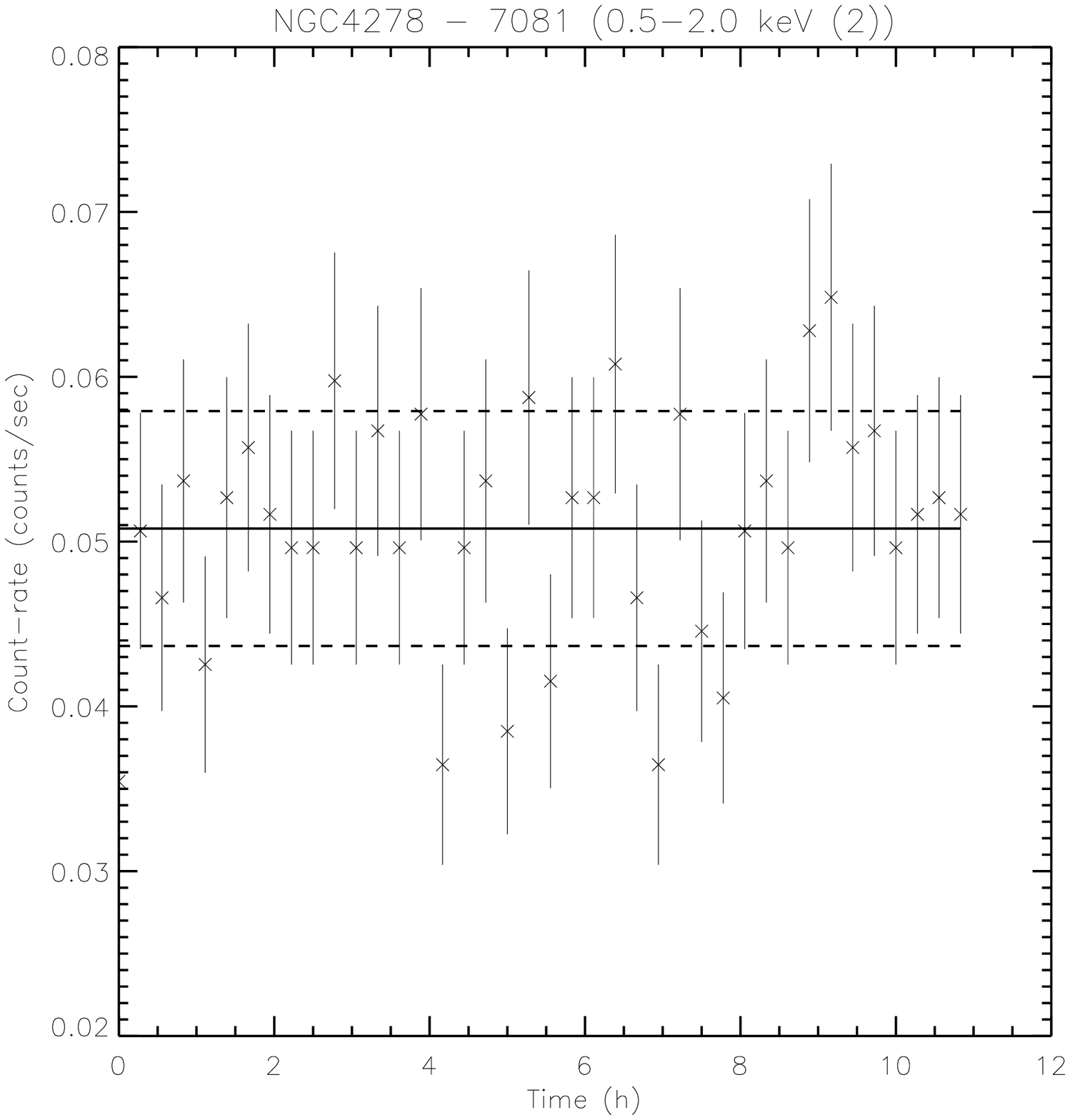}}
\subfloat{\includegraphics[width=0.30\textwidth]{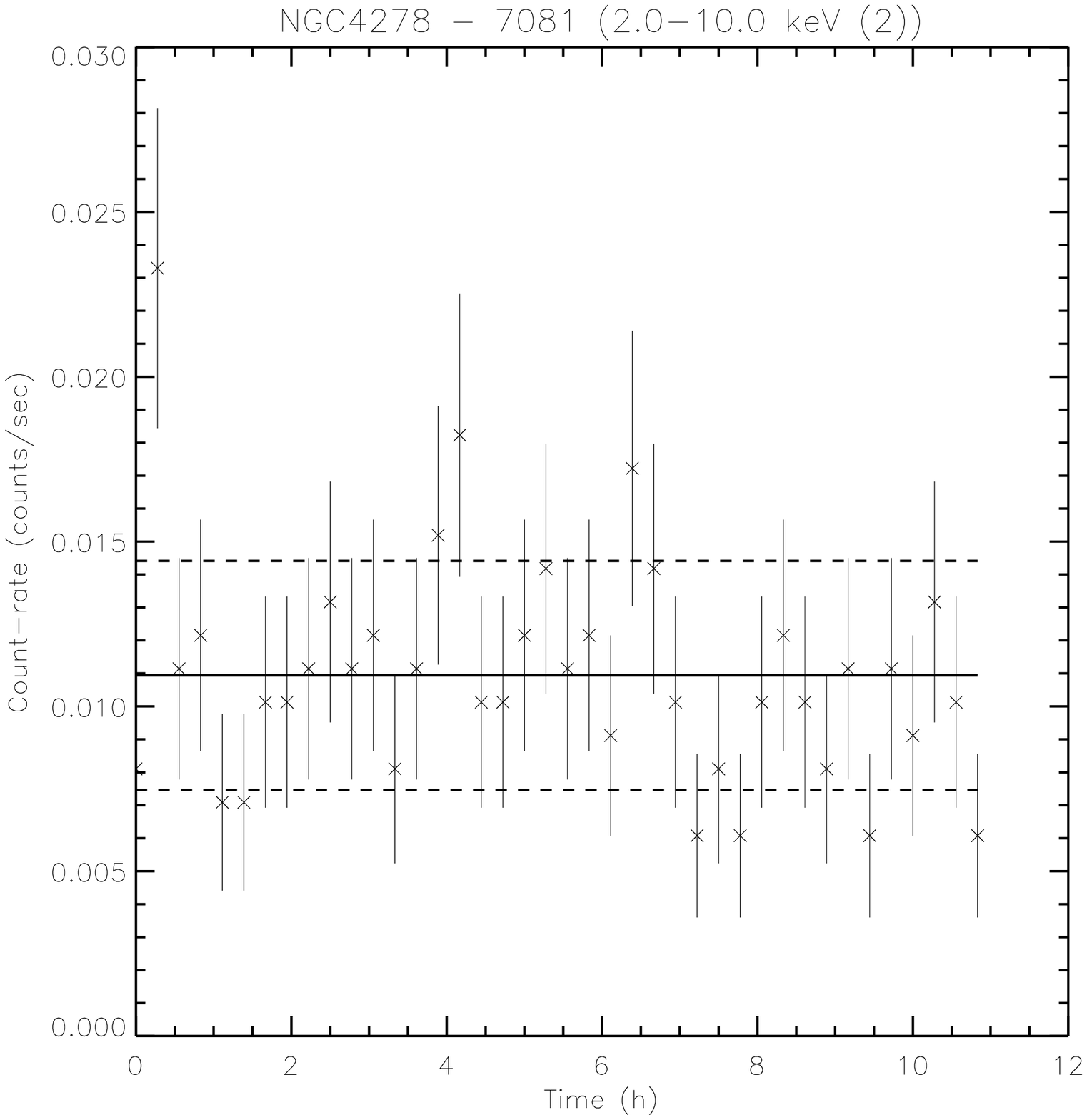}}
\subfloat{\includegraphics[width=0.30\textwidth]{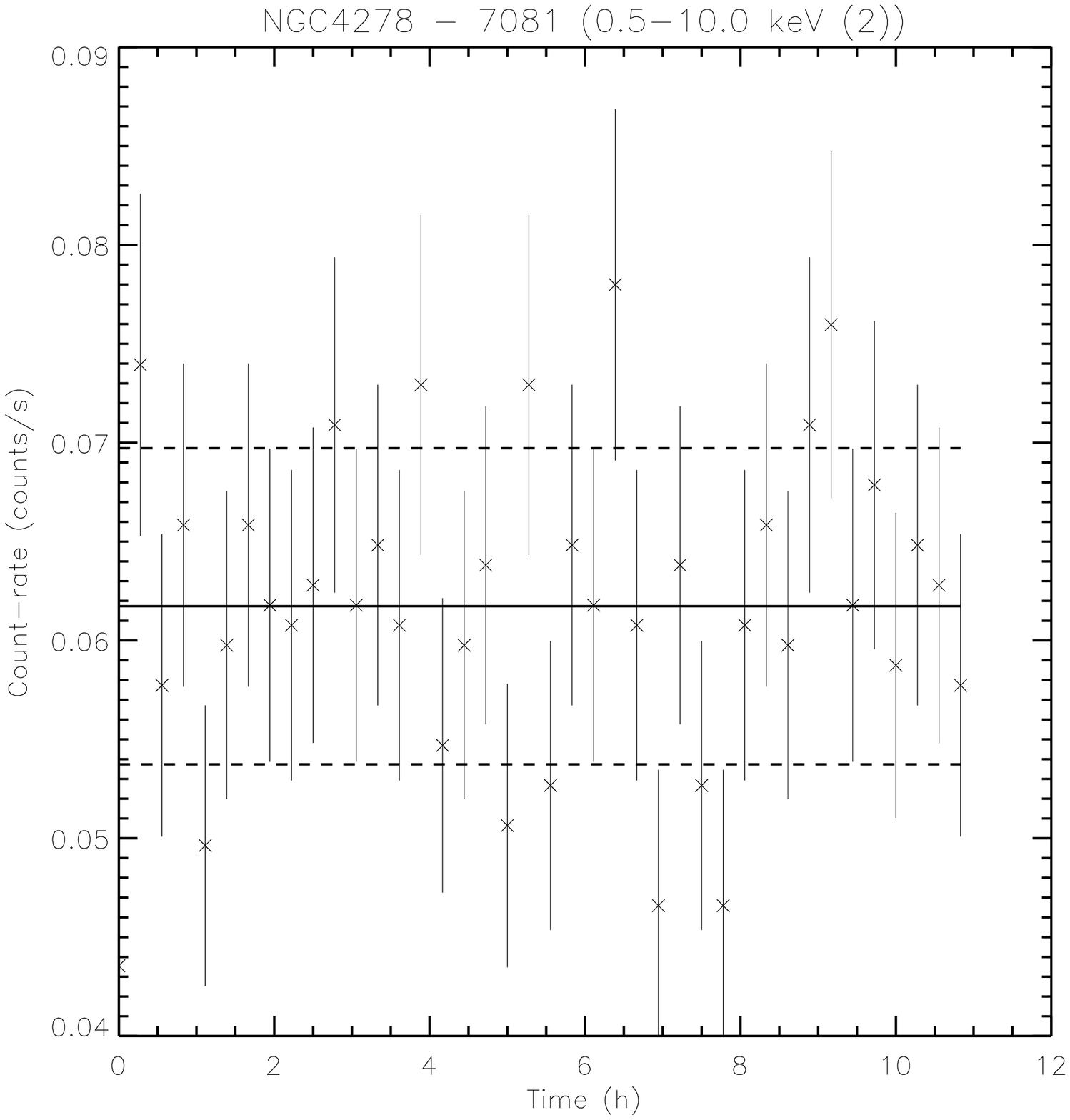}}
\caption{Light curves of NGC\,4278 from \emph{Chandra} data. Note that ObsID. 7077 and 7081 are divided in two segments.}
\label{l4278}
\end{figure}

\begin{figure}[H]
\setcounter{figure}{7}
\centering
\subfloat{\includegraphics[width=0.30\textwidth]{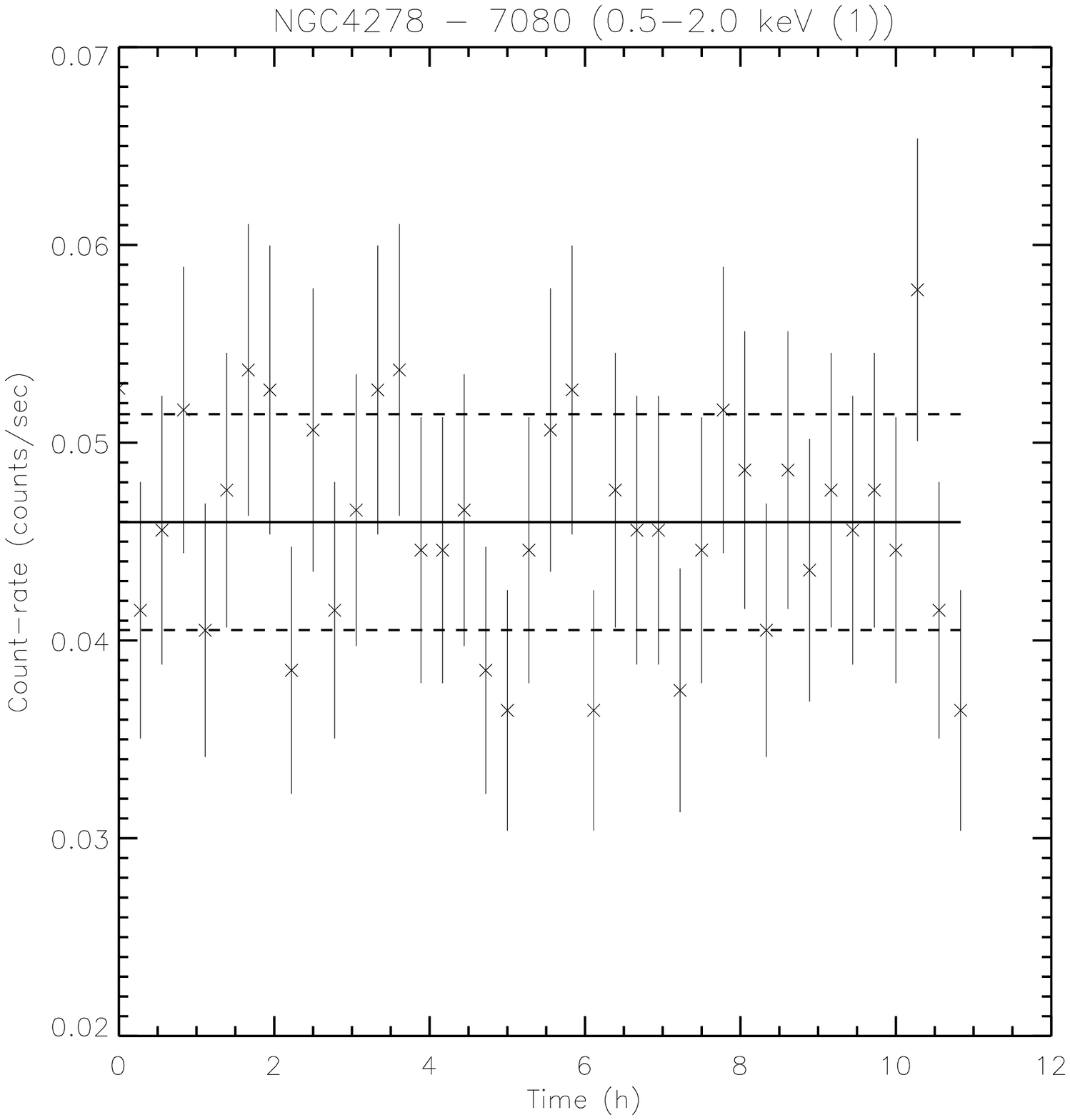}}
\subfloat{\includegraphics[width=0.30\textwidth]{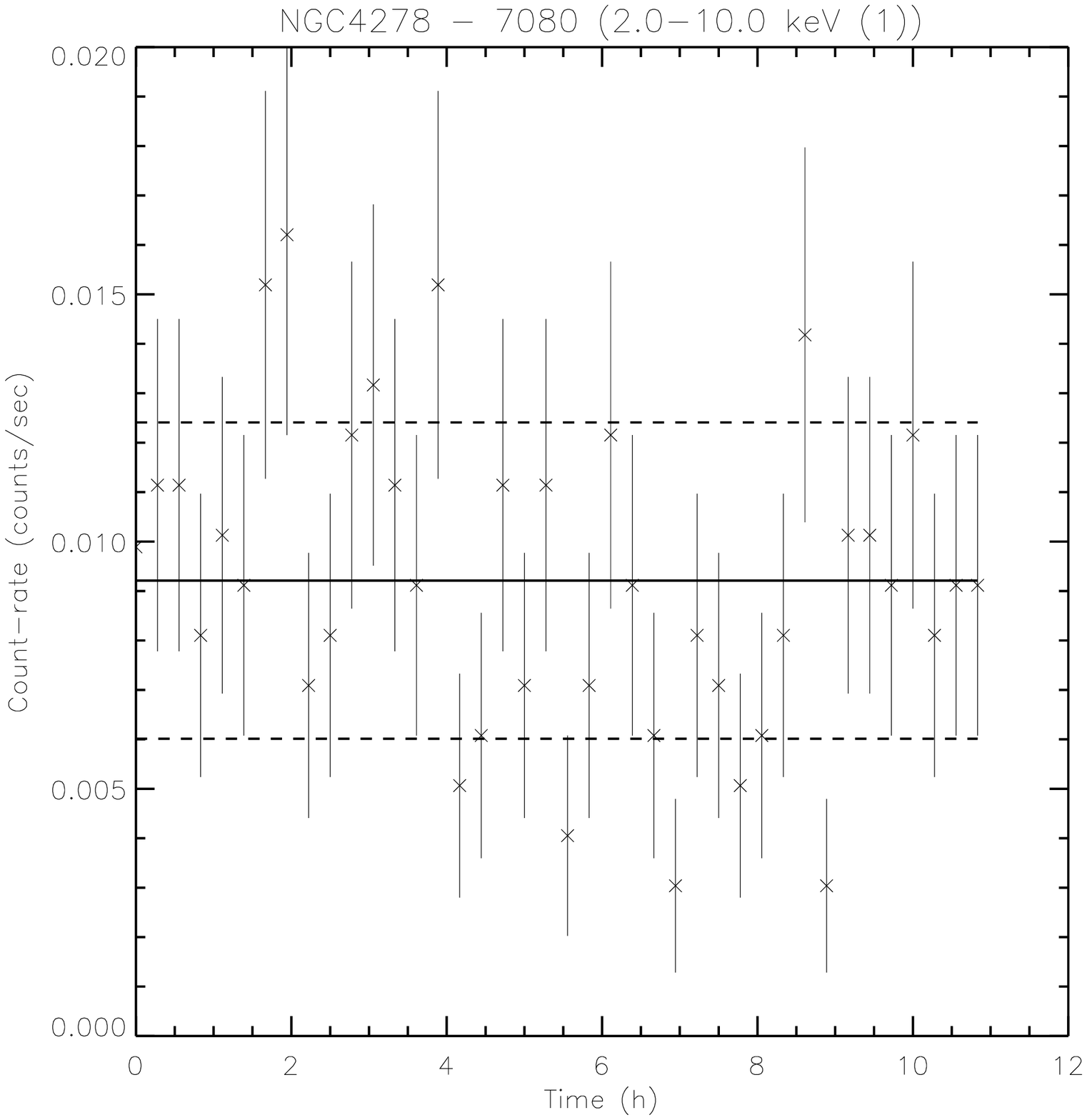}}
\subfloat{\includegraphics[width=0.30\textwidth]{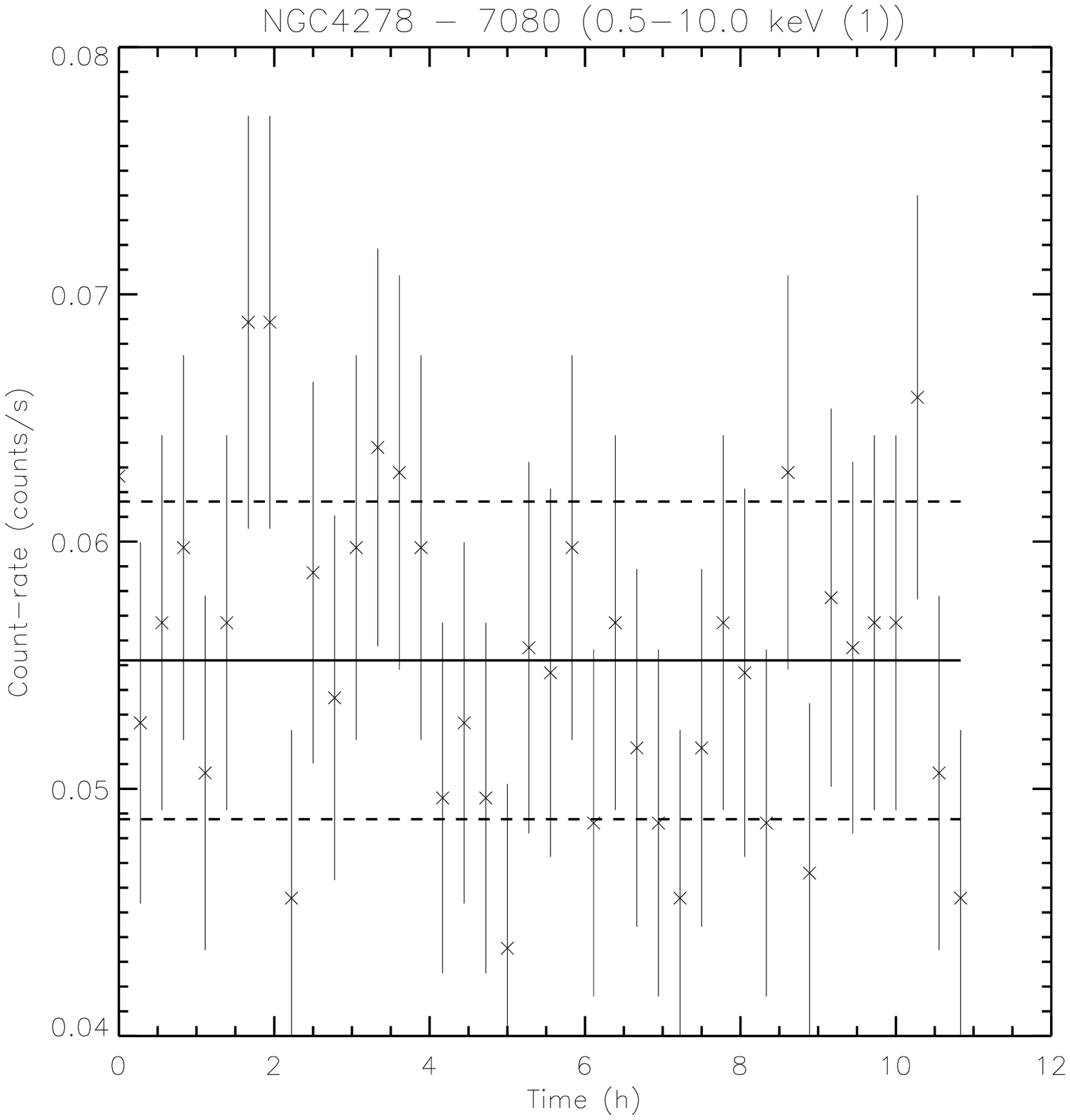}}
\caption{(Cont.)}
\end{figure}

\begin{figure}[H]
\centering
\subfloat{\includegraphics[width=0.30\textwidth]{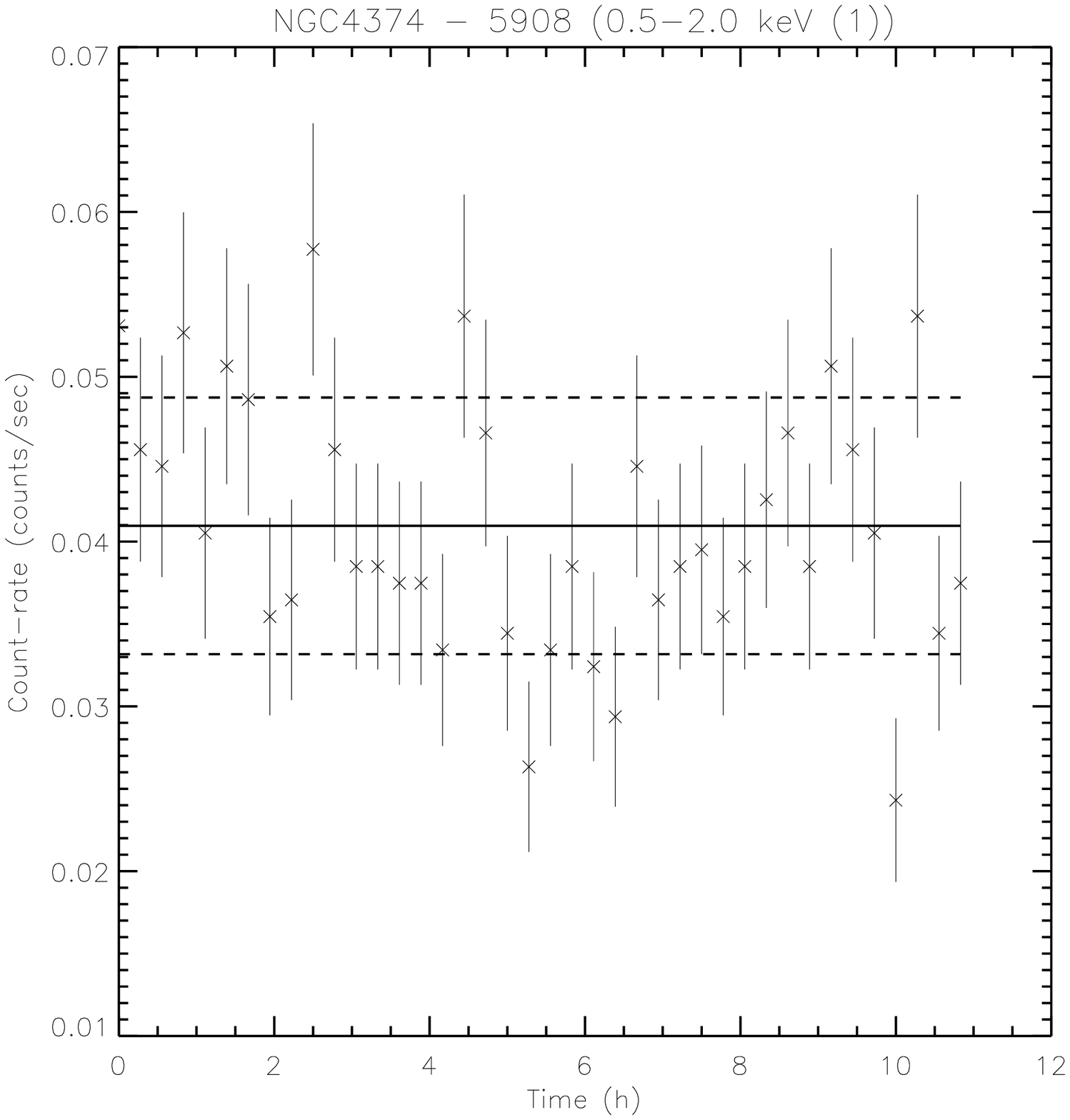}}
\subfloat{\includegraphics[width=0.30\textwidth]{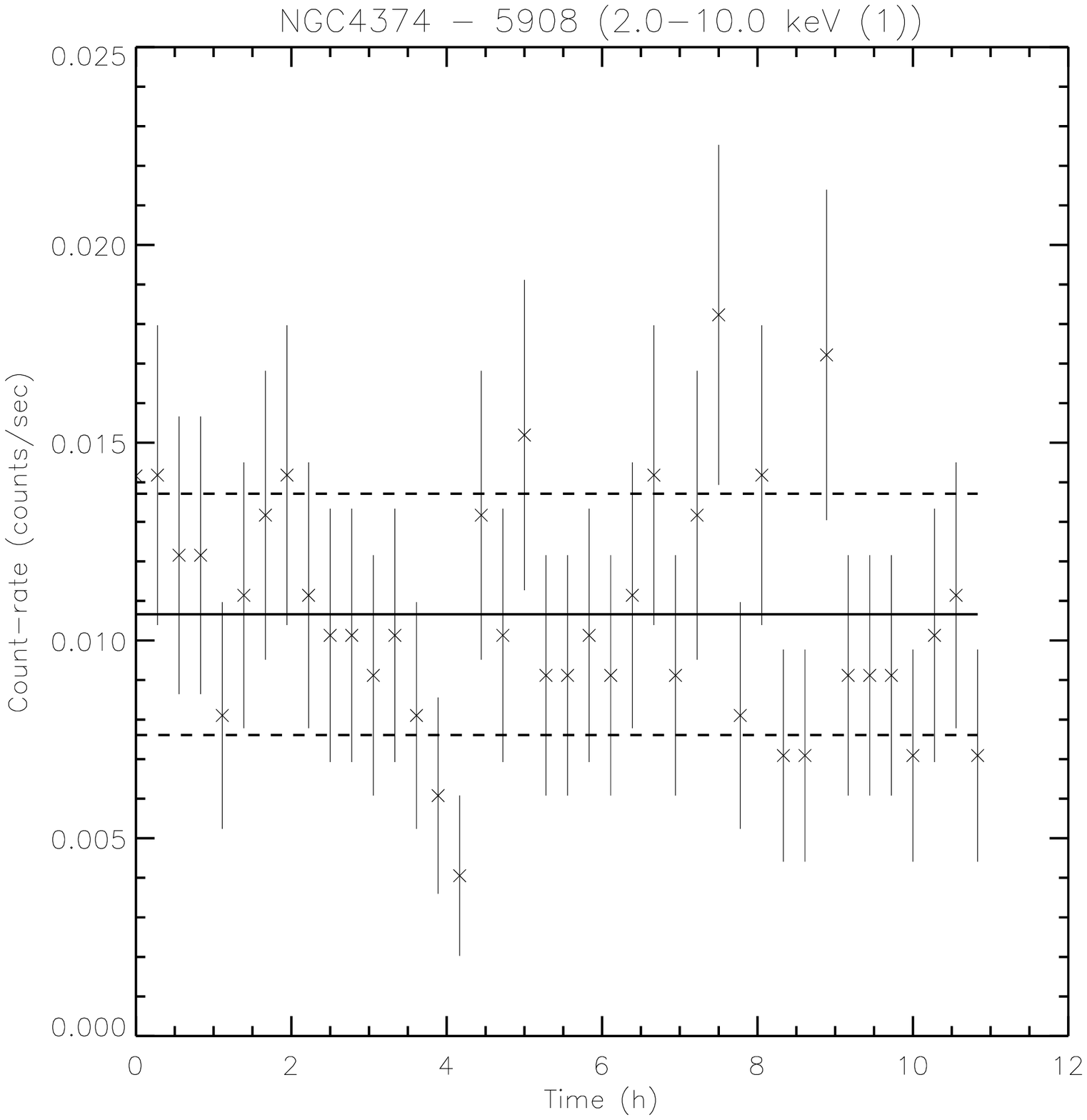}}
\subfloat{\includegraphics[width=0.30\textwidth]{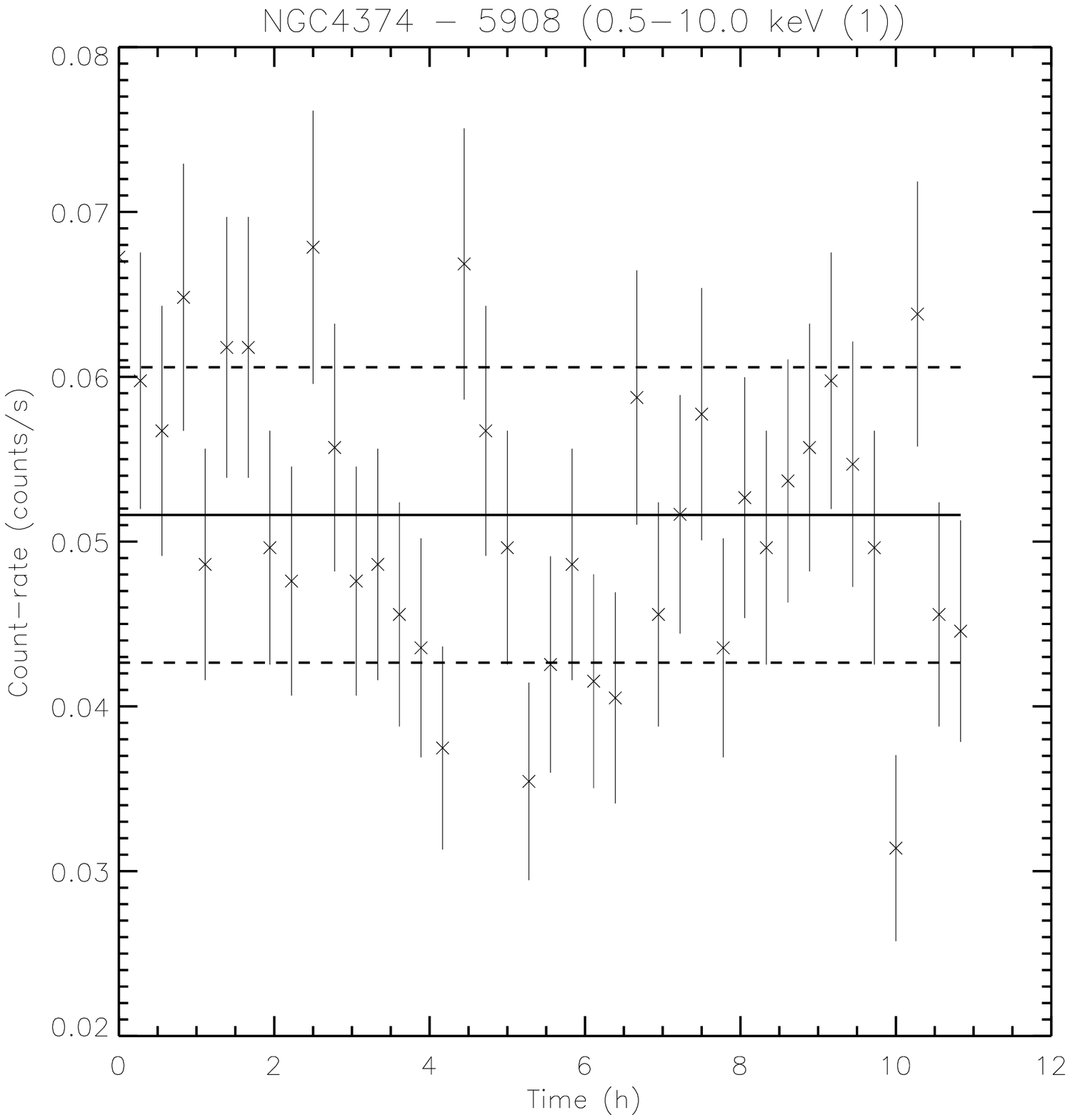}}

\subfloat{\includegraphics[width=0.30\textwidth]{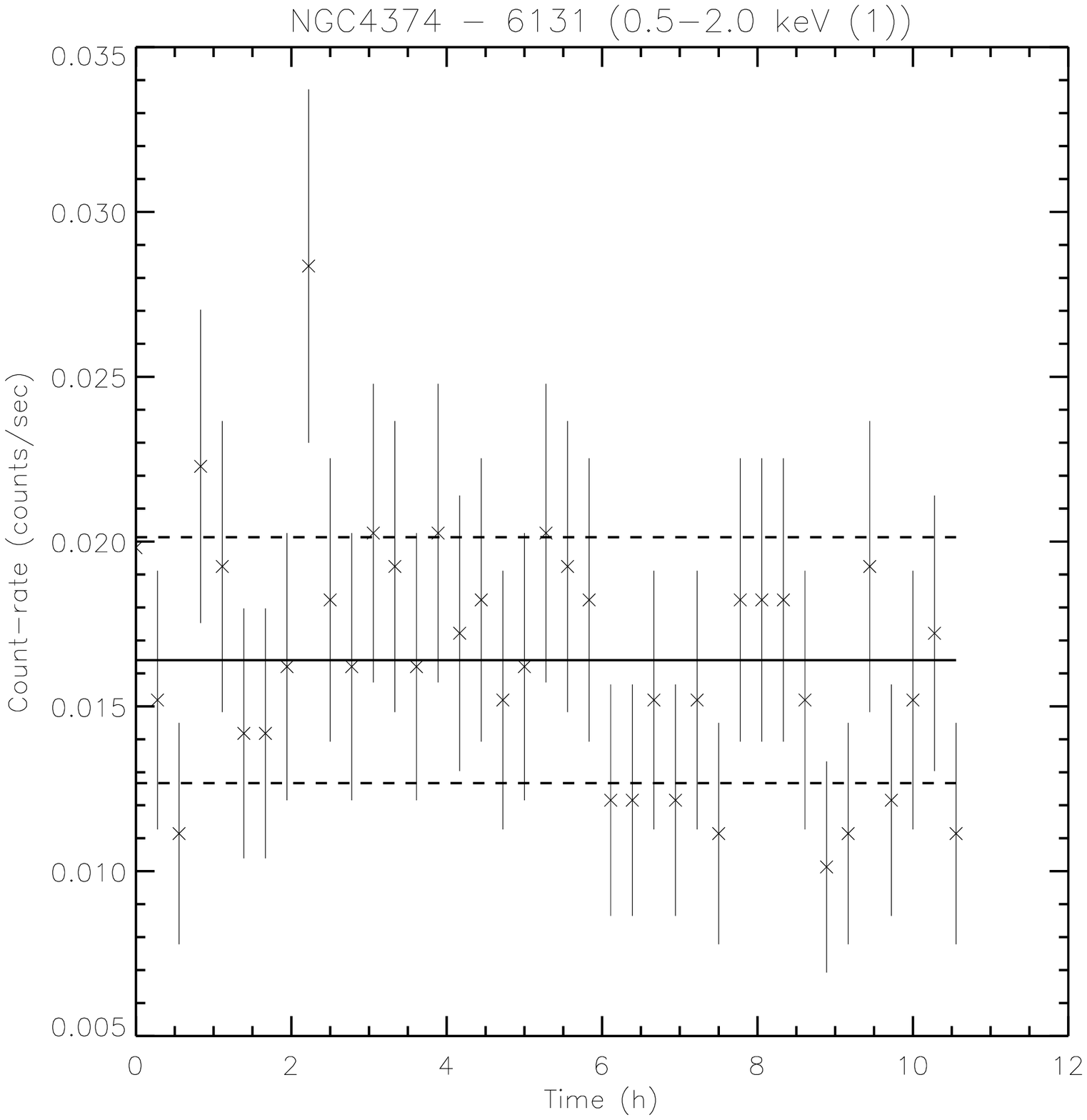}}
\subfloat{\includegraphics[width=0.30\textwidth]{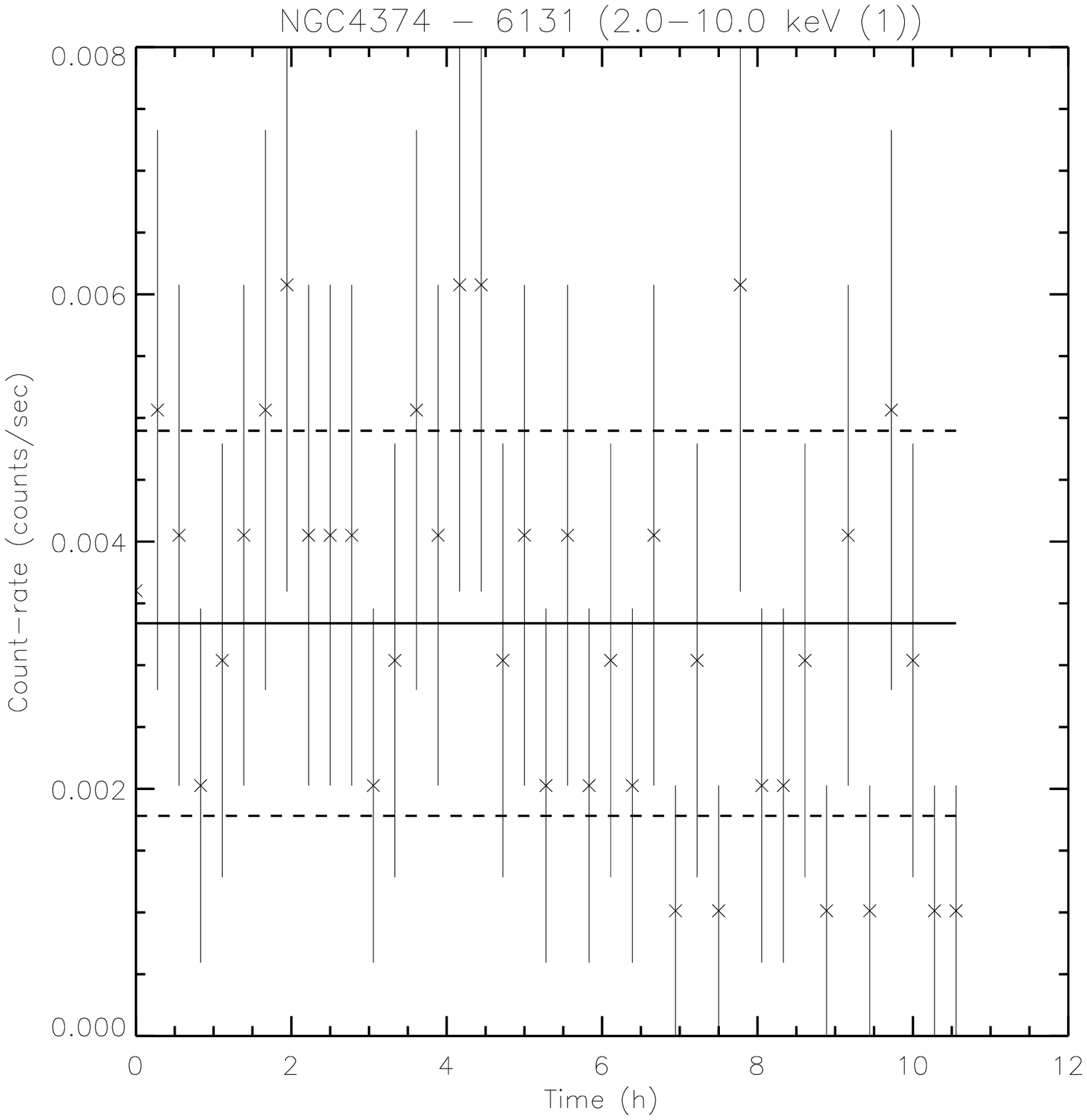}}
\subfloat{\includegraphics[width=0.30\textwidth]{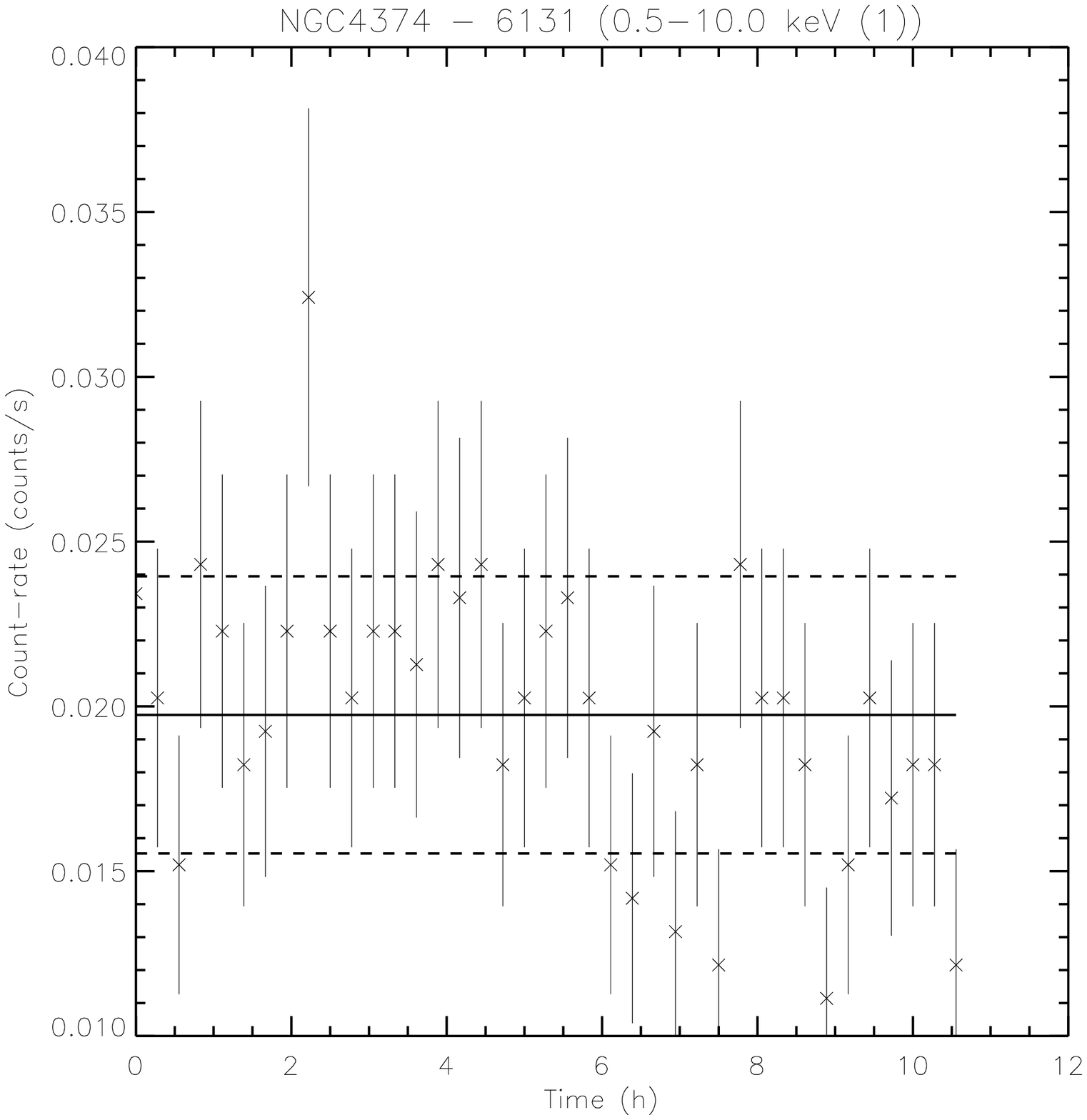}}
\caption{Light curves of NGC\,4374 from \emph{Chandra} data.}
\label{l4374}
\end{figure}

\begin{figure}[H]
\centering
\subfloat{\includegraphics[width=0.30\textwidth]{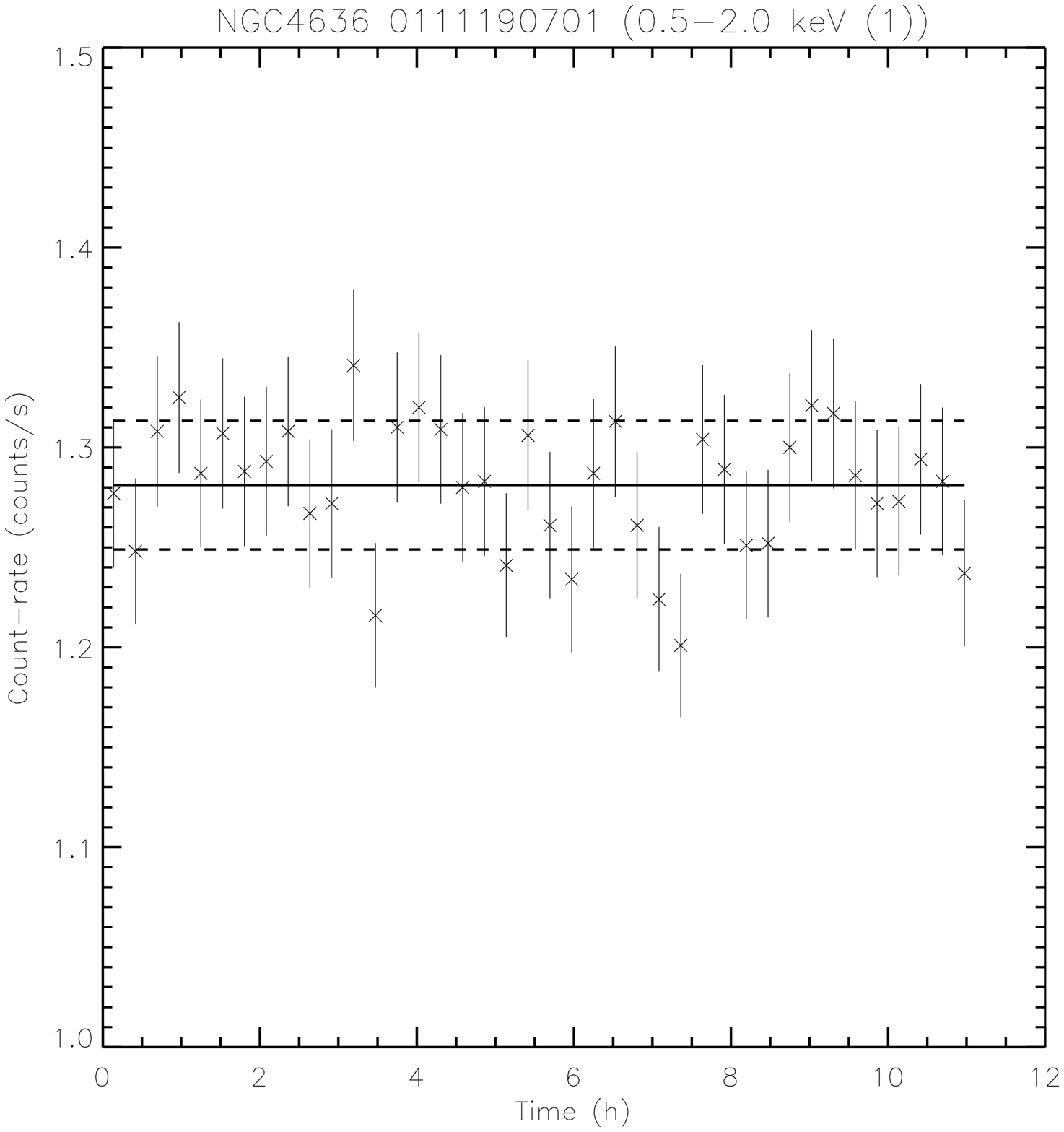}}
\subfloat{\includegraphics[width=0.30\textwidth]{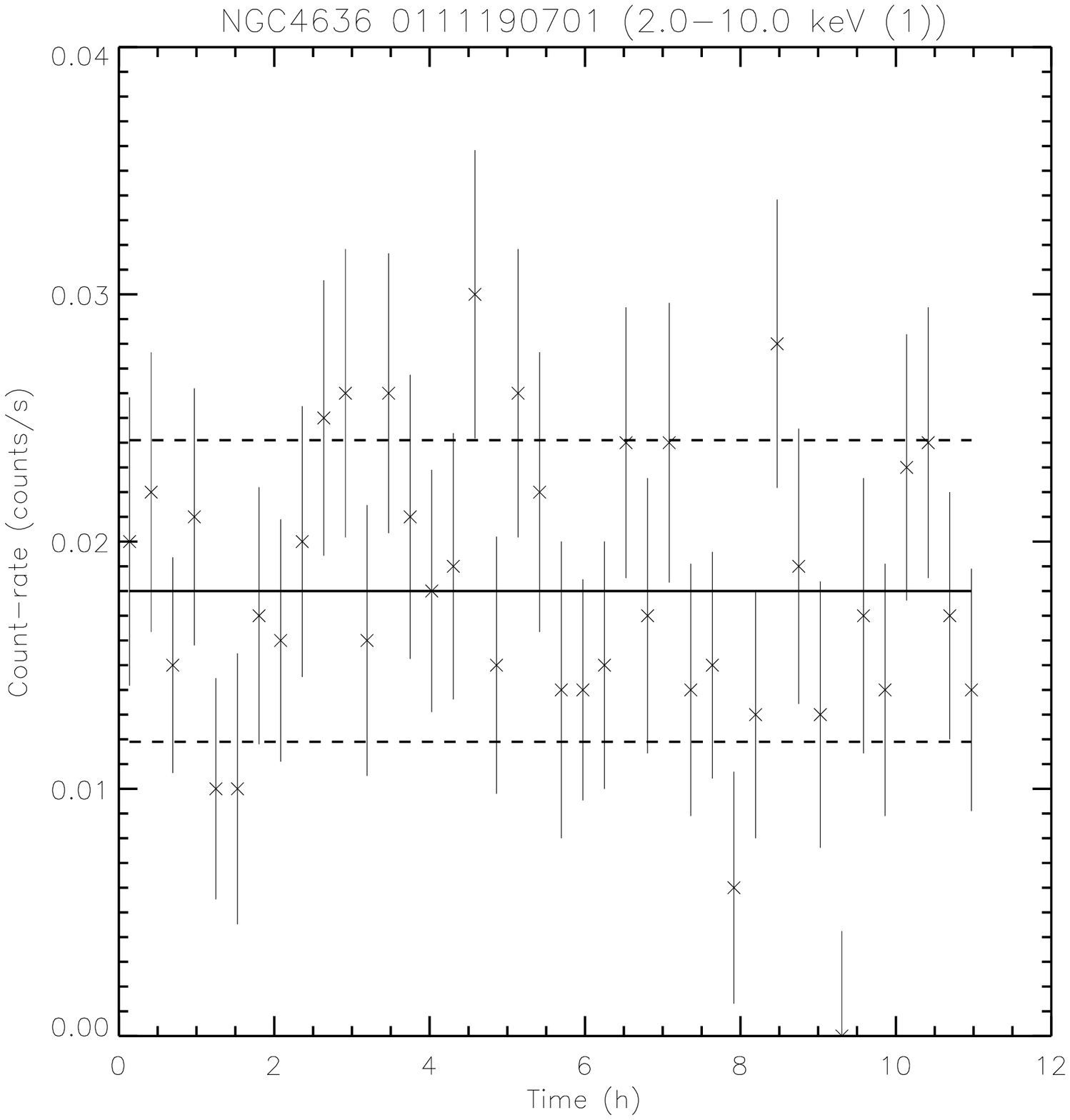}}
\subfloat{\includegraphics[width=0.30\textwidth]{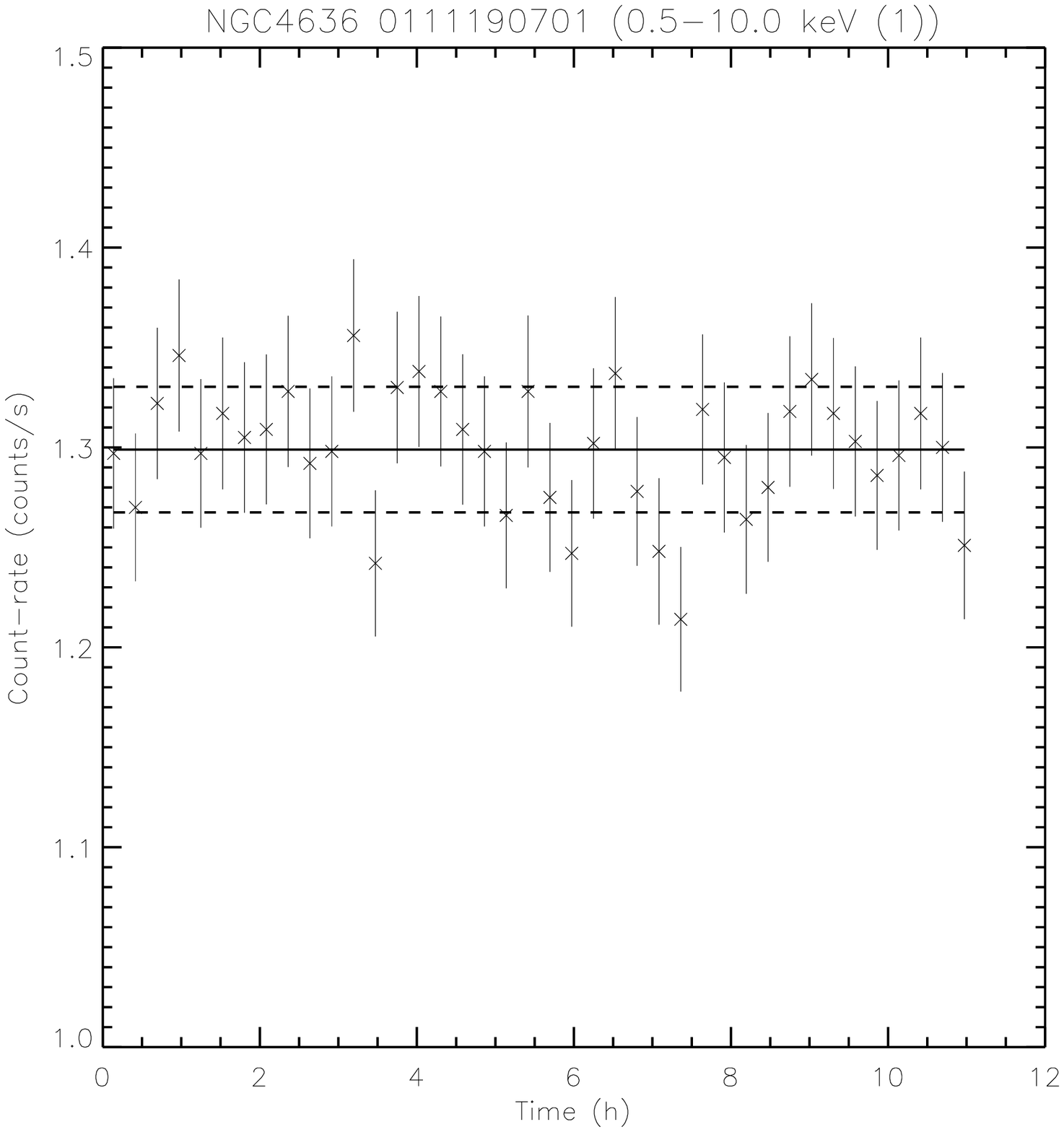}}
\caption{Light curves of NGC\,4636 from \emph{XMM}--Newton data.}
\label{l4636}
\end{figure}

\begin{figure}[H]
\centering
\subfloat{\includegraphics[width=0.30\textwidth]{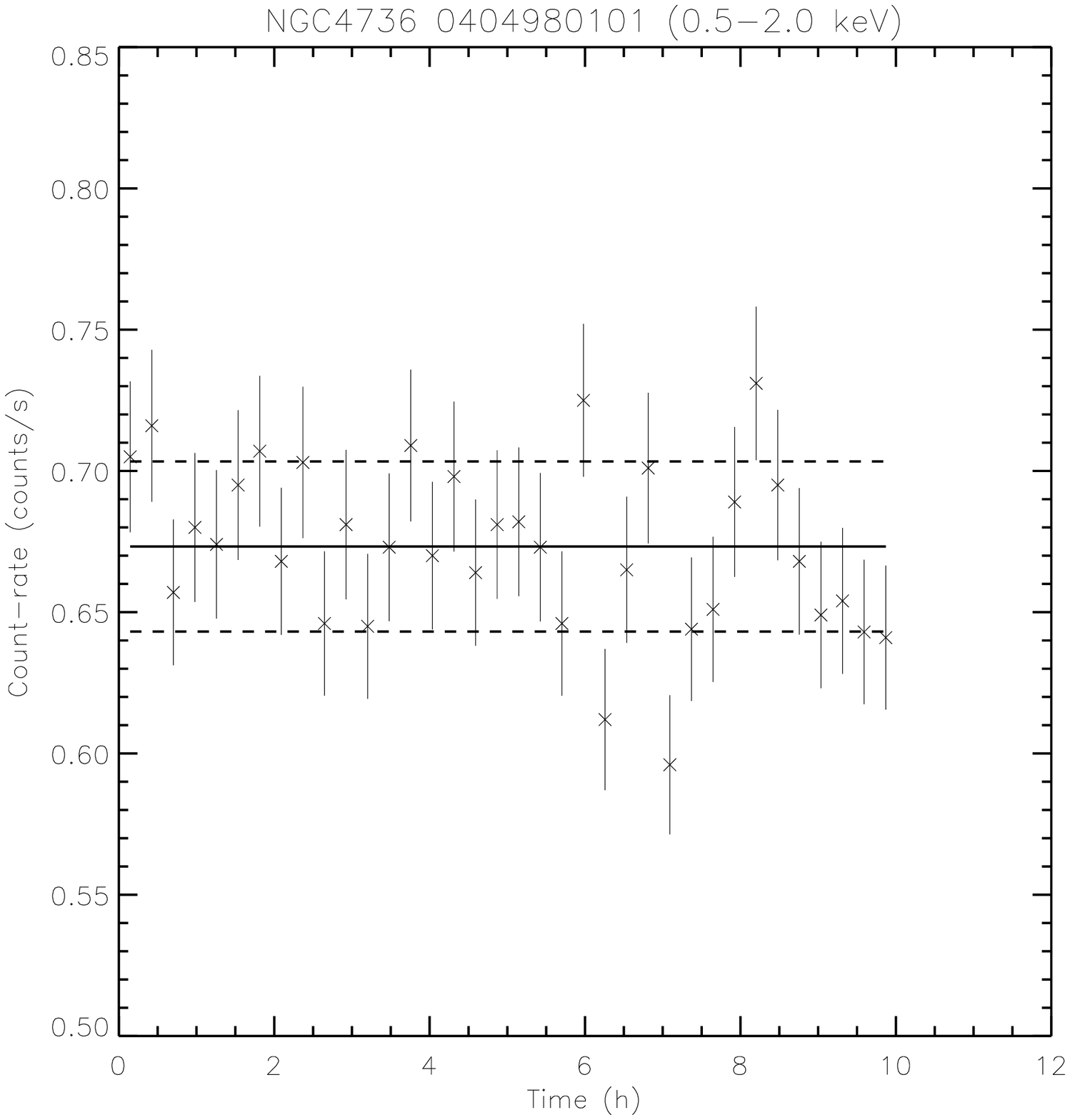}}
\subfloat{\includegraphics[width=0.30\textwidth]{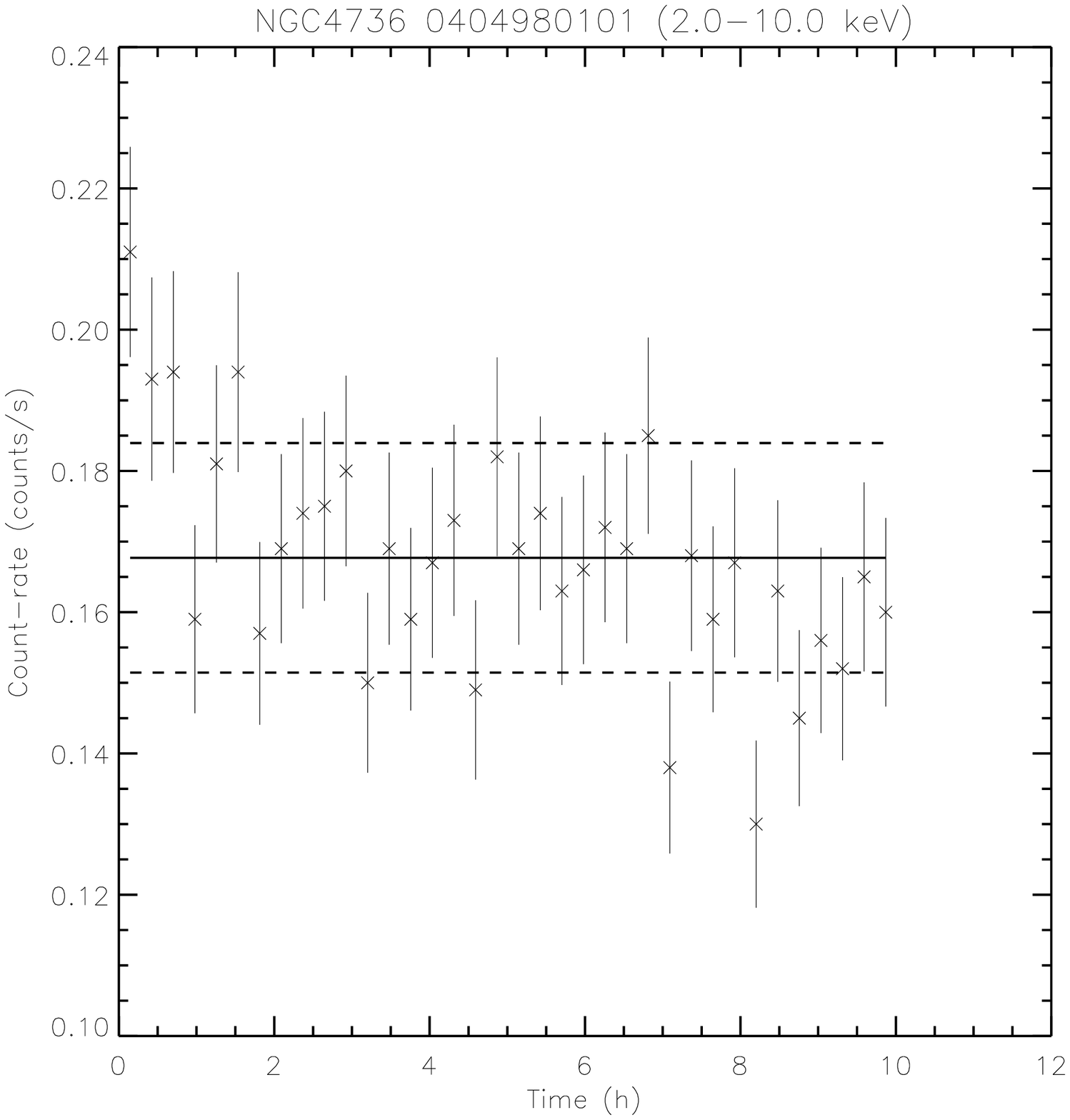}}
\subfloat{\includegraphics[width=0.30\textwidth]{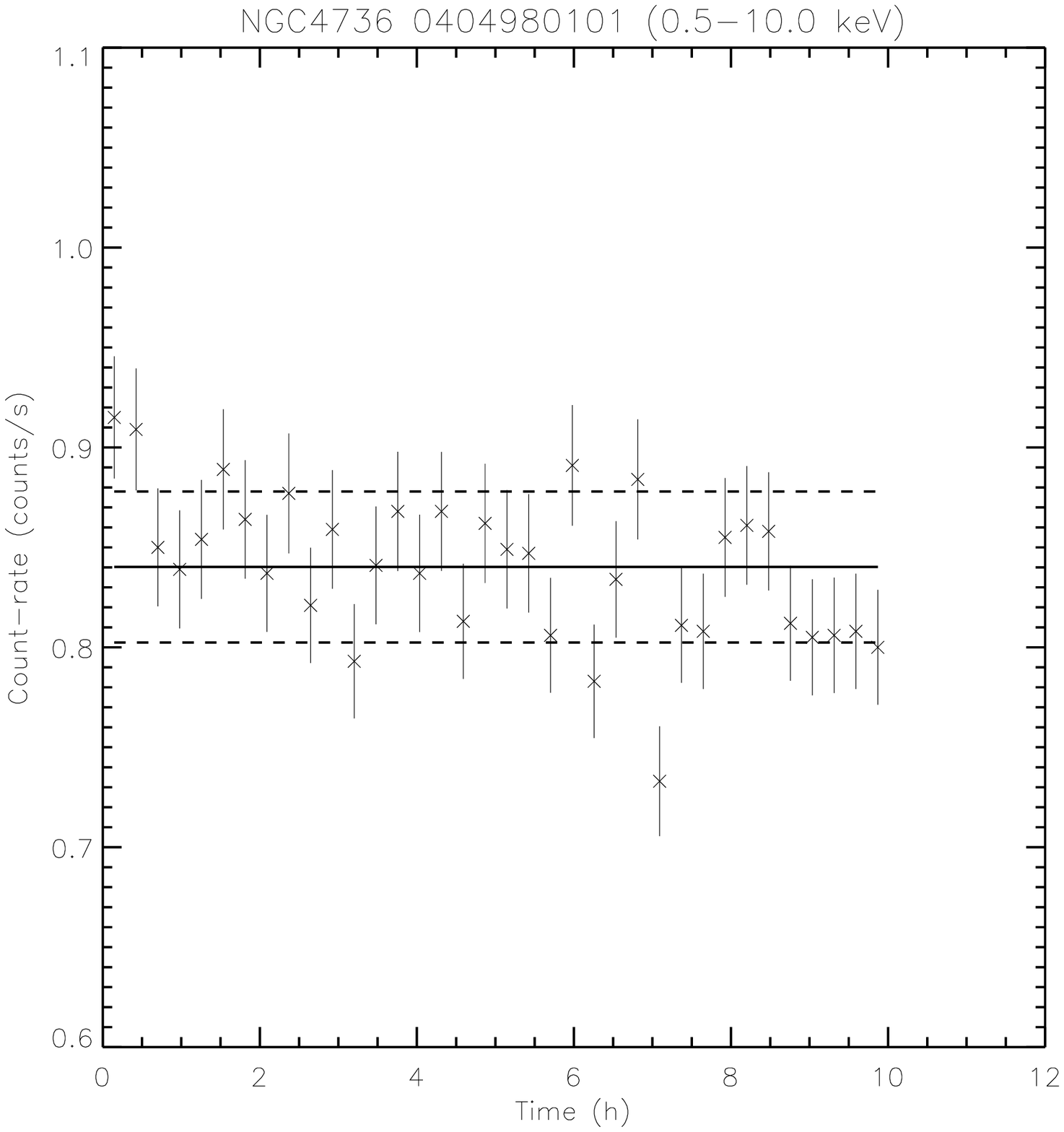}}

\subfloat{\includegraphics[width=0.30\textwidth]{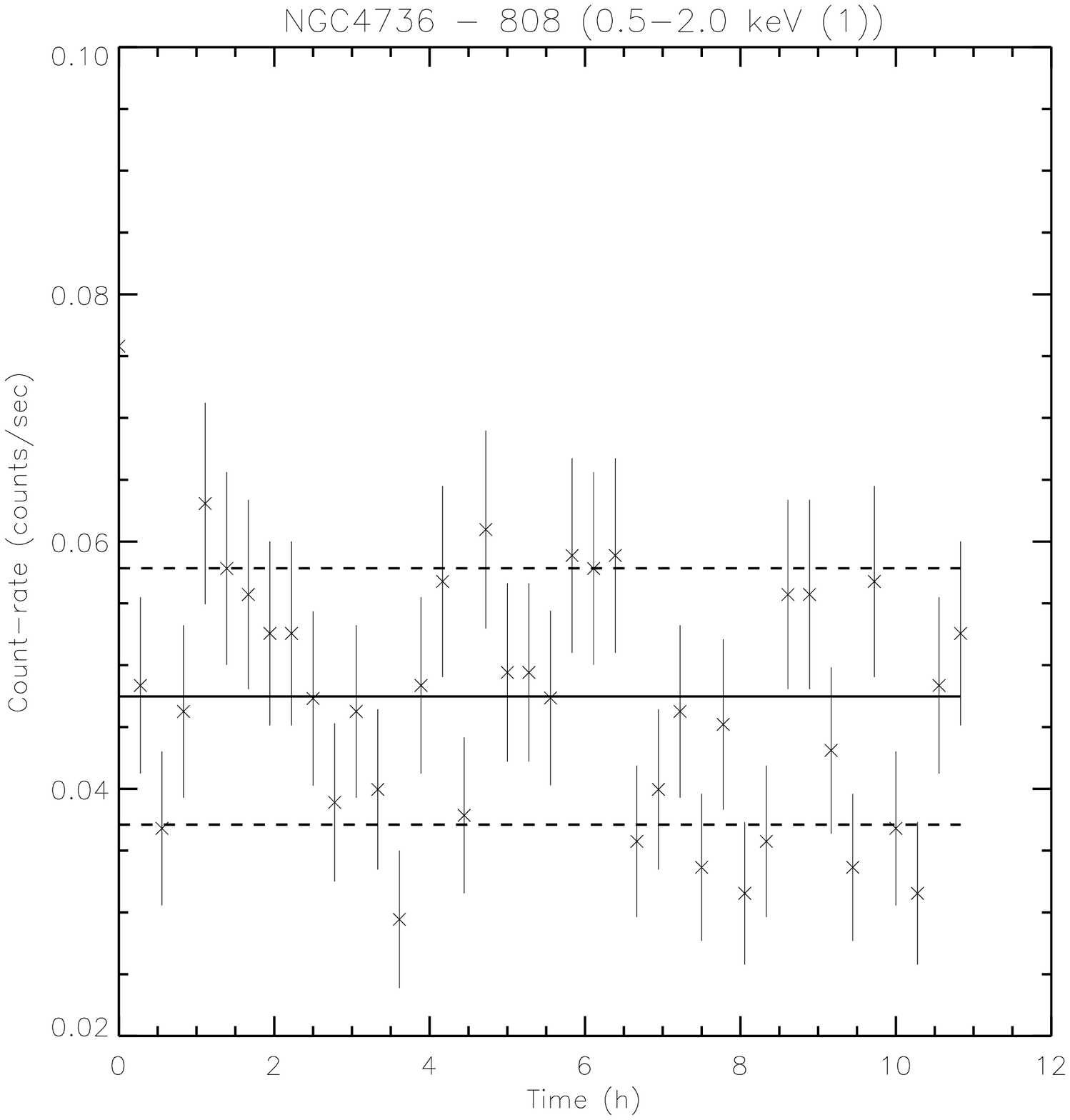}}
\subfloat{\includegraphics[width=0.30\textwidth]{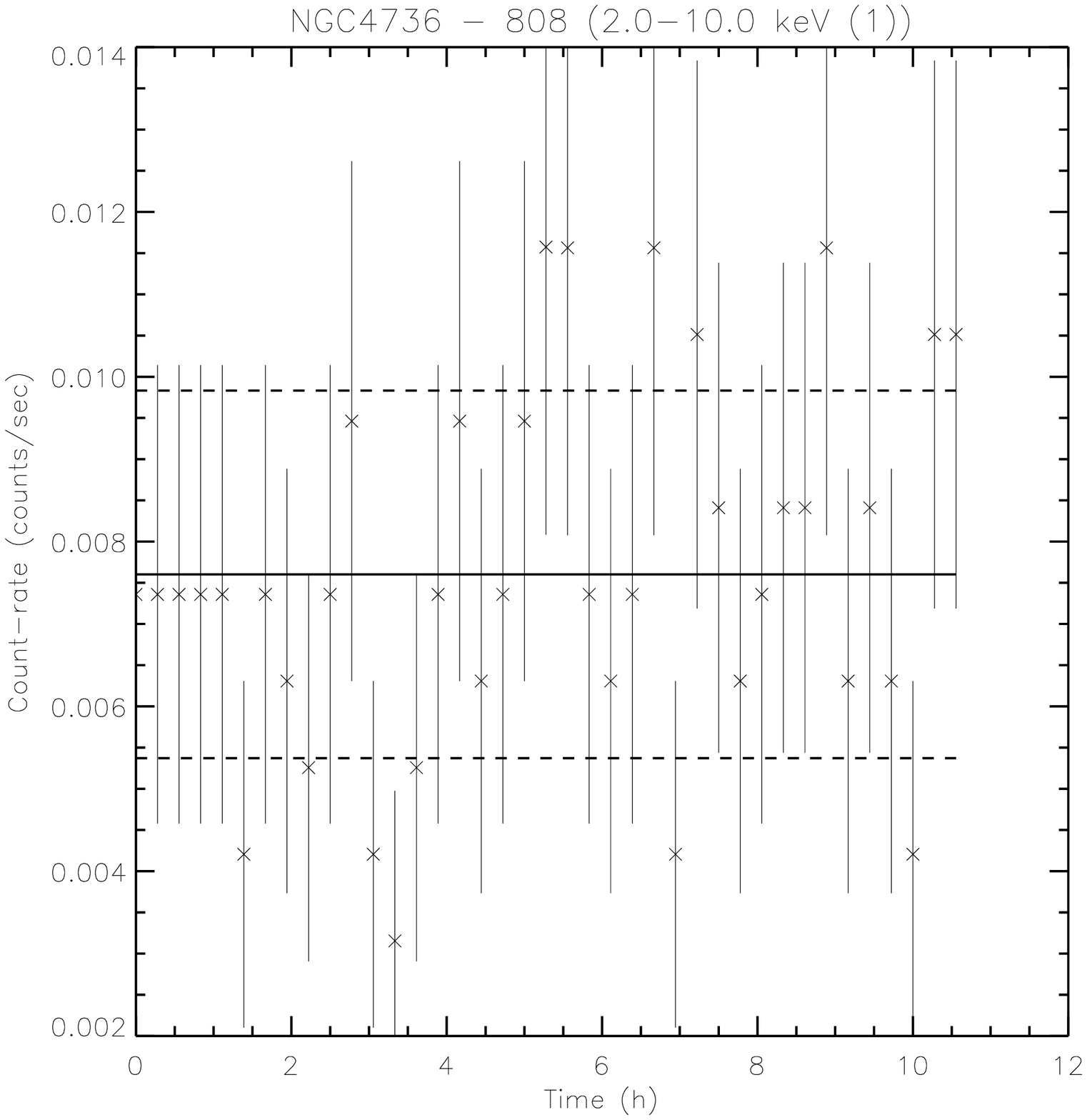}}
\subfloat{\includegraphics[width=0.30\textwidth]{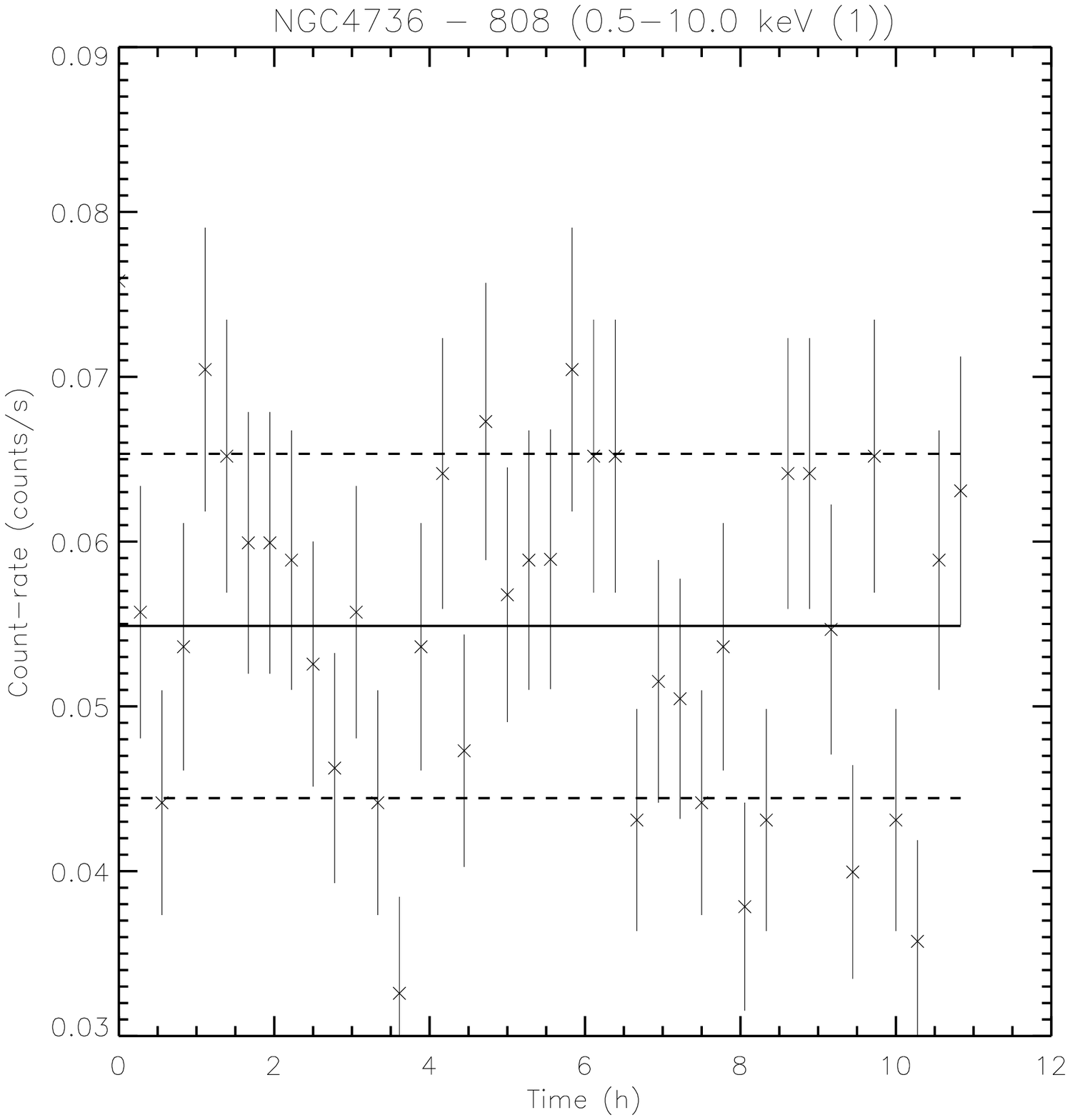}}
\caption{Light curves of NGC\,4736 from \emph{XMM}--Newton (up) and \emph{Chandra} (down) data.}
\label{l4736}
\end{figure}

\begin{figure}[H]
\centering
\subfloat{\includegraphics[width=0.30\textwidth]{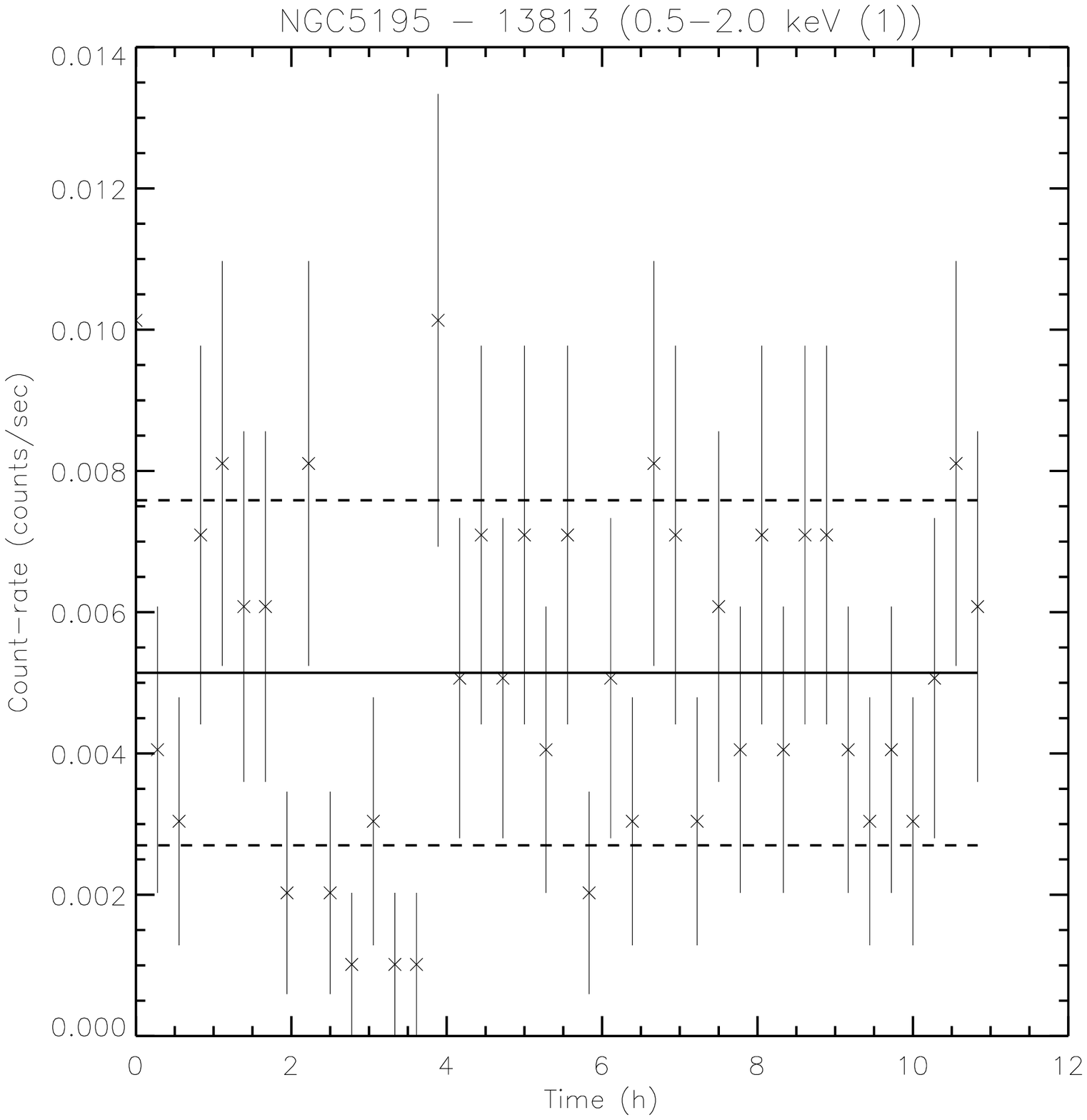}}
\subfloat{\includegraphics[width=0.30\textwidth]{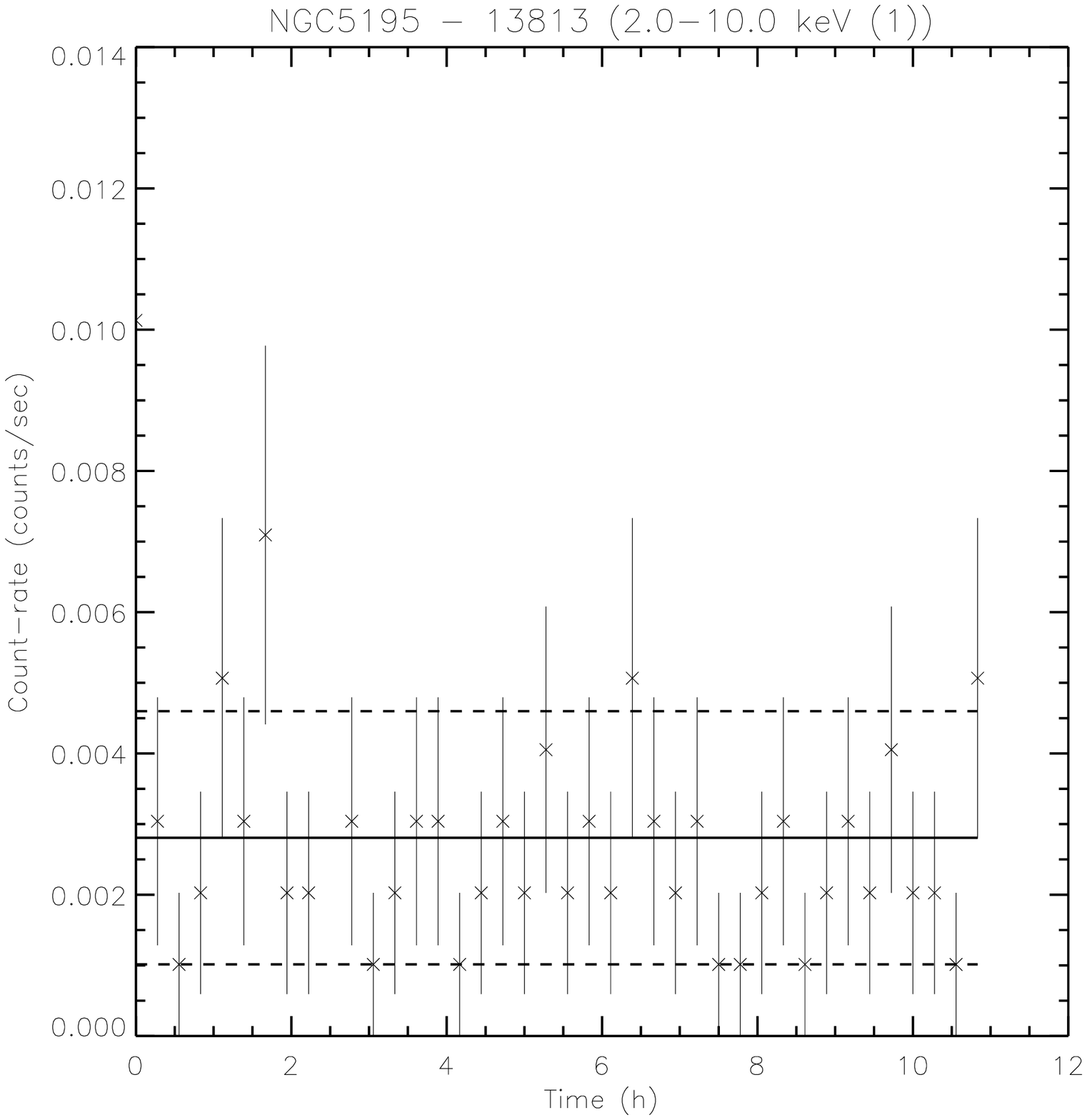}}
\subfloat{\includegraphics[width=0.30\textwidth]{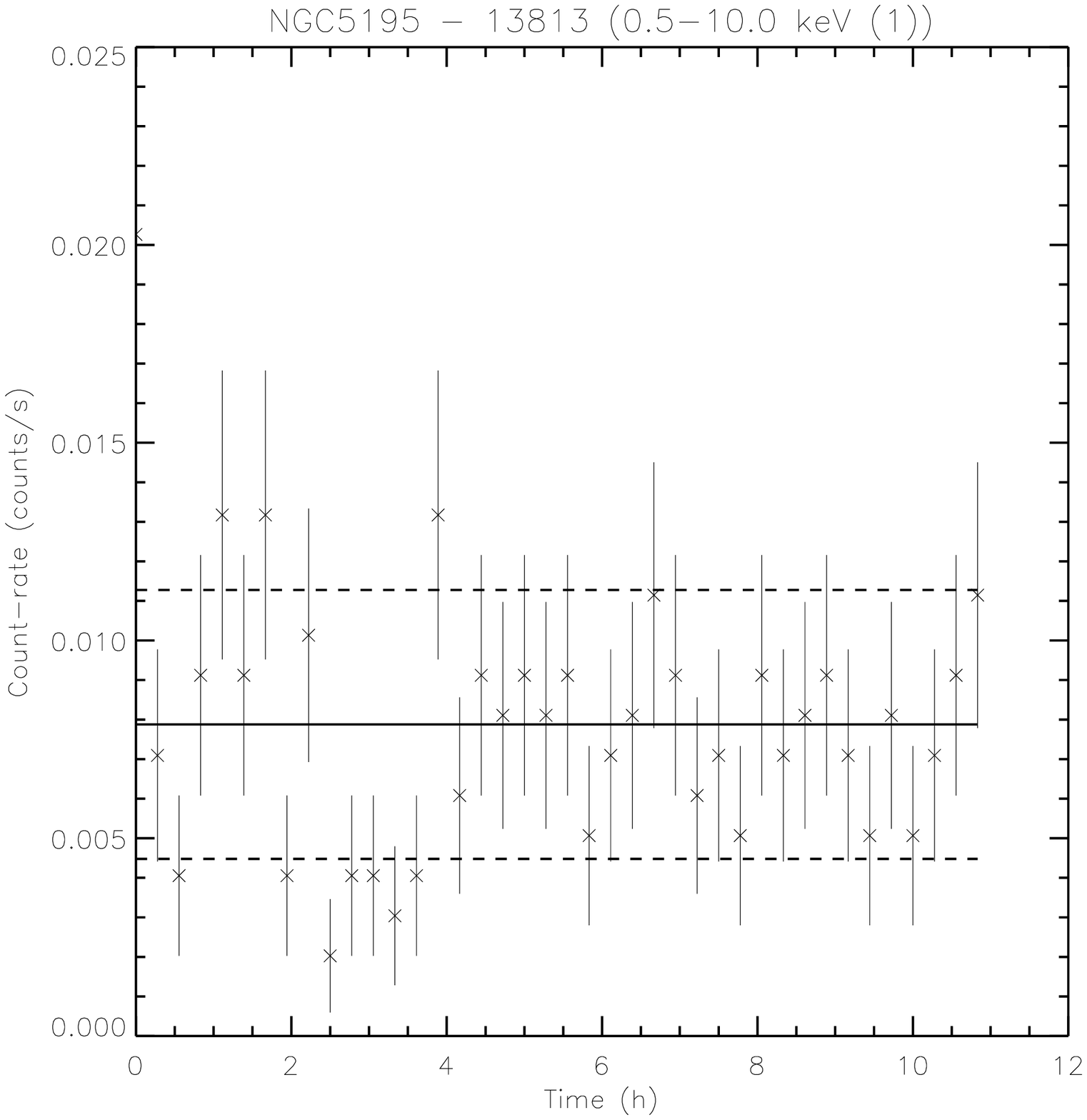}}

\subfloat{\includegraphics[width=0.30\textwidth]{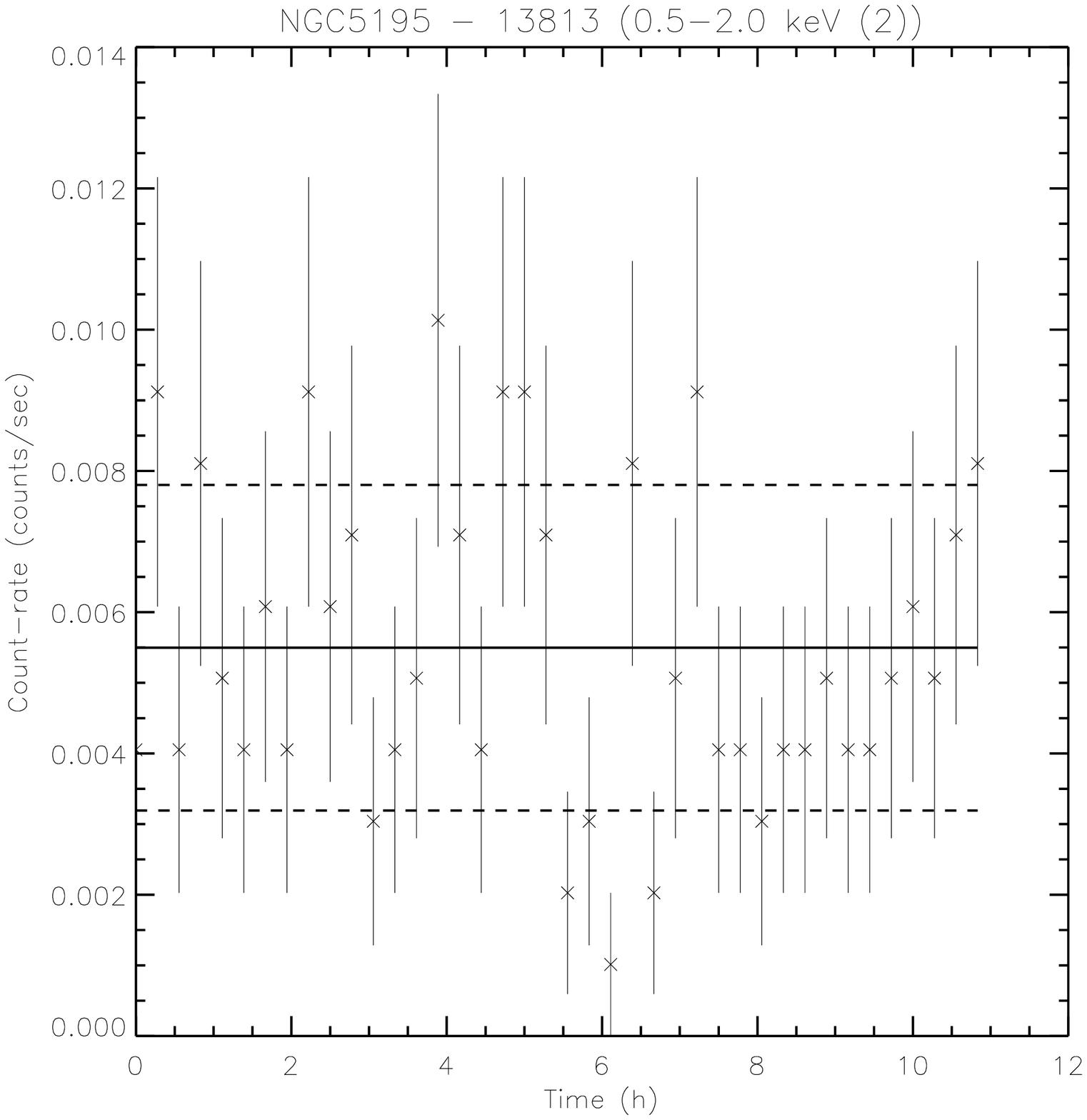}}
\subfloat{\includegraphics[width=0.30\textwidth]{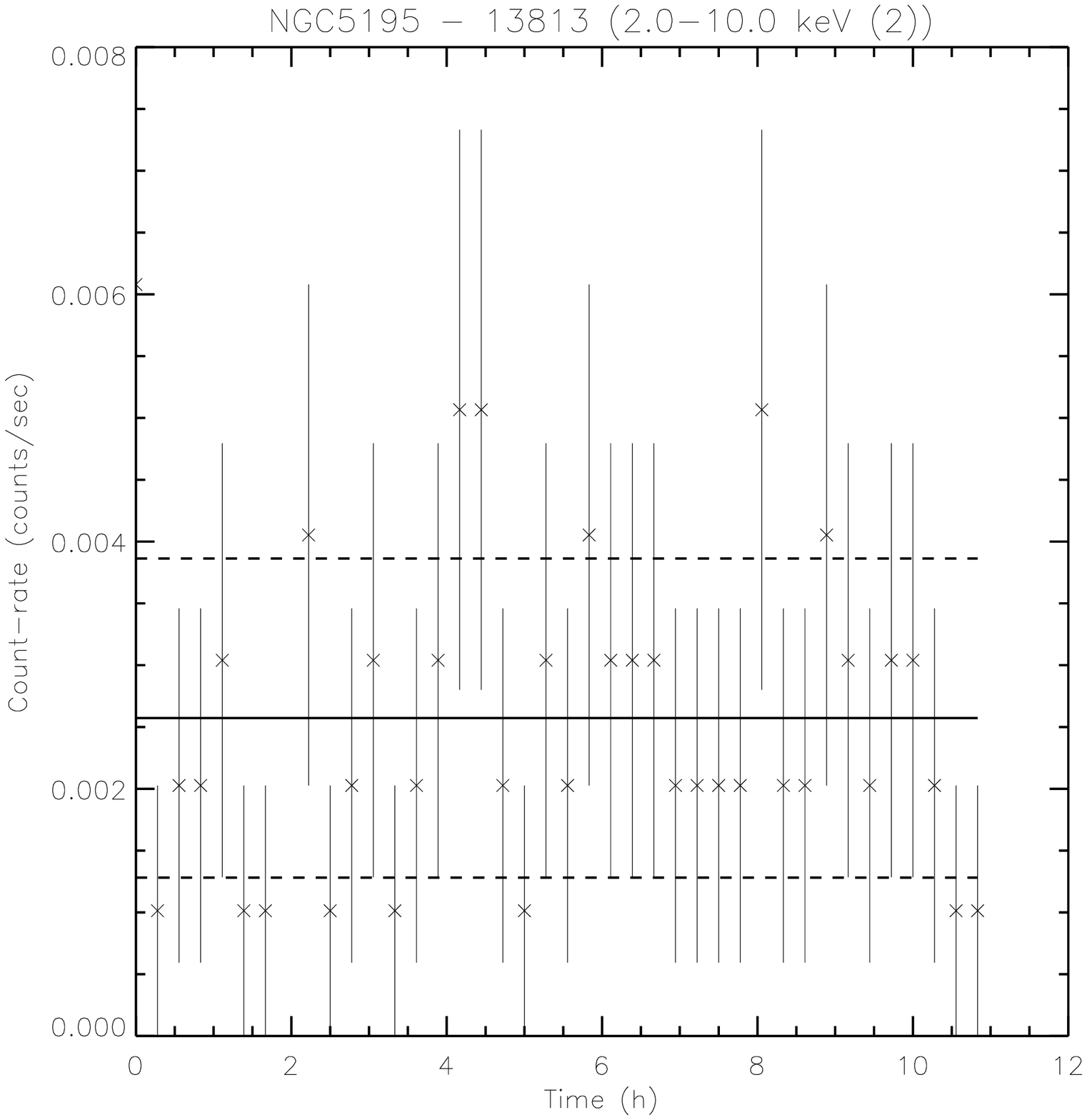}}
\subfloat{\includegraphics[width=0.30\textwidth]{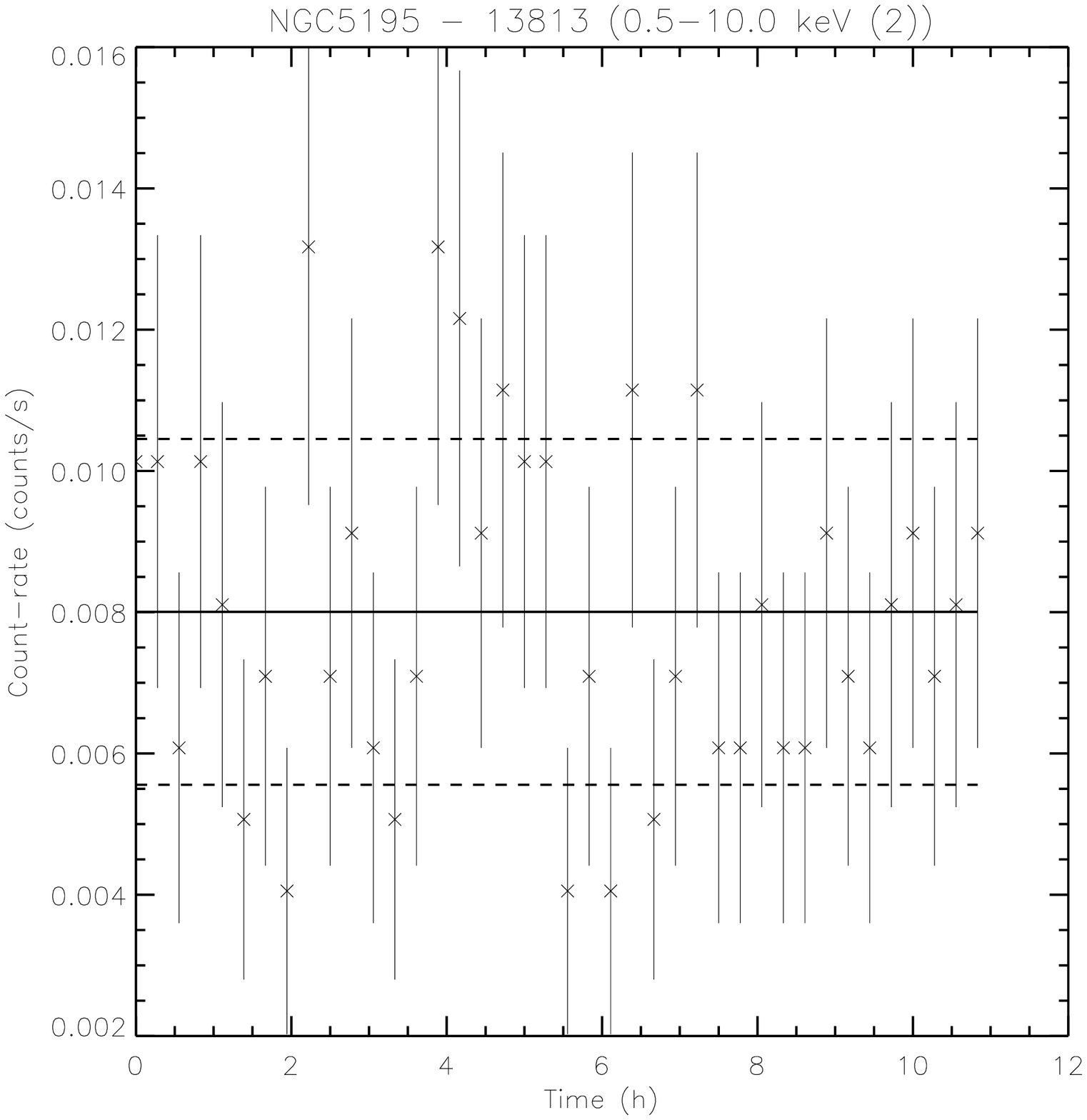}}

\subfloat{\includegraphics[width=0.30\textwidth]{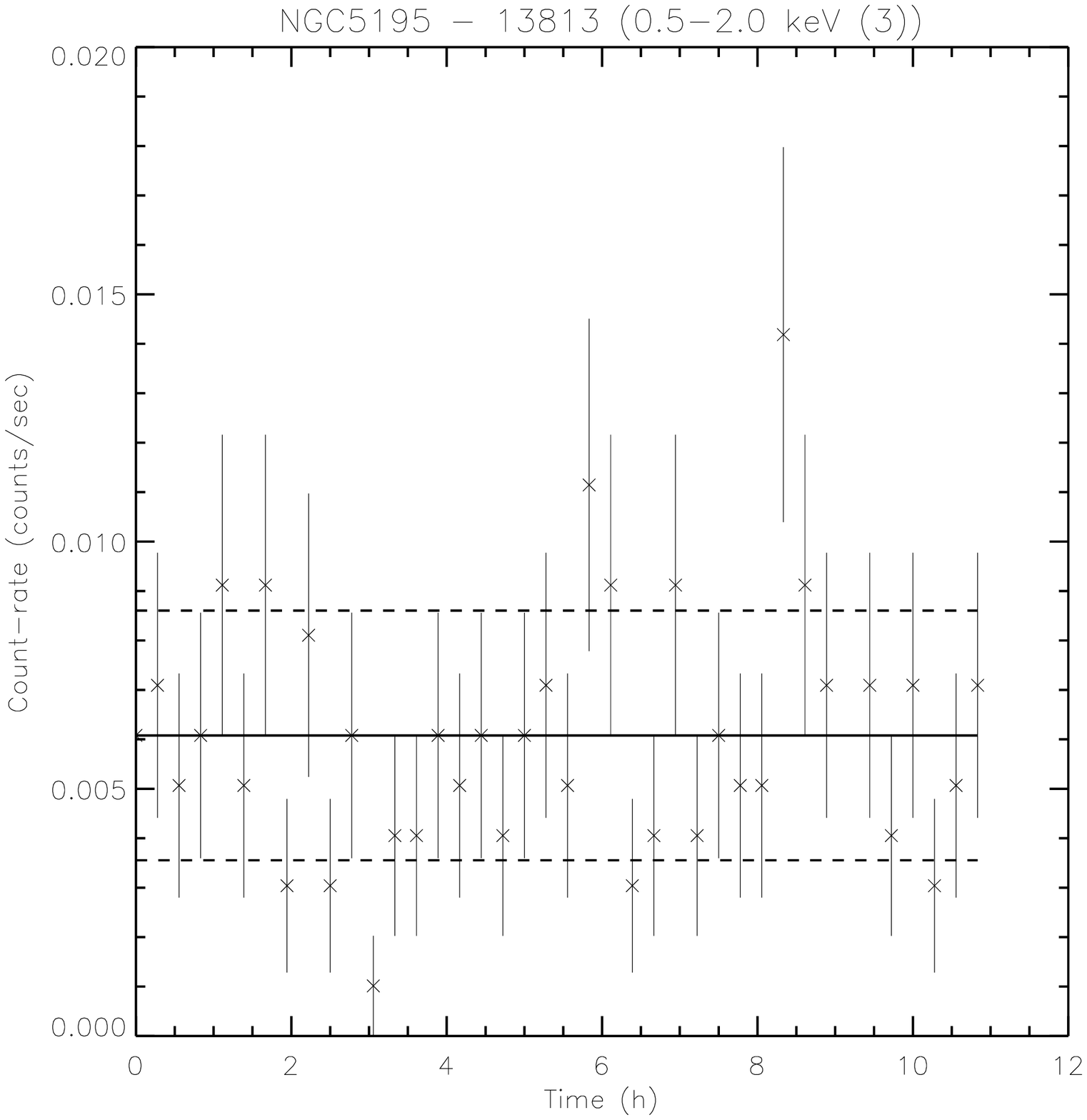}}
\subfloat{\includegraphics[width=0.30\textwidth]{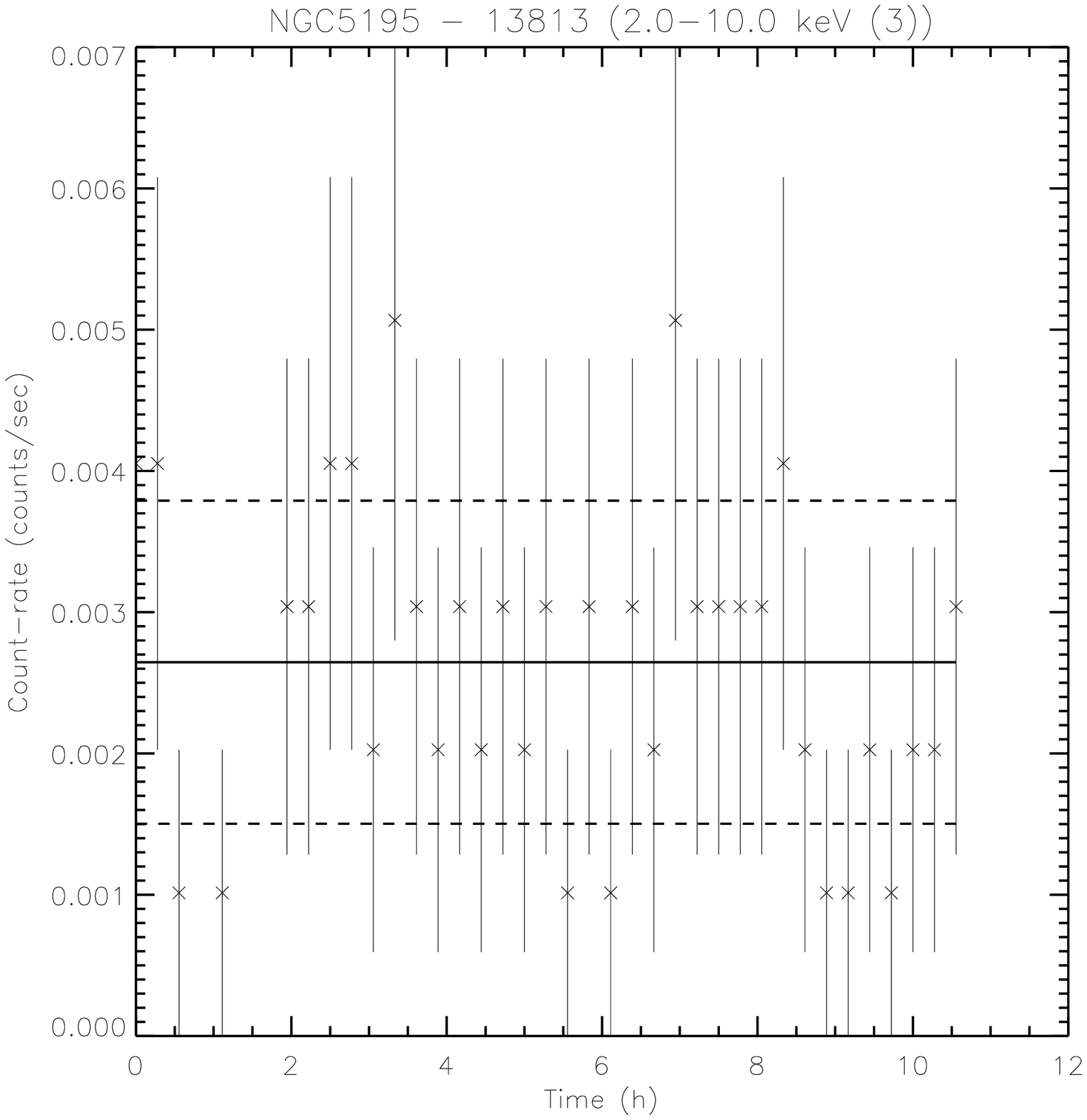}}
\subfloat{\includegraphics[width=0.30\textwidth]{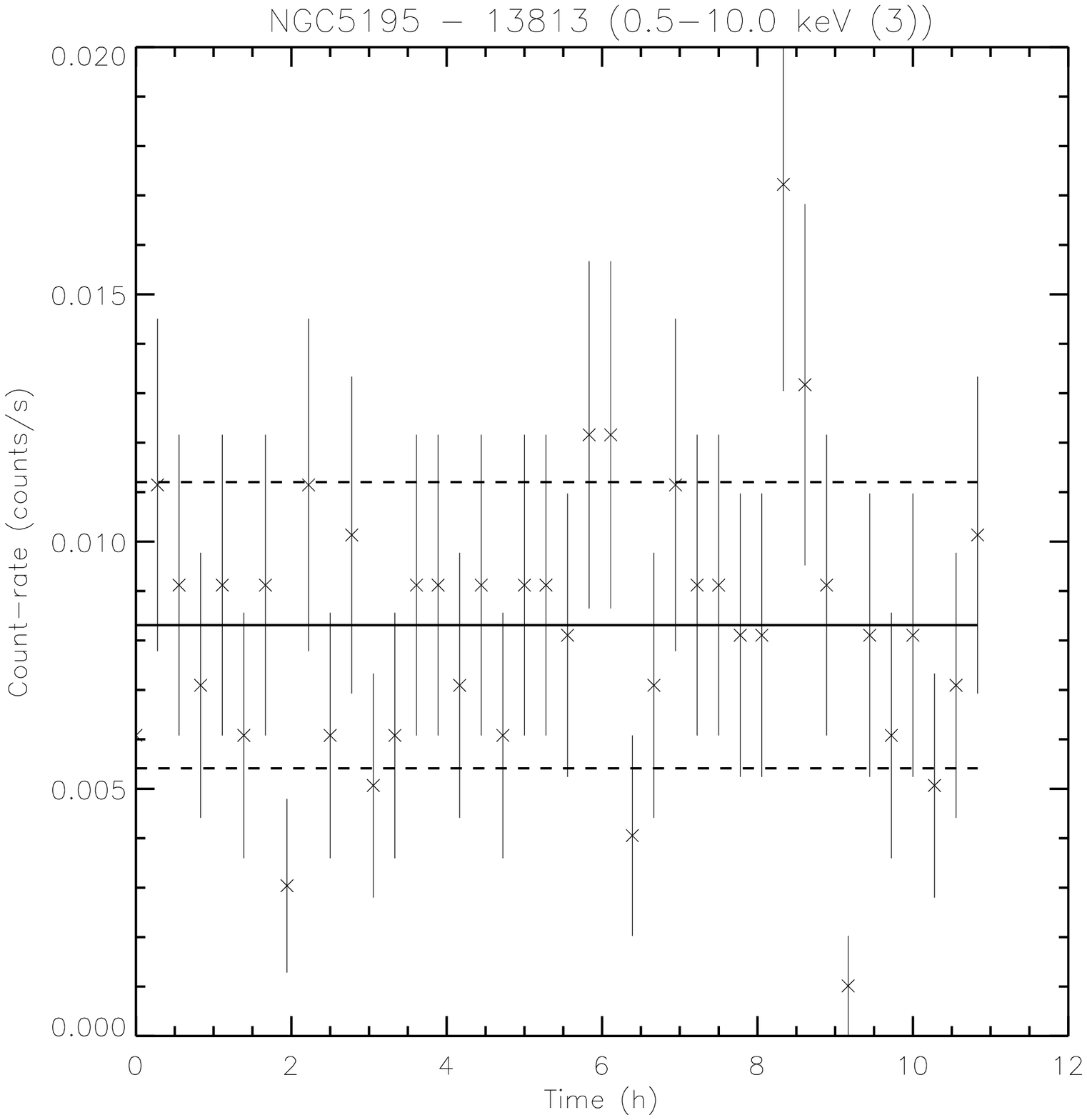}}

\subfloat{\includegraphics[width=0.30\textwidth]{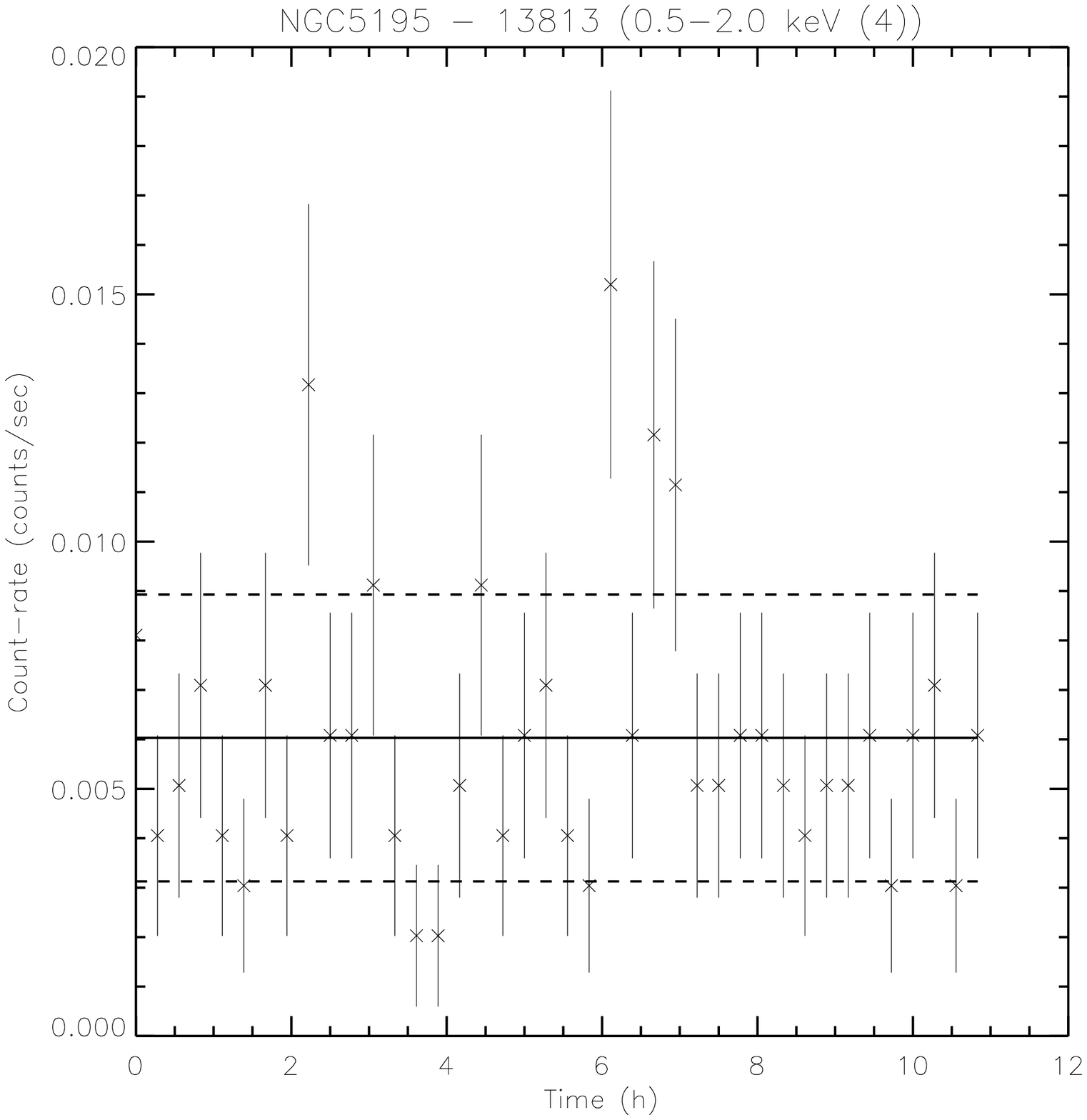}}
\subfloat{\includegraphics[width=0.30\textwidth]{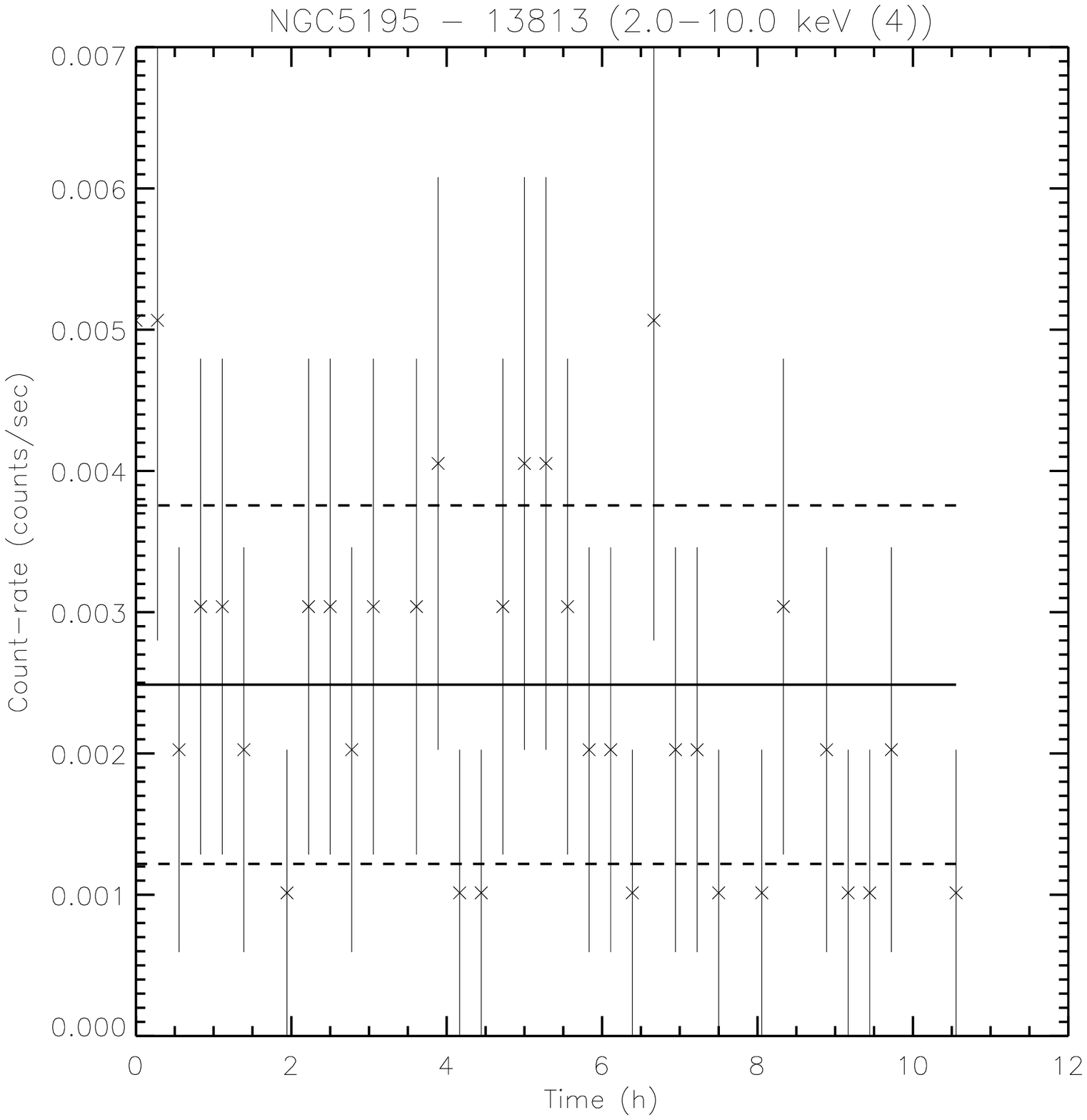}}
\subfloat{\includegraphics[width=0.30\textwidth]{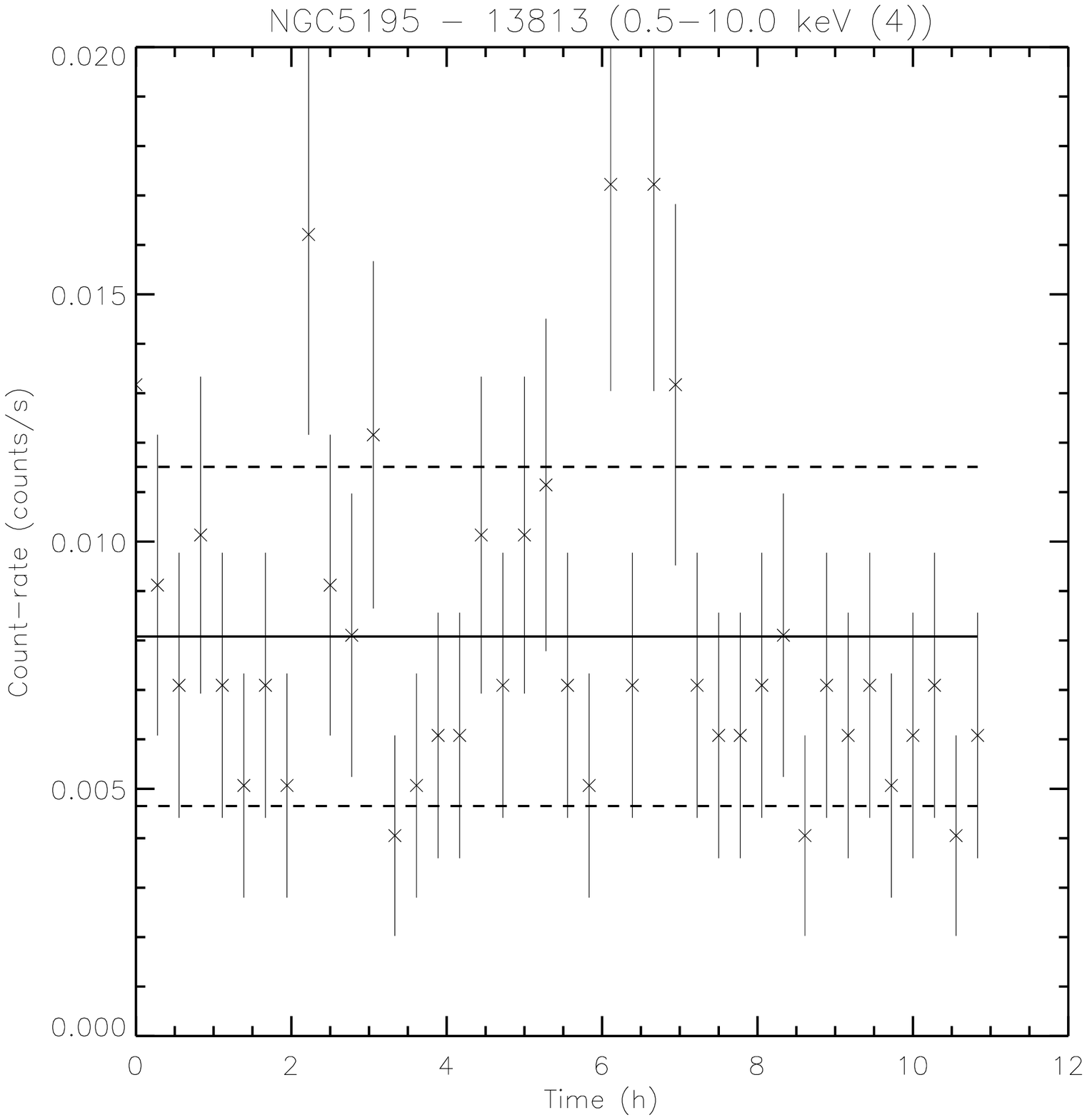}}

\caption{Light curves of NGC\,5195 from \emph{Chandra} data. Note that ObsID. 13813 is divided in four segments and ObsID. 13812 in three segments.}
\label{l5195}
\end{figure}

\begin{figure}[H]
\setcounter{figure}{11}
\centering
\subfloat{\includegraphics[width=0.30\textwidth]{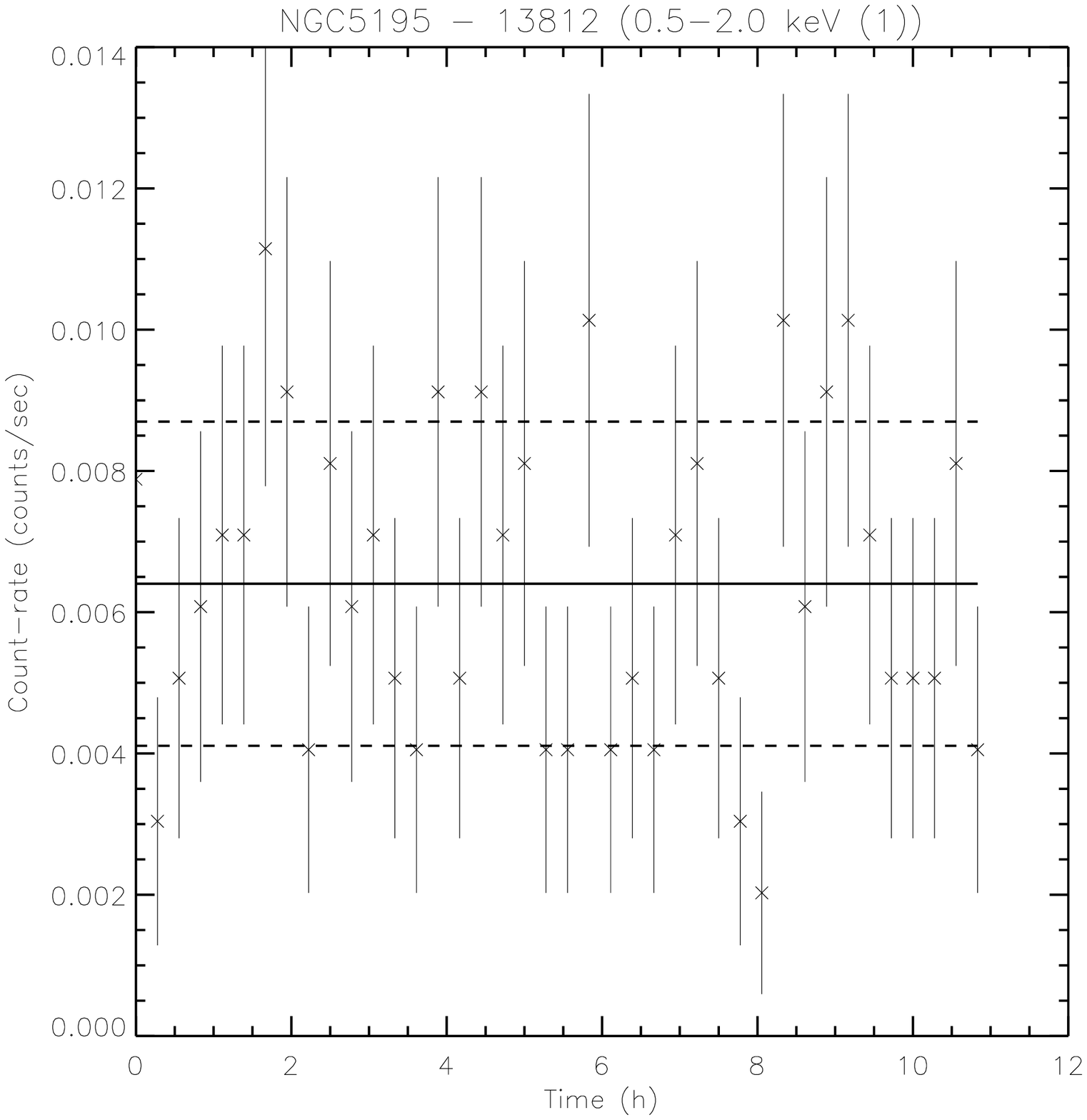}}
\subfloat{\includegraphics[width=0.30\textwidth]{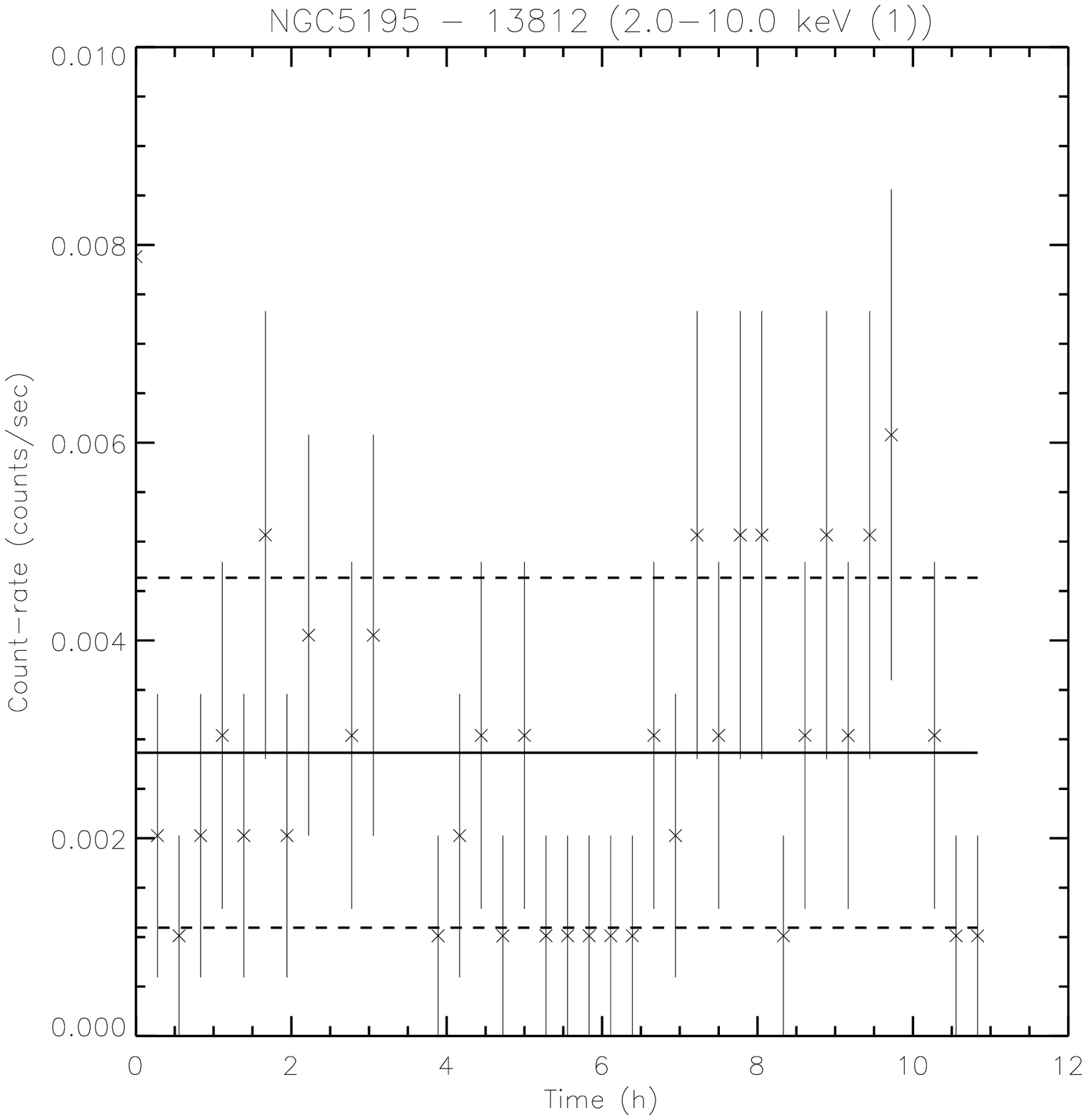}}
\subfloat{\includegraphics[width=0.30\textwidth]{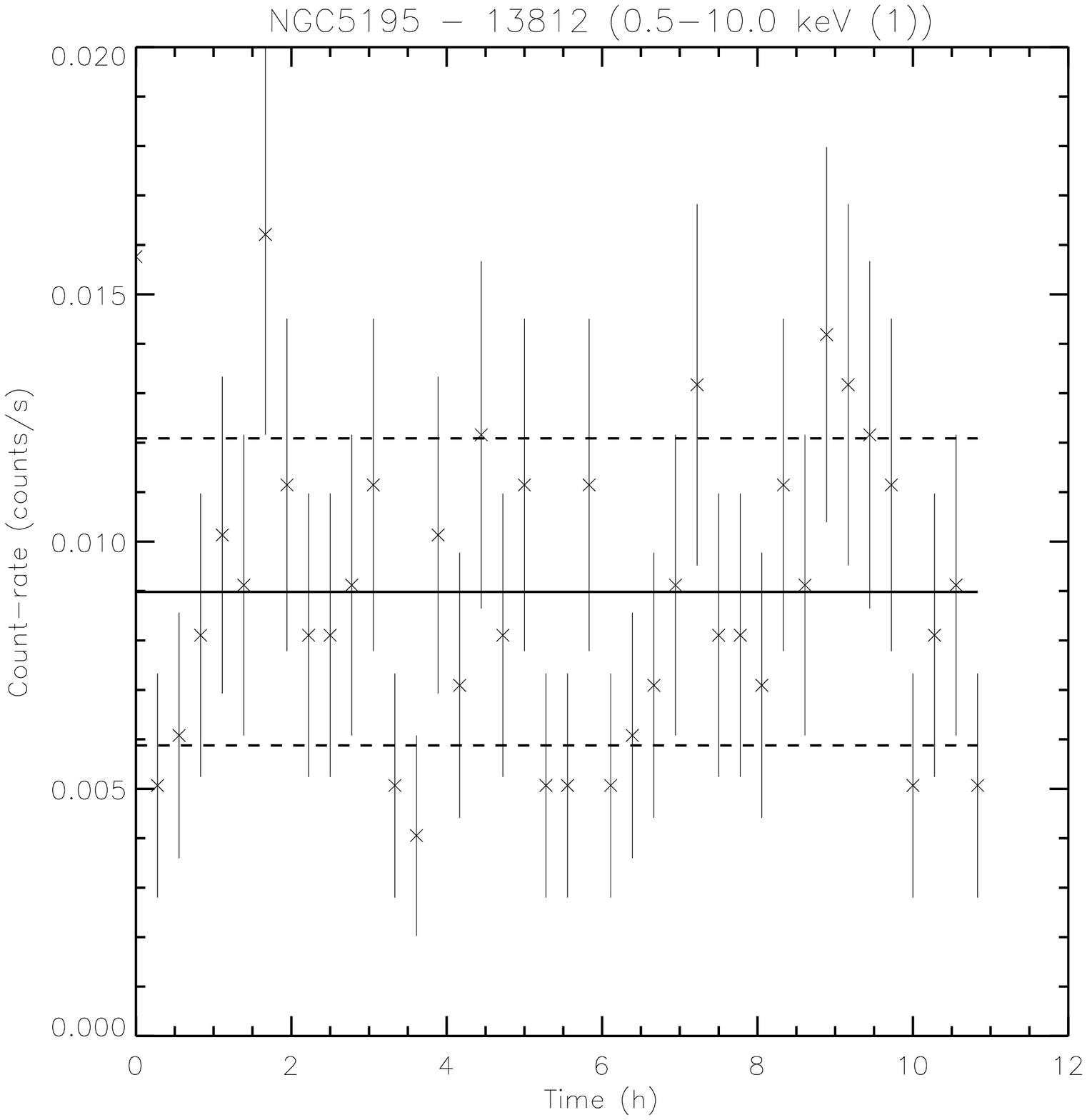}}

\subfloat{\includegraphics[width=0.30\textwidth]{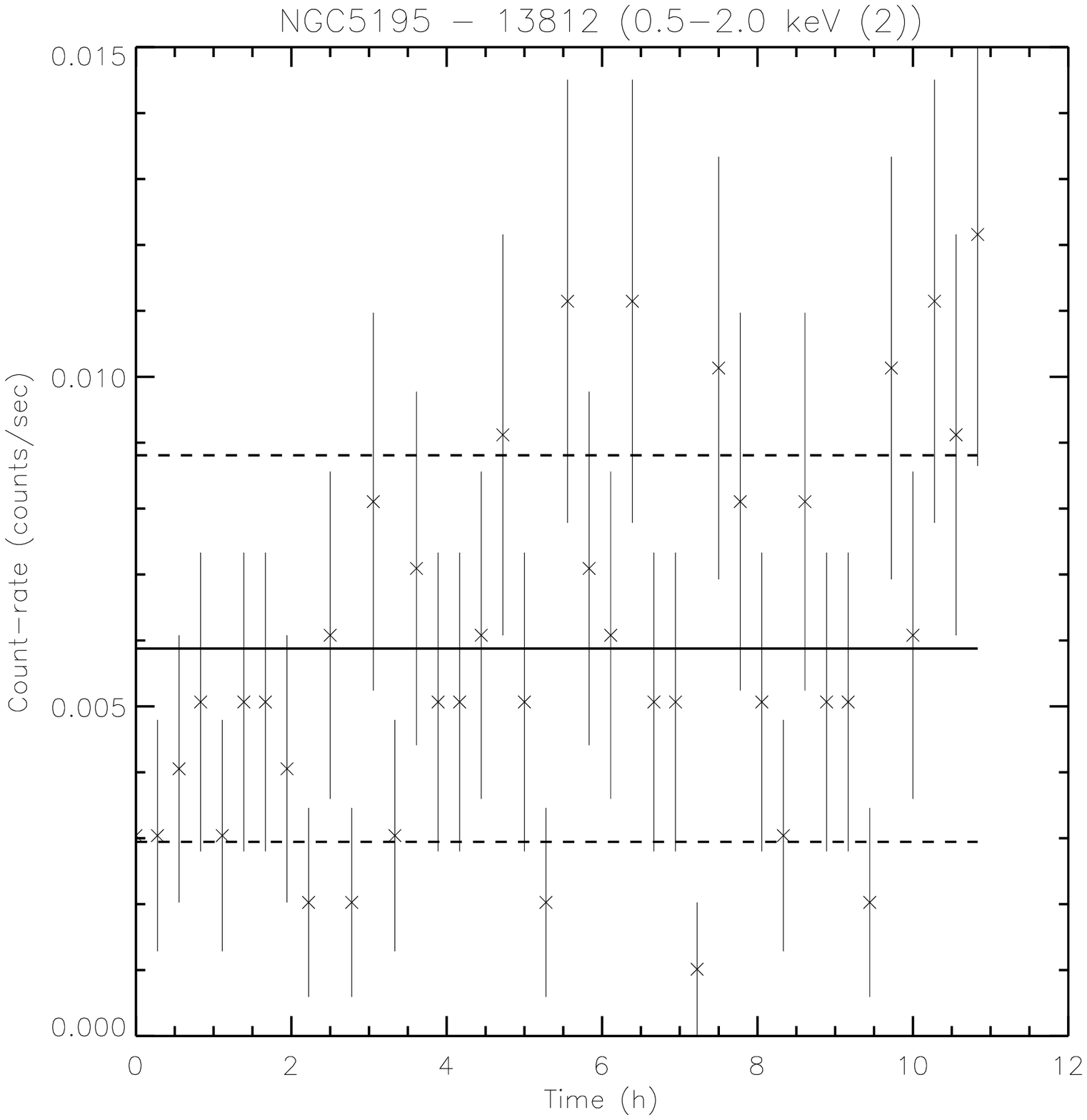}}
\subfloat{\includegraphics[width=0.30\textwidth]{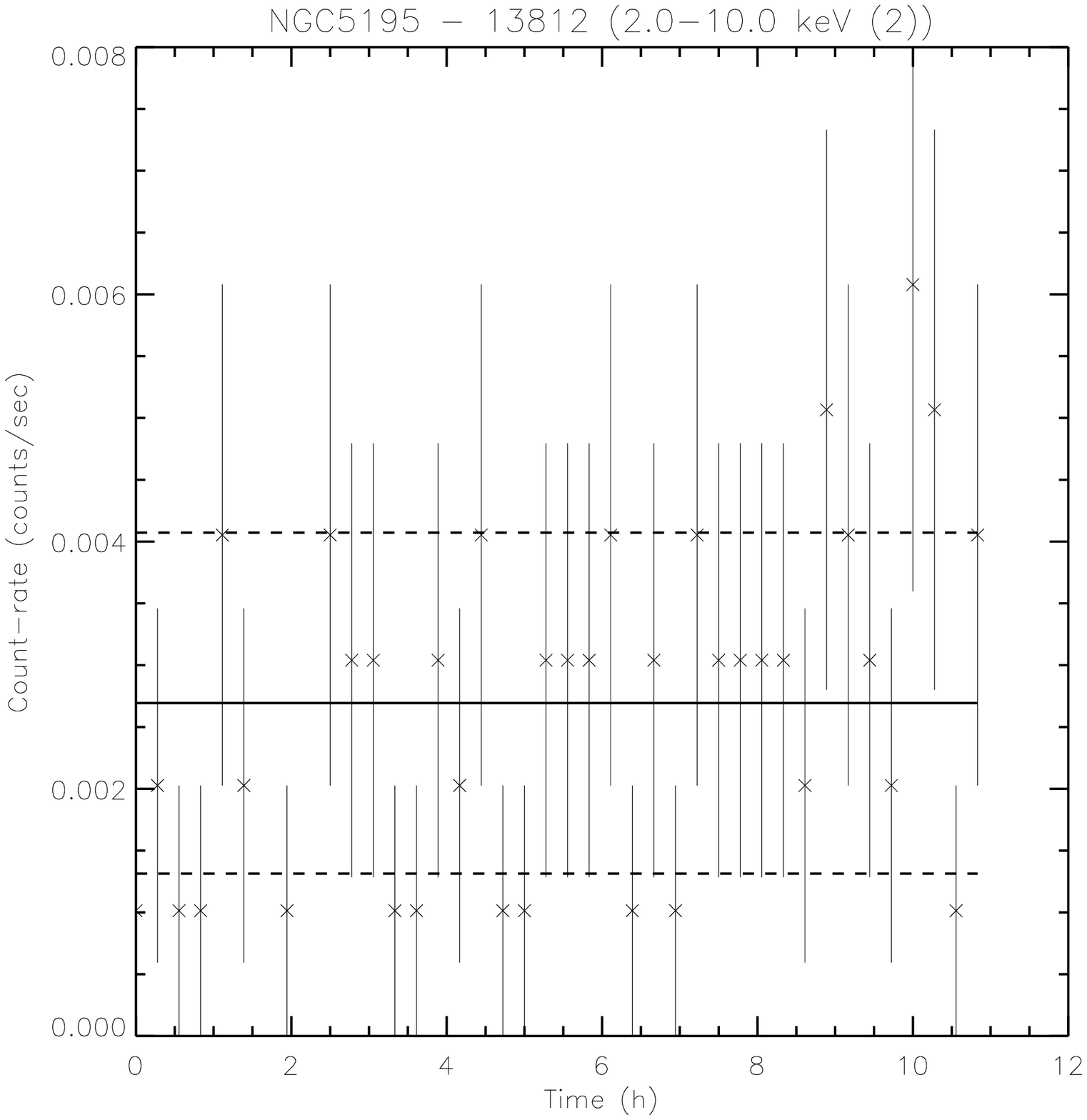}}
\subfloat{\includegraphics[width=0.30\textwidth]{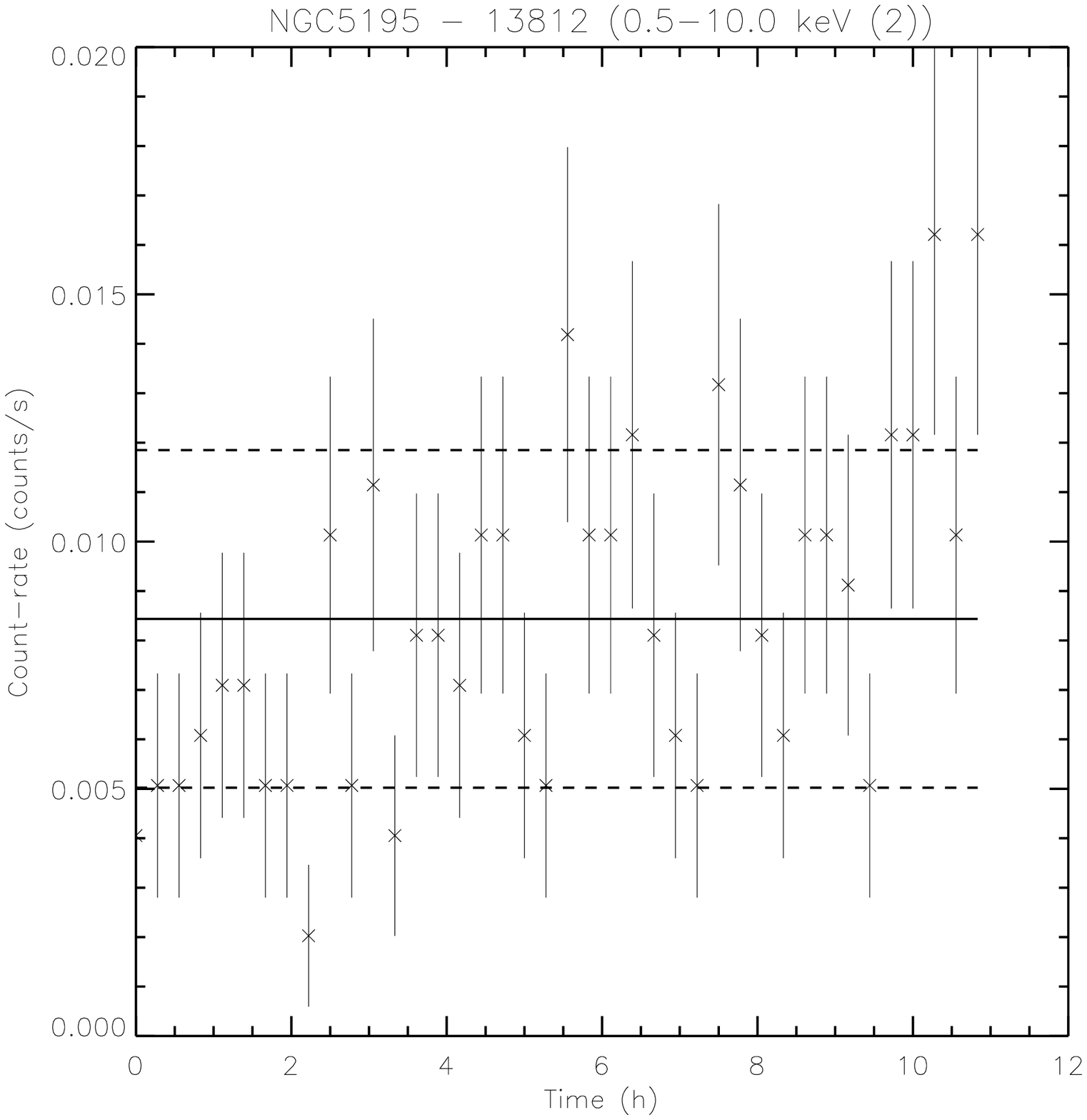}}

\subfloat{\includegraphics[width=0.30\textwidth]{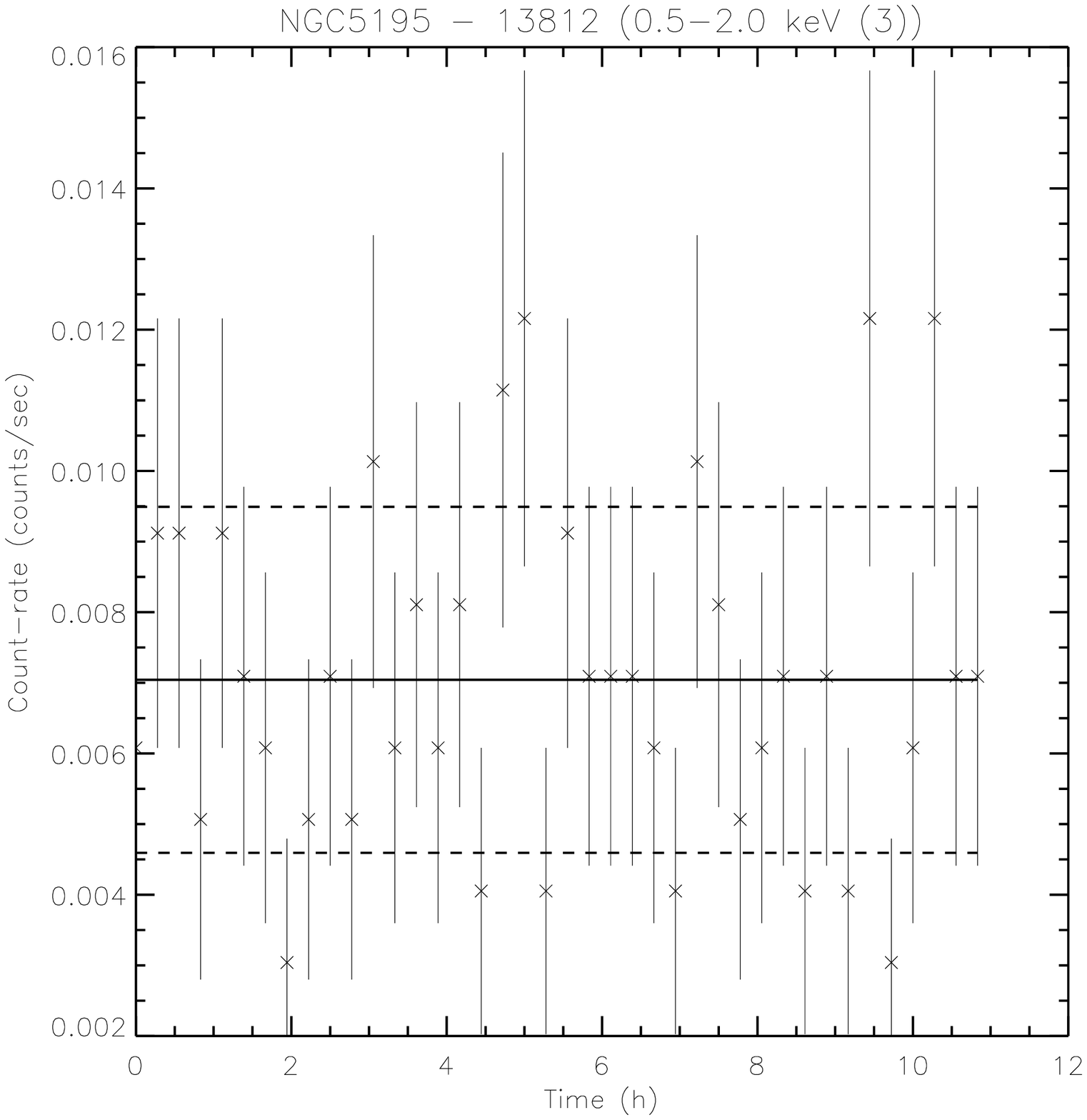}}
\subfloat{\includegraphics[width=0.30\textwidth]{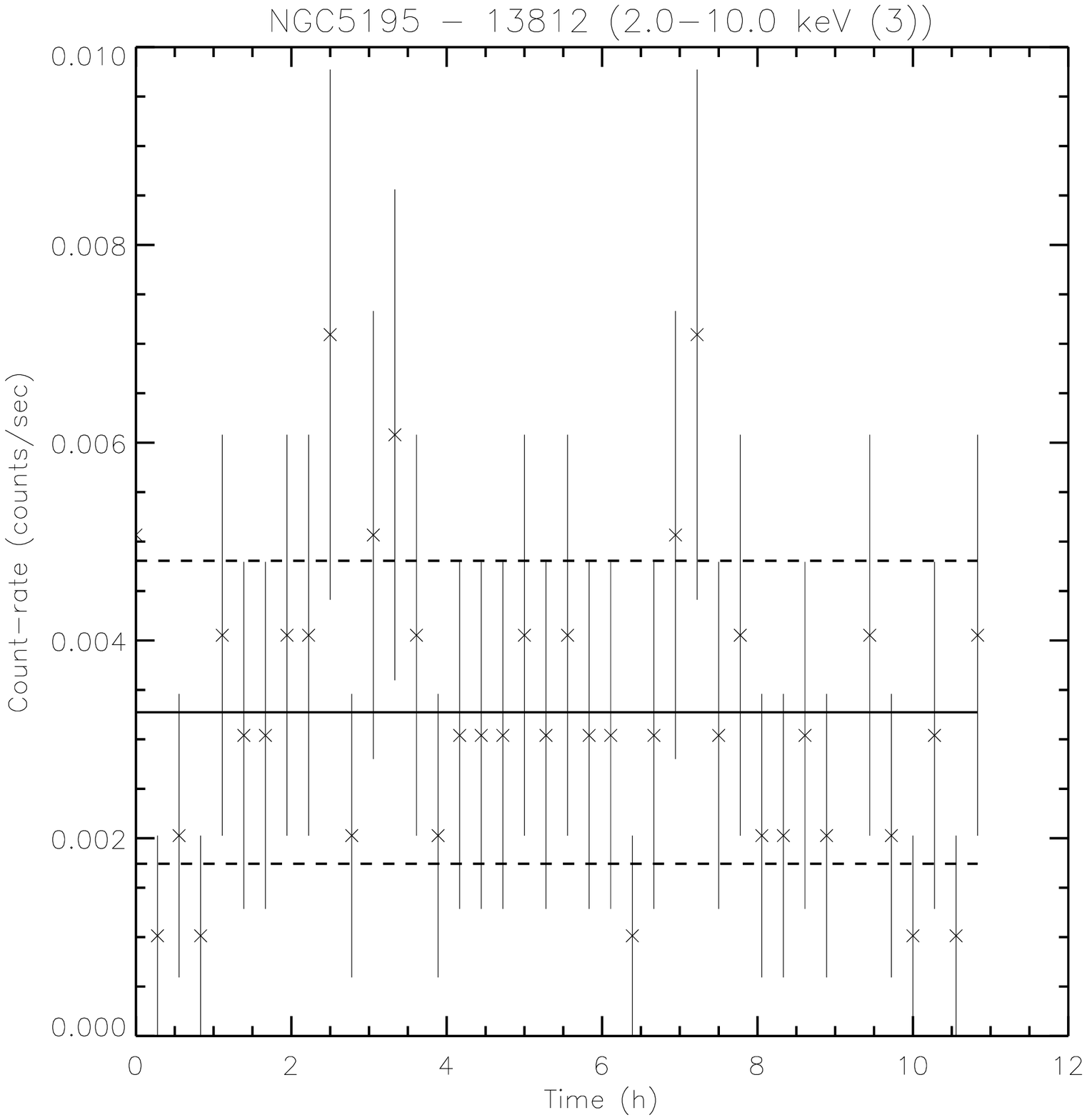}}
\subfloat{\includegraphics[width=0.30\textwidth]{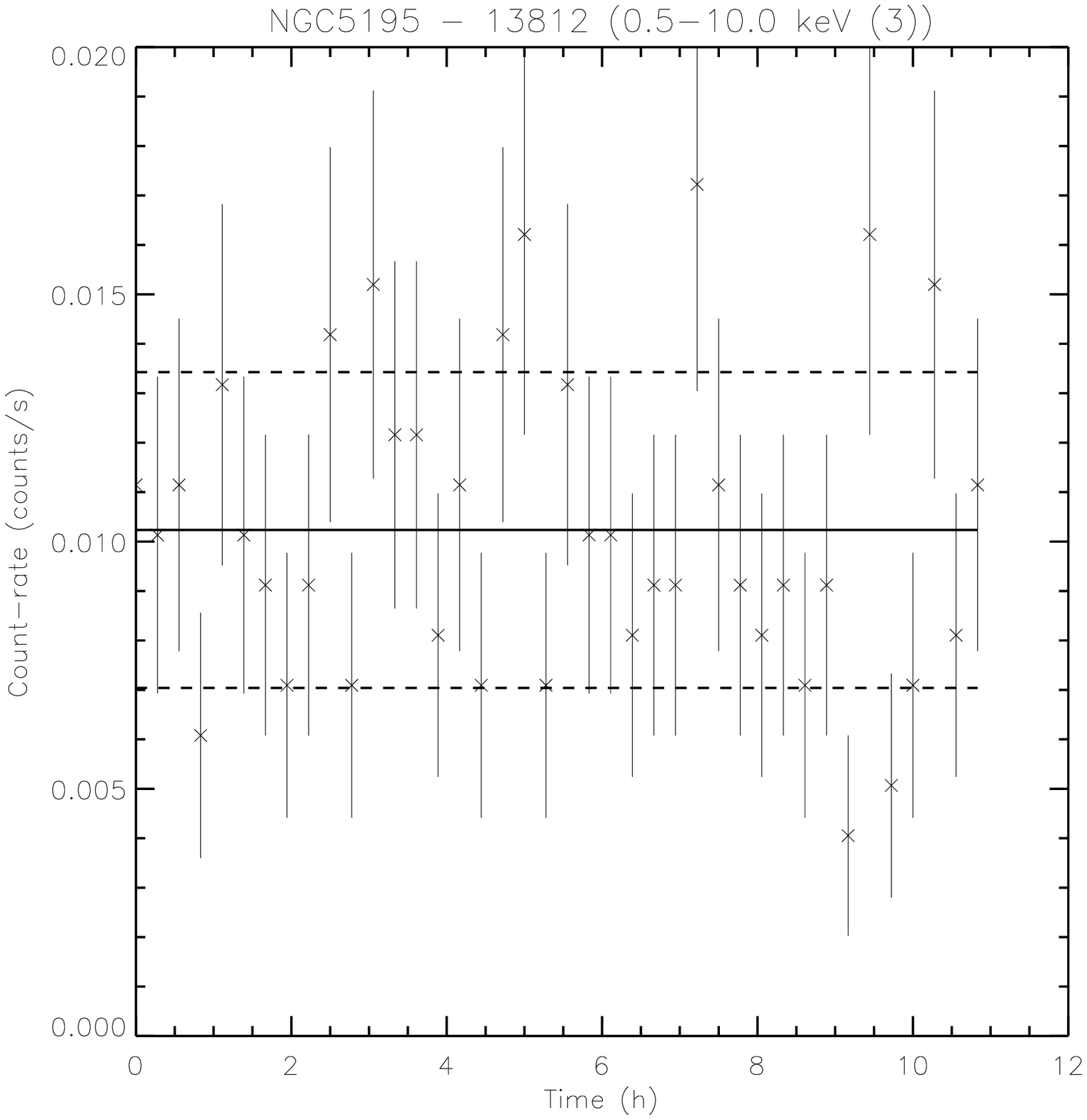}}
\caption{(Cont.)}
\end{figure}

\begin{figure}[H]
\centering
\subfloat{\includegraphics[width=0.30\textwidth]{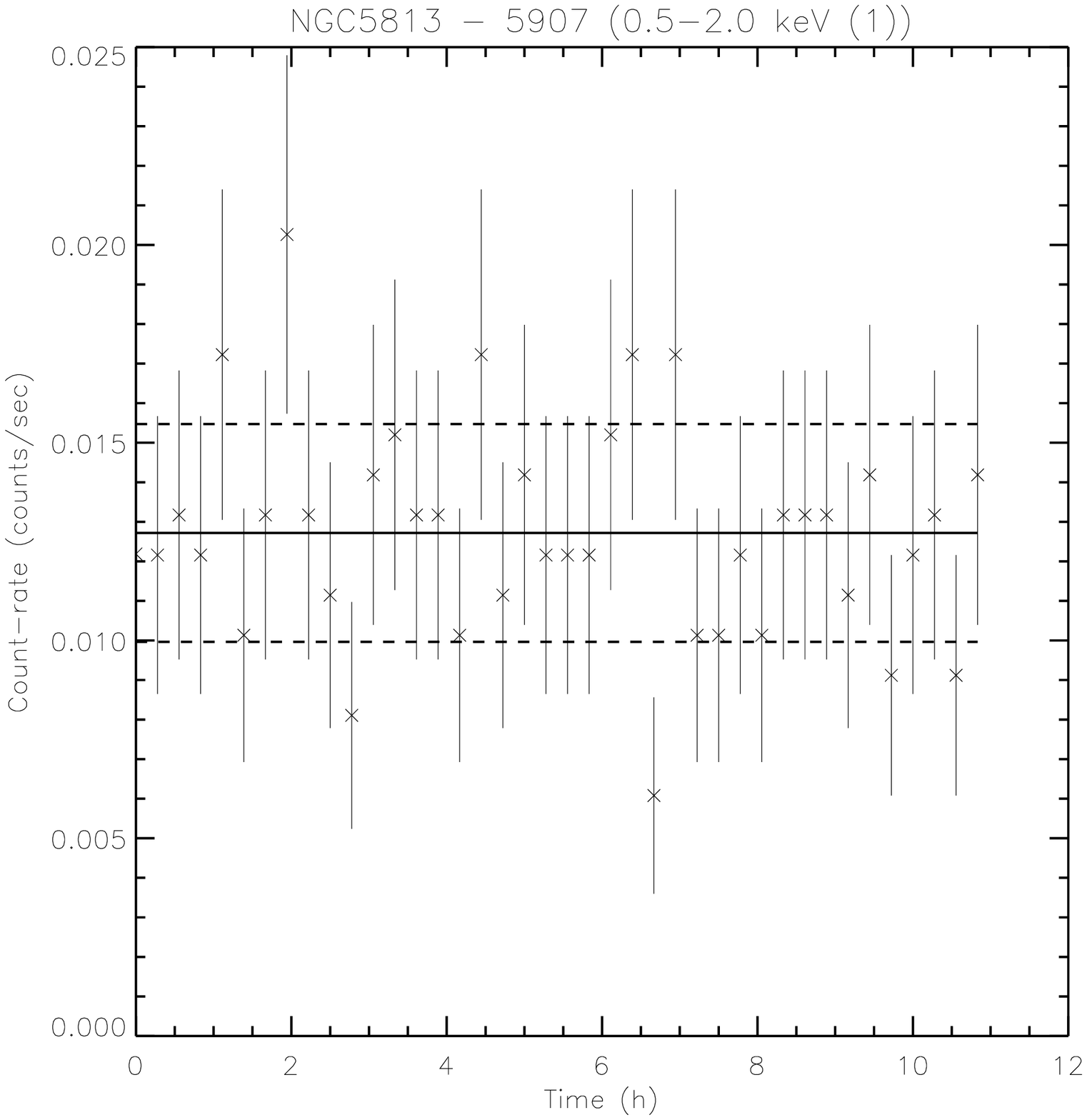}}
\subfloat{\includegraphics[width=0.30\textwidth]{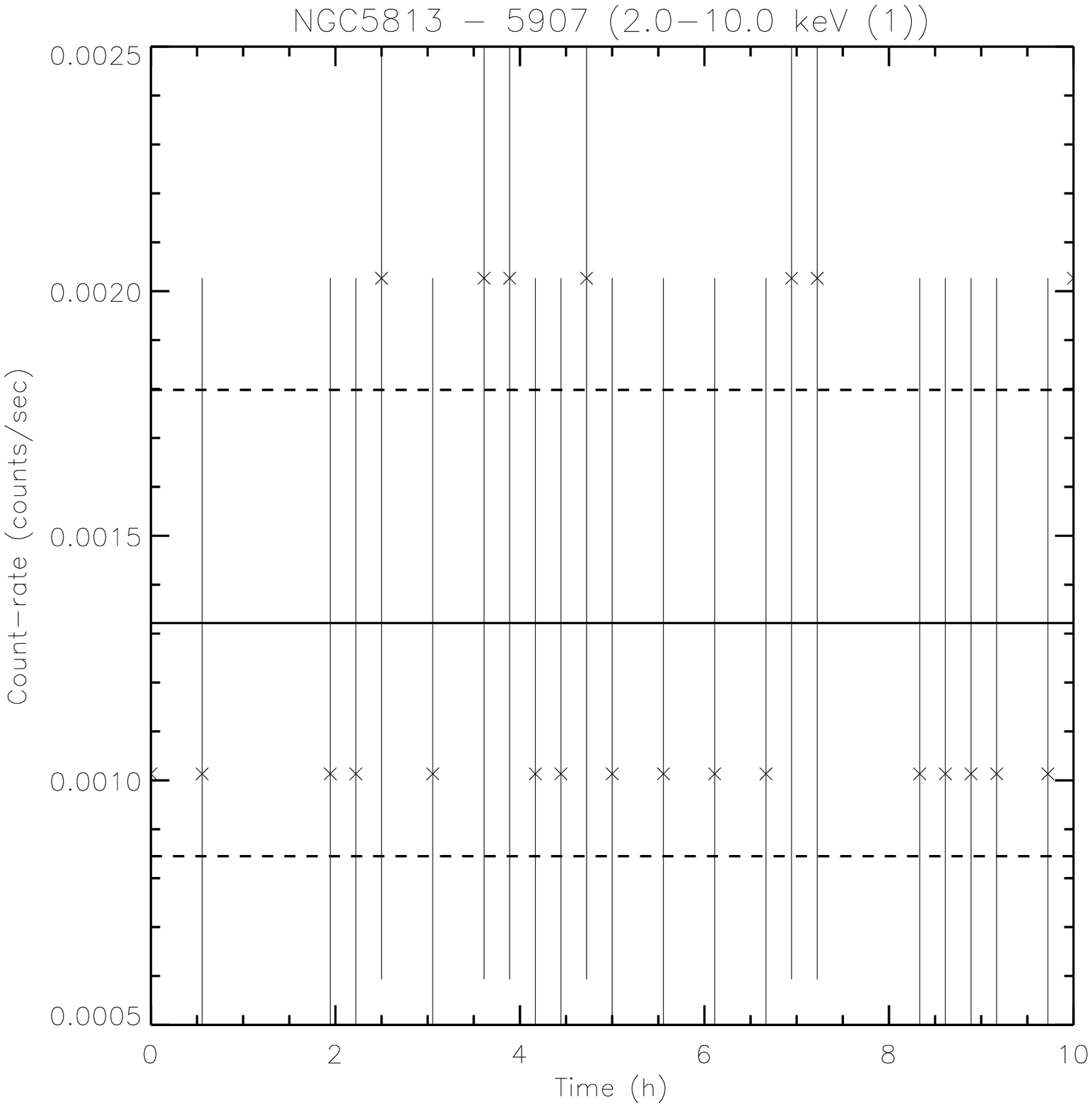}}
\subfloat{\includegraphics[width=0.30\textwidth]{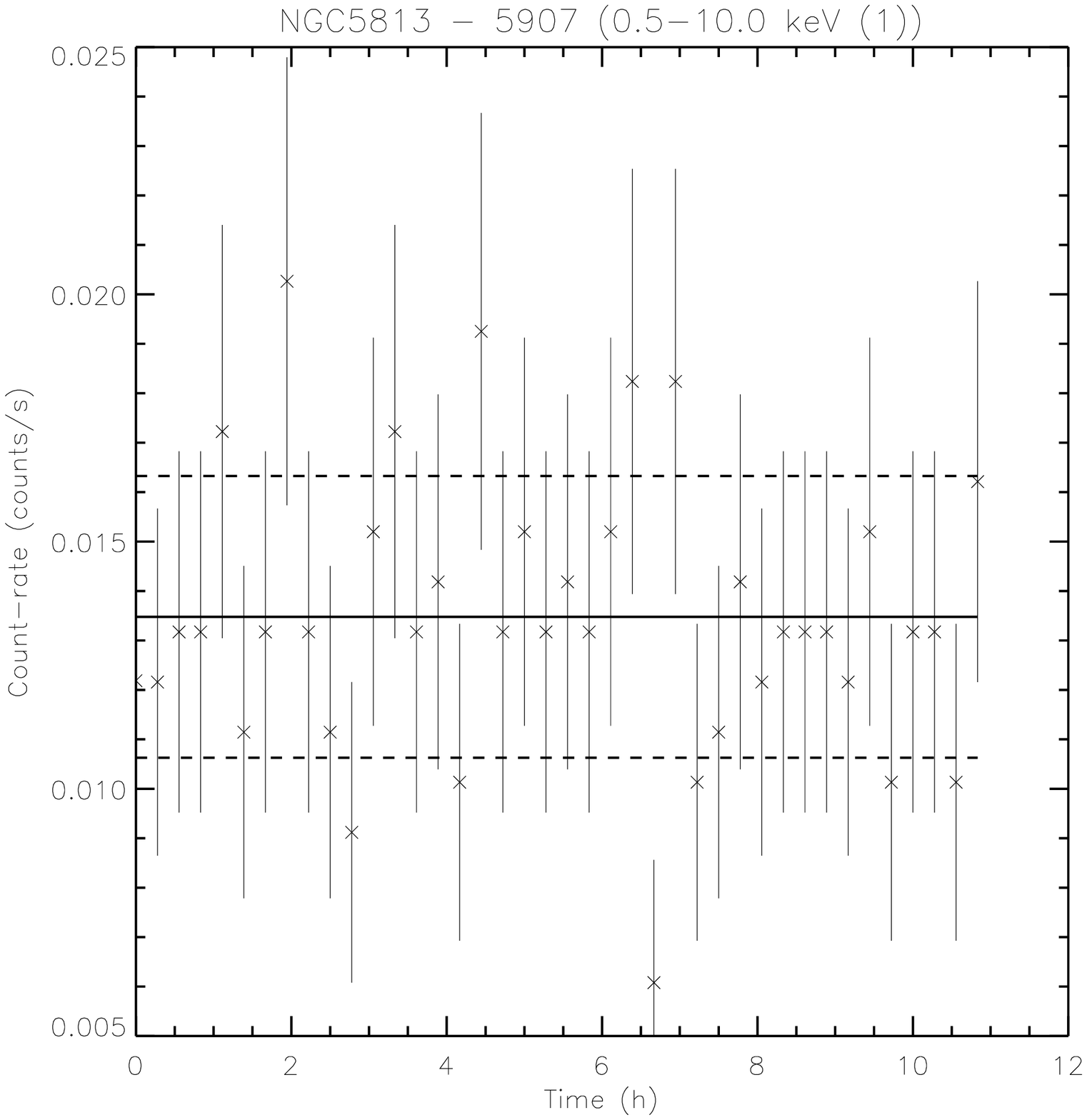}}

\subfloat{\includegraphics[width=0.30\textwidth]{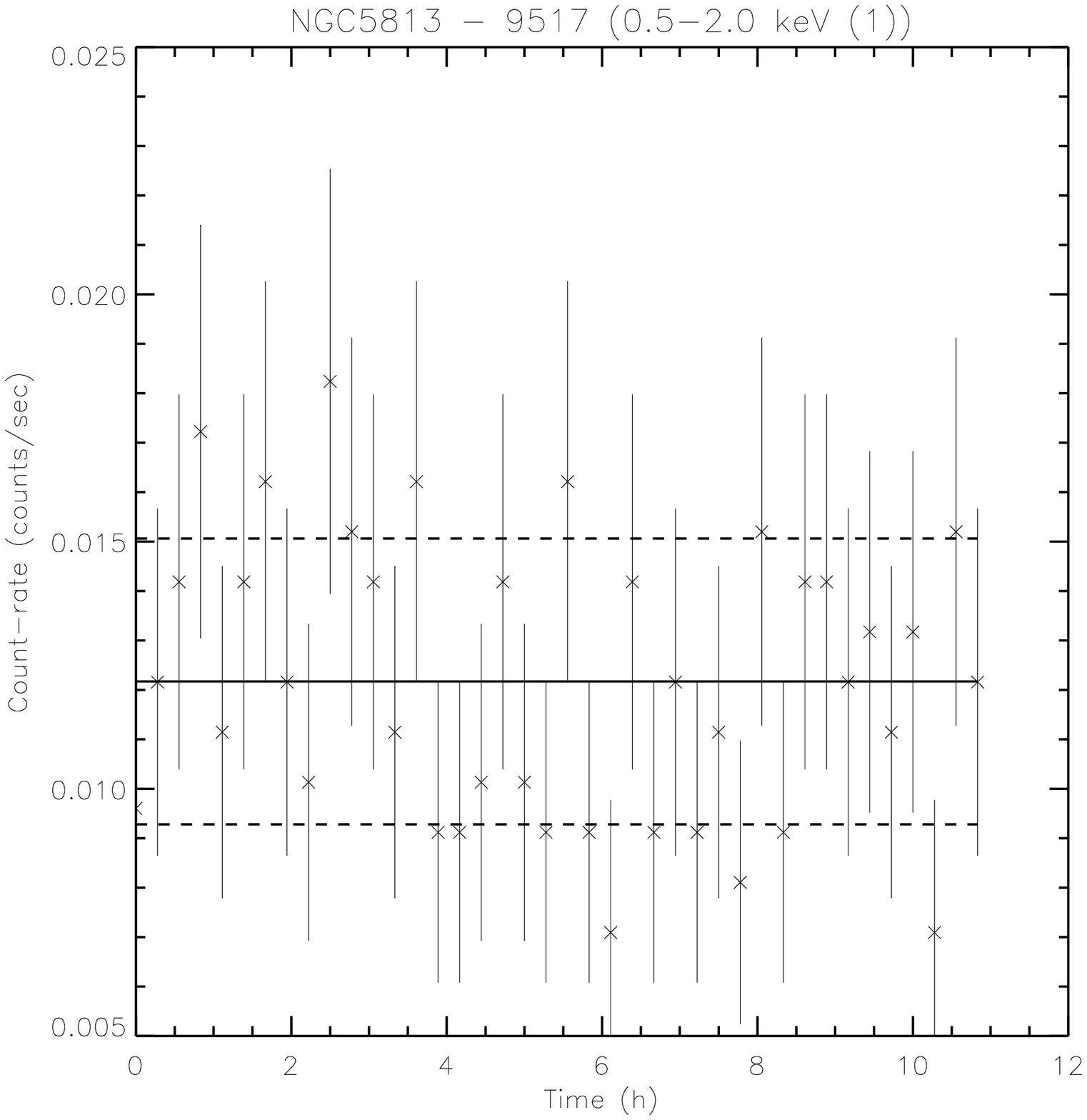}}
\subfloat{\includegraphics[width=0.30\textwidth]{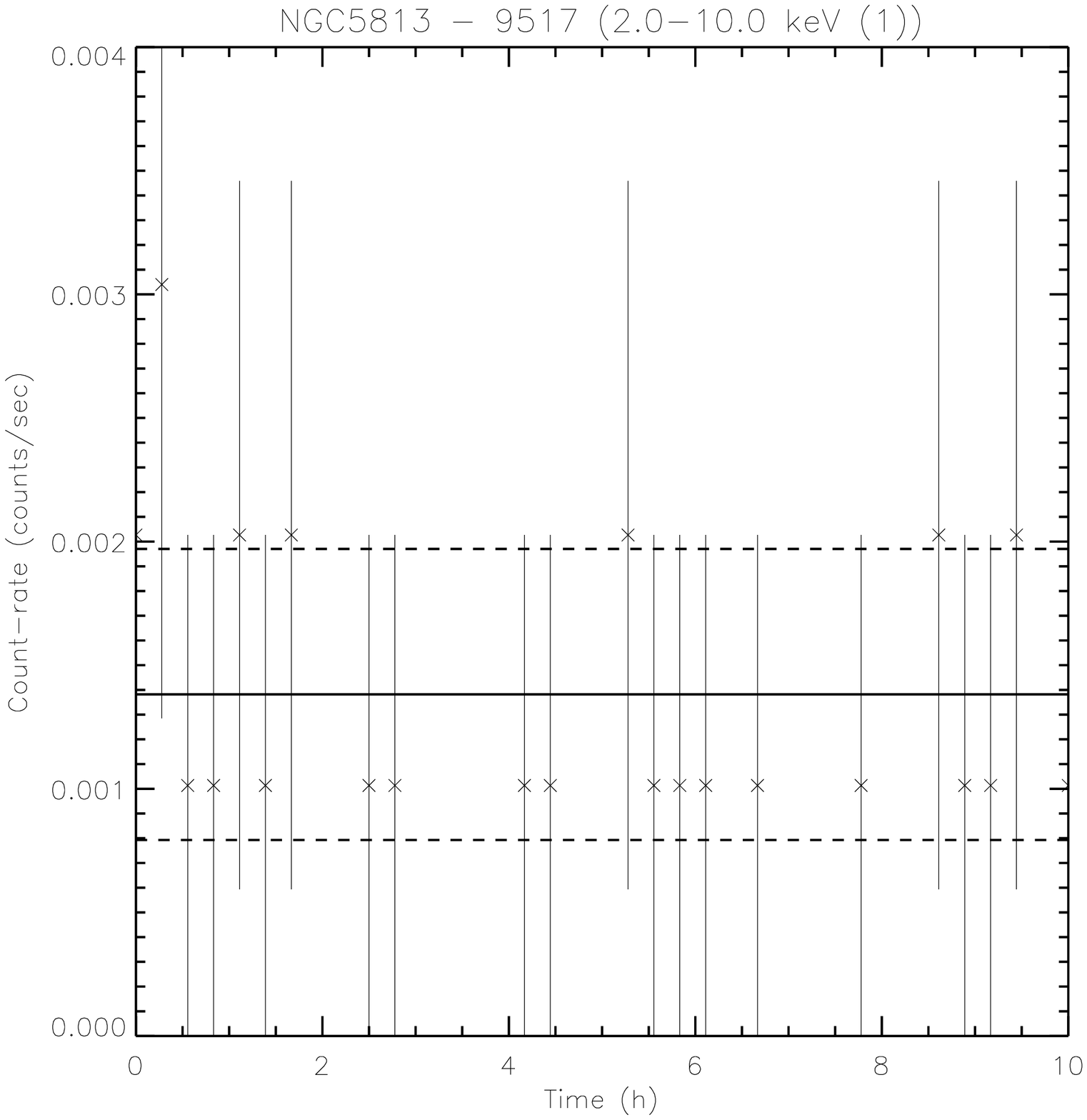}}
\subfloat{\includegraphics[width=0.30\textwidth]{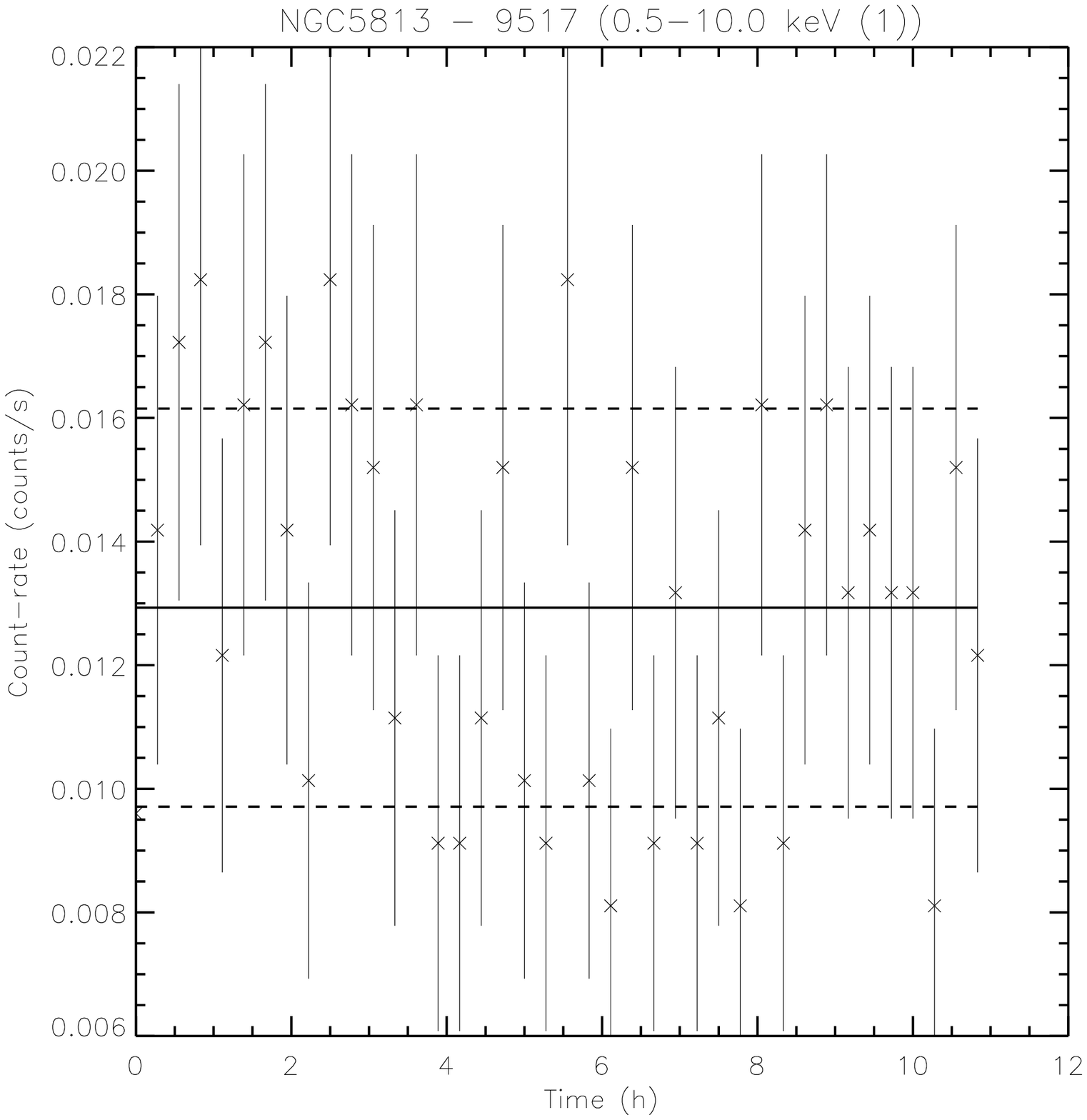}}

\subfloat{\includegraphics[width=0.30\textwidth]{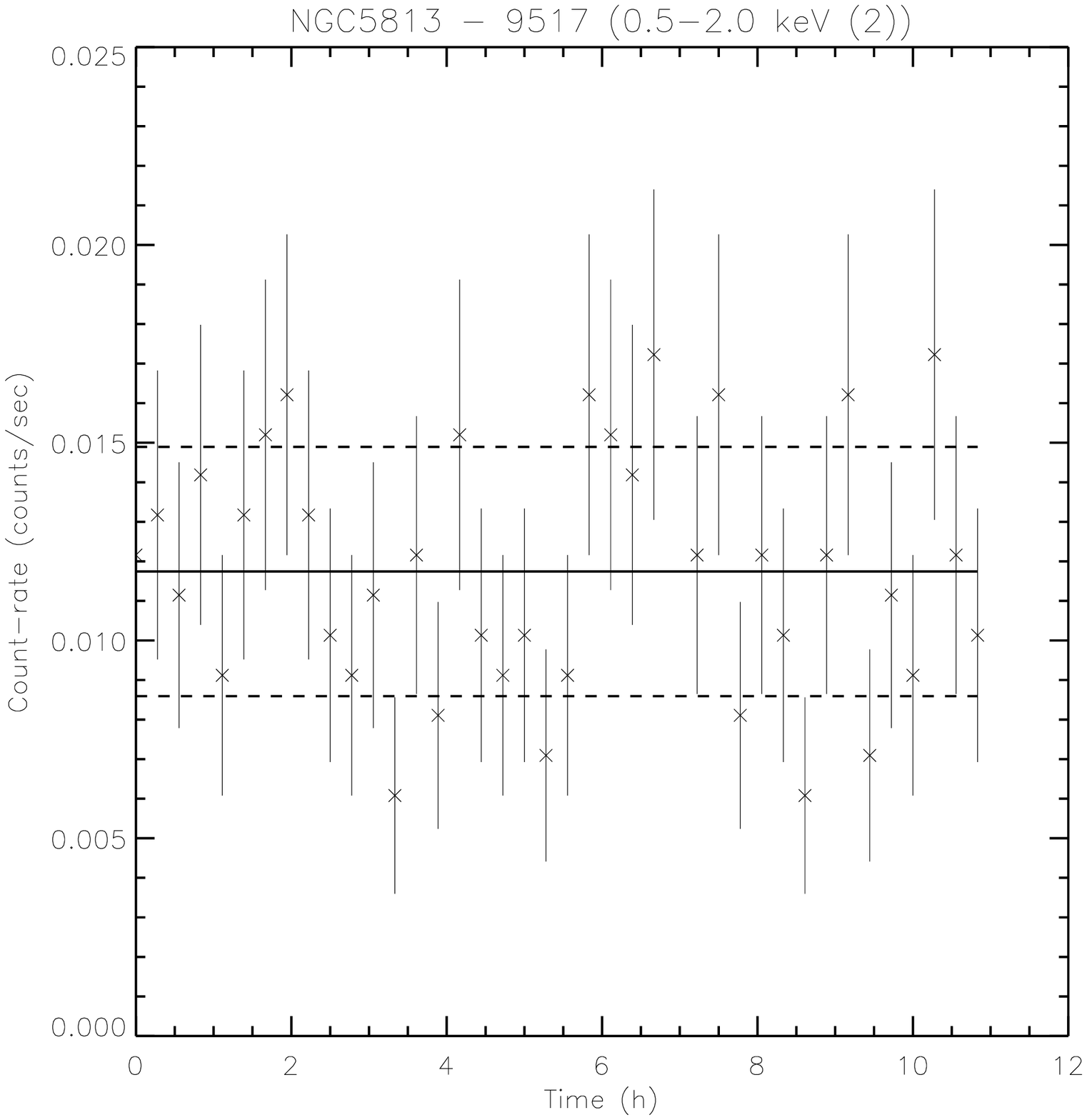}}
\subfloat{\includegraphics[width=0.30\textwidth]{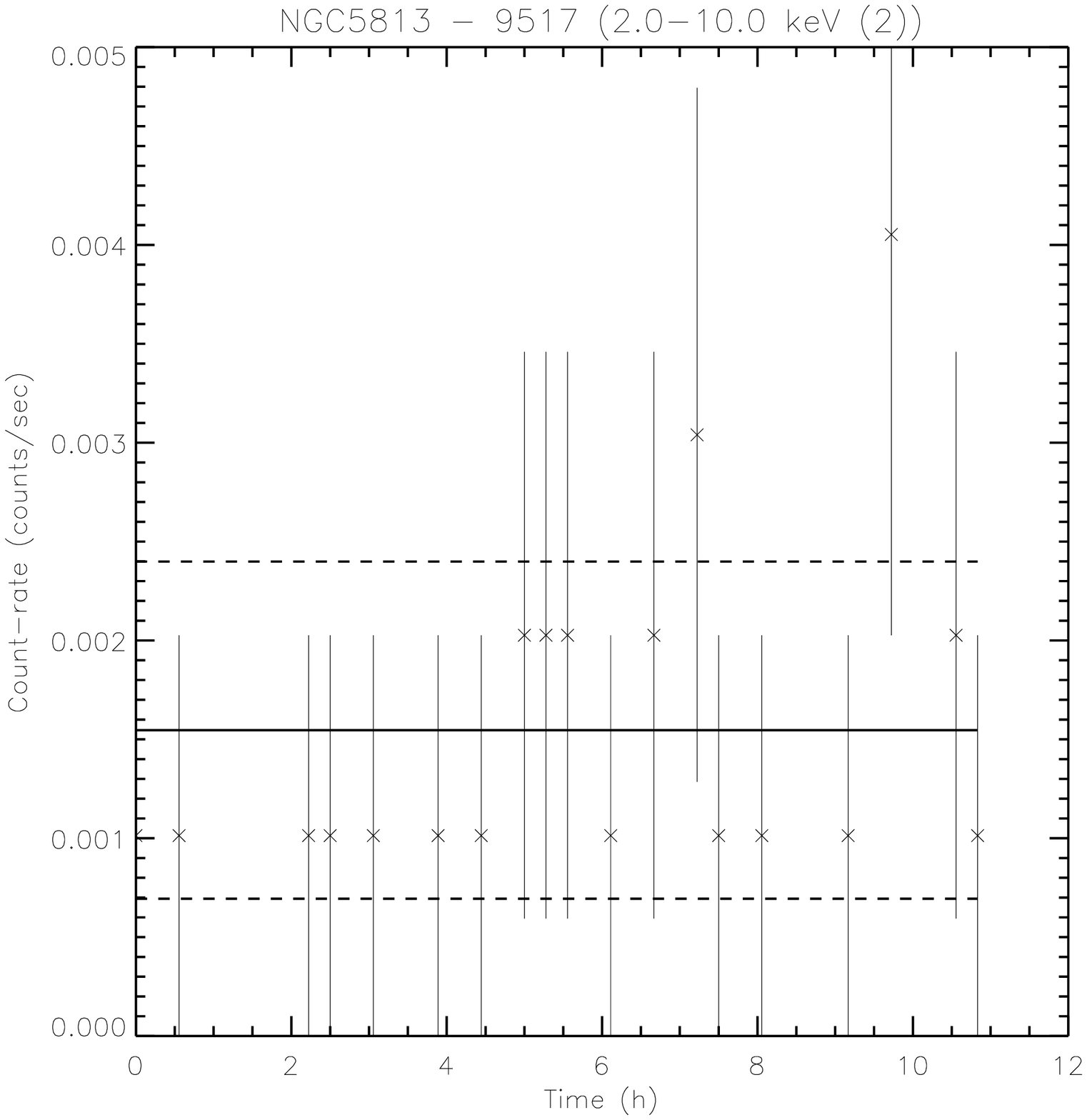}}
\subfloat{\includegraphics[width=0.30\textwidth]{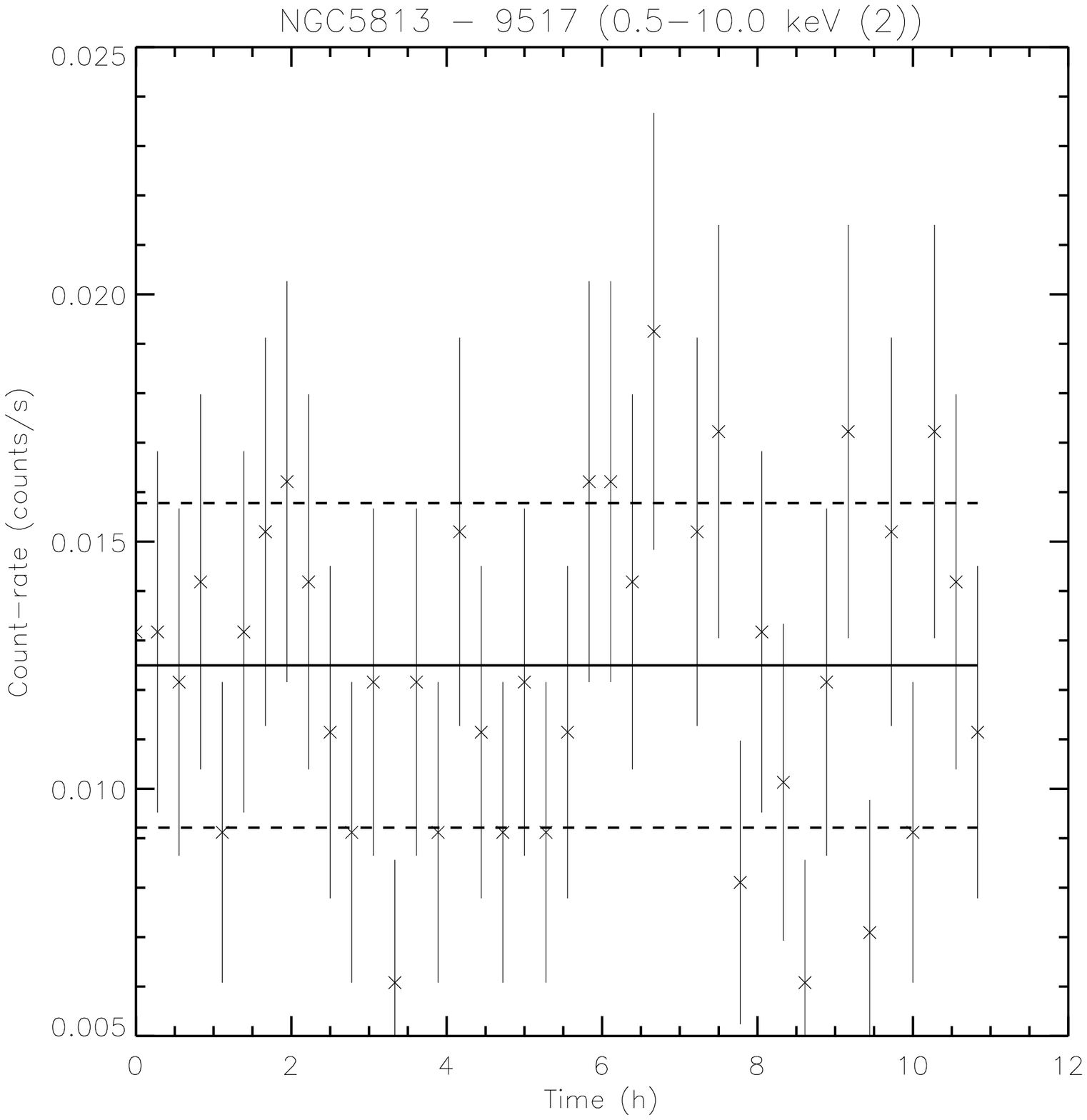}}

\subfloat{\includegraphics[width=0.30\textwidth]{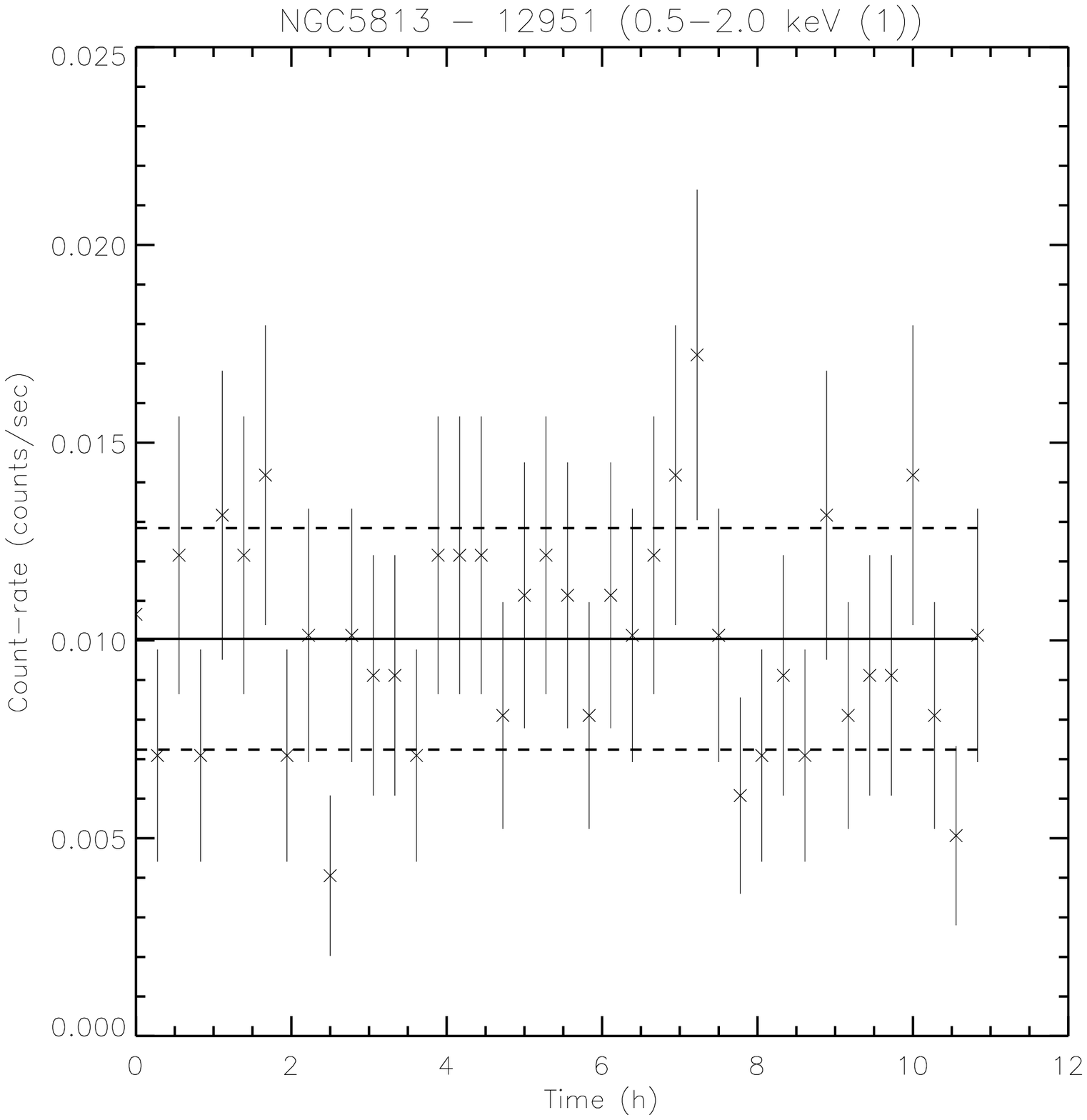}}
\subfloat{\includegraphics[width=0.30\textwidth]{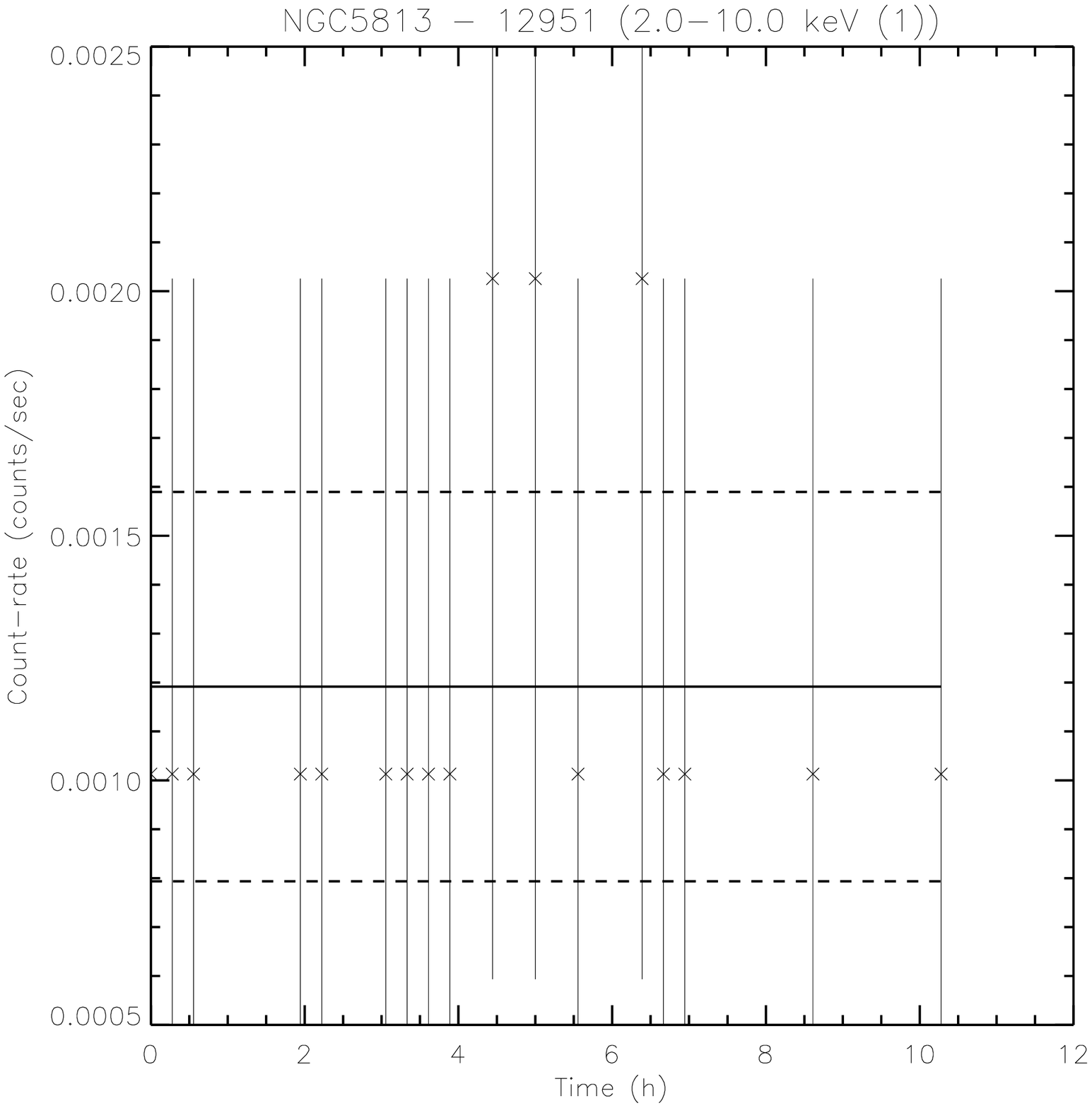}}
\subfloat{\includegraphics[width=0.30\textwidth]{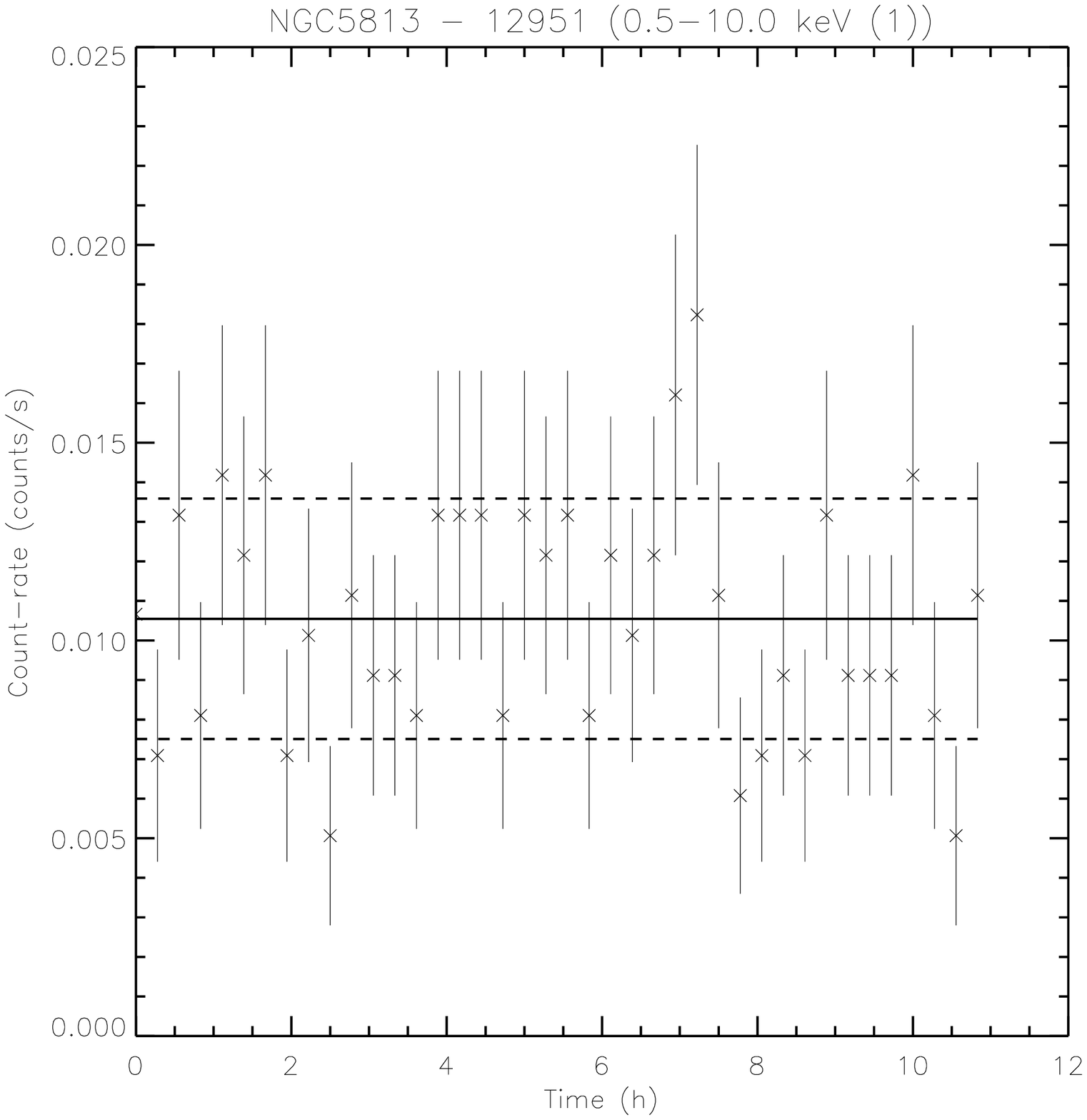}}
\caption{Light curves of NGC\,5813 from \emph{Chandra} data. Note that ObsID. 9517 and 13253 are divided in two segments}
\label{l5813}
\end{figure}

\begin{figure}[H]
\setcounter{figure}{12}
\centering
\subfloat{\includegraphics[width=0.30\textwidth]{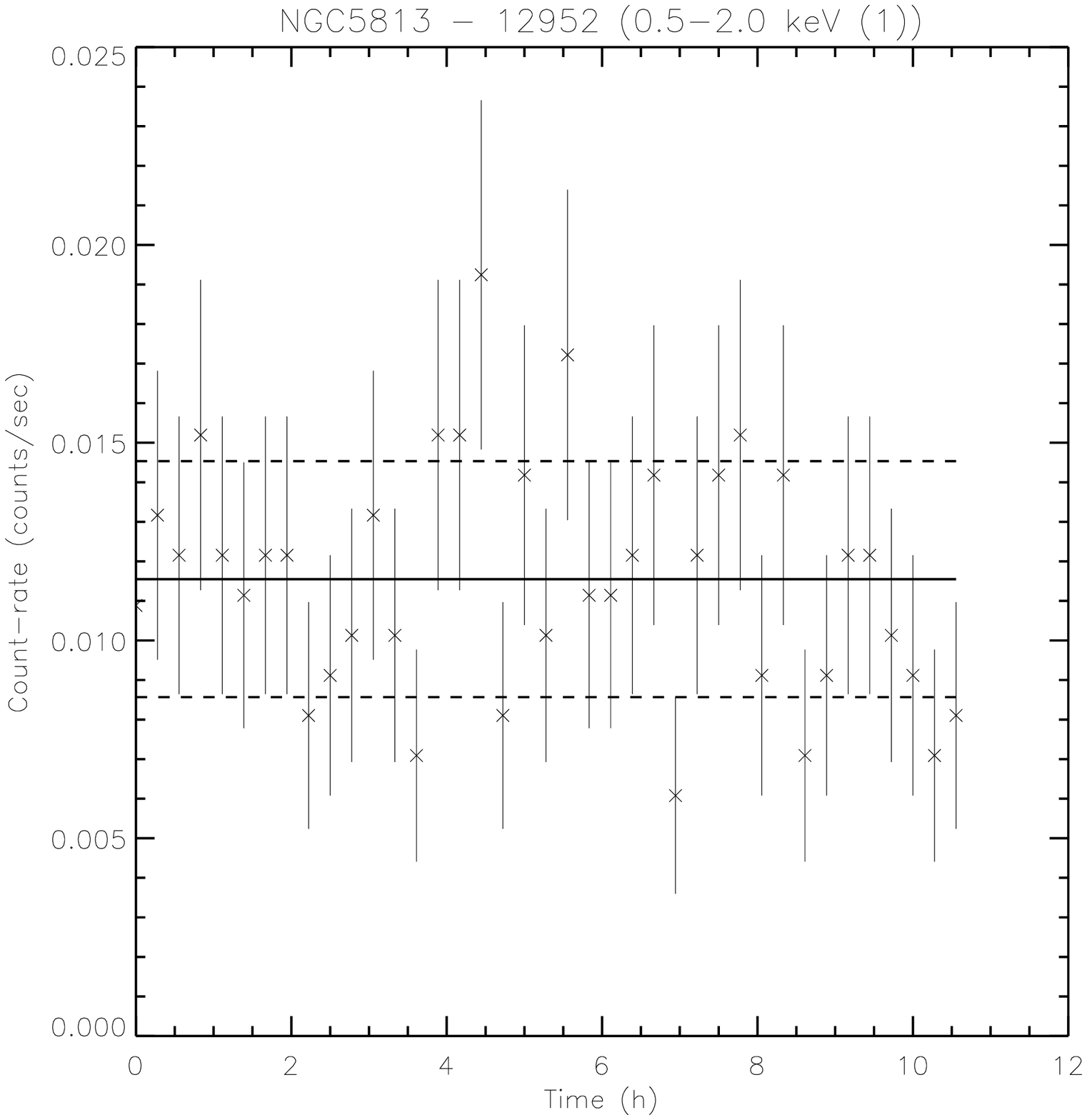}}
\subfloat{\includegraphics[width=0.30\textwidth]{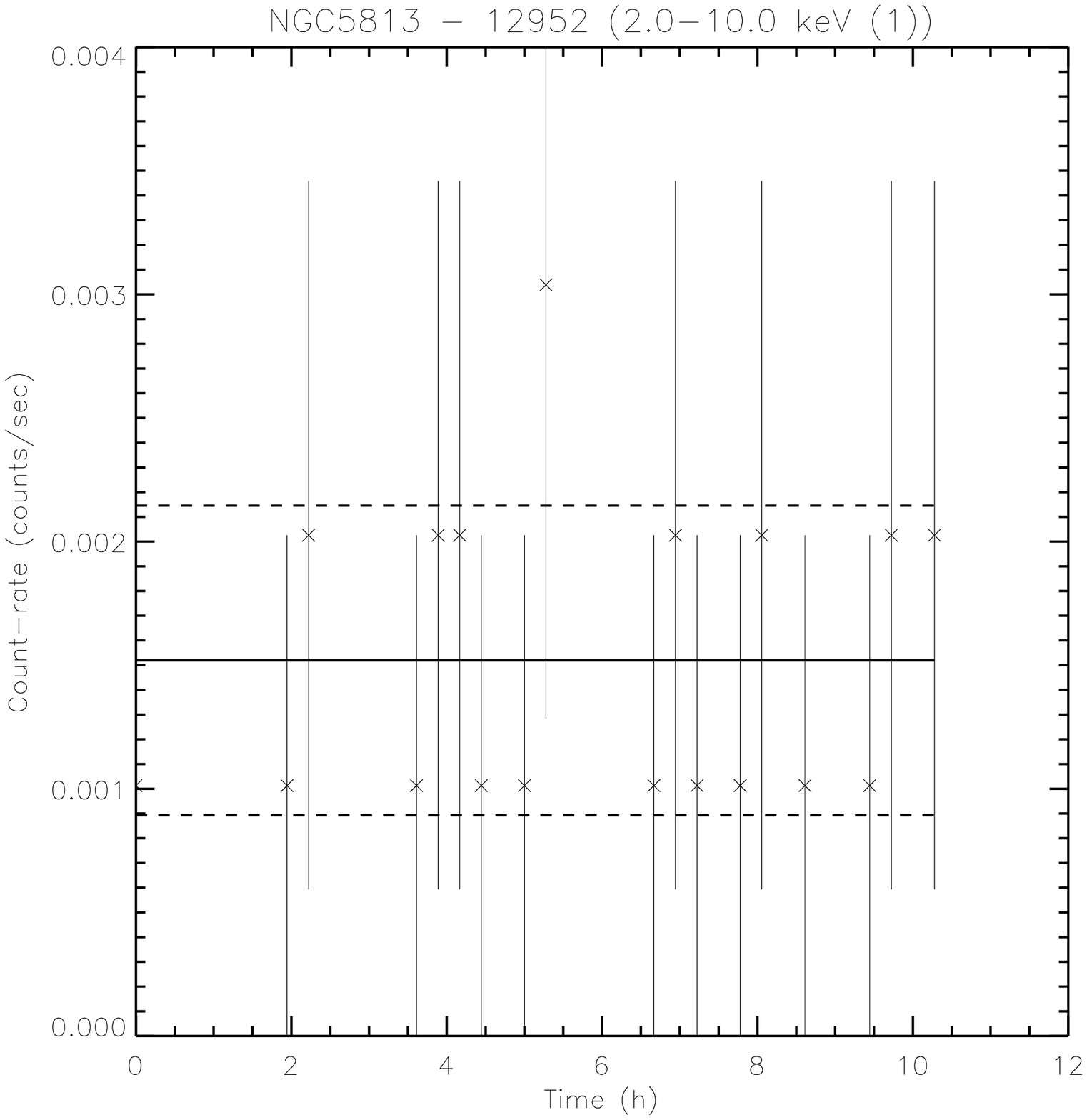}}
\subfloat{\includegraphics[width=0.30\textwidth]{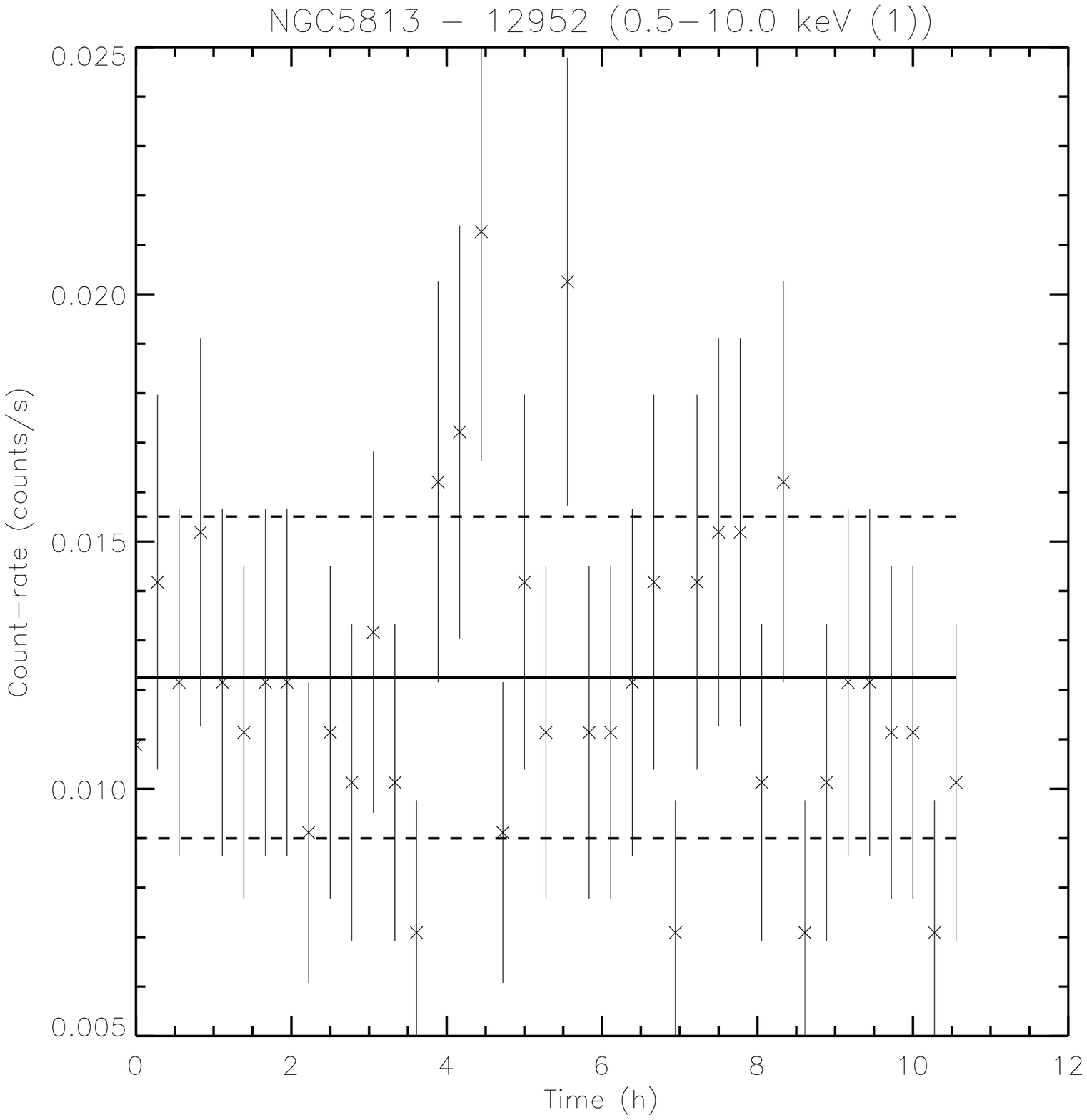}}

\subfloat{\includegraphics[width=0.30\textwidth]{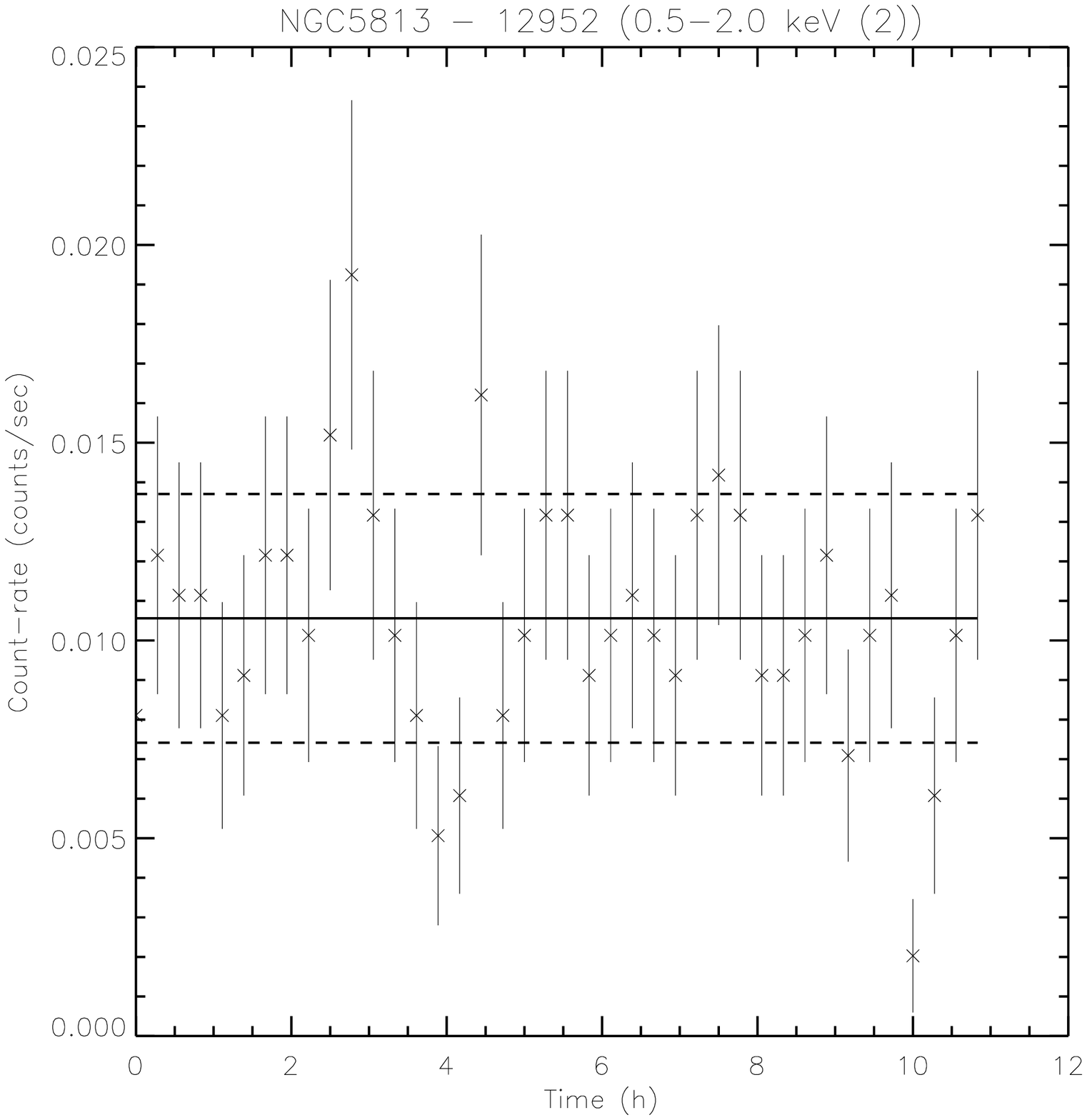}}
\subfloat{\includegraphics[width=0.30\textwidth]{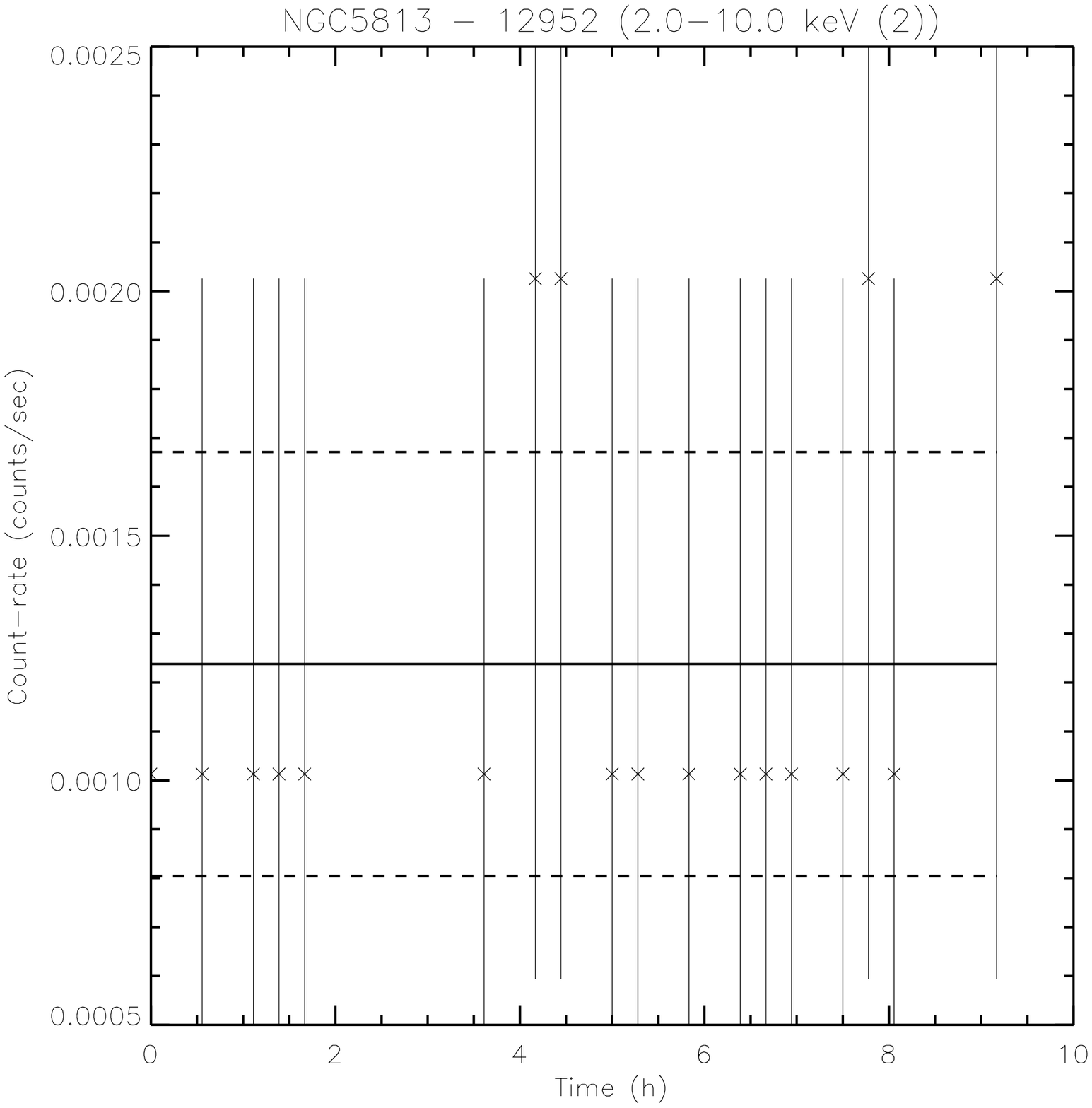}}
\subfloat{\includegraphics[width=0.30\textwidth]{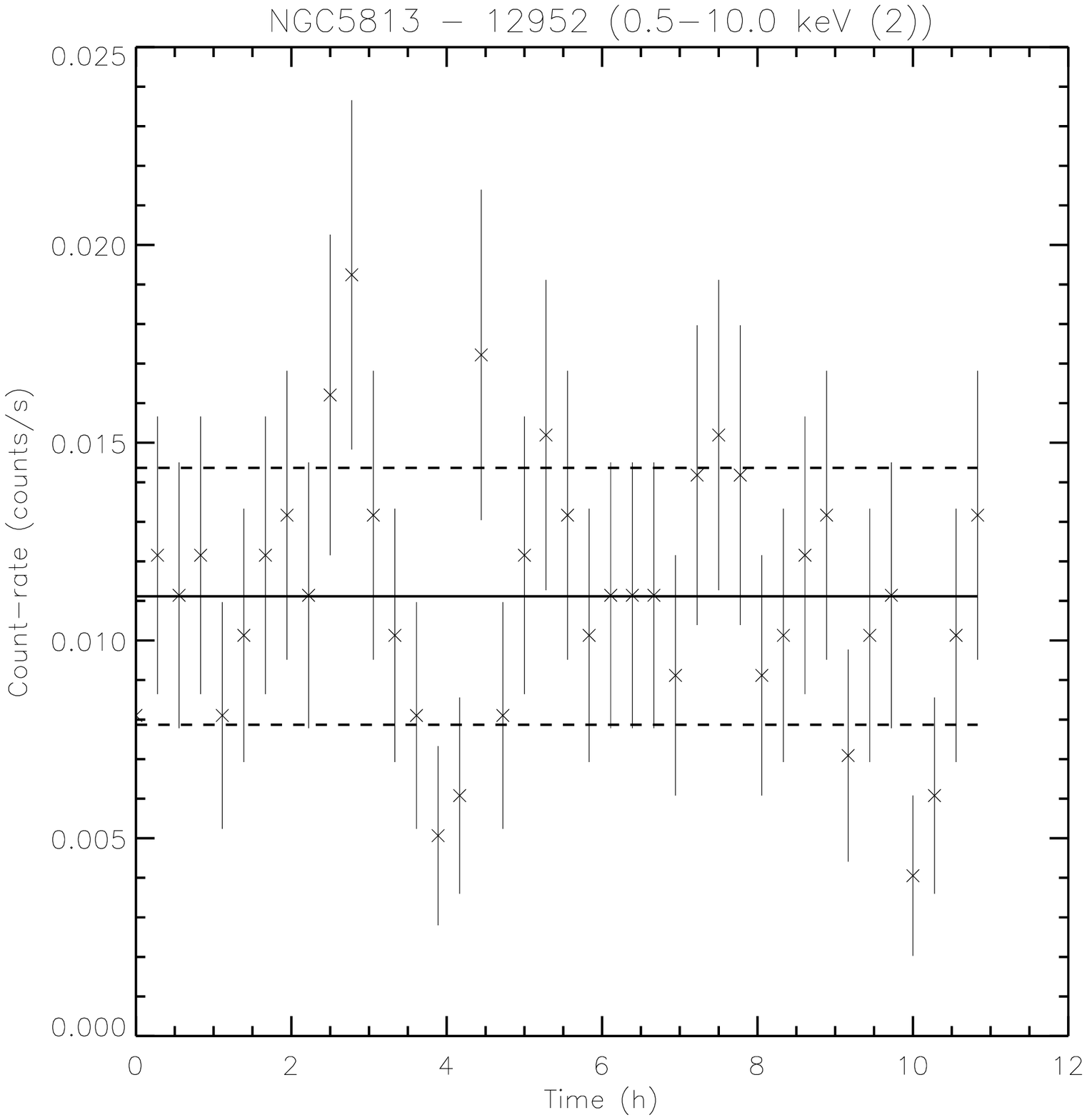}}

\subfloat{\includegraphics[width=0.30\textwidth]{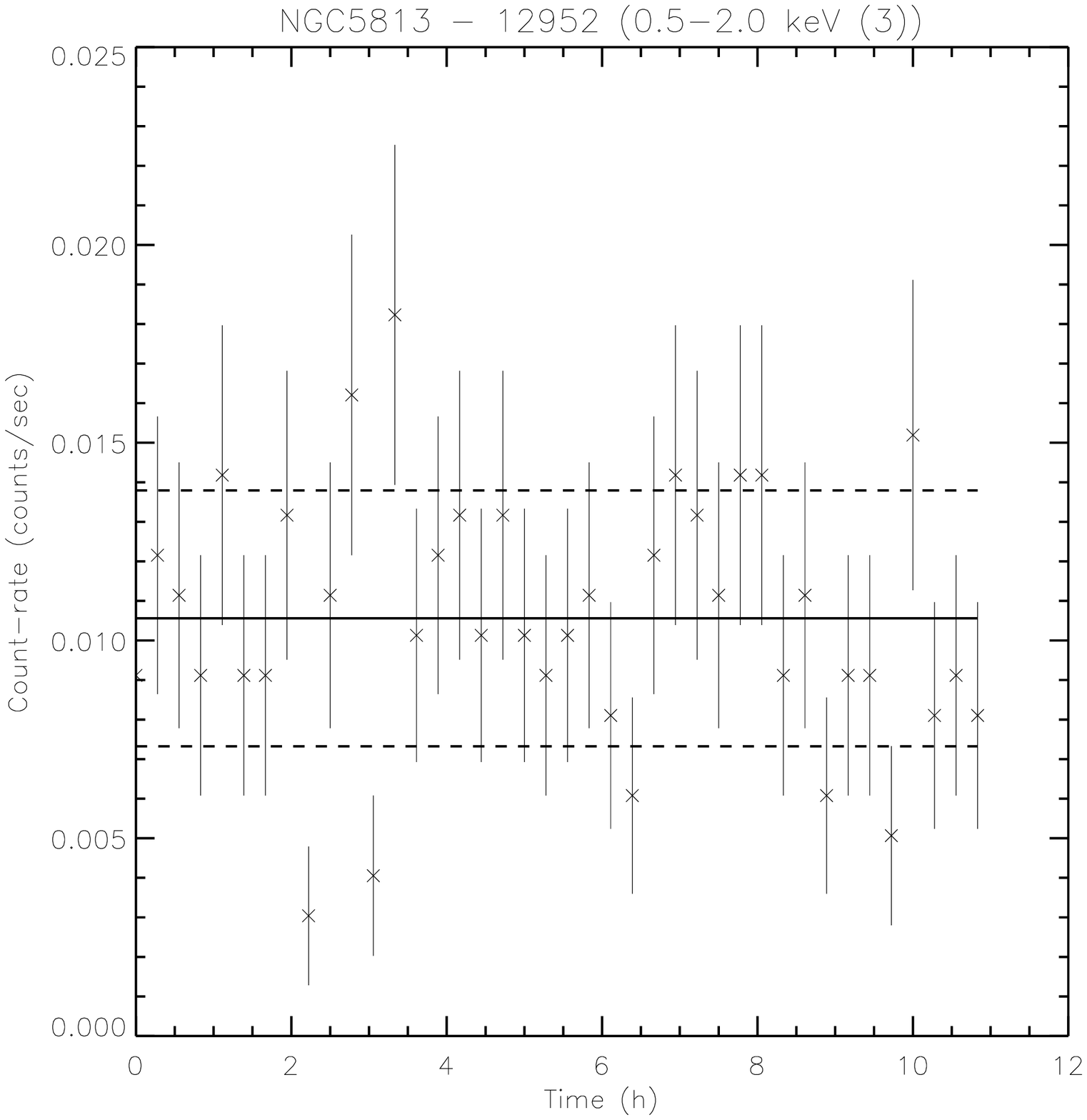}}
\subfloat{\includegraphics[width=0.30\textwidth]{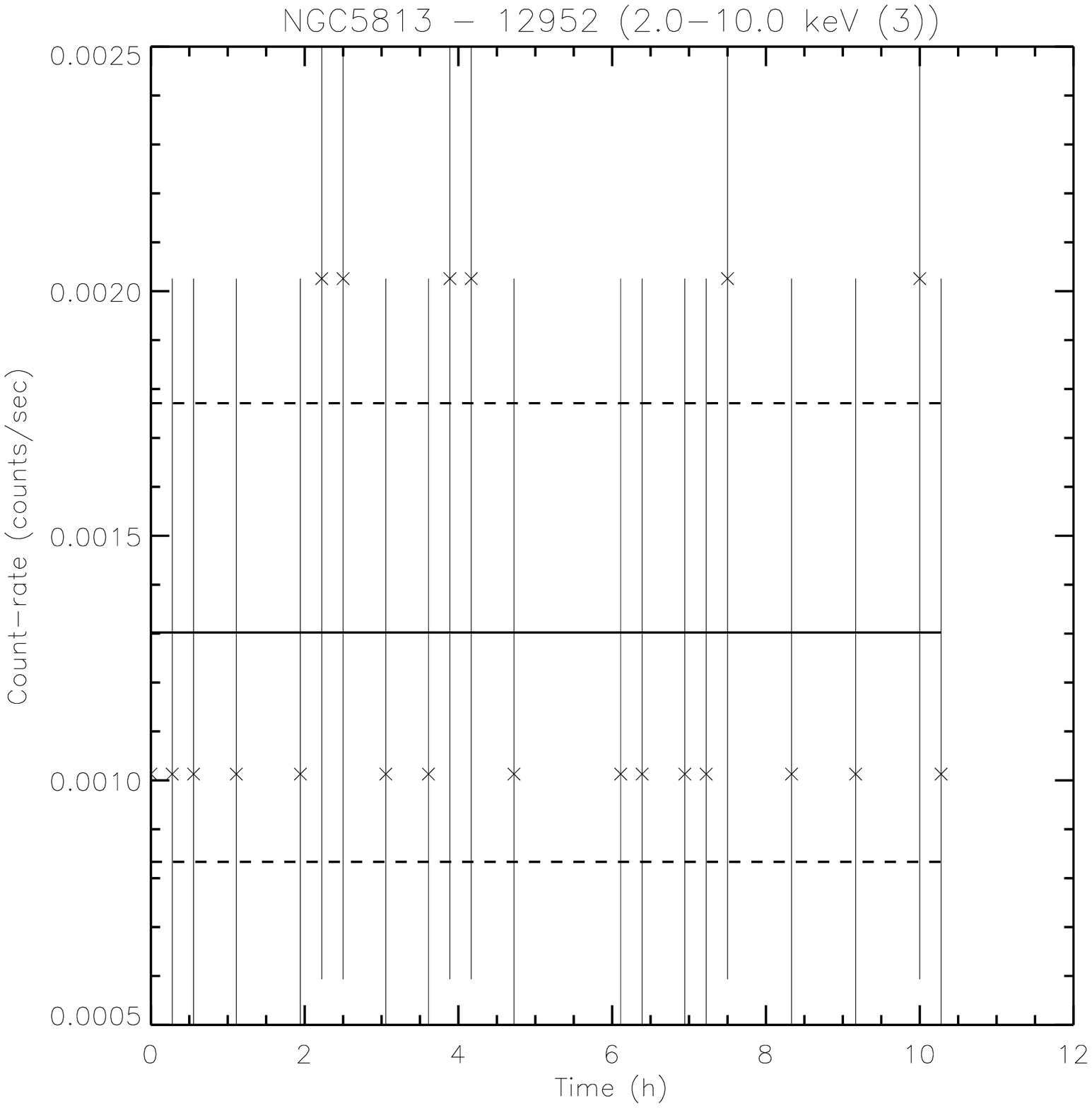}}
\subfloat{\includegraphics[width=0.30\textwidth]{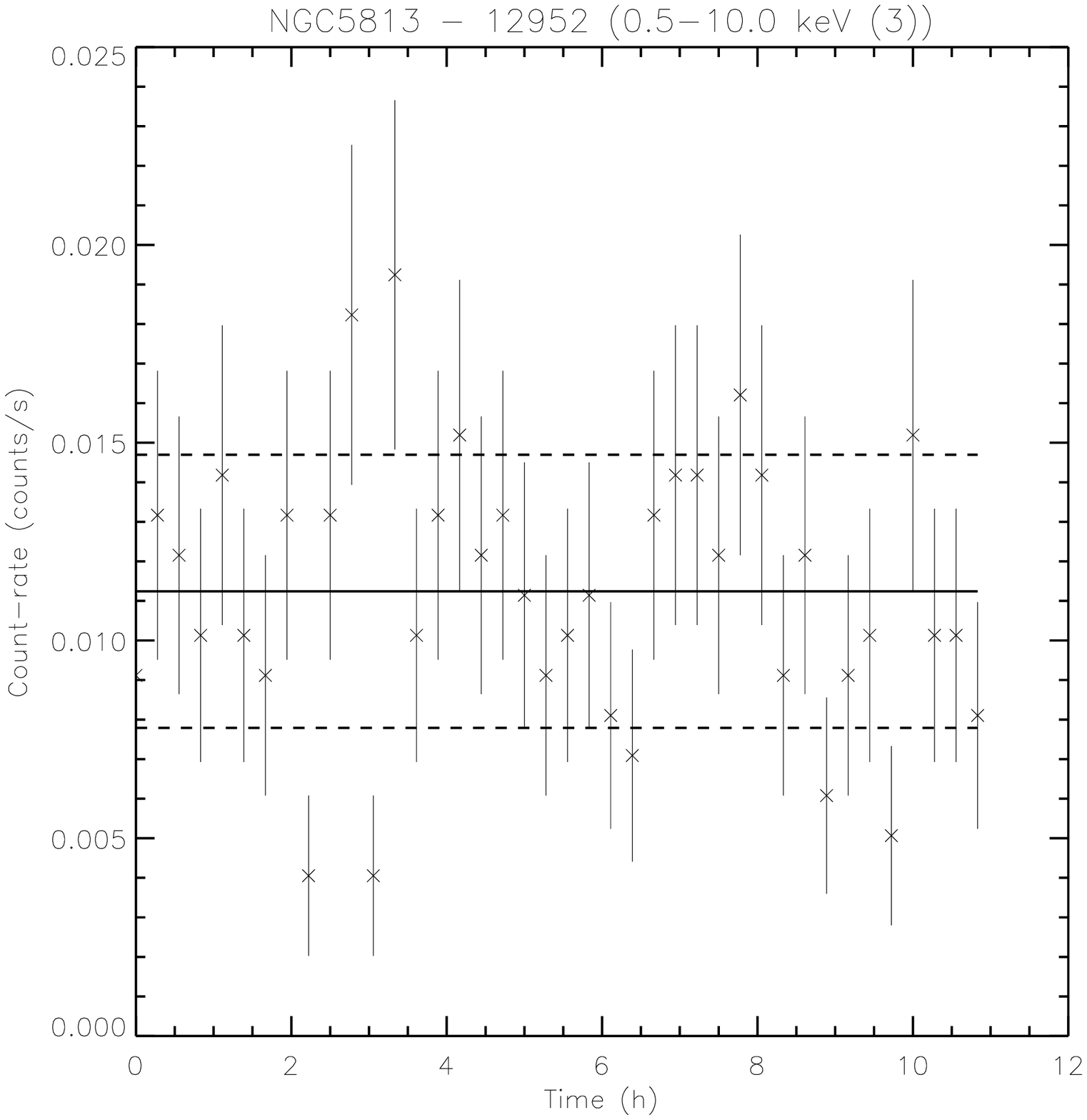}}

\subfloat{\includegraphics[width=0.30\textwidth]{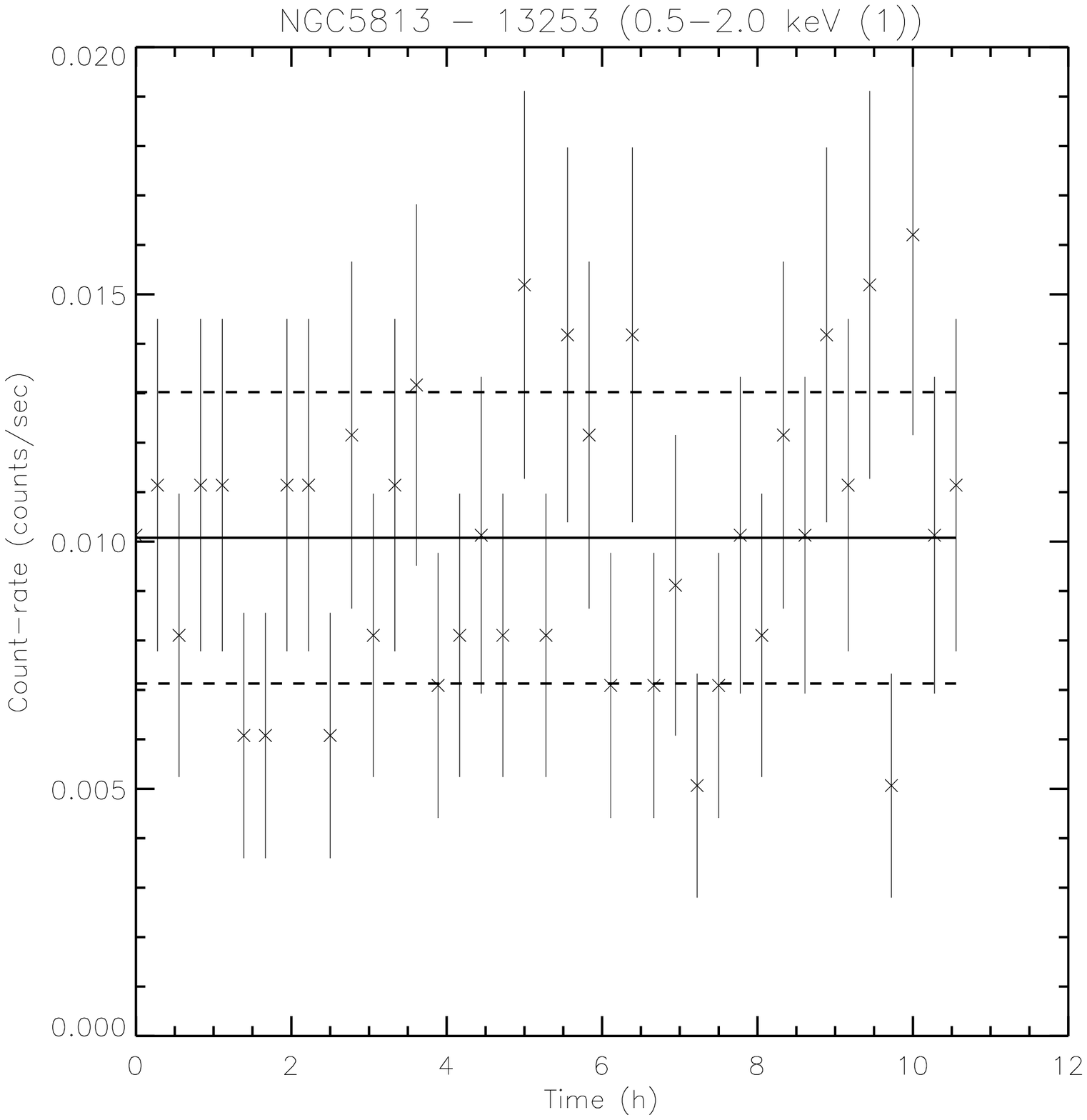}}
\subfloat{\includegraphics[width=0.30\textwidth]{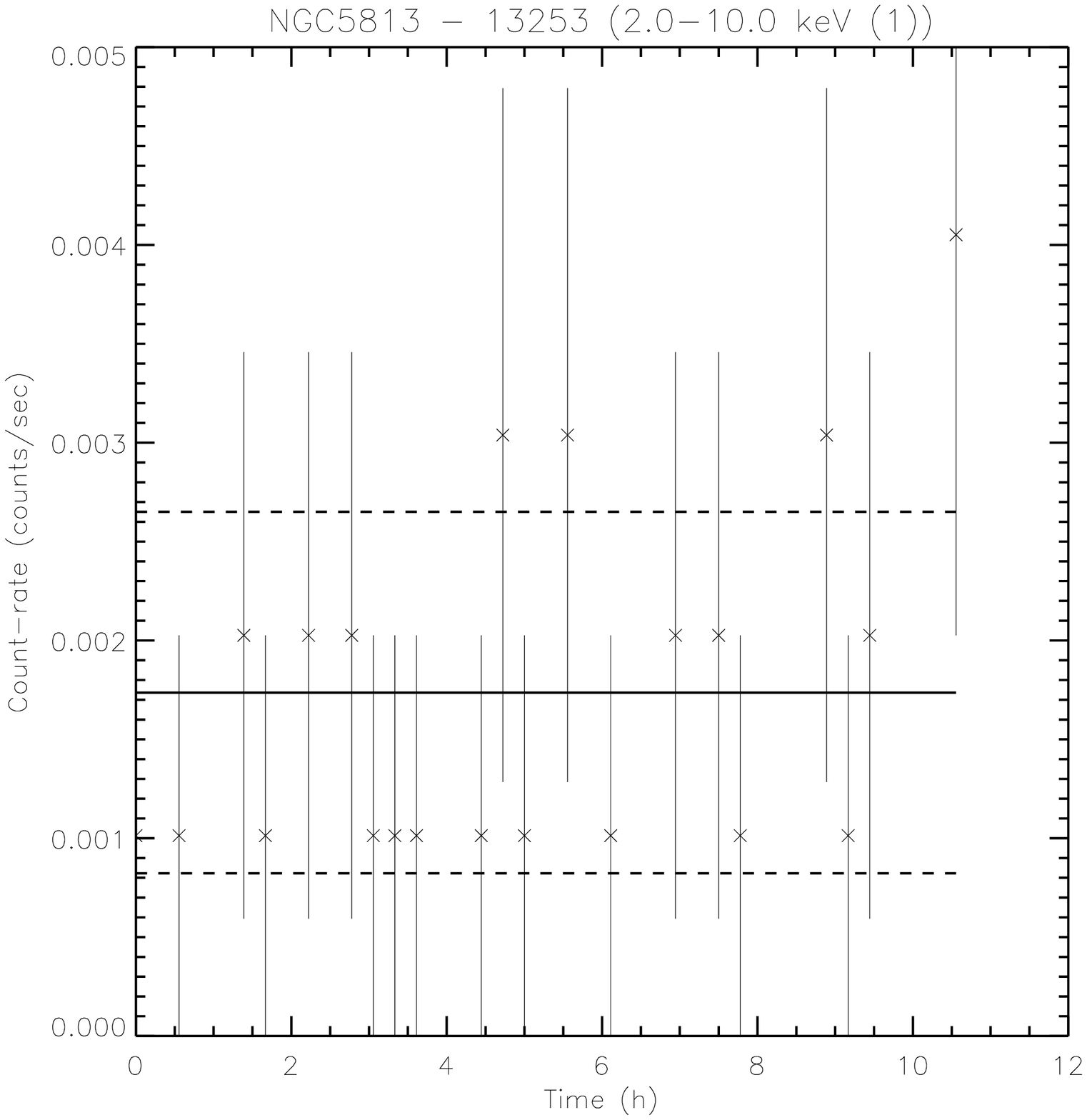}}
\subfloat{\includegraphics[width=0.30\textwidth]{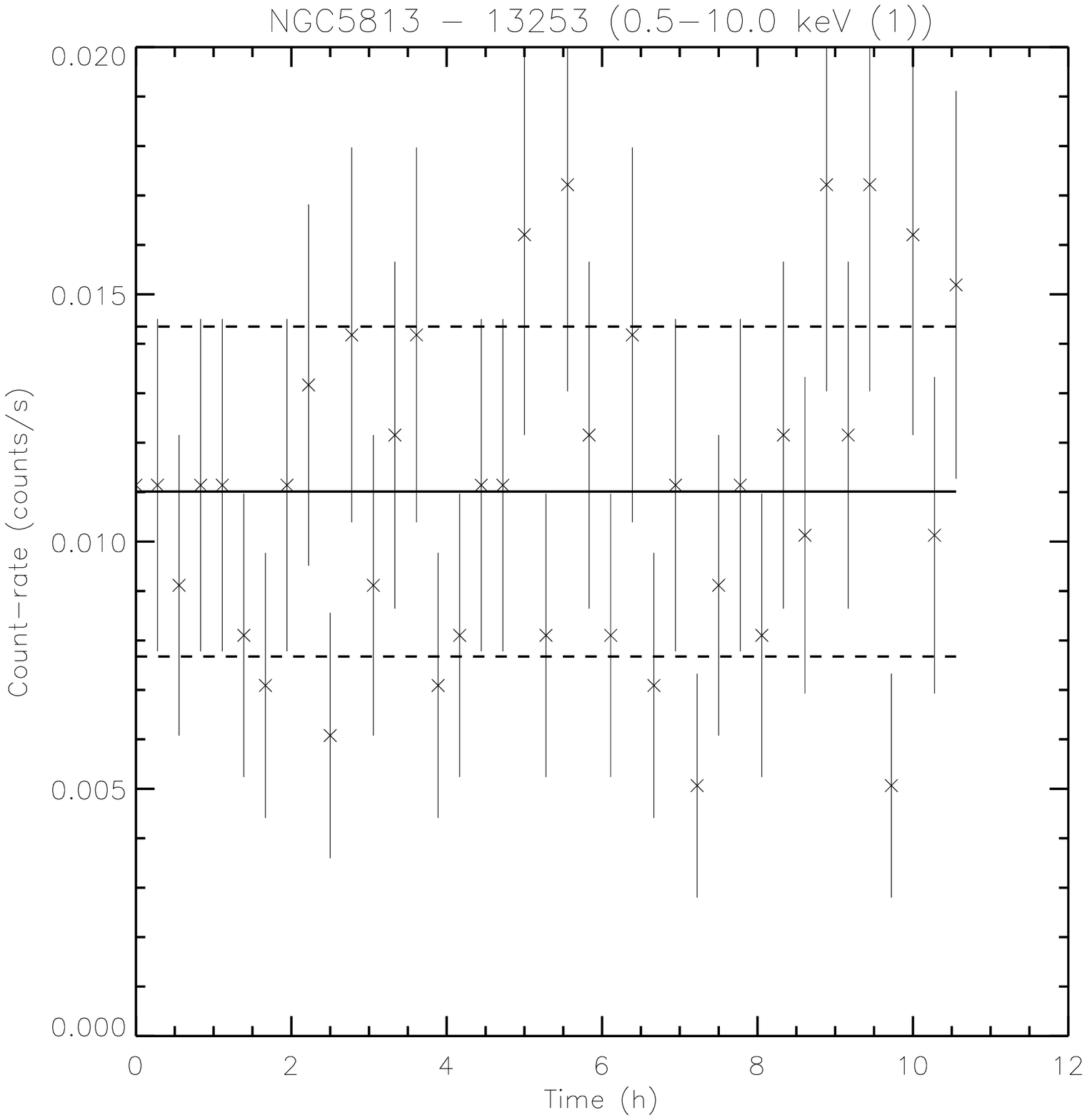}}
\caption{(Cont.)}
\end{figure}

\begin{figure}[H]
\setcounter{figure}{12}
\centering
\subfloat{\includegraphics[width=0.30\textwidth]{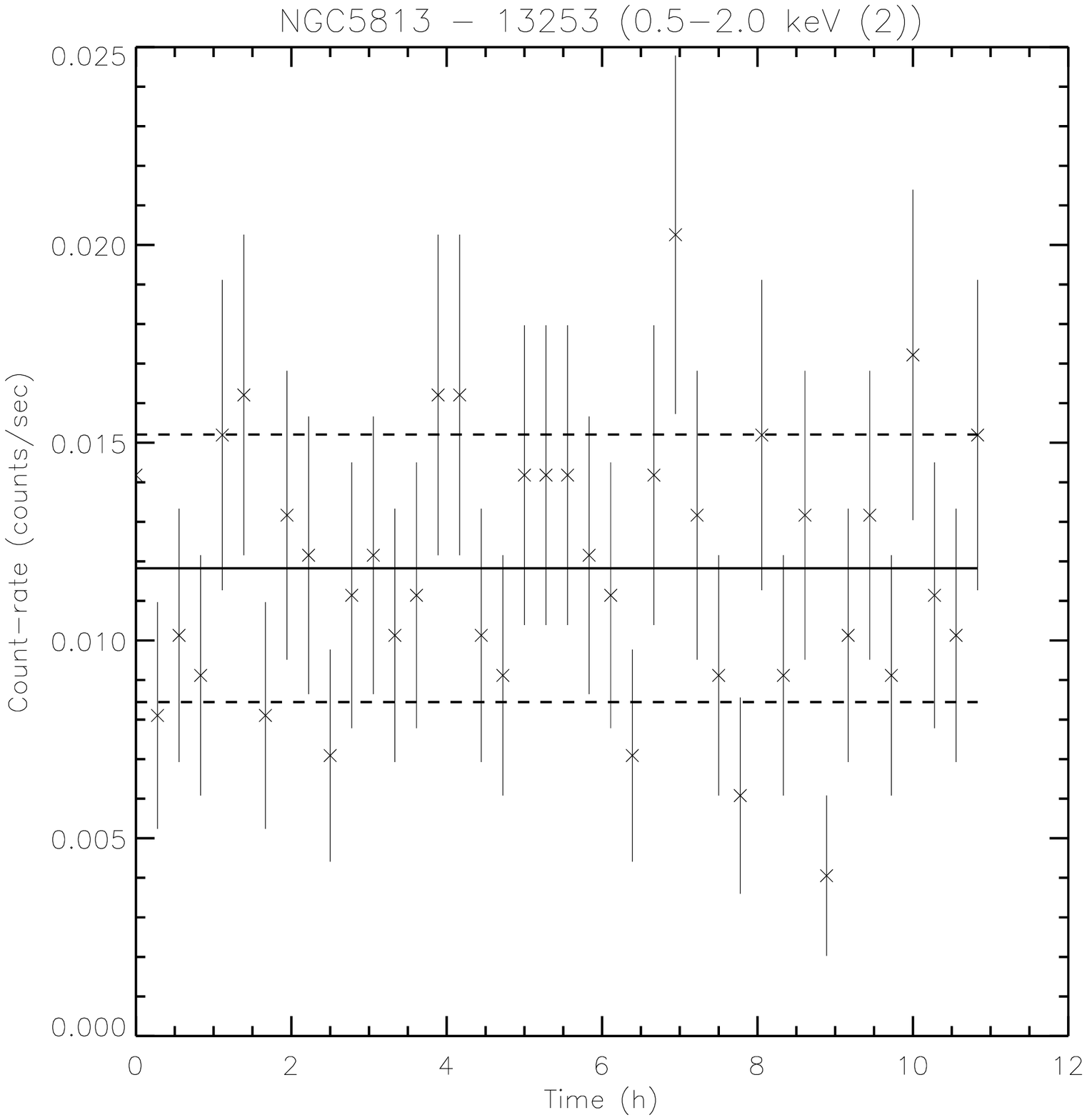}}
\subfloat{\includegraphics[width=0.30\textwidth]{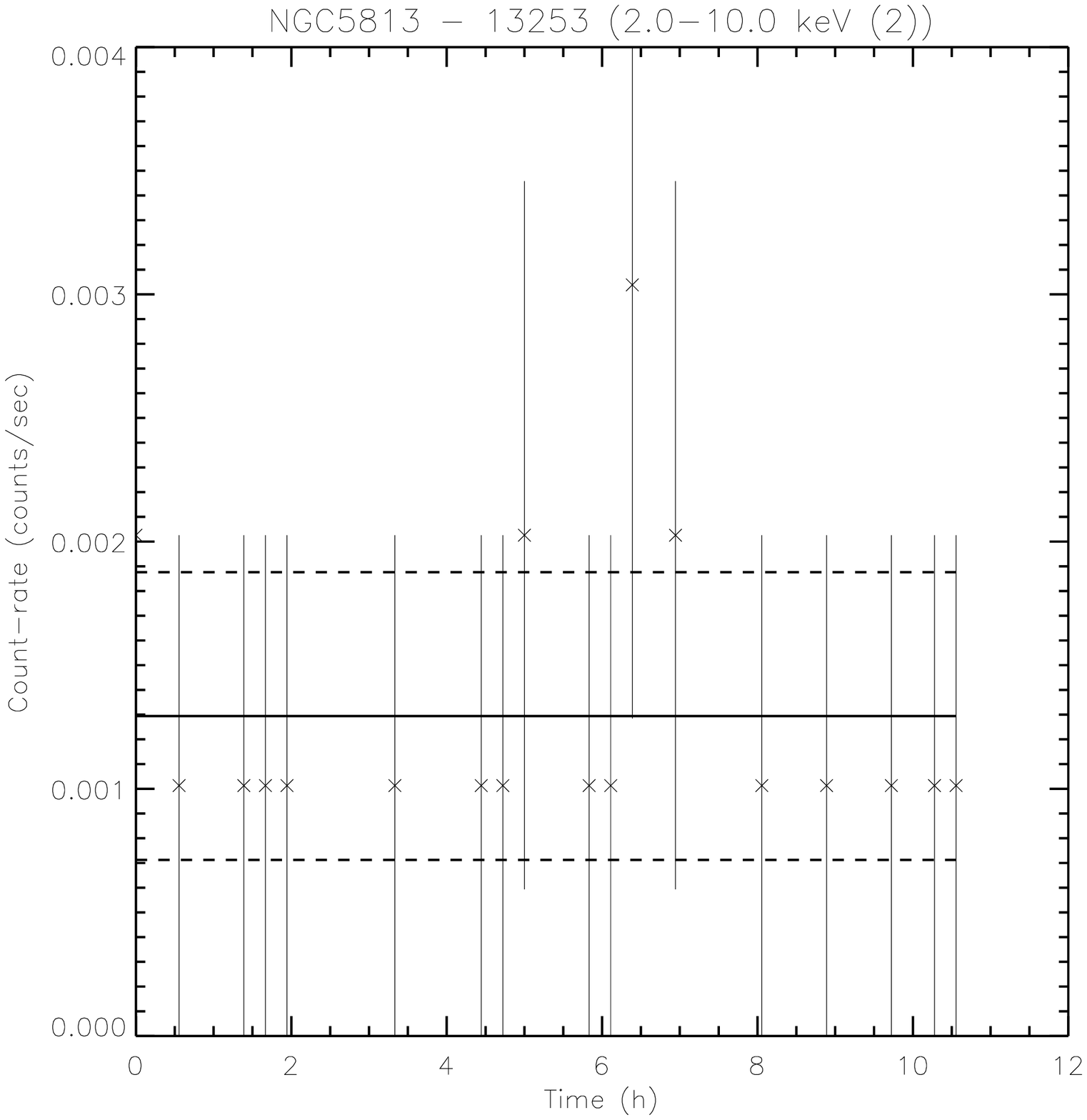}}
\subfloat{\includegraphics[width=0.30\textwidth]{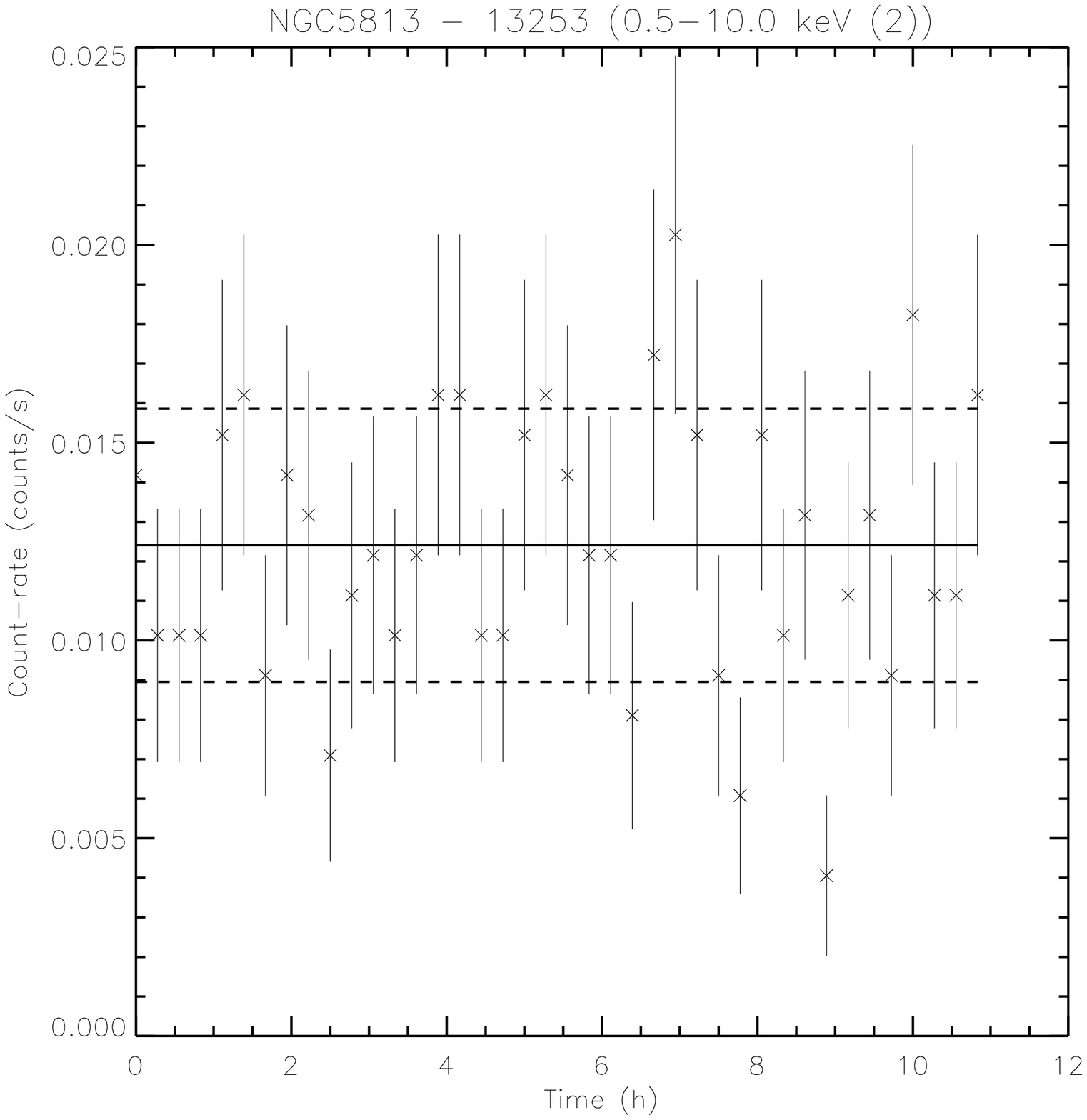}}

\subfloat{\includegraphics[width=0.30\textwidth]{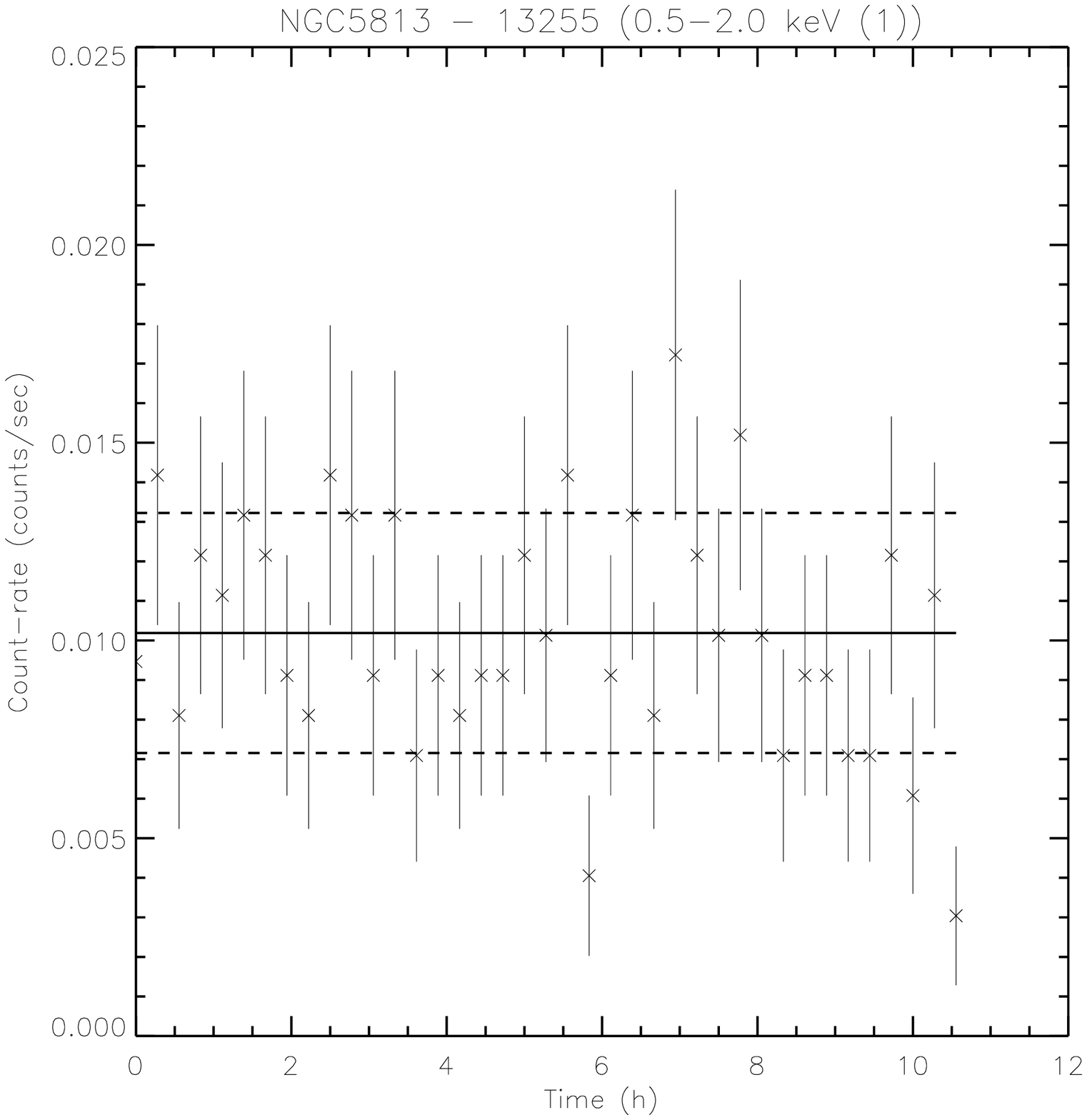}}
\subfloat{\includegraphics[width=0.30\textwidth]{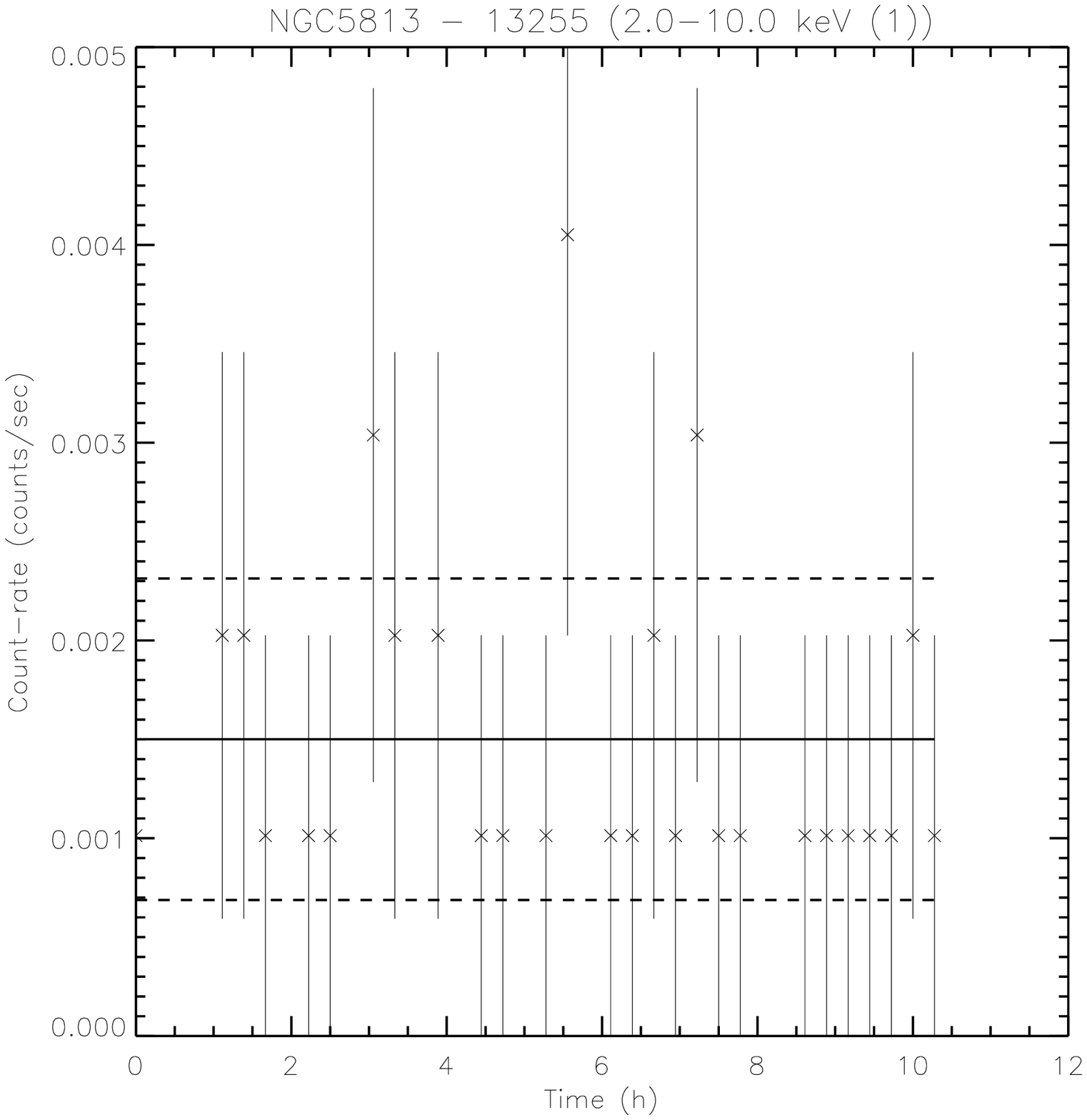}}
\subfloat{\includegraphics[width=0.30\textwidth]{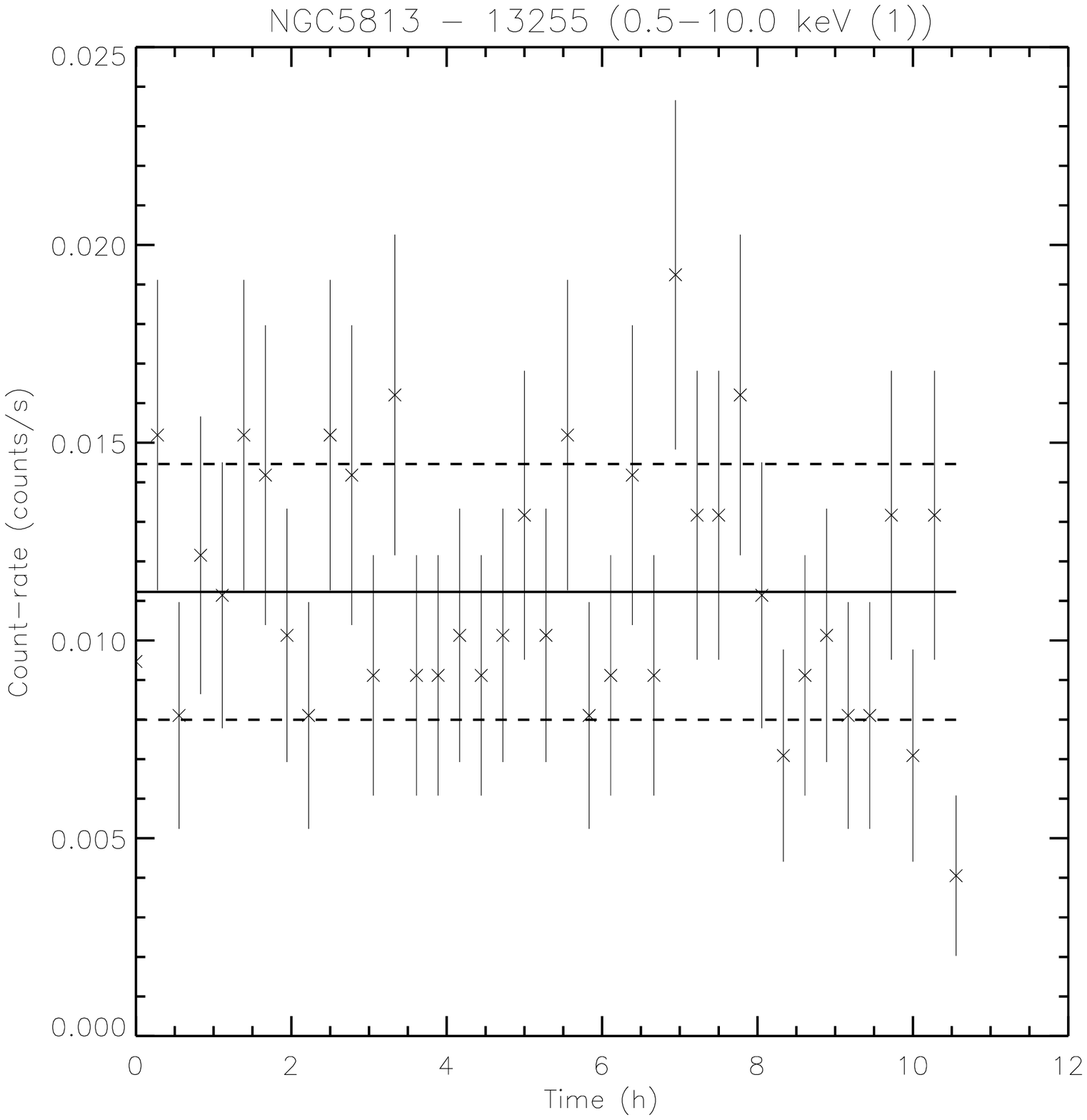}}
\caption{(Cont.)}
\end{figure}

\end{document}